\documentstyle[12pt]{article}
\def\Journal#1#2#3#4{{#1} {\bf #2}, #3 (#4)}

\def\NPB{{\em Nucl. Phys.} B}

\def\PLB{{\em Phys. Lett.}  B}
\def\PRL{{\em Phys. Rev. Lett.} }
\def\PRD{{\em Phys. Rev.} D}
\def\ZPC{{\em Z. Phys.} C}
\def\PRC{{\em Phys. Rep.} C}

\def\JP{{\em J. of Physics}}
\def\EPJC{{\em Eur. Phys. J. }C}
\def\CPC{{\em Comp. Phys. Commun. }}
\def\JPG{{\em J. Phys.} G}  
\def\PR{{\em Phys. Rep. }}

\def\ra{\rightarrow}

\def\xb{x_{Bj}}

\def\be{\begin{equation}}
\def\ee{\end{equation}}
\def\bea{\begin{eqnarray}}
\def\eea{\end{eqnarray}}

\def\lsim{\:\raisebox{-0.5ex}{$\stackrel{\textstyle<}{\sim}$}\:}
\def\gsim{\:\raisebox{-0.5ex}{$\stackrel{\textstyle>}{\sim}$}\:}
\def\gev{GeV }
\def\eg{{\it{e.g. }}}
\def\ie{{\it{i.e. }}}
\def\g2{GeV$^2$}
\def\psr{{pseudorapidity }}
\def\bi{\bibitem}
\begin{document}
\titlepage
\vspace*{-1.5cm}
\begin{flushright}
IFT 99/15\\
{\large \bf hep-ph/0011083} \\
JULY 2000
\end{flushright}
\vspace*{1.cm}
\centerline{\bf {\huge {{Survey of present data on  photon structure
functions}}}}
\vskip 0.5cm 
\centerline{\bf {\huge {and resolved photon processes}}}
\vskip 2.5cm
\centerline{\Large Maria Krawczyk and Andrzej Zembrzuski}                 
\centerline{Institute of Theoretical Physics, Warsaw University, Poland}

\vskip 0.5cm
\centerline{and}
\vskip 0.5cm
\centerline{\Large Magdalena Staszel}
\centerline{Institute of Experimental Physics, Warsaw University, Poland}
\vskip 3cm
\begin{abstract}
Present data on the partonic content  of the photon
from LEP, TRISTAN and HERA accelerators
are reviewed and the essential aspects of the underlying ideas 
and methods are pointed out. Results of the unpolarized photon 
structure function
$F_2^{\gamma}$
from DIS$_{e\gamma}$ experiments and on large $p_T$  
jet production processes in the resolved $\gamma \gamma$ collisions
 are presented for both  real and  virtual 
photons. The results of  analysis of the hadronic final state 
accompanying the DIS$_{e\gamma}$ measurements, showing some discrepancies
with the Monte Carlo models, are collected together 
and presented as a separate issue.
Also results on the DIS$_{e\gamma}$ with leptonic final states are shown.
The results from resolved real and virtual photon processes at HERA
collider based on the single and double jet events, also charged particles 
and prompt photons, are presented.
In the context of virtual photon processes the data 
for forward jet and forward particle production are included.
In addition a short presentation of the recent data on 
 the heavy quark content of the photon is given.
Related topics --  the polarized (spin dependent)
structure functions for the real and virtual photon,
 the structure function of the electron 
and the photonic content of the proton
are also shortly mentioned.
\end{abstract}

\newpage
\tableofcontents
\newpage
\section{Introduction}
\subsection{The concept of the ``structure of the photon''}
The photon is one of the basic elementary particles.  This
 ``horrendous-looking hypothesis of light-quanta''    
 introduced in 1905 by Einstein has been  based on the Planck's 
concept of quanta of energy dating from 1900. 
The photon properties, as follows from Quantum Electrodynamics (QED), 
are well known: the photon is a  massless, chargeless object with a 
pointlike coupling to the charged  fundamental particles. As such, it was 
used  in the past and  is still used  at  present as an ideal tool for  
probing structure of more complicated objects,  for example hadrons, 
 acting as  a microscope with the resolution given by its 
wavelength.

The  concept  of ``the structure of the light-quanta'' which has appeared 
later and which is the topic of this work, makes people confused even today. 
However it is very natural  in the quantum field theory, since 
a photon, as any  elementary particle, can fluctuate into 
various states consisting of leptons, quarks, $W^{\pm}$ bosons, hadrons.... 
A transition to a particular  observed state  can be realized as a physical process 
due to the interaction with another object(s), for example, ... 
 {\sl  ``through an interaction with a 
Coulomb field the photon could materialize as a pair of electrons,
$\gamma\rightarrow e^+e^-$. Although not usually thought of in these terms, 
this phenomenon was the earliest manifestation of photon structure''} \cite{vdm}.

Similarly the process $\gamma\rightarrow q\bar{q}$ leads to the
``hadronic (partonic) structure of the photon'' 
in the perturbative Quantum Chromodynamics (pQCD).
This notion offers an effective way to describe 
hard and  semihard processes involving photons.
The hadronic properties of the photon become apparent also in the soft 
photon-hadron interactions, where the similarity between photon and 
vector mesons $\rho,\omega,\phi$ interaction is observed.
Probing the ``structure'' of the photon can be performed in an 
analogous way as testing the  structure of other objects - 
once a hard probe is at our disposal, the partonic structure can be 
revealed.  Often the  probe  happens to be just another photon 
(...``an ideal tool''). 
Sometimes this leads to additional complications 
and even confusion when the partonic content  of the photon is discussed.

Contrary to the nucleon case, the Quark-Parton Model (QPM or PM) gives 
the definite predictions for the hadronic structure 
functions $F_{1,2}^{\gamma}$ for the real photon,  and for the related 
parton densities in the photon. 
They exhibit a logarithmic dependence on the $Q^2$ scale.
The leading QCD corrections to these quantities 
can be summed up using the evolution equations, which, due to existence 
of  the inhomogeneous (PM) term,  can be solved in principle without any 
input, leading to the so called asymptotic solution.
Despite this fact evolution equations are solved usually in an
 analogous  way as for the nucleon, \ie  by introducing some experimental 
or model input, to  cure the singular behaviour present 
in the asymptotic solution.
The gross features of the parton densities in the photon
present already in the PM,
as the scaling violation and large quark densities at large $x_{Bj}$,
are unique for the photon (in general for gauge bosons),
therefore they  are treated as  one of the basic tests of QCD.
This holds to even larger extent  for  the structure functions of the
virtual photon. Here the existence of the second scale, the ``mass'' of 
the virtual photon, makes it possible, at least in principle,  to obtain
the QCD predictions in the input-independent way. 
The QCD  description of 
all basic processes induced by the  real 
and virtual photons is now available with  the NLO accuracy.

Basic ideas related to the considered topic are discussed
in review articles, \eg \cite{vdm}, \cite{rev}. 

\subsection{Recent measurements}

There are two basic types of inclusive processes where the structure
of photon is tested in existing experiments:
\begin{itemize}
\item the deep inelastic scattering, DIS$_{e\gamma}$
and DIS$_{e\gamma^*}$, 
where the structure functions for a real and 
virtual photon  are measured,
\item
the large $p_T$ jet (and particles) production in ${\gamma}p$
and ${\gamma}{\gamma}$ collisions,
where individual quark and gluon densities in the photon
may be probed (the so called resolved photon processes).
\end{itemize}

The early experiments of the first type were performed
at PETRA and  PEP $e^+e^-$ colliders almost twenty years ago,
starting with the first measurement by the PLUTO
Collaboration \cite{early} in 1981. Recent 
results on the structure function $F^{\gamma}_2$, 
for a real $\gamma$, based on few years runs at LEP1, \ie at the 
CM energy $\sim M_Z$, are now available in published or preliminary 
form. Also new data taken at LEP 1.5 (CM energy 130-136 GeV)
and at LEP2 (161-172, 183, 189 and 202 GeV) have been or are being 
analysed.
Final analyses appeared from the  
TRISTAN collider (CM energy $\sim$ 60 GeV).
\footnote{In principle the measurement
 could be performed also at the SLC collider (at energy $\sim$~91~GeV),
but the luminosity is too low for a precise analysis of two-photon 
processes.} Altogether the
measurements cover a wide range of the (average) $Q^2$ from 0.2 to 
$\sim 390-400$ GeV$^2$ \footnote{The first high $Q^2$ results were
obtained by the AMY Collaboration \cite{amy2}} 
allowing  to test the scaling violation pattern.
The range of the $x_{Bj}$ 
variable extends from $\sim$ 0.001, where the possible rise of 
the structure function 
of the real photon can be observed, to 0.98.
In earlier analyses the charm (sometimes also bottom) quark
contribution to $F^{\gamma}_2$ was subtracted,
later on it was included, although with a different treatment  than 
for the light quarks (only the QPM term).  Recently the  first dedicated  
measurement of $F^{\gamma}_{2,c}$ has been performed at LEP2
 for 0.0014$<x_{Bj}<$0.87 and 5$<Q^2<$100 GeV$^2$.

The  measurement of the resolved real photon processes, 
\ie production of large $p_T$ jets (also particles), started  a decade later.
First evidence for the resolved photon process was reported in 1992 by the AMY 
collaboration at TRISTAN \cite{amy3}.
Data have been and are still 
being taken    in ${\gamma}{\gamma}$ collisions at the abovementioned 
$e^+e^-$ machines and in  the photoproduction processes at the $e^{\pm}p$ 
collider HERA (with the CM energy $\sim$ 300 - 320  GeV). The latter one  
gave first results on the effective parton density (the light quark 
distributions combined with the gluon distributions) 
and on the gluonic content of the photon, showing the rise at small $x$,
a part of the momentum of the photon taken by a parton. 
The probed range of $x$ is from 0.05-0.1 to 1, where $x$ equal to 1 
corresponds to the direct interaction of the photon.
The hard scale in such processes, $\tilde{Q}$, 
is usually given by  $p_T$, and the range 
of the measured   $p_T^2$ $(=\tilde Q^2)$ extends from  
$\sim$ 20 up to 1000 GeV$^2$. 
There appeared also first data on the prompt photon production,
and on the  $c$ and $b$  quark content of the photon, both from $e^+e^-$  
and $e^{\pm}p$ collisions.

An impressive progress  has been observed during the last year in the 
measurements of the ``structure of the virtual photon''. 
Fifteen  years after the first measurement of the $F_2^{\gamma^*}$
by PLUTO Collaboration {\cite{pluto}},
a new measurement has just been performed  at LEP.
A large amount of data  concerning the resolved virtual photon
comes from the $ep$ collider HERA, with the virtualities of the photon 
 spanning  the range from 0.1 to 85 GeV$^2$. 
Here an effective parton  distribution for a virtual photon 
has been extracted from the dijet production.
Here also a new type of analysis related to the forward jet production 
in events with large $Q^2$ (DIS$_{ep}$-type  events) is being performed, 
with the aim to test the 
underlying dynamics of the QCD radiation processes at the small $x_{Bj}$
in the proton.
The resolved virtual photon subprocesses contribute here as well and 
they should be included when describing the forward jet production.

The discrepancies between the data and Monte Carlo (MC) 
models  in the hadronic final 
states in the photon-induced processes are observed by many 
collaborations.
They are especially visible in the DIS$_{e\gamma}$ experiments,
where there is  a  need of unfolding, \ie reconstruction of basic 
kinematical variable 
$x_{Bj}$ from the visible  energy of hadronic system $W$.
The improvement in the unfolding has been obtained recently, still the 
dependence on the chosen MC used in the analysis
cannot be avoided \cite{mariak}.
Similar situation, \ie the lack of proper description of hadronic final state, 
occurs also at the $ep$ collider HERA. It was found there that 
implementation of the modified transverse momenta distribution of the partons 
in the MC programs helps to describe the data.
The same modified MC models have been used recently in analysis of the $e^+e^-$ data
from the  LEP and TRISTAN colliders.  The observed
discrepancies may also be a signal of a new 
production mechanism not included in the standard
perturbative QCD analysis, as for example the multiple interaction. 
This mechanism, when   added, usually improves description of the data, both at
the $e^+e^-$ and $ep$ colliders. The situation is still not clear and it 
enforces the advanced study of various aspects of hadron production in
the DIS$_{e\gamma}$ setup (see \eg \cite{miller,cart,finch-lep,mariak})
and in the photon-induced jet production \cite{butter-jets}. 

\subsection{The aim of the survey}
The aim of this survey is to  present  our knowledge on the structure
of the photon, as it stands  AD 2000. We chose to use the data as a 
guide through the field. We collected  
 main results from  precision measurements for the unpolarized real 
and virtual photons\footnote{So far there is no data 
for the polarized photon structure functions nor for the individual 
polarized parton densities.}, performed during 
last few years at  LEP, TRISTAN and HERA 
colliders\footnote{ We refer to data taken up to 1998, and papers that 
appeared till June 2000.}. In order to show  
the status and  also the  progress achieved 
in this rapidly  expanding    field of research in high energy physics,
we present the   very recent (mainly published) results, and also some  of the 
previous measurements  playing the role of the milestones in the 
field.\footnote{A comprehensive collection of the earlier data can be found in 
\cite{whalley}; for the new results see \cite{rev}h, j.}
  
The survey is a collection of short presentations
of the various experimental  data without much detail on the experimental
setup. Only basic information on the type of measurements for each entry,
used Monte Carlo generators and, wherever 
possible, comparison with the theoretical predictions are given.  
This allows to compare different results in an easy way,
and also to show   various attempts to improve
the Monte Carlo description  of the data.
Although the collection of the experimental
results is the main body of the
survey, we found it  important to point out the essential aspects of the 
analyses and to  mention also some new theoretical ideas and future 
measurements.

This survey is an extension of our first attempt \cite{rev}i, 
which covered the data which appeared up to early 1998 
when the first measurements of 
some  quantities related to the structure of photon have
appeared, and also some  problems in describing  data
have been recognized.
The present overview contains much more data, and it reports new types 
of analyses undertaken to  solve the  observed problems  (
already with some positive results).

\subsection{The content of the survey}
 
We start with the results of measurements of the hadronic 
structure function $F_2^{\gamma}$ of the unpolarized real photons
in DIS$_{e\gamma}$ experiments.
The results of analysis of the hadronic final state 
accompanying these measurements
are collected together and presented as a separate issue.
We discuss also results on double-tag hadronic events (DIS$_{e\gamma^*}$),
and single and double-tag leptonic events.
The results from resolved real and virtual photon processes at HERA
collider based on the single and double jet events, and in addition
on charged particles 
and prompt photons, are presented.
In the context of virtual photon processes the HERA data 
for forward jet and forward particle production are included.
In addition a short presentation of the very recent data on 
 the heavy quark content
of the photon is given to signal the new type of measurements.
Related topics - the polarized photon
structure functions, 
the structure function of the electron and the photonic content of the proton -
are also shortly mentioned. In the Appendix we give a short description 
of the existing parton parametrizations for the real and 
the virtual photon.
Bibliography contains, besides references to the data we describe, 
references to MC generators and
the comprehensive list of theoretical papers divided into groups 
according to the topics.

We will not discuss the total cross sections, either for  real or virtual 
photons, nor diffraction phenomena. The data 
from the dedicated studies of the structure of jets and  remnant jets are omitted, 
 a compromise between the completeness of 
the picture and the   size of the presentation.

Each of the topics presented in the survey is preceded  by a 
short review of the
basic theoretical ideas, the description of 
the  specific - for the discussed topic - 
 type of analysis, used MC programs, the  quality of the description 
in terms of the MC and  theoretical   predictions, and finally the  notation.
(In different types of measurements and in analyses made by different 
groups, different notation is used. We try to keep
a consistent notation throughout the survey. The double role played
by the virtual photon - serving as a probe of the structure of the
target, and in some cases being a target itself, which may be resolved
by final large $p_T$ hadrons/jets, causes additional problems, \eg
$Q^2$ versus $P^2$ notation for DIS$_{ep}$ events at HERA.)
In each entry we present individual measurements by 
the short description of results together with  some representative
figures, as well as  comments/conclusions mainly quoted 
from original publications. (Standard names of the Monte Carlo generators
 and the acronyms of  the parton parametrizations are used in the descriptions, see references and Appendix, respectively.)
We also give tables with the measured values of the hadronic structure function
$F_2^{\gamma}$ and for parton densities
for the real and virtual photon.

We denote each entry by the name of the collaboration in alphabetic order,
the publication year (trying to keep the chronological order)
and the reference number listed in
 the bibliography,
and in addition by the name of the collider.
The conference presentations or submitted  papers are labelled with 
 {\bf conf} in the header of the entry.
By adding a small letter we distinguish among the entries published or
presented by the
same collaboration in the same year, in the alphabetic order 
corresponding to their appearance in the text, \eg  {\bf DELPHI 96a}. 

\newpage
\section{Partonic content of the real photon\label{sec2}}
The deep inelastic scattering on the $\gamma$ , DIS$_{e\gamma}$,
with the hadronic final state produced in the single-tag events,
corresponding to the $\gamma^*\gamma$
collision at $e^+e^-$ colliders,
is the main source of information on the partonic content 
of the real photon. For the unpolarized $e^+e^-$ case, the hadronic structure
functions $F_{1, 2}^{\gamma}$ can be introduced to describe this 
process, in analogous way as for the proton in the DIS$_{ep}$ experiments.

Contrary to the proton case, the (Quark) Parton Model gives definite 
predictions for $F_{1, 2}^{\gamma}$,  based on the fully calculable
QED process
$\gamma^*\gamma\rightarrow q\bar{q}$. 
These functions,
and related quark densities in the photon, depend logarithmically
on $Q^2$. The Parton Model contribution together with  QCD 
corrections constitute the  
inhomogeneous evolution equations which can be solved without any extra 
input, leading to the definite QCD predictions both for $x_{Bj}$ and $Q^2$
dependence. This was why   the photon structure functions 
were treated at the beginning as a unique test of QCD,
allowing also, in principle, for an extraction of $\Lambda_{QCD}$.
The pointlike (called also asymptotic) solution obtained in this way
suffers, however, from the
singularities at small $x_{Bj}$, calling for an additional 
contribution of nonperturbative (hadronic) origin. Even after 
proper implementation of the hadronic contribution
the specific properties of $F_{2}^{\gamma}$, namely the large quark density
at large $x_{Bj}$ and logarithmic rise with $Q^2$ remain, providing 
a very well defined (a ``must'') test of QCD.

First measurements of $F_{2}^{\gamma}$ were performed at PETRA and 
PEP in early eighties \cite{whalley}. Recently new data from LEP (LEP1 with the CM
energy $\sqrt{s}\sim$ 90 GeV, LEP 1.5 with 130-136 GeV and LEP2 with 161-202 GeV) 
and TRISTAN (with the CM energy 58 GeV)
$e^+e^-$ colliders have appeared. They reach very small $x_{Bj}\sim 10^{-3}$ 
(at small $Q^2$), and also very high $<Q^2>\sim 400$ GeV$^2$ regions.
In sec. \ref{sec222} 
we collect these  results for the hadronic structure function 
$F_{2}^{\gamma}$ of the real photon. 

In extracting $F_{2}^{\gamma}$ the unfolding of the true kinematical
variables is a real challenge. Specific problems which occur here,
special data analyses and related experimental results
are discussed in the separate section (sec. \ref{sec23}). 
The contributions due to the resolved, both target and probe, photons
happen to be  also of relevance for proper unfolding of
$F_2^{\gamma}$, see below. 
On top of it the target is always slightly virtual, which influences 
the extraction of $F_2^{\gamma}$. 
The problem how to correct for the offshellness of the target
photon is easier to handle 
for a leptonic structure function $F_2^{\gamma (QED)}$,
based on the process $\gamma^* \gamma \ra l^+ l^-$, see corresponding
results in sec. \ref{sec4}.

The description of the heavy quark content of the
photon deserves a comment. At very high scale ($\sim Q^2$)
the c and b-quark, or even t-quark play a role of another (massless) parton
in the photon. At the present relatively low scales it is often enough
to use just the (Q)PM description of the massive c (and b) - quark contribution  
to $F_{2}^{\gamma}$.
In early measurements heavy quark contributions were even subtracted from data,
as they are not given by the pointlike solution of the QCD
evolution equation.
It is only recently that the charm part of $F_{2}^{\gamma}$, $F_{2, c}^{\gamma}$, was measured at LEP in the dedicated experiment.
These and other very recent results on the heavy quark content of the photon 
are shortly commented in sec. \ref{sec5}.

The other source, of growing importance, of information 
on the partonic content of the 
real photon is the hard particle(s) or jet production
in $\gamma \gamma$ collision at the $e^+e^-$ colliders 
or $\gamma p$ collision at $ep$ HERA collider ($\sqrt s \sim$ 300-320 GeV). 
These hard processes were observed first at TRISTAN and HERA.
Here, as in a typical hard 
 hadron-hadron process, the individual partons of parent particles
(\eg two photons in $e^+e^-$ or photon and proton in the
photoproduction at $ep$ collision)
interact and the corresponding parton densities in the photon
may be extracted from 
data. The collection of the results for these {\sl resolved} real
photon processes is presented in sec. \ref{sec24}.
For corresponding results for a virtual photon see sec. \ref{sec3}.

Neither the structure functions of the polarized real photon nor
the corresponding resolved polarized photon processes have been
measured so far; sec. \ref{sec61} contains a short note on this topic.

\subsection{Source of the  real photons} 
In the present experiments 
there are no beams of high energy {\sl real} photons. Therefore testing 
of the ``structure'' of the real photon is performed at the $e^+e^-$ 
collisions, both in the DIS$_{e\gamma}$ and the resolved $\gamma\gamma$
(with both photons real)
processes, or in photoproduction at $e^{\pm}p$ colliders.
In all these cases the flux of 'initial
real' photons  arises from the electron or positron beams. 
The flux of these (almost) real photons in present experimental setups 
for unpolarized electron (positron) beams can be 
approximated by the Weizs\"acker - Williams (WW, sometimes called also 
the Equivalent Photon Approximation, EPA) formula \cite{ww,rev}:
\bea
f_{\gamma/e}(z)={{\alpha}\over{2\pi}}{{1+(1-z)^2}\over{z}}\ln {{P^2_{max}}
\over{P^2_{min}}}-{{2 (1-z)}\over{z}}(1-{{P^2_{min}}\over{P^2_{max}}}),
\label{eq1}
\eea
where $z$ denotes the fraction of the electron energy taken
by the photon. $P^2_{min}=m_e^2z^2/(1-z)$
with $m_e$ denoting  the electron mass, and $P^2_{max}$ stands for the
maximal value of the photon virtuality
appropriate for the experimental setup\footnote{The critical 
discussion of this approach for the 
${\gamma}{\gamma}$ processes in $e^+e^-$ collisions at LEP 2
can be found in \cite{schu}.}.
Eq. \ref{eq1}  gives the typical soft bremsstrahlung spectrum. Note 
that in $e\gamma$ or $\gamma \gamma$ 
options planned for  the future  Linear Colliders, where the real photons
would be produced in the Compton backscattering process, 
the corresponding flux is expected to be much harder \cite{tel}.
Here also a flux of photons with definite polarization can be obtained
in an easy way.

\subsection{DIS$_{e\gamma}$ experiments \label{sec22}} 

\begin{figure}[ht]
\vskip -1.5cm\relax\noindent\hskip 2.cm
       \relax{\includegraphics{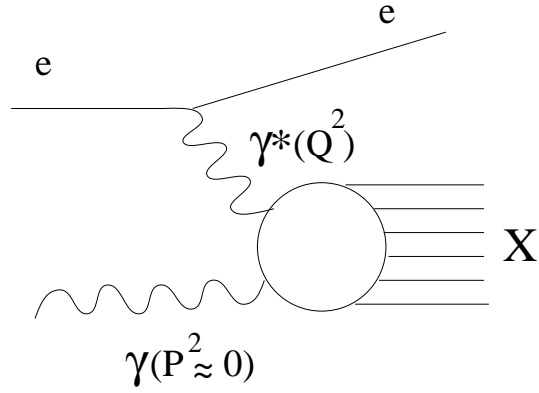}}
\vspace{6.2cm }
\caption{{\small\sl Deep inelastic scattering
on the (almost) real photon $e{\gamma}~(P^2=-p^2\approx0)\ra eX$. The photon 
probe has a large virtuality $Q^2=-q^2>$0.}}
\label{fig:jerzy1}
\end{figure}

In this section we consider the standard DIS$_{e\gamma}$ measurements 
in unpolarized $e{\gamma}$ collisions based on the process 
(fig.~\ref{fig:jerzy1}):
\be
{e(k)\gamma(p)} \ra e(k') {\ }hadrons, \label{eg}
\ee
at $e^{+}e^-$ colliders (single tagged events) \cite{rev}.
Here the target is a real (in practice   almost real, 
with $p^2\sim 0$) photon, radiated by 
 the initial electron\footnote{For simplicity we will use the symbol $e$
for both electron and positron.}. 

In the DIS$_{e\gamma}$ process (\ref{eg}) the real photon
with four-momentum $p$, and the virtuality  $P^2=-p^2=0$, 
is probed by the highly virtual photon with four-momentum $q=k-k'$, 
where  $k$  and $k'$  are the four-momenta of the initial and final
electron, and a large virtuality $Q^2=-q^2>$0.\footnote{The positive 
square  of the four-momentum of the exchanged particle   
will be called its virtuality.}

The differential cross section for the process  (\ref{eg}) is given by:
\begin{eqnarray}
\label{si}
{{d\sigma^{e { \gamma} \ra e X}}\over{dx_{Bj}dy}}=
{{4\pi \alpha^2 (2p\cdot k)}\over
{Q^4}}{[(1-y)F_2^{\gamma}+x_{B_j}y^2F_1^{\gamma}]} \\
={{2\pi \alpha^2 (2p\cdot k)}\over
{Q^4}}{[(1+(1-y)^2)F_2^{\gamma}-y^2F_L^{\gamma}]}, \label{si2} 
\end{eqnarray}
with the relation among the \underline{transverse} $F_T^{\gamma}$ 
(= $F_1^{\gamma}$) and \underline{longitudinal} $F_L^{\gamma}$  structure 
functions: $F_2^{\gamma}=F_L^{\gamma}+2x_{Bj}F_1^{\gamma}$ \cite{f2gdef}.
The standard DIS variables are defined by  
\be
x_{Bj}={Q^2 \over {2p \cdot q}}     
,~~~~~~y={{p\cdot q}\over {p\cdot k}}, \label{xy} 
\ee
where $x_{Bj}$ is a standard Bjorken variable, while $y$,
in case of a massive target (\eg proton) denotes 
the scaled energy of the exchanged photon in the target rest frame.
In practice, in $DIS_{e\gamma}$ measurements the variable $y$ is small and 
the cross section (\ref{si}, \ref{si2}) is effectively saturated by 
$F_2^{\gamma}$. 
Note that  in $e^+e^-$  collisions there is always a small 
off-shellness of the photon target, \ie $p^2=-P^2\neq0$, and there may 
appear in addition the \underline{third structure function}, $F_3^{\gamma}$ 
\cite{f2gdef}.
After integration over the azimuthal angle between two scattering 
planes of the scattered $e^+$ and $e^-$ 
it  disappears however from the cross section in the single-tag 
measurements, since it 
enters the cross section as the term $F_3^{\gamma} \cos (2\bar {\phi})$).
See sec. \ref{sec3} for a discussion on virtual photon measurements where
$P^2$ is kept far from zero.
Discussion on the additional structure functions
measured for a leptonic  final state can be found in sec. \ref{sec4}.

\subsubsection{Theoretical description}
In the (Quark) Parton Model (denoted as QPM\footnote{Usually in  QPM
quarks are treated as massive.} or PM) one assumes that 
the hadronic 
final state in eq.~(\ref{eg}) is due to the production of quark pairs:
$q_i$ and $\bar q_i$ ($i$=1,2...$N_f$, with $N_f$ being
the  number of quark flavours) with the 
fractional charge $Q_i$. The  $F_2^{\gamma}$ is obtained by the 
integration over the four-momentum of the outgoing quarks with 
respect to the target photon direction. The full (QPM or the lowest 
order) expression for $F_2^{\gamma}$, keeping the terms with quark 
mass $m_{q_i}$, is given by the Bethe-Heitler formula \cite{rev}: 
\begin{eqnarray}
\nonumber{F}_2^{\gamma}={{{\alpha}}\over{{\pi}}}N_c\sum_{i=1}
^{N_f} Q_i^4 x_{Bj}
[ (-1+8x_{Bj}(1-x_{Bj})-x_{Bj}(1-x_{Bj}){{4m_{q_i}^2}\over{Q^2}}){\beta}\\
{+[x_{Bj}^2+(1-x_{Bj})^2+x_{Bj}(1-3x_{Bj}){{4m_{q_i}^2}\over{Q^2}}
-x_{Bj}^2{{8m_{q_i}^2}\over{Q^2}}]\ln{{1+{\beta}}\over{1-{\beta}}} ] }
\label{mh},
\end{eqnarray}
where the quark velocity $\beta$, in the $\gamma^* \gamma$ CM system,
 is given by
\begin{eqnarray}
{\beta}=\sqrt{1-{{4m_{q_i}^2}\over s}}=\sqrt{1-{{4m_{q_i}^2x_{Bj}}\over
 {Q^2(1-x_{Bj})}}}.
\label{beta}
\end{eqnarray}
The squared energy of the $\gamma^* \gamma$ collision, $s$, equal to the
 square of the invariant mass of the hadronic system, $W^2$, is given by
\begin{eqnarray}
s=W^2={{Q^2}\over x_{Bj}}(1-x_{Bj}).
\label{w2}
\end{eqnarray}
In the limit of $s$ well above threshold, \ie for $\beta \approx 1$, 
which for fixed $x_{Bj}$ (not too small and not too close to 1) 
corresponds to the  Bjorken limit, one gets
\be
\ln{{1+{\beta}}\over{1-{\beta}}}
\approx\ln{{Q^2(1-x_{Bj})}\over{m_{q_i}^2x_{Bj}}}.
\ee
The structure function $F_2^{\gamma}$ can be approximated in such 
case by
\begin{eqnarray} 
F_2^{\gamma}= 
{{\alpha }\over{\pi}}N_c\sum^{N_f}_{i=1} 
Q_i^4 x_{Bj}[ [x_{Bj}^2+(1-x_{Bj})^2]\ln{W^2\over m^2_{q_i}}
+8x_{Bj}(1-x_{Bj})-1]
\label{f2}
\end{eqnarray}
and it can be used to define the quark densities in the photon:
\begin{eqnarray} 
F_2^{\gamma} =x_{Bj}\sum^{2N_f}_{i=1} Q_i^2 q_i^{\gamma}(x_{Bj},Q^2).
\label{f2q}
\end{eqnarray}  
In the above formulae $N_c$=3 denotes the number of colors and $Q_i$
- the fractional quark charge. In the 
last  equation a (natural) assumption, that quark and antiquark 
distributions in photon  are the same, has been  introduced. Note that
\begin{itemize}
\item
$ F_2^{\gamma}$ is calculable in the (Q)PM, in contrast to the structure 
function of hadrons, \eg the nucleon structure function $ F_2^{N}$.
\item
$F_2^{\gamma}$ is proportional to $\alpha$, the fine structure 
coupling constant ($\approx$ 1/137). There is an overall logarithmic 
dependence on the energy scale squared $W^2$ (or $Q^2$, see 
eq.~(\ref{w2})). A large value of $F_2^{\gamma}$, or in other words 
a large quark density, is predicted for large $x_{Bj}$.
\item  $F_2^{\gamma}$ is a sum of the quark contributions, each 
containing the factor $Q_i^4$, so the individual
quark density is proportional to $Q^2_i$ (see eqs. (\ref{f2}, \ref{f2q})).
\end{itemize}
The longitudinal structure function $F_L^{\gamma}$ is not zero in 
PM, in contrast to the corresponding function for hadrons, and it 
is scale invariant:
\be 
F_L^{\gamma}={{\alpha}\over{\pi}}N_c\sum_{i=1}^{N_f}Q^4_i
x_{Bj}[4x_{Bj}
(1-x_{Bj})].
\ee
Note that also the third structure function is scale invariant:  
\be
F_3^{\gamma}={{{\alpha}}\over{{\pi}}}N_c\sum_{i=1}^{N_f}Q^4_i
x_{Bj}[-x_{Bj}^2].
\ee
In the leading logarithmic approximation (LLA) the Callan-Gross 
relation $F_2^{\gamma}=2x_{Bj} F_1^{\gamma}$ holds as in the case 
of hadrons. 

In this approximation the PM formula for the quark density 
is given by
\be
q_i^{\gamma}(x_{Bj},Q^2)|_{PM}^{LL}
={{\alpha }\over {2\pi}}N_c Q_i^2 [ x_{Bj}^2+
(1-x_{Bj})^2]\ln{{Q^2}\over {\Lambda_{QCD}^2}} . 
\ee
Since the LL contribution corresponds to the on-shell quarks one 
can treat $x_{Bj}$ in the above formula as equal to the part of 
four-momentum of the initial photon carried by the quark;  this latter 
variable is usually  denoted by $x_{\gamma}$ (or simply $x$). Note 
also that in the above formula instead of a quark mass there appears 
the QCD scale $\Lambda_{QCD}$. Therefore this expression allows to 
describe all the light quark (light as compared to the scale $W^2$)
contributions to $F_2^{\gamma}$ in a universal way. Heavy quark 
contributions should be treated separately, according to the QPM 
(eqs.~\ref{mh} and \ref{beta}), as long as $Q^2$ is not considerably
larger than $m_q^2$ \cite{rev}.

An additional $Q^2$ dependence will appear in $F_2^{\gamma}$ and further in 
$q^{\gamma}_i$ due to the QCD corrections. Here also the 
gluonic content of the photon, $G^{\gamma}$, appears.
Large logarithmic corrections  $\sim$ ln$Q^2$ can be described by 
the Dokshitzer-Gribov-Lipatov-Altarelli-Parisi (DGLAP) type of
equations \cite{dglap,dewitt} or by other 
techniques \cite{rev,witten,smith,bb,uw} in the 
leading logarithm approximation
LLA, in the next-to-leading logarithm approximation NLLA, 
or with a higher accuracy (NNLLA,...). 

The inhomogeneous DGLAP equations for the (massless) parton densities in the real photon 
can be represented in the following way 
(below $q_i^{\gamma}$ is  used for both quarks and antiquarks):
\begin{eqnarray}
\nonumber
{{{\partial q_i^{\gamma}(x,Q^2)}\over{\partial lnQ^2}}=
{{{\alpha}}\over{2{\pi}}}Q^2_i P_{q \gamma}
{+{{{\alpha}_s}\over{2{\pi}}}\int^1_x{{dy}\over y}
[P_{qq}({x\over y})q_i^{\gamma}(y,Q^2))+
P_{qg}({x\over y})G^{\gamma}(y,Q^2)]}}
\end{eqnarray}
\begin{eqnarray}
{{\partial G^{\gamma}(x,Q^2)}\over{\partial lnQ^2}}=
0+{{{{\alpha}_s}\over{2{\pi}}}\int^1_x{{dy}\over y}
[P_{gq}({x\over y})\sum_{i=1}^{2N_f} q_i^{\gamma}(y,Q^2)+
P_{gg}({x\over y})G^{\gamma}(y,Q^2)]} \label{DGLAPeqs}
\end{eqnarray}
with the standard splitting functions $P_{qq},P_{qg},P_{gq},P_{gg}$
and in addition with the function 
\be
P_{q{\gamma}}
=N_c[x^2+(1-x)^2],
\ee
describing the splitting of the photon into quarks (from eq. 14).

The dominant LL ($Q^2$) contribution in a general solution of the
eqs. (\ref{DGLAPeqs}) (corresponding to the one-loop expressions
for $\alpha_s$ and for the splitting functions)
 is coming from the strong ordering
of the transverse momenta of radiated gluons along the chain between
the target and the probe particles.

Note that it is possible to solve the above equations, based on 
the pointlike couplings only, without the initial
conditions, assuming the LL ($Q^2$) or NLL ($Q^2$) behaviour of
the solution \cite{witten}. 
Obtained in this way the, so called, {\underline {asymptotic 
solutions}}, which give definite predictions for both the $x_{Bj}$
and $Q^2$ dependence of the $q_i^{\gamma}$, $G^{\gamma}$ and $F_2^{\gamma}$.
The logarithmic dependence of this solutions on the QCD parameter 
$\Lambda_{QCD}$ would,
in principle, allow to extract this basic parameter  from  $F_2^{\gamma}$ data.
However these pointlike solutions have a singular behaviour  at small $x$ 
\cite{uw}, and  in practice while solving the  equations (14) 
the initial conditions have to be assumed from a model (\eg VMD) or 
taken from measurements, at some (low) 
$Q^2_0$ scale \cite{grg}.
This way  nonperturbative contributions enter, which regularize
somehow the unphysical behaviour of the purely pointlike solution.

At this stage the parton parametrizations
for the photon, can be  constructed.   
For light quarks and a gluon the procedure is 
based on eqs. (\ref{DGLAPeqs}), while 
treatment of the heavy quarks depends on the scale.
At present energies
(or $Q^2$ scales) 
for $c,b$ quark contributions  the QPM formula (eq.~\ref{mh}), 
with possible QCD corrections,
 is applied close to the threshold, see also sec. \ref{sec5}.
In the Appendix some details
of existing parton parametrizations for a real photon are given.

In pre-QCD times, the hadronic structure of the photon has been {\sl solely}
attributed to the vector meson component ($\rho, \omega, \phi$) 
in the real photon. Therefore for the matrix element  of the photon 
between \eg the nucleon states,  the following representation 
by the corresponding matrix elements for the $\rho$ meson current  
was assumed \cite{bernstein}:
\be
<N\mid J_{\mu}\mid N>=-{{m_{\rho}^2}\over{g_{\rho}}}{1\over{q^2-m^2_{\rho}}}
<N\mid {\rho}_{\mu}\mid N>+...
\ee

That was the basic assumption of the 
Vector Meson Dominance (VMD) model or,
if higher than $\phi$ vector meson states were included, 
of the Generalized VMD (GVMD)\footnote{denoted also VDM and
GVDM} \cite{bernstein,vdm}. 
Once a photon ``becomes'' a hadron, the cross section for
$\gamma\gamma\rightarrow hadrons$ is dominated by {\sl soft}
exchange processes, as for any other hadron-hadron collisions 
({\sl soft} VMD contribution). This mechanism is also present in 
DIS$_{e\gamma}$ case, leading to
nonperturbative (or hadronic)
contribution to the measured structure function $F_2^{\gamma}$,
as well as to $q^{\gamma}$ or   $G^{\gamma}$ densities.

It is clear that the VMD not only models the {\sl soft} 
photon-induced 
processes, but, being built  into the parton densities,  
it is also important in the description of the {\sl hard} large $p_T$ 
processes  involving resolved photons.
For the general discussion of various components of the photon,
 with less conventional
description of the photon-induced processes and the photon structure
functions, see ref. \cite{sas2} and the Appendix.
\newline

Parton distributions of the photon do not fulfill the, standard for hadrons, 
``momentum'' sum rule formulated for quarks and gluons.
The reason for this is that $\gamma$ can interact directly, being
a parton itself.
Various attempts to introduce a usefull momentum sum rule for the 
photon were undertaken, see \cite{sum}, \cite{sas}
and \cite{gal}.

Finally we would like to mention  specific problems
when  counting the order of the perturbation (leading logarithms)
in the higher order calculations
for the photon-induced processes. 
This is already seen in eq. \ref{mh}, where the QPM formula
contains both LL and NLL terms, both being, however,  of the QED origin.
The presence of the  inhomogeneous term in the  evolution equations 
for the quark densities in the photon (~\ref{DGLAPeqs})
leads to further complications in the interpretation of what is meant by
the   LO and the  NLO QCD analysis. 
For a recent discussion, see \cite{Chyla,vogt}, and also 
comments in secs. \ref{sec241} and \ref{sec245}.

For future discussion (sec. \ref{sec34}) let us mention other than 
the DGLAP approaches 
to the deep inelastic scattering processes (for any target). They may
become important if the 
region of small $x_{Bj}$ is tested in DIS experiments. In such case large
logarithms $\sim {\rm ln}(1/x_{Bj})$ appear, and summing them to all orders
becomes necessary. The corresponding evolution equation,
Balitsky-Fadin-Kuraev-Lipatov equation (BFKL),  leading to LL ($1/x_{Bj}$)
terms does not rely on the strong ordering in the transverse momenta
along the chain, allowing for the parton emission with 
 large transverse momenta
\cite{bfkl}\footnote{The NLO BFKL \cite{fadlip}
calculations demonstrate large corrections to the LO BFKL results.}.
The combined approach was also proposed 
(Catani-Ciafaloni-Fiorani-Marchesini equation)\cite{ccfm}.
Note that BFKL and CCFM approaches should apply for any target as the DGLAP
does, since
they correspond to the QCD chain to be attached to the target particle.

\subsubsection{Measurements of $F^{\gamma}_2$ \label{sec222}}

Measurements of the 
$F_2^{\gamma}(x_{Bj},Q^2)$ for the real photon base on the 
single tagged events 
at $e^+e^-$ colliders (\ie DIS$_{e\gamma}$ events, see eq. (\ref{eg})
and fig. \ref{fig:jerzy1}). The bulk of the hadrons is produced at 
small angles to the incoming $e^+e^-$ beams, and remains undetected. 
Due to the particle losses  
in the experiment  not $x_{Bj}$ but the quantity $x_{vis}$ is measured,
\be
x_{vis}={Q^2 \over {Q^2+W_{vis}^2}},
\ee
where $W_{vis}$ is the invariant mass of the visible hadronic system.
Since the target photon has always the nonzero  
virtuality $P^2$, the measured quantity is in fact 
\be
x_{vis}={Q^2 \over {Q^2+W_{vis}^2+P^2}}.
\ee
In order to reconstruct $x_{Bj}$ from a measured value for $x_{vis}$,
an unfolding procedure is needed.
To this end, in some analyses the related variables, $W_{rec},x_{rec}$, 
are also used, see \eg {\bf L3 98a, OPAL 2000}. 
They are reconstructed from the measured
four-momenta of electrons with the constraints from the {\sl transverse}
momentum conservation. Even using  the above technique, 
the value of $W_{rec}$ is still generally smaller than the
 true value due to \eg energy loss in the forward regions. 
Therefore  new variables {\ $W_{cor}$ and $x_{cor}$}
are formed to correct for this fact,  see {\bf OPAL 2000}.  
The very important improvement in the description of the data
has been achieved due to the recently proposed   two-dimensional unfolding,
with the energy in the forward cone used as the second variable 
\cite{cart}. It was used in {\bf ALEPH 99b,conf},
and {\bf OPAL 2000} leading to much
 smaller errors \cite{mariak, finch-lep}.
 The proper inclusion 
of the nonzero virtuality
 of the target photon\footnote{The procedure ``is not well defined if the 
photon is not pointlike'' \cite{mariak-pryw}.} should lead to 
the further improvement \cite{mariak}
(see sec. \ref{sec4} where this effect for 
the leptonic structure functions
is discussed).
The unfolding  can be performed using the  traditional method based on 
the linear scale or using the  approach with the logarithmic scale,
 especially useful for extracting $F_2^{\gamma}$ in the small $x_{Bj}$ 
region \cite{paris94}. 

In order to perform reliable unfolding of the structure function 
$F_{2}^{\gamma}$ from the cross section for
$\gamma^* \gamma \ra hadrons$, the corresponding events  are
grouped into classes of similar topologies (see discussion in 
\cite{amy2}, \cite{miller} 
and \eg {\bf TOPAZ 94}, {\bf{OPAL 97a}}, {\bf DELPHI 96b,conf}),
which roughly  coincide with   the contributions from  QPM, VMD 
and RPC (RPC - Resolved Photon Contribution).
First analyses applied the FKP approach, with the pointlike QCD 
contribution assumed for light quarks, QPM description of the 
heavy quarks, and VMD 
contribution for the {\sl soft} hadronic production. The $p_T^0$ parameter
separates the pointlike and the hadronic ($\rho$-type)
configuration of the quarks in the target photon
(at the first vertex level). The additional 
parameter $p_T^{min}$,  in the FKP approach of the same order as $p_T^0$, 
is introduced as a cutoff for the perturbative $2\rightarrow 2$ 
subprocesses QCD calculations.
More details on the modelling of the hadronic final state can be found in 
sec. \ref{sec23}.

Nowadays the Monte Carlo generators
HERWIG and PYTHIA, adapted for DIS$_{e\gamma}$ 
in 1995 (see \eg {\bf OPAL 97a}),
 can describe these contributions (using the cutoff $p_T^{min}$) 
with any of the existing  parton parametrizations for the photon.
It is no longer necessary for the resolved $\gamma$ term 
to fit an empirical $p_T^0$ parameter
to the data before unfolding, as in the parton density in the photon both 
nonperturbative (\eg \`a~la VMD) 
and the perturbative contributions are included.
The PYTHIA and modified HERWIG (``HERWIG + power law-$p_T$'') 
programs with broader $p_t$ distribution 
$dp_t^2/p_t^2 + \tilde{p}_{t0}^2$ \footnote{The parameter
 $\tilde p_{t0} \sim$ 0.66 GeV should not be 
confused with $p_T^0$ separating the soft and hard hadronic processes
within the FKP approach.}
 originally introduced 
to improve description of $\gamma p$ processes ({\bf ZEUS 95c}), 
seem to describe $\gamma\gamma$ data 
 better than the default versions (\eg {\cite{lauber}}).
On the other hand, the TWOGAM generator, based on QPM + VMD + RPC
contributions is successfully used to describe the newest
$F_2^{\gamma}$ data although with a large $p_T^{min}$ parameter,
$\sim$ 2.3 - 3.5 GeV, instead of the default value 1.8 GeV ({\bf L3 99a}).
The general purpose generator PHOJET based on the Dual Parton
Model is also used successfully in the $F_2^{\gamma}$  measurements 
(with $p_T^{min}$ = 2.5 GeV).
The large dependence on the chosen Monte Carlo model leads to 
large errors in the $F_2^{\gamma}$ data points, or even to
two sets of data points from the same event sample ({\bf L3 98a}).
The multi-dimensional unfolding method reduces the dependence on MC models, 
as it was found in very recent analyses ({\bf OPAL 2000} 
and {\bf ALEPH 99b,conf}), see sec. \ref{sec23}.

The $F_2^{\gamma}$  measurements at small
$x_{Bj}$ are presented in {\bf DELPHI 96a},  {\bf OPAL 97b, OPAL 2000}
and {\bf L3 98a}, with the lowest value  $x_{Bj}$ (center of bin value) 0.001 
(0.0022). The low $Q^2$  measurements at LEP ($<Q^2> \sim$ 2 GeV$^2$)
 were reported in
 {\bf OPAL 97b, L3 98a, OPAL 2000} while
the largest $<Q^2>$=400 GeV$^2$ data are  discussed in 
{\bf DELPHI 98,conf}. 
Note also that in the earlier measurements the $c$ (and $b$) quark 
contribution was usually subtracted from the structure function 
$F_2^{\gamma}$ (\eg {\bf OPAL 94}). For the dedicated measurement of 
$F_{2,c}^{\gamma}$ reported in 1999 by OPAL
collaboration, see  sec. \ref{sec5}.

The general features of $F_2^{\gamma}$, as far as $x_{Bj}$ and $\log Q^2$
dependences are concerned, agree with the theoretical expectations,
although the precision of the data does not allow in many cases 
to distinguish between existing parton parametrizations
and clarify the small $x_{Bj}$ behaviour of $F_2^{\gamma}$.  
The parton densities which give, at present, the best description of the
$F_2^{\gamma}$ data are GRV and SaS1D \cite{mariak}.

As we have already pointed out, the study of the hadronic final state  
together with the possibly resolved (real and virtual) photon contribution
became part of the measurements of the structure function 
for the photon at $e^+e^-$ colliders
(details of the studies of the hadronic final state will be presented
in the next section).
 Here we would only like to stress 
that some discrepancies have been found for certain distributions,
\eg the hadronic energy flow.
This fact is included in the estimation of 
the uncertainty of the measured function $F_2^{\gamma}$.

We start the presentation  of the $F_2^{\gamma}$ data from  the LEP collider, 
then the TRISTAN data  are discussed.
Figures and 
tables with corresponding numbers for the $F_{2}^{\gamma}$ as a
function of $x_{Bj}$ at fixed $<Q^2>$, and also for $F_{2}^{\gamma}$ averaged
over relevant $x_{Bj}$ ranges are given.
 Collective figures of
$F_2^{\gamma}$ versus $x_{Bj}$ and $F_2^{\gamma}$ versus $Q^2$
containing all existing results
are presented at  the end of this section.

\newpage
\centerline{\bf \huge DATA}                                                
~\newline\newline
$\bullet${\bf {ALEPH 99a \cite{finch99} (LEP 1)}}\\
Data on $F_2^{\gamma}$ for $x_{Bj}$ from
0.005 to 0.97 and $Q^2$ between 6 and 3000~GeV$^2$,
in three $Q^2$ bins, were  collected in the period 1991-95. 
An analysis of the hadronic final state was 
performed using the QPM+VMD
model and the ``HERWIG 5.9 + power law $p_t$''
involving GRV LO parton parametrizations (see next 
section for details).

\begin{figure}[ht]
\vskip 0.in\relax\noindent\hskip 0.6cm
       \relax{\includegraphics{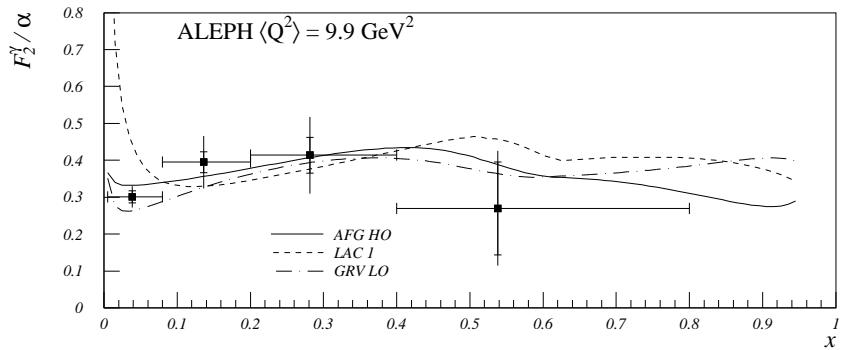}}
\vspace{0.cm}
\caption{\small\sl The unfolded structure function 
$F_2^{\gamma}/\alpha$ as a function of $x_{Bj}$ for $<Q^2>$=9.9 GeV$^2$.
Comparison with predictions of different parton 
parametrizations: AFG HO (solid line), LAC 1(dashed line), GRV LO 
(dot-dashed line)
(from \cite{finch99}).}
\label{fig:finch4}
\end{figure}

\begin{figure}[ht]
\vskip 0.cm\relax\noindent\hskip 0.6cm
       \relax{\includegraphics{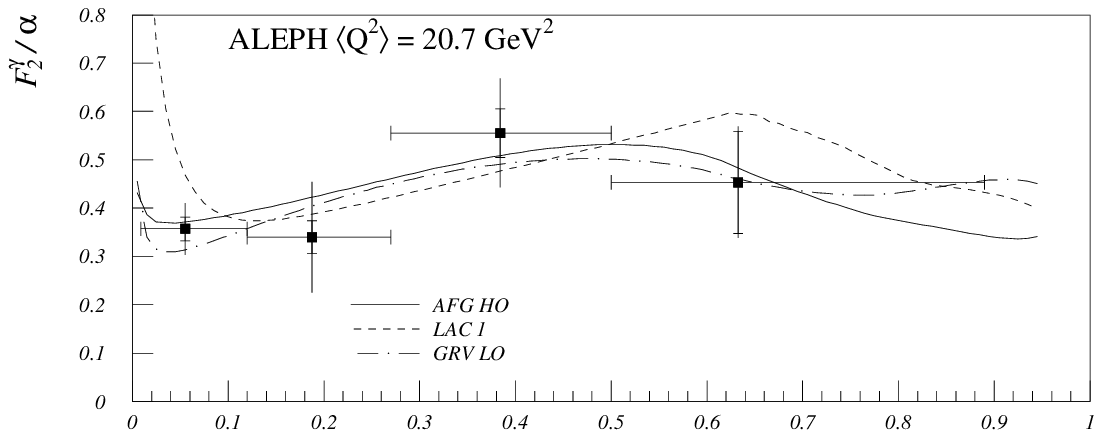}}
\vspace{0.cm}
\caption{\small\sl The same as in fig. \ref{fig:finch4}
for $<Q^2>$=20.7 GeV$^2$ (from \cite{finch99}).}
\label{fig:finch3}
\end{figure}

\begin{figure}[ht]
\vskip 0.cm\relax\noindent\hskip 0.6cm
       \relax{\includegraphics{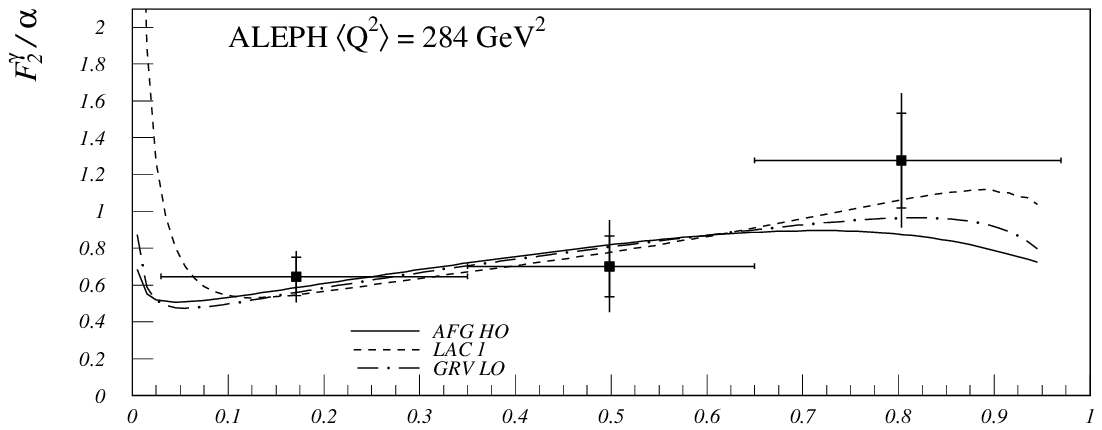}}
\vspace{0.cm}
\caption{\small\sl The same as in fig. \ref{fig:finch4}
for $<Q^2>$=284 GeV$^2$ (from \cite{finch99}).}
\label{fig:finch2}
\end{figure}

The unfolded results for $F_2^{\gamma}$ as a function of $x_{Bj}$ 
are presented in table \ref{table1} and in 
figs.~\ref{fig:finch4}, \ref{fig:finch3}, \ref{fig:finch2},
where the comparison with predictions of the parton parametrizations:
AFG HO, LAC 1 and GRV LO is made.
The values of $F_2^{\gamma}$ averaged over $x_{Bj}$ (for 0.1 $<x_{Bj}<$ 0.6) 
for three values of $<Q^2>$ are given in table \ref{table2}.

\begin{table}[ht]
\caption{}
\label{table1}
$$
\begin{array}{|c|c|c|}
\hline
<Q^2>&x_{Bj}&F_2^{\gamma}/\alpha\\
~[GeV^2]~&&(stat. + syst.)\\
\hline
~~~9.9~~~&0.005 - 0.08&0.30\pm0.02\pm0.02\\
         &0.08 - 0.2&0.40\pm0.03\pm0.07\\
         &0.2 - 0.4&0.41\pm0.05\pm0.09\\
         &0.4 - 0.8&0.27\pm0.13\pm0.09\\
\hline
~~20.7~~~&0.009 - 0.120&0.36\pm0.02\pm0.05\\
         &0.12 - 0.27&0.34\pm0.03\pm0.11\\
         &0.27 - 0.50&0.56\pm0.05\pm0.10\\
         &0.50 - 0.89&0.45\pm0.11\pm0.05\\
\hline
~~~284~~~&0.03 - 0.35&0.65\pm0.1\pm0.09\\
         &0.35 - 0.65&0.70\pm0.16\pm0.19\\
         &0.65 - 0.97&1.28\pm0.26\pm0.26\\
\hline
\end{array}
$$
\end{table}
\vspace*{4.5cm}
\vspace*{4.cm}
\vspace*{4.cm}

\begin{table}[ht]
\caption{}
\label{table2}
$$
\begin{array}{|c|c|}
\hline
<Q^2>&<F_2^{\gamma}/\alpha>\\
~[GeV^2]~&(tot)\\
\hline
9.9 &0.38\pm 0.05\\
\hline
20.7&0.50\pm 0.05\\
\hline
284&0.68\pm 0.12\\
\hline
\end{array}
$$
\end{table}

The $Q^2$ dependence of $F_2^{\gamma}$ (from the $0.1 < x_{Bj} < 0.6$
range) has also been measured (not shown) and  a good agreement
with the {\bf OPAL 97c} measurements was found.

The $F_2^{\gamma}$ data were compared to different parametrizations
and the $\chi^2$ test was performed. LAC1 and 2 and WHIT 4,5 and 6,
all containing a large gluon density at low $x$, show large $\chi^2$.

~\newline
Comment: {\it ``Comparisons to parametrized parton density functions
show that those containing a large gluon content are
inconsistent with the data.''}
\newline\newline
$\bullet${\bf {ALEPH 99b,conf \cite{aleph99conf} (LEP 2)}}\\
The measurement of $F_2^{\gamma}$ at $\sqrt s$=183 GeV based on the 1997
data was performed using a two-dimensional unfolding method 
(the principle of maximum entropy) \cite{cart}.
 The two $Q^2$ ranges: 7-24 GeV$^2$
and 17-200 GeV$^2$ were studied. In the analysis ``HERWIG 5.9 + power law 
$p_t$'' with the GRV LO and SaS1D parton parametrizations and in addition
the  PHOJET model were used.  

The hadronic final state analysis
was  performed (see next section for details) and it was found that
the unfolding procedure reduces the 
dependence on Monte Carlo models, and that
 ``PHOJET is in poor agreement with the data for the higher $Q^2$ range''.

\begin{figure}[ht]
\vskip 0.in\relax\noindent\hskip 2.5cm
       \relax{\includegraphics{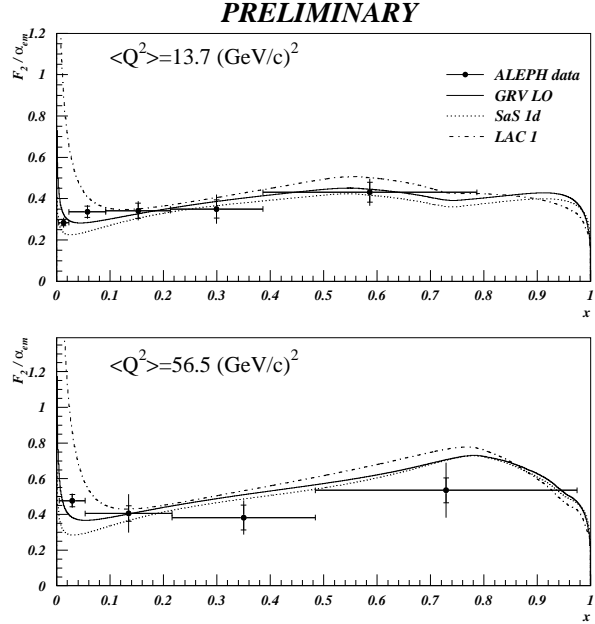}}
\vspace{-1.cm}
\caption{\small\sl The unfolded structure function 
$F_2^{\gamma}/\alpha$ as a function of $x_{Bj}$ for $<Q^2>$=13.7 GeV$^2$
and 56.5 GeV$^2$.
Comparison with predictions of different parton 
parametrizations: GRV LO (solid line), SaS1D (dotted line), LAC1
(dot-dashed line)
(from \cite{aleph99conf}).}
\label{fig:aleph99conf_5}
\end{figure}

The unfolded results for $F_2^{\gamma}$ as a function of $x_{Bj}$ 
are presented in fig.~\ref{fig:aleph99conf_5}
for $<Q^2>$=\newline
13.7 GeV$^2$ and 56.5 GeV$^2$ and in table \ref{table1a}
(only for lower $Q^2$).

\vspace*{9cm}
\vskip -0.5cm
\begin{table}[ht]
\caption{}
\label{table1a}
$$
\begin{array}{|c|c|c|}
\hline
<Q^2>&x_{Bj}&F_2^{\gamma}/\alpha\\
~[GeV^2]~&&(stat. + syst.)~in ~\% \\
\hline
~~~13.7~~~&0.003 - 0.023&0.28\pm5.3\pm6.3\\
         &0.023 - 0.092&0.34\pm7.5\pm5.3\\
         &0.092 - 0.213&0.34\pm10.8\pm8.0\\
         &0.213 - 0.386&0.35\pm13.6\pm15.7\\
         &0.386 - 0.786&0.43\pm11.1\pm10.5\\
\hline
\end{array}
$$
\end{table}
\vskip -0.5cm
~\newline
Comment: {\it The new unfolding method was used in the analysis. It ``leads to
smaller statistical errors and a reduced model dependence compared to 
one-dimensional procedure''. The  GRV LO parametrization
describes the $F_2^{\gamma}$ data well. At low $x$, 
``the SaS1D parametrization is found to be lower than data, 
whereas the LAC 1 parametrization is considerably higher...''.}

\begin{figure}[hc]
\vskip 0.0in\relax\noindent\hskip 0.cm
       \relax{\includegraphics{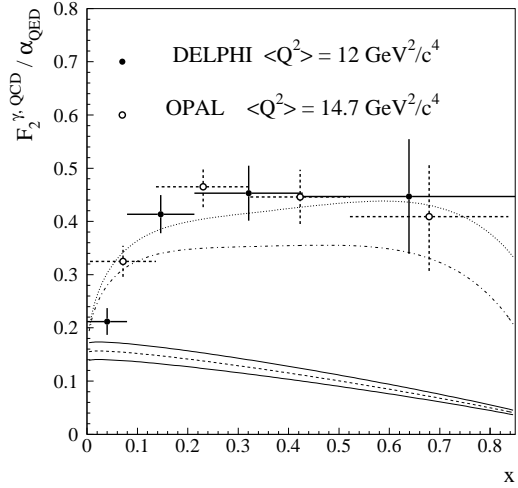}}
\vspace{0.cm}
\caption{\small\sl The unfolded $F_2^{\gamma}/\alpha$ 
for the light quarks together with {\bf OPAL 94} data. The curves show the  
sum of the GVMD model prediction multiplied  
by the threshold factor 1-$x_{Bj}$ and the prediction
of the FKP parametrization,
with the parameter $p_T^0$ equal
0.1 GeV (upper line) and 0.5 GeV (lower line). The bottom  
curves show separately the GVMD contribution with different target masses
(from \cite{delphi2}).}
\label{fig:delphi7}
\end{figure}

\vskip -0.5cm
~\newline\newline
$\bullet${\bf {DELPHI 96a \cite{delphi2} (LEP 1) }}\\
Data on $F_2^{\gamma}$ were taken in the period 1991-93 for $Q^2$ between  
4 and 30 GeV$^2$ ($<Q^2>$ = 12 GeV$^2$) and for $x_{Bj}$ down to 0.003. 
The so called   $F_2^{\gamma (QED)}$ was  also measured
for nonzero virtuality $P^2$ of the
target photon (see sec. \ref{sec42} for  details).  Estimated  target photon 
virtuality $P^2$   was used in the unfolding
of  $F_2^{\gamma}$ 
\footnote{It was found that although  $<P^2>$=0.13
GeV$^2$, a fixed value of 0.04 GeV$^2$ fits the data better.}.
The TWOGAM event generator  was used to simulate QPM events,
another event generator was 
used to  obtain the QCD correction (LL) to the pointlike
contribution for light quarks in the FKP approach.
 The  GVMD  and the pointlike (FKP)  contributions were studied 
with $p_T^0$ = 0.1 and 0.5 GeV. 
Results for $F_2^{\gamma}$ for the light quarks  
are presented in fig.~\ref{fig:delphi7} and in  table \ref{table3}. 

\vspace*{6.3cm}
\begin{table}[ht]
\caption{}
\label{table3}
$$
\begin{array}{|c|c|c|}
\hline
<Q^2>&x_{Bj}&F_2^{\gamma}/\alpha\\
~[GeV^2]~&&(stat. + syst.)\\
\hline
~~~12~~~&0.003 - 0.080&0.21\pm0.03\pm0.06\\
&0.080 - 0.213&0.41\pm0.04\pm0.05\\
&0.213 - 0.428&0.45\pm0.05\pm0.05\\
&0.428 - 0.847&0.45\pm0.11\pm0.10\\
\hline
\end{array}
$$
\end{table}

The averaged value of  $F_2^{\gamma}/\alpha$ over the $x_{Bj}$ 
range between 0.3 and 0.8 was extracted:
$$
\begin{array}{|c|c|}
\hline
<Q^2>&<F_2^{\gamma}/\alpha>\\
~[GeV^2]~& \\
\hline
~12~&0.45 \pm0.08\\
\hline
\end{array}
$$

\begin{figure}[ht]
\vskip 7.in\relax\noindent\hskip -2.4cm
       \relax{\includegraphics{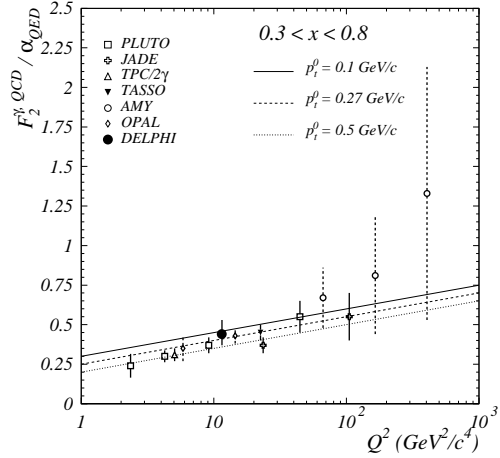}}
\vspace{-12.cm}
\caption{\small\sl $F_2^{\gamma}/\alpha$ averaged over $x_{Bj}$ 
between 0.3 and 0.8 as a function of $Q^2$. The curves 
show the FKP parametrization 
predictions for different values of 
the parameter $p_T^0$  (from \cite{delphi2}).}
\label{fig:delphi9}
\end{figure}

For comparison with other measurements of the $Q^2$ dependence of
the $F_2^{\gamma}$, see fig.\ref{fig:delphi9}.
\vspace*{-0.5cm}

\begin{figure}[ht]
\vskip 0cm\relax\noindent\hskip -3.cm
       \relax{\includegraphics{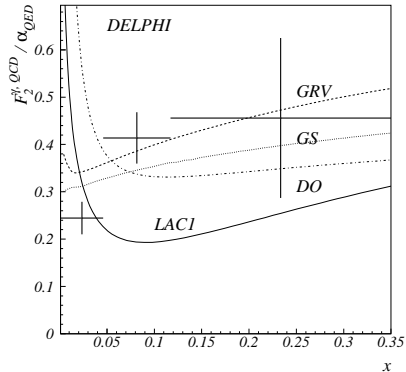}}
\vspace{0cm}
\caption{\small\sl The unfolded $F_2^{\gamma}/\alpha$ for $<Q^2>=$   
12 GeV$^2$ as a function of $x_{Bj}$ with
the LO parametrizations LAC1, GS, DO, 
and GRV (from \cite{delphi2}).}
\label{fig:delphi10}
\end{figure}

The study of the $F_2^{\gamma}$ behaviour at $<Q^2>$=12 GeV$^2$ in 
the low $x_{Bj}$ domain leads to the
following results (fig.~\ref{fig:delphi10} and table \ref{table4}):
\begin{table}[ht]
\caption{}
\label{table4}
$$
\begin{array}{|c|c|c|}
\hline
<Q^2>&x_{Bj}&F_2^{\gamma}/\alpha\\
~[GeV^2]~&&(stat. + syst.)\\
\hline
12&0.003 - 0.046&0.24\pm0.03\pm0.07\\
&0.046 - 0.117&0.41\pm0.05\pm0.08\\
&0.117 - 0.350&0.46\pm0.17\pm0.09\\
\hline
\end{array}
$$
\end{table}
\vspace*{4.3cm}
~\newline
Comment: {\it The effect of the nonzero $P^2$ was included in the 
unfolding of $F_2^{\gamma}$. 
No rise of $F_2^{\gamma}$ at small {$x_{Bj}$} has been found.
The GRV  and GS leading order parametrizations of the 
quark density in the photon are in agreement with the data.}

~\newline
$\bullet${\bf {DELPHI 96b,conf \cite{delphi} (LEP 1)}}\\
The measurement  of the photon structure function
 $F_2^{\gamma}$ at $<Q^2>$ = 13 and 106 GeV$^2$ 
(data collected in the years 1991-95) for $x_{Bj}$ down to 0.003
together with a study  of the hadronic final state (see next section) 
is reported. Three types of subprocesses have been considered: 
QPM (direct term with $m_q\neq 0$ for all quarks), 
soft hadronic GVMD or VMD (the TPC/2$\gamma$-type parametrization)
and the {\sl resolved} target
photon contribution (RPC) with  GS2 parton parametrization.
In addition the effects of the resolved probe
photon were studied.
The unfolding was done in the linear, and, for better sensitivity to
low-$x_{Bj}$, logarithmic scale in $x_{Bj}$.

\begin{figure}[ht]
\vskip 0.in\relax\noindent\hskip 1.5cm
       \relax{\includegraphics{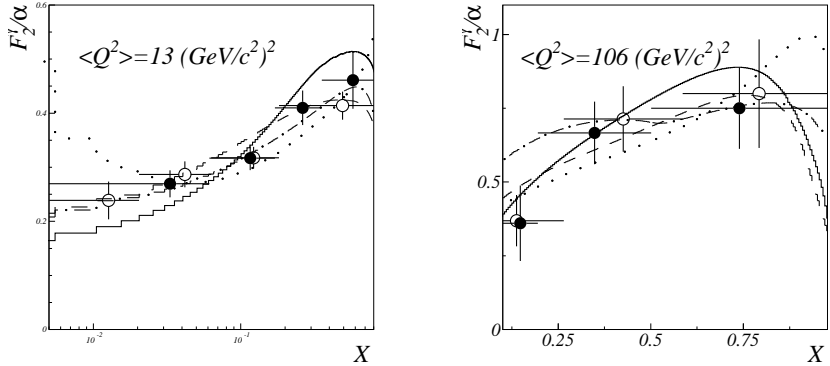}}
\vspace{0.ex}
\caption{\small\sl Unfolded $F_2^{\gamma}/\alpha$ as a function of $x_{Bj}$
 compared with 
QPM+GVMD+RPC (with parton parametrization GS2) (solid line) and
predictions of the  SaS4 (dashed line), GS2 (dash-dotted)
and GRV3 (dotted) parametrizations (from \cite{delphi}).} 
\label{fig:tyap96}
\end{figure}

Fig.~\ref{fig:tyap96} shows the $x_{Bj}$ dependence of $F_2^{\gamma}$
for two $Q^2$ samples, 
compared with the predictions of various parton parametrizations.

\vspace*{4.5cm}

The value of $F_2^{\gamma}/\alpha$ averaged in the range 
$0.3 <x_{Bj}<0.8$ was extracted for $Q^2$ = 13 and 106 GeV$^2$
(see table \ref{table5}).

\begin{table}[ht]
\caption{}
\label{table5}
$$
\begin{array}{|c|c|}
\hline
<Q^2>&<F_2^{\gamma}/\alpha>\\
~[GeV^2]~& (stat.+syst.)\\
\hline
~13&0.38~\,\pm0.031\pm0.016\\
106\,&0.576\pm0.081\pm0.076\\
\hline
\end{array}
$$
\end{table}
~\newline
Comment: {\it  The importance of the final hadronic 
state topology was noticed and the resolved photon contribution
was included. The study of a
linear and logarithmic unfolding was performed.}
\newpage
~\newline
$\bullet${\bf {DELPHI 97a,conf \cite{tyapkin} (LEP 1, LEP 2)}}\\
A study of $F_2^{\gamma}$ in the $Q^2$ range between
3 and 150 GeV$^2$, based on data from the 1994-95 runs for energies
around the $Z^0$ mass and from 1996 for energies between 161
and 172 GeV is presented. An analysis of the hadronic final state is performed
using the TWOGAM Monte Carlo
program, where QPM (with $N_f$ = 4), soft hadronic
VMD and RPC (with the GS2 parton parametrization, both single and double 
resolved photon terms)
parts are included (see also next section).

\begin{figure}[hc]
\vskip 0in\relax\noindent\hskip -3.5cm
       \relax{\includegraphics{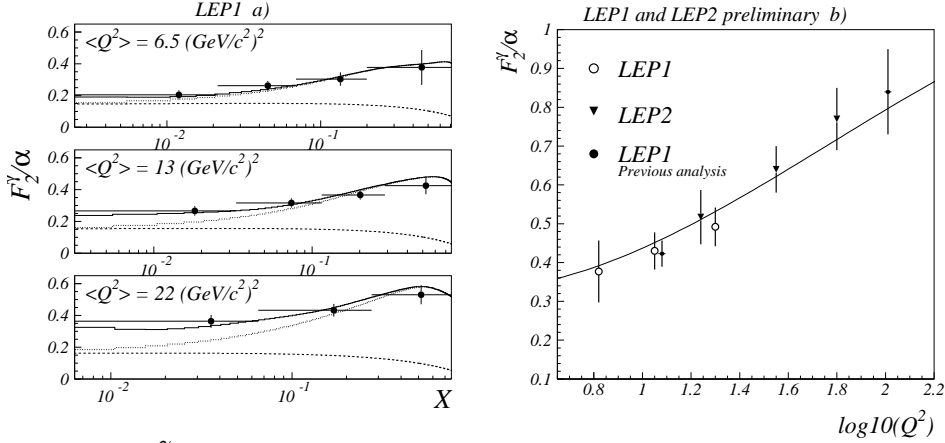}}
\vspace{0cm}
\caption{\small\sl a) $F_2^{\gamma}/\alpha$ versus $x_{Bj}$ from the DELPHI 
experiment based on the LEP 1 data for three values of $<Q^2>$: 6.5, 13, and
22 GeV$^2$. The solid line corresponds to QPM+GVMD+RPC 
(with with parton parametrization GS2), dotted - QPM+GVMD and dashed - GVMD.
 b) $F_2^{\gamma}/\alpha$ 
averaged over $x_{Bj}$ from 0.3 to 0.8 as a function of $Q^2$.
Results obtained from the LEP1 and LEP2 data are shown together with
results of a previous analysis of the LEP1 data (from \cite{tyapkin}).}
\label{fig:tyapkin4}
\end{figure}

The unfolded results for $F_2^{\gamma}$ as a
function of $x_{Bj}$ and $Q^2$ are presented in 
figs.~\ref{fig:tyapkin4}a and~\ref{fig:tyapkin4}b, respectively.\\

\vspace*{5cm}
~\newline
Comment: {\it 
``There is some indication of deviations of TWOGAM predictions for 
the data in the high $Q^2$ region.''}

\begin{figure}[ht]
\vskip 0.cm\relax\noindent\hskip 0.cm
       \relax{\includegraphics{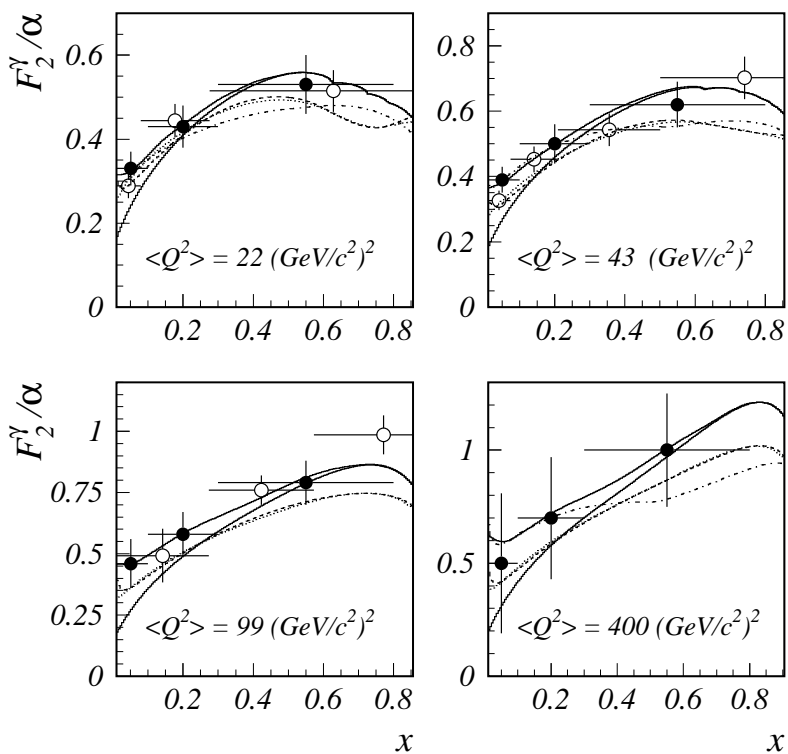}}
\vspace{0.cm}
\caption{\small\sl 
The DELPHI structure function $F_2^{\gamma}/\alpha$ as a function of 
$x_{Bj}$ at $<Q^2>$= 22, 43, 99 and 400 ~GeV$^2$.
Comparison with the parton parametrizations: GS (dash - dotted line),
GRV (dashed line) and SaS (dotted line); TWOGAM prediction is the
solid line (from \cite{tyappriv}).}
\label{fig:f2typ}
\end{figure}

~\newline
$\bullet${\bf {DELPHI 98,conf \cite{delphi99,tyappriv} (LEP 2) }}\\
Data on $F_2^{\gamma}$ were taken in the period 1996-1998 for $Q^2$ between  
10 and 1000 GeV$^2$ (for the  energy 163-188 GeV).
The three components were applied: the QPM for all quarks simulated by
the TWOGAM generator, the soft hadronic 
VMD contribution and RPC for single and double resolved photon processes, 
and a good description was
obtained (see also next section). The data are shown in table 
\ref{table6} and in fig. \ref{fig:f2typ}.

\begin{figure}[ht]
\vskip 0.cm\relax\noindent\hskip 0.cm
       \relax{\includegraphics{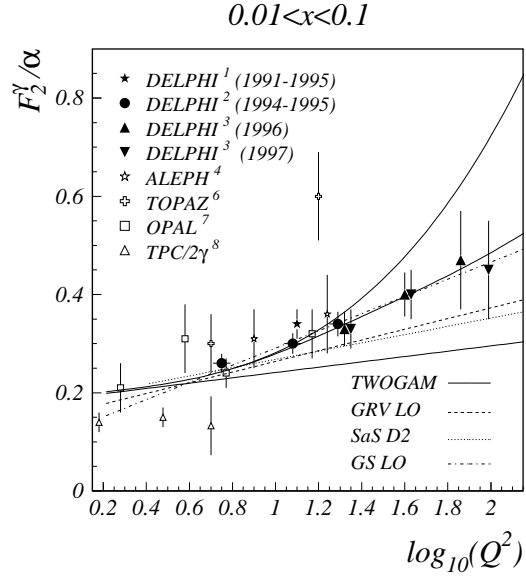}}
\vspace{0.cm}
\caption{\small\sl The DELPHI results for $F_2^{\gamma}/\alpha$
as a function of $Q^2$ for $0.01<x_{Bj}<0.1$ compared with other
data (from \cite{tyappriv}).}
\label{fig:f2q1typ}
\end{figure}

\begin{figure}[ht]
\vskip 0.cm\relax\noindent\hskip 0.cm
       \relax{\includegraphics{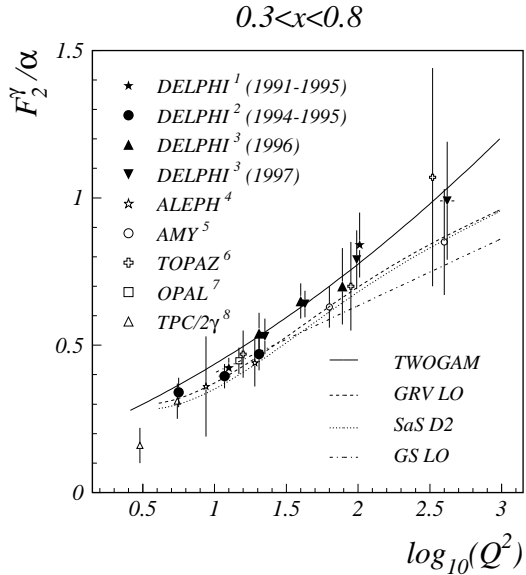}}
\vspace{0.cm}
\caption{\small\sl The DELPHI results for $F_2^{\gamma}/\alpha$
as a function of $Q^2$ for $0.3<x_{Bj}<0.8$ compared with other
data (from \cite{tyappriv}).}
\label{fig:f2qtyp}
\end{figure}

\begin{table}[ht]
\caption{}
\label{table6}
$$
\begin{array}{|c|c|c|}
\hline
<Q^2>&x_{Bj}&F_2^{\gamma}/\alpha\\
~[GeV^2]~&&(stat. + syst.)\\
\hline
~~~21~~~&0.01 - 0.1&0.33\pm0.01\pm0.03\\
         &0.1 - 0.3&0.41\pm0.03\pm0.02\\
         &0.3 - 0.8&0.51\pm0.05\pm0.04\\
\hline
~~42~~~&0.01 - 0.1&0.41\pm0.01\pm0.03\\
         &0.1 - 0.3&0.48\pm0.02\pm0.02\\
         &0.3 - 0.8&0.59\pm0.03\pm0.04\\
\hline
~~99~~~&0.01 - 0.1&0.45\pm0.06\pm0.02\\
         &0.1 - 0.3&0.52\pm0.05\pm0.03\\
         &0.3 - 0.8&0.73\pm0.05\pm0.03\\
\hline
~~~400~~~&0.01 - 0.1&0.5\pm0.3\pm0.1\\
         &0.1 - 0.3&0.70\pm0.2\pm0.2\\
         &0.3 - 0.8&1.0\pm0.1\pm0.3\\
\hline
\end{array}
$$
\end{table}
\vspace*{9.cm}
\newpage
The $Q^2$ dependence was studied in two $x_{Bj}$ ranges:
$${\rm for }~~0.01 <x_{Bj}<0.1{\hspace{0.5cm}} 
F_2^{\gamma}/\alpha=(0.12\pm0.02)+(0.073\pm0.009) \ln Q^2,$$
$${\rm while~for }~~0.3 <x_{Bj}<0.8{\hspace{0.5cm}} 
F_2^{\gamma}/\alpha=(0.09\pm0.05)+(0.14\pm0.02) \ln Q^2.$$
The data for $F_2^{\gamma}/\alpha$ as a function of $Q^2$ 
are presented in figs. \ref{fig:f2q1typ}, \ref{fig:f2qtyp}
in comparison with the Monte Carlo predictions and LO calculations.
A {\sl suppression factor} was introduced to describe the high $Q^2$ data.
\vspace*{7.2cm}
\vspace*{7cm}
\newpage
~\newline
Comment: {\it ``Prediction for the RPC component tends to have
too high cross section for the high $Q^2$'' and the low $x_{Bj}$.
``For the first time TWOGAM has been tuned to
 get a good agreement in the region of high $Q^2$''.}

~\newline
$\bullet${\bf {L3 98a \cite{l3_f2_98} (LEP 1)}}\\
The $F_2^{\gamma}$ data for 1.2 $\le Q^2\le $ 9.0 GeV$^2$ collected in years
1991-95 are reported. Two bins: $<Q^2>$ = 1.9 GeV$^2$ in the $x_{Bj}$ 
range 0.002 $< x_{Bj}<0.1$ and $<Q^2>$ = 5 GeV$^2$ in the $x_{Bj}$ 
range 0.005 $< x_{Bj}<0.2$ were analysed.
To improve the measurement of $W_{vis}$ the kinematics of the tagged electron 
was included (the {\sl transverse momentum} conservation).
In the analysis three Monte Carlo generators: PHOJET 1.05c 
($p_T^{min}$ = 2.5 GeV), HERWIG 5.9 and TWOGAM ($p_T^{min}$ = 2.3 GeV) have been used.
They led to different results for $x_{Bj}$, $p_T$ distributions 
and  the energy flow (for more details see next section).

\begin{figure}[hc]
\vskip 0in\relax\noindent\hskip 3.cm
       \relax{\includegraphics{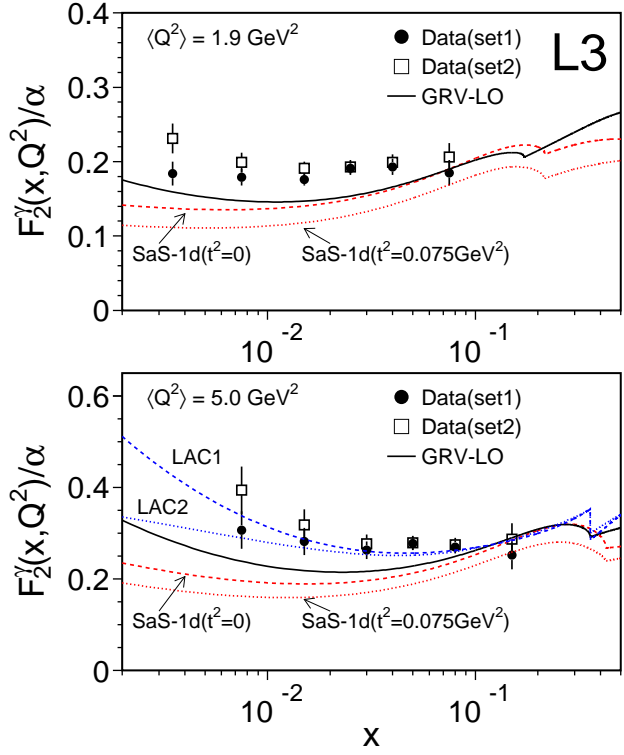}}
\vspace{0cm}
\caption{\small\sl Two sets of data for  $F_2^{\gamma}/\alpha$ 
versus $x_{Bj}$ for two values of $<Q^2>$: 1.9 and
5 GeV$^2$. Set 1 (2) corresponds to the unfolding based on 
PHOJET (TWOGAM). The solid line corresponds to GRV LO, dashed - 
SaS1D (for $P^2$=0, $P^2$ is denoted here as $t^2$) and dotted - SaS1D
($P^2$=$t^2$=0.075 GeV$^2$) parton distributions. 
For the $Q^2$=5 GeV$^2$ in addition the LAC1 and LAC2
predictions are shown (from \cite{l3_f2_98}).}
\label{fig:l3_f2_98_6}
\end{figure}

Two (!) data sets for $F_2^{\gamma}$ based on two Monte Carlo 
analyses: the PHOJET and TWOGAM, are given.
They differ up to 14\% at high $x_{Bj}$ and up to 28 \% at low $x_{Bj}$.
The results are presented in  
fig.~\ref{fig:l3_f2_98_6} and in table \ref{table7}
(the values of $F_2^{\gamma}/{\alpha}$
are given at the centre of the $x_{Bj}$ bin).

\begin{table}[hb]
\caption{}
\label{table7}
$$
\begin{array}{|r|r|c|c|}
\hline
\,\,<Q^2>~~\!\!&x_{Bj}~~~~~~&F_2^{\gamma}/{\alpha}(set 1)&
F_2^{\gamma}/{\alpha}(set 2)\\  
~[GeV^2]~~~&&(stat. + syst.)&(stat. + syst.)\\
\hline
~~~~1.9~~~~~&0.002-0.005&0.184\pm0.009 \pm 0.013&0.231\pm0.011 \pm 0.016\\
            &0.005-0.010&0.179\pm0.007 \pm 0.009&0.199\pm0.008 \pm 0.010\\
            &0.010-0.020&0.176\pm0.006 \pm 0.006&0.191\pm0.007 \pm 0.006\\
            &0.020-0.030&0.191\pm0.008 \pm 0.004&0.193\pm0.008 \pm 0.004\\
            &0.030-0.050&0.193\pm0.008 \pm 0.007&0.199\pm0.008 \pm 0.007\\
            &0.050-0.100&0.185\pm0.007 \pm 0.015&0.206\pm0.008 \pm 0.017\\
            
\hline
~~~5~~~~~   &0.005-0.010&0.307\pm0.021 \pm 0.035&0.394\pm0.027 \pm 0.045\\
            &0.010-0.020&0.282\pm0.014 \pm 0.027&0.318\pm0.016 \pm 0.031\\
            &0.020-0.040&0.263\pm0.011 \pm 0.015&0.277\pm0.012 \pm 0.016\\
            &0.040-0.060&0.278\pm0.013 \pm 0.007&0.279\pm0.013 \pm 0.007\\
            &0.060-0.100&0.270\pm0.012 \pm 0.008&0.275\pm0.012 \pm 0.008\\
            &0.100-0.200&0.252\pm0.011 \pm 0.029&0.287\pm0.013 \pm 0.032\\
\hline
\end{array} 
$$
\end{table}
\vspace*{9.4cm}

\begin{figure}[hc]
\vskip 0in\relax\noindent\hskip 3.cm
       \relax{\includegraphics{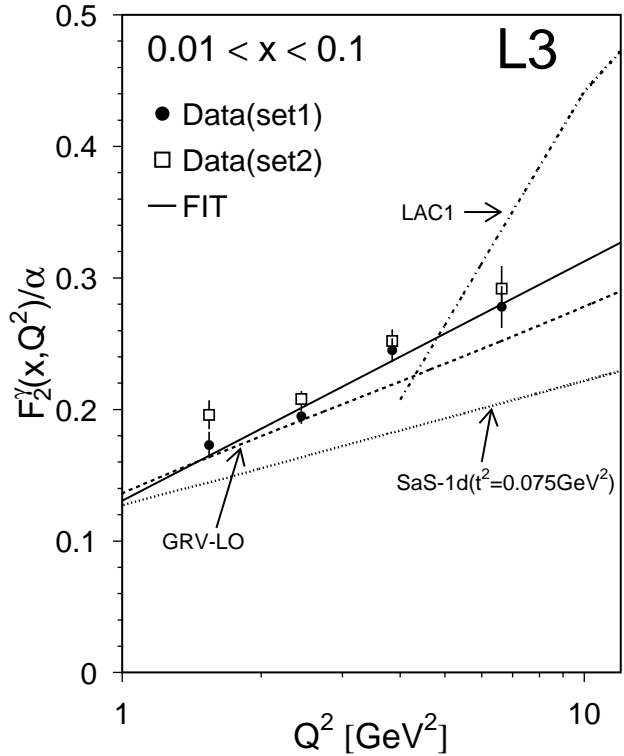}}
\vspace{0cm}
\caption{\small\sl Two sets of data for  $F_2^{\gamma}/\alpha$
 as a function of $Q^2$.
 Set 1 (2) corresponds to the unfolding based on 
PHOJET (TWOGAM). The solid line corresponds to the fit to set 1,
dashed line denotes the GRV LO parton parametrization prediction, dotted - 
SaS1D ($P^2$=0.075 GeV$^2$, $P^2$ is denoted here as $t^2$)
and dot-dashed corresponds to  LAC1  (LAC2 is 
similar to LAC1) (from \cite{l3_f2_98}).}
\label{fig:l3_f2_98_7}
\end{figure}

The $Q^2$ - dependence was also studied; results for $0.01<x_{Bj}<0.1$ 
are presented in table \ref{table8} (see also fig.~\ref{fig:l3_f2_98_7}),
for the two sets of data.
\begin{table}[ht]
\caption{}
\label{table8}
$$
\begin{array}{|r|c|c|}
\hline
\,\,<Q^2>~~\!\!&F_2^{\gamma}/{\alpha}(set 1)&F_2^{\gamma}/{\alpha}(set 2)\\  
~[GeV^2]~~~&(stat. + syst.)&(stat. + syst.)\\
\hline
~~~~1.5~~~~~&0.173\pm0.004 \pm 0.009&0.196\pm0.005 \pm 0.010\\
~~~~2.4~~~~~&0.195\pm0.005 \pm 0.004&0.208\pm0.005 \pm 0.004\\
~~~~3.8~~~~~&0.245\pm0.006 \pm 0.007&0.252\pm0.006 \pm 0.007\\
~~~~6.6~~~~~&0.278\pm0.009 \pm 0.013&0.292\pm0.009 \pm 0.014\\
\hline
\end{array} 
$$
\end{table}

The following $Q^2$ - dependence  was found for set1:
$$F_2^{\gamma}(Q^2)/\alpha=(0.131\pm0.012\pm0.021)+
(0.079\pm0.011\pm0.009)\ln(Q^2/{\rm (GeV)}^2.$$

\vspace*{9.2cm}

The $F_2^{\gamma}$ data are not corrected for the virtuality $P^2$ of the 
target photon since this correction is model dependent:
7.5\% for GVMD and 10-20\% for SaS1D.
As estimated by PHOJET and TWOGAM (QPM in TWOGAM),
$<P^2>=0.075$ (0.087) GeV$^2$ \cite{mariak-pryw}. 
\newline
\newline
Comment: {\it ``Because of the large discrepancies with the data, we
do not use HERWIG for the results.''
``At low values of $x_{Bj}$ the data are above 
the prediction of the GRV LO and SaS1D models (...). 
The LAC model can reproduce the $x_{Bj}$ behaviour of $F_2^{\gamma}$
 at $<Q^2>$=5 GeV$^2$ but it predicts too fast a rise of $F_2^{\gamma}$ 
as a function of $\ln Q^2$''. \\
Because of the significant differences in results for
 $F_2^{\gamma}$
obtained using different Monte Carlo 
for the first time  two different data sets
were 'derived' from the same sample of events.}

~\newline
$\bullet${\bf {L3 99a  \cite{l3_q2_98} (LEP 2)}}\\
Results of the measurements of 
 $F_2^{\gamma}$ at the CM energy 183 GeV  
in the interval  $9<Q^2<30$ GeV$^2$,  
 $x_{Bj}$ from 0.01 to 0.5 are presented (data collected in 1997).
Three Monte Carlo generators were used:  TWOGAM ($p_T^{min}$=3.5 GeV)
and PHOJET 1.05c ($p_T^{min}$=2.5 GeV) with JAMVG 
(the latter used for the charm quark contribution, 
as in PHOJET the charm quark is treated 
as massless). The $p_T^0$ cutoff in TWOGAM needed to
describe data is 3.5 GeV, since the $p_T^{min}$ = 2.3 GeV used in
previous analysis ({\bf L3 98a}) ``produces a too large cross section
for hard processes in the high $Q^2$ region \cite{chlin}''. The hadronic
final state is reasonably  described by the used Monte Carlo generators.
Comparison was made with the SaS1D, GRV LO, GRV HO and LAC1 predictions.
The values of $F_2^{\gamma}/\alpha$ are not corrected for $P^2\ne 0$.

\begin{figure}[hc]
\vskip 0in\relax\noindent\hskip 3.cm
       \relax{\includegraphics{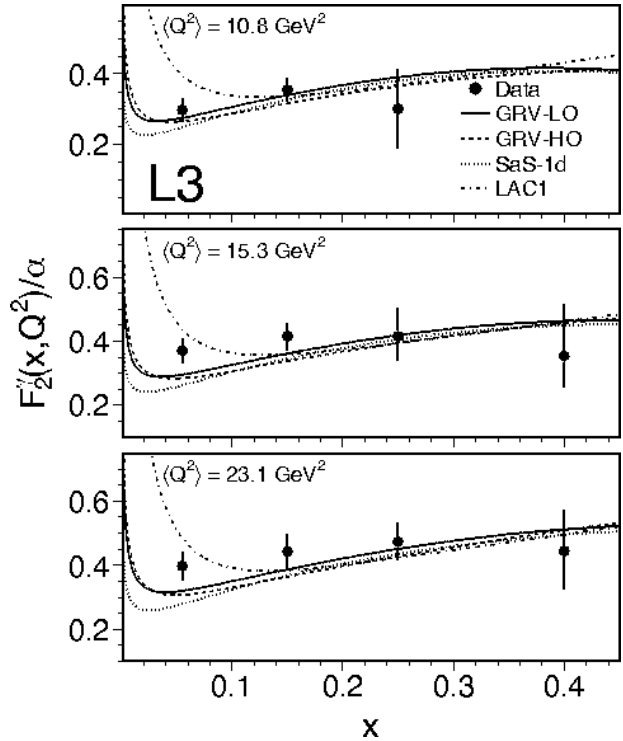}}
\vspace{0cm}
\caption{\small\sl $F_2^{\gamma}/\alpha$ as a function of $x_{Bj}$ 
for three $Q^2$ bins unfolded with PHOJET (plus JAMVG). The data are 
compared with predictions of the GRV LO, GRV HO, 
SaS1D and  LAC1 parton parametrizations (from \cite{l3_q2_98}).}
\label{fig:l3_q2_98_3}
\end{figure}

The results for $F_2^{\gamma}/\alpha$ obtained using PHOJET
are given in table \ref{table9} and in fig.~\ref{fig:l3_q2_98_3}.

\begin{table}[ht]
\caption{}
\label{table9}
$$
\begin{array}{|r|r|c|}
\hline
\,\,<Q^2>~~\!\!&x_{Bj}~~~~~~&F_2^{\gamma}/{\alpha}\\  
~[GeV^2]~~~&&(stat. + syst.)\\
\hline
~~~~10.8~~~~~&0.01-0.1&0.30\pm0.02 \pm 0.03\\
            &0.1-0.2&0.35\pm0.03 \pm 0.02\\
            &0.2-0.3&0.30\pm0.04 \pm 0.10\\
\hline
~~~15.3~~~~~&0.01-0.1&0.37\pm0.02 \pm 0.03\\
            &0.1-0.2&0.42\pm0.04 \pm 0.01\\
            &0.2-0.3&0.42\pm0.05 \pm 0.05\\
            &0.3-0.5&0.35\pm0.05 \pm 0.08\\
\hline            
~~~23.1~~~~~&0.01-0.1&0.40\pm0.03 \pm 0.03\\
            &0.1-0.2&0.44\pm0.04 \pm 0.04\\
            &0.2-0.3&0.47\pm0.05 \pm 0.02\\
            &0.3-0.5&0.44\pm0.05 \pm 0.11\\
\hline
\end{array} 
$$
\end{table}
\vspace*{9.4cm}

\begin{figure}[hc]
\vskip 0in\relax\noindent\hskip 3.cm
       \relax{\includegraphics{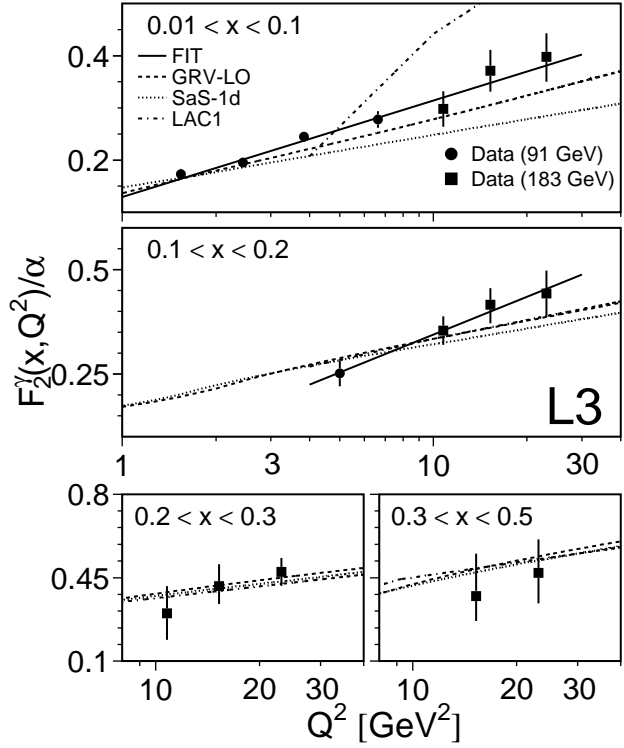}}
\vspace{0cm}
\caption{\small\sl $F_2^{\gamma}/\alpha$ as a function of $Q^2$ 
for different $x_{Bj}$ regions from LEP1 and LEP2 data
(the unfolding based on PHOJET with JAMVG). 
The solid line corresponds to the fit,
the GRV LO parton parametrization prediction
is denoted as a dashed line, SaS1D - dotted,  
and dot-dashed line corresponds to  the LAC1 prediction
(from \cite{l3_q2_98}).}
\label{fig:l3_q2_98_4}
\end{figure}

The $Q^2$ dependence of  $F_2^{\gamma}$ was studied as well, 
results are shown in fig.~\ref{fig:l3_q2_98_4}.\newline

\vspace*{9.5cm}

The $Q^2$ - dependence was found to be, for 0.01$<x_{Bj}<0.1$:
$$F_2^{\gamma}(Q^2)/\alpha=(0.13\pm0.01\pm0.02)+
(0.08\pm0.009\pm0.009)\ln(Q^2/{\rm GeV)}^2,$$
and for 0.1$<x_{Bj}<0.2$:
$$F_2^{\gamma}(Q^2)/\alpha=(0.04\pm0.008\pm0.008)+
(0.13\pm0.03\pm0.03)\ln(Q^2/{\rm GeV)}^2.$$

\noindent
Comment: {\it High $p_T^{min}$ cutoff (3.5 GeV) in TWOGAM is needed.\\
``The values of $F_2^{\gamma}/\alpha$
are not corrected for the fact that $P^2$ is not strictly equal to zero''.\\
``The agreement  between data and the different Monte Carlo predictions is good
except at pseudorapidity values, $\eta>$3 and $\eta<-1$'',
note however that these regions contribute negligibly, when $F_2^{\gamma}$
is extracted.

``At low $x_{Bj}$, the rise of the  $F_2^{\gamma}$ (with $Q^2$)
is larger than predicted
by the GRV and SaS1D models, thus requiring a modification 
of the gluon density in the photon structure function $F_2^{\gamma}$.''}

~\newline
$\bullet${\bf {L3 2000  \cite{l3_highq2} (LEP 1)}}\\
The measurement of the structure function for the real 
photon was performed based on  1991-1995 data
for large $Q^2$, between 40 to 500 GeV$^2$ with $<Q^2>$=120 GeV$^2$.  
The virtual photon structure function was studied as well, see sec. \ref{sec32}.

The $F_2^{\gamma}$ data from single-tag events were obtained using the JAMVG
generator modelling the QPM with $N_f$=4, the PHOJET 1.05c with a cutoff 
$p_T^{min}$=2.5 GeV and TWOGAM with $p_T^{min}$=3.5 GeV 
generating three processes: QPM, 
VMD and QCD resolved photon contributions.
The hadronic final state was investigated  (see next section for details),
and it is  properly described by JAMVG (QPM) model for $x_{Bj}>$0.5
and by PHOJET for smaller $x_{Bj}$.

\begin{figure}[hc]
\vskip 0in\relax\noindent\hskip 1.8cm
       \relax{\includegraphics{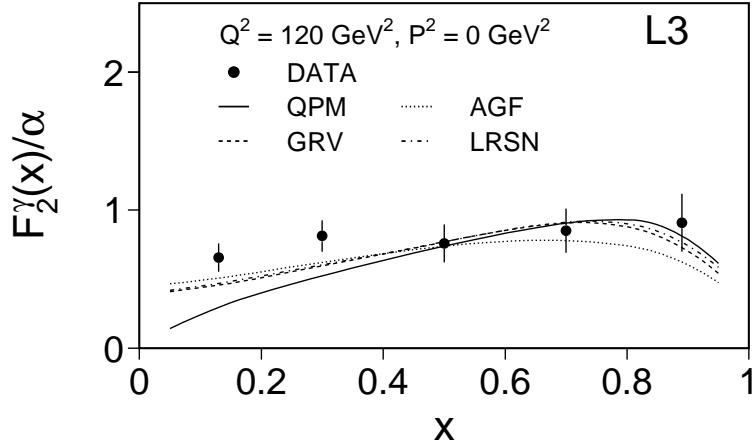}}
\vspace{-0.7cm}
\caption{\small\sl   $F_2^{\gamma}/\alpha$
 as a function of $x_{Bj}$ for $<Q^2>$=120 GeV$^2$
 from single-tag events.
The solid line corresponds to the QPM prediction, 
the GRV LO parton parametrization  prediction
is denoted as a dashed line, AGF - dotted  
line, and dot-dashed line corresponds to the LRSN prediction
(from \cite{l3_highq2}).}
\label{fig:l3_highq2_6a}
\end{figure}

The $F_2^{\gamma}$ data are presented in table \ref{table_l3high} 
and fig.~\ref{fig:l3_highq2_6a},
where the results are compared with the predictions of QPM and 
of QCD calculations, the latter being needed
to describe the $x_{Bj}$ region  below 0.5.  

\begin{table}[ht]
\caption{}
\label{table_l3high}
$$
\begin{array}{|r|r|c|}
\hline
\,\,<Q^2>~~\!\!&x_{Bj}~~~~~~&F_2^{\gamma}/{\alpha}\\  
~[GeV^2]~~~&&(stat. + syst.)\\
\hline
~~~120~~~~~&0.05-0.20&0.66\pm0.08\pm0.06\\
            &0.20-0.40&0.81\pm0.08\pm0.08\\
            &0.40-0.60&0.76\pm0.12\pm0.07\\
            &0.60-0.80&0.85\pm0.14\pm0.08\\
            &0.80-0.98&0.91\pm0.19\pm0.09\\
\hline
\end{array} 
$$
\end{table}
\vspace*{6.5cm}

\begin{figure}[hc]
\vskip 0in\relax\noindent\hskip 1.9cm
       \relax{\includegraphics{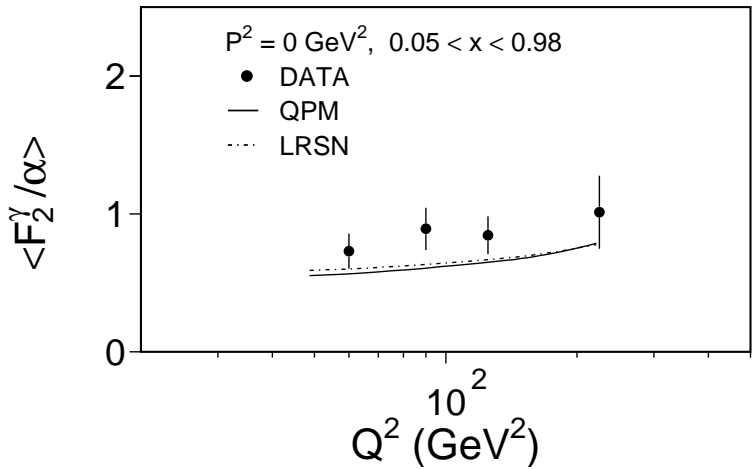}}
\vspace{-0.7cm}
\caption{\small\sl The $Q^2$ dependence of   $F_2^{\gamma}/\alpha$
 averaged over $x_{Bj}=0.05-0.98$
for single-tag data. Comparison with the QPM prediction (solid line)
 and the  LRSN NLO calculation (dashed line)
(from \cite{l3_highq2}).}
\label{fig:l3_highq2_8a}
\end{figure}

The averaged values of $<F_2^{\gamma}/\alpha>$ as a function of $<Q^2>$
are presented in table \ref{table_l3_highq2_av} for two $x_{Bj}$
ranges and in fig.~\ref{fig:l3_highq2_8a}.

\begin{table}[ht]
\caption{}
\label{table_l3_highq2_av}
$$
\begin{array}{|c|c|c|c|}
\hline
<Q^2>&<F_2^{\gamma}/\alpha>&<F_2^{\gamma}/\alpha>\\
~[GeV^2]~& (stat.+syst.)&(stat.+syst.)\\
&x_{Bj}=0.05-0.98&x_{Bj}=0.3-0.8\\
\hline
60&0.73\pm0.11\pm0.07&0.66\pm0.09\pm0.06\\
90&0.89\pm0.13\pm0.09&0.79\pm0.14\pm0.08\\
125&0.85\pm0.11\pm0.09&0.88\pm0.12\pm0.08\\
225&1.01\pm0.25\pm0.10&1.18\pm0.22\pm0.11\\
\hline
\end{array}
$$
\end{table}
\vspace*{6.5cm}

The virtuality of the photon is estimated to be $P^2$=0.014 GeV$^2$.

~\newline
Comment: {\it ``The structure function $F_2^{\gamma}(x)$ of real 
photons shows an excess at low $x_{Bj}$ over QPM and over several 
QCD calculations.''}

~\newline
$\bullet${\bf {OPAL 94 \cite{opal94} (LEP 1)}}\\
The measurement of $F_2^{\gamma}$ at $<Q^2>$=5.9 and 14.7 GeV$^2$
was performed using data from the period 1990-92.
The new Monte Carlo TWOGEN program was used to generate
events according to chosen formula for $F_2^{\gamma}(x, Q^2, P^2)$
or $F_2^{\gamma}(x, Q^2)$. 
The VMD with two different distributions of $q\bar{q}$ pair
(peripheral and QPM - type) and pointlike QCD contributions 
for light quarks (in the FKP approach) were studied.
The cutoff parameter $p_T^0$ separating these contributions 
was determined to be 0.27$\pm 0.10$ GeV for both $Q^2$ values.
The Monte Carlo model with this parameter value was then used 
to unfold $F_2^{\gamma}$. See also sec. \ref{sec23}.
 
\begin{figure}[ht]
\vskip 0.in\relax\noindent\hskip 1.45cm
       \relax{\includegraphics{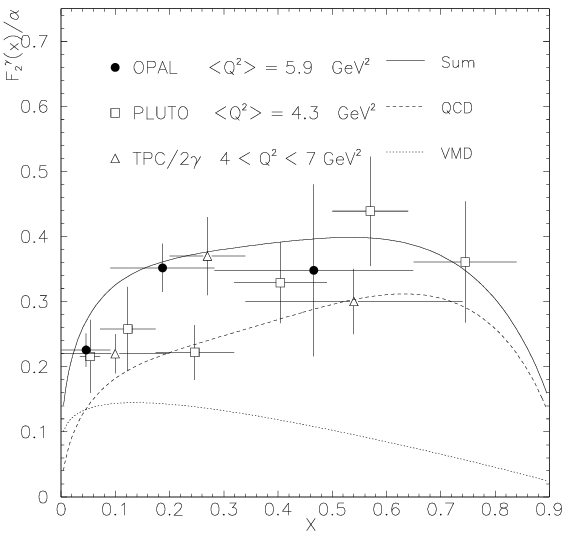}}
\vskip -0.2in\relax\noindent\hskip 7.85cm       
       \relax{\includegraphics{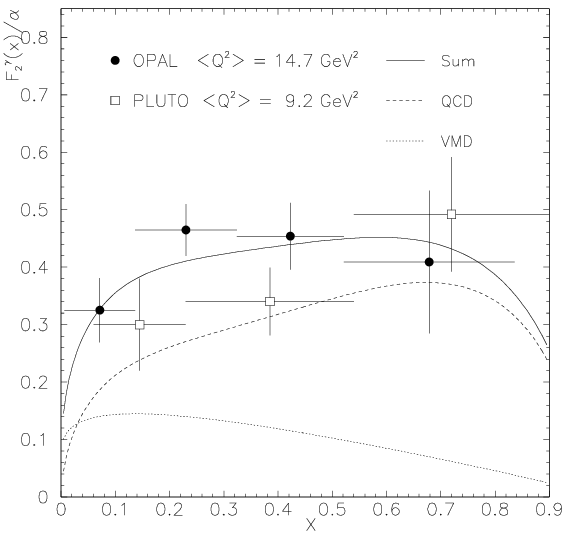}}
\vspace{-0.4cm}
\caption{\small\sl The results for unfolded charm subtracted
$F_2^{\gamma}/\alpha$ shown and compared with 
other experiments at similar $<Q^2>$ values. 
The curves show contributions 
of (left) VMD (dots), QCD-based model (dashes) and their sum (line) for 
 $<Q^2>$=5.9 GeV$^2$, (right)$<Q^2>$=14.7 GeV$^2$ 
(from \cite{opal94}).}
\label{fig:opal94a}
\end{figure}

Results  for unfolded $F_2^{\gamma}$,
with the charm contribution calculated using QPM, compared with 
the PLUTO and TPC/2$\gamma$
data are presented in fig.~\ref{fig:opal94a}. 
\newline

\vspace*{5.5cm}

~\newline
Comment: {\it In the analysis a simple formula
$F_2^{\gamma}/\alpha = 0.2 (1-x)$ was applied to describe the VMD contribution.
The TWOGEN with a QPM formula for $F_2^{\gamma}(x, Q^2, P^2)$
was used to obtain the correction for the $P^2\ne 0$. \newline
``No increase of $F_2^{\gamma}(x)$ is observed'' at small $x_{Bj}$.}

~\newline
$\bullet${\bf {OPAL 97a \cite{opal2} (LEP 1) }}\\
The measurement of $F_2^{\gamma}$ was done for two samples,
$6<Q^2<30$ GeV$^2$ and  $60<Q^2<400$~GeV$^2$, using all data 
at the $Z$ peak (years 1990-95) with new type of unfolding
as compared to {\bf OPAL 94}. The detailed 
analysis of the hadronic final states was performed and sizeable 
discrepancies with the expectations 
 were found especially at low $x_{Bj}$ (see also next section).
The influence of the choice of different Monte Carlo generators 
(HERWIG 5.8d, PYTHIA 5.718, also for comparisons and systematic
checks the F2GEN) on the unfolded $F_2^{\gamma}$
was studied. The dependence of $F_2^{\gamma}$ on the $P^2$ values 
was estimated based on the SaS1D parametrization.

\begin{figure}[ht]
\vskip 0.in\relax\noindent\hskip 0.in
       \relax{\includegraphics{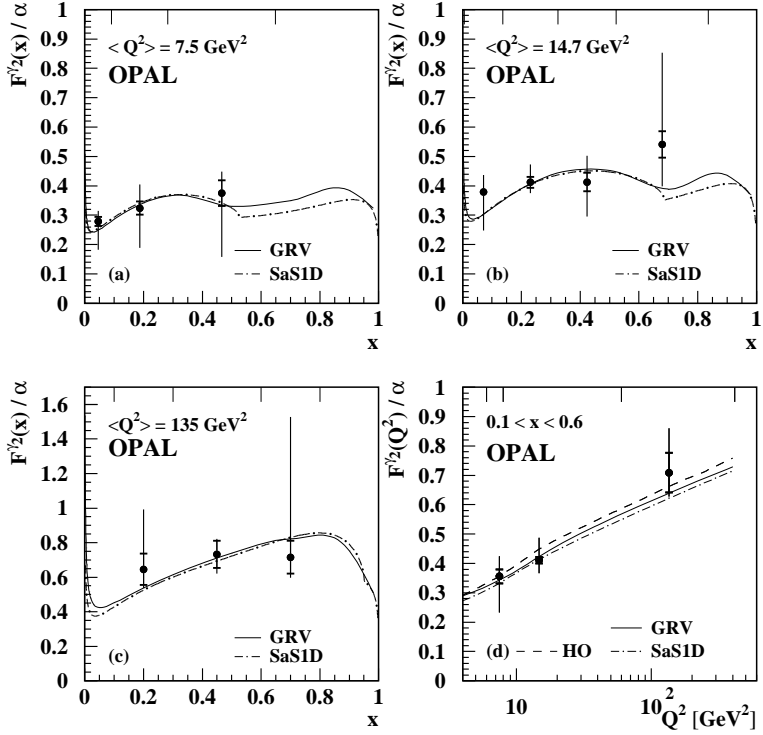}}
\vspace{0.ex}
\caption{\small\sl The $F_2^{\gamma}/\alpha$ data from the OPAL experiment for
the number of flavours $N_f$=4. Curves in (a)-(d) show predictions
of the GRV and SaS1D parametrizations (from \cite{opal2}).}
\label{fig:scan1}
\end{figure}

The results for  unfolded $F_2^{\gamma}$ are presented in  
fig.~\ref{fig:scan1}a,b,c and in table \ref{table10}
(the value of $F_2^{\gamma}/{\alpha}$
is given at the centre of the $x_{Bj}$ bin).

\begin{table}[ht]
\caption{}
\label{table10}
$$
\begin{array}{|r|r|c|}
\hline
\,\,<Q^2>~~\!\!&x_{Bj}~~~~~~&F_2^{\gamma}/{\alpha}\\  
~[GeV^2]~~~&&(stat. + syst.)\\
\hline
~~~~7.5~~~~~&0.001-0.091&0.28\pm0.02^{+0.03}_{-0.10}\\
            &0.091-0.283&0.32\pm0.02^{+0.08}_{-0.13}\\
            &0.283-0.649&0.38\pm0.04^{+0.06}_{-0.21}\\
\hline
~~~14.7~~~~~&0.006-0.137&0.38\pm0.01^{+0.06}_{-0.13}\\
            &0.137-0.324&0.41\pm0.02^{+0.06}_{-0.03}\\
            &0.324-0.522&0.41\pm0.03^{+0.08}_{-0.11}\\
            &0.522-0.836&0.54\pm0.05^{+0.31}_{-0.13}\\
\hline
~~135.0~~~~~&0.100-0.300&0.65\pm0.09^{+0.33}_{-0.06}\\
            &0.300-0.600&0.73\pm0.08^{+0.04}_{-0.08}\\
            &0.600-0.800&0.72\pm0.10^{+0.81}_{-0.07}\\
\hline
\end{array} 
$$
\end{table}
\vspace*{8.4cm}

The measurement of $F_2^{\gamma}/{\alpha}$ as a function of $Q^2$
averaged over the $x_{Bj}$ range, 
0.1$<x_{Bj}<$0.6, gives
results shown in fig.~\ref{fig:scan1}d and in table \ref{table11}.
\begin{table}[ht]
\caption{}
\label{table11}
$$
\begin{array}{|r|c|}
\hline
\,\,<Q^2>~~\!\!&<F_2^{\gamma}/{\alpha}>\\ 
~[GeV^2]~~~&(stat. + syst.)\\
\hline
~~~~7.5~~~~~ &0.36\pm0.02^{+0.06}_{-0.12}\\
~~~14.7~~~~~&0.41\pm0.01^{+0.08}_{-0.04}\\
~~135.0~~~~~&0.71\pm0.07^{+0.14}_{-0.05}\\
\hline
\end{array} 
$$
\end{table}

The slope $d(F_2^{\gamma}/{\alpha})/dlnQ^2$ is measured to be 
$0.13^{+0.06}_{-0.04}$.
~\newline\newline
Comment: {\it New measurements of $F_2^{\gamma}(x,Q^2)$ are presented,
``allowing for the first time for uncertainties in the description of 
the final state by different Monte Carlo models''.\\
The $F_2^{\gamma}$ is not correced for $P^2$.
Large discrepancies between the hadronic energy flow
data and Monte Carlo simulations are observed at low $Q^2$ for
x$_{vis} < 0.1$,
when the results are presented versus pseudorapidity or 
azimuthal angle (see next section).}
\newline\newline
$\bullet${\bf{OPAL 97b \cite{bech} (LEP 1)}}
~\newline
The measurement of $F_2^{\gamma}$ (years 1993-94)
was done for $1.1<Q^2<2.5$ GeV$^2$
  and $2.5<Q^2<6.6$ GeV$^2$ (average $Q^2$=1.86 and 3.76 GeV$^2$)
 as a function of $x_{Bj}$, reaching
the lowest,  at that time, measured (center of bin on a logarithmic scale)
 value: $x_{Bj}$=0.004. 
For a better sensitivity on the  low $x_{Bj}$ 
region the unfolding procedure was performed
on a logarithmic scale. Final state topology
was analysed as well (using the HERWIG 5.18d, PYTHIA 5.722, F2GEN 
- both with the pointlike and the peripheral $q\bar{q}$ distributions generators).
The GRV LO and SaS1D parton parametrizations were used in the 
analysis. Charm contribution is treated differently in different
MC programs.
Discrepancy between the data (hadron energy flow)
and results from the Monte Carlo generators, as well as
between different models  was found for low $x_{vis}<$0.05 
(see sec. \ref{sec23}).

\begin{figure}[ht]
\vskip 0.in\relax\noindent\hskip 1.2cm
       \relax{\includegraphics{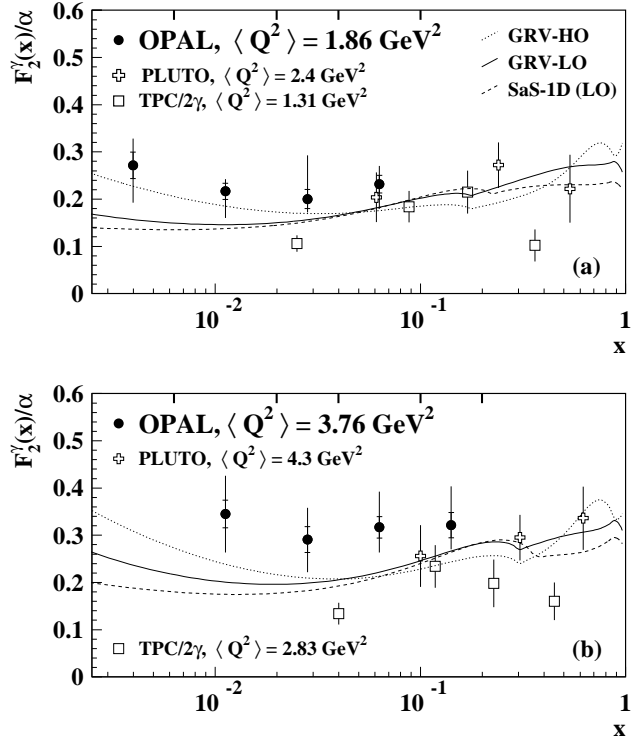}}
\vspace{0.1cm}
\caption{\small\sl The OPAL $F_2^{\gamma}/{\alpha}$ data
(circles) as a function of $x_{Bj}$ for 
 $<Q^2>$= 1.86 GeV$^2$ (a) and $<Q^2>$=
3.76 GeV$^2$ (b). Also data from 
PLUTO (crosses) for $<Q^2>$= 2.4, 4.3 GeV$^2$ 
and TPC/2$\gamma$ (squares) for $<Q^2>$=1.31, 
2.83 GeV$^2$ are shown. The curves show predictions of
the GRV HO (dots), GRV LO (line), and SaS1D (dashed) parametrizations.
The range of the $x_{Bj}$-bins of the OPAL results are marked at the tops of 
figures (from \cite{bech}).}
\label{fig:bech2}
\end{figure}

The obtained values of $F_2^{\gamma}/{\alpha}$  
are given in table \ref{table12} and shown in 
fig.~\ref{fig:bech2} together with early measurements from 
PLUTO and TPC/2$\gamma$. 
\begin{table}[ht]
\caption{}
\label{table12}
$$
\begin{array}{|r|r|c|}
\hline
\,\,<Q^2>~~\!\!&x_{Bj}~~~~~~&F_2^{\gamma}/{\alpha}\\  
~[GeV^2]~~~&&(stat. + syst.)\\
\hline
~~~~1.86~~~~~&0.0025-0.0063&0.27\pm0.03^{ +0.05}_{-0.07}\\
            &0.0063-0.020&0.22\pm0.02^{ +0.02}_{-0.05}\\
            &0.020-0.040&0.20\pm 0.02^{ +0.09}_{-0.02}\\
            &0.040-0.100&0.23\pm 0.02^{ +0.03}_{-0.05}\\
\hline
~~~~3.76~~~~~&0.0063-0.020&0.35\pm0.03^{ +0.08}_{-0.08}\\
            &0.020-0.040&0.29\pm0.03^{ +0.06}_{-0.06}\\
            &0.040-0.100&0.32\pm 0.02^{ +0.07}_{-0.05}\\
            &0.100-0.200&0.32\pm 0.03^{ +0.08}_{-0.04}\\
\hline
\end{array} 
$$
\end{table}
\vspace*{9.cm}
~\newline
Comment: {\it No correction for $P^2\neq$0 was made. GRV HO is consistent 
with the low-x OPAL results in the lower 
$Q^2$ bin, but at  higher $Q^2$
it underestimates the low x OPAL
data. GRV LO and SaS1D describe the unfolded
results worse than GRV HO. Shapes of measured $F_2^{\gamma}$
are flat within the errors, but a small rise in the low x region is 
not excluded.}

~\newline
$\bullet${\bf{OPAL 97c \cite{LP291} (LEP 2)}}
\newline
New data on $F_2^{\gamma}$
from LEP 2 at the CM energies 161-172 GeV were
collected in 1996 in two samples $Q^2$=6-20 GeV$^2$ ($0.004<x_{Bj}<0.76$) 
and  $Q^2$=20-100 GeV$^2$ ($0.012<x_{Bj}<0.94$). 
Also the  distribution
of the final hadronic energy flow was studied using
HERWIG 5.9, PYTHIA 5.718 and F2GEN, see next section.
The LEP1 {\bf OPAL 97a} data were used for comparison.
For unfolding the Monte Carlo 
 based on HERWIG program with the GRV parton parametrization was used.
Two types of binning were performed, with $<Q^2>$ = 9, 14.5, 30 and 59 GeV$^2$,
and with $<Q^2>$ = 11 (combined samples 9 and 14.5 GeV$^2$)
and 41 GeV$^2$ (30 and 59 GeV$^2$).

\begin{figure}[ht]
\vskip 0.in\relax\noindent\hskip 2.5cm
       \relax{\includegraphics{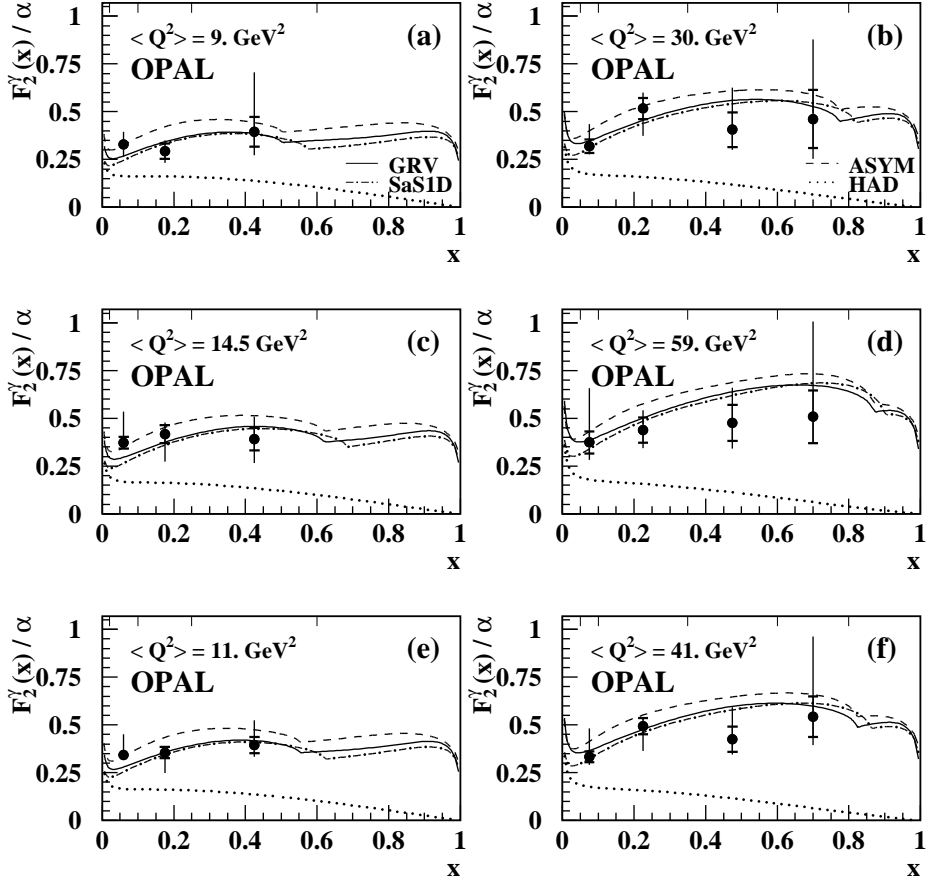}}
\vspace{0.ex}
\caption{\small\sl The $x_{Bj}$-dependence
of $F_2^{\gamma}/{\alpha}$ for different values of
$<Q^2>$: 9, 30, 14.5, 59, 11, and 41 GeV$^2$. For comparison
the GRV LO (solid line) and SaS1D (dot-dashed line)
parametrizations are shown;
dotted line represents the hadronic component 
and the dashed one the (augmented) asymptotic solution (ASYM)  
(from \cite{LP291}).}
\label{fig:nisius1}
\end{figure}

\begin{figure}[hb]
\vskip 0.cm\relax\noindent\hskip 0.5cm
       \relax{\includegraphics{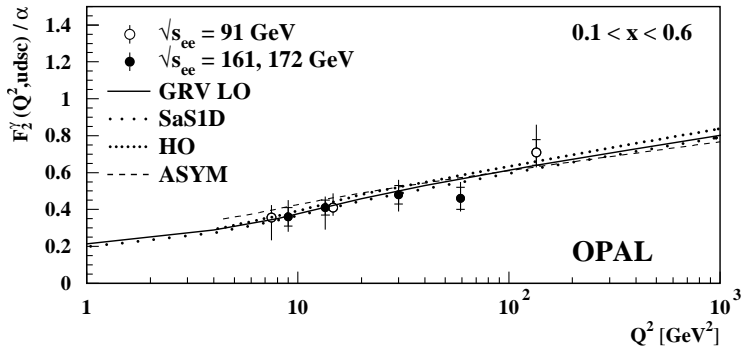}}
\vspace{0.1cm}
\caption{\small\sl 
The $Q^2$- dependence of $F_2^{\gamma}/{\alpha}$  
averaged over 0.1$<x_{Bj}<$0.6, 
for the energy 91 GeV (open circles) and  for
the energies 161, 172 GeV (full circles). Predictions of the QCD
calculation are shown by the lines: 
solid (GRV LO), dotted (SaS1D), and double-dotted
(HO, based on the GRV HO parametrization for light quarks);  dashed
line corresponds to the (augmented) asymptotic solution (ASYM)
(from \cite{LP291}).} 
\label{fig:nisius2}
\end{figure}

The unfolded $F_2^{\gamma}/{\alpha}$ as a function
of $x_{Bj}$ and $Q^2$ are presented in  
figs.~\ref{fig:nisius1} and \ref{fig:nisius2} and in tables 
\ref{table13}, \ref{table14}.

\begin{table}[ht]
\caption{}
\label{table13}
$$
\begin{array}{|r|r|c|}
\hline
\,\,<Q^2>~~\!\!&x_{Bj}~~~~~~&F_2^{\gamma}/{\alpha}\\  
~[GeV^2]~~~&&(stat. + syst.)\\
\hline
~~~~9~~~~~&0.02-0.1&0.33\pm0.03^{ +0.06}_{-0.06}\\
            &0.100-0.250&0.29\pm0.04^{ +0.04}_{-0.05}\\
            &0.250-0.600&0.39\pm 0.08^{ +0.30}_{-0.10}\\
\hline
~~~~14.5~~~~~&0.02-0.1&0.37\pm0.03^{ +0.16}_{-0.01}\\
            &0.100-0.250&0.42\pm0.05^{ +0.04}_{-0.14}\\
            &0.250-0.600&0.39\pm 0.06^{ +0.10}_{-0.11}\\
\hline
~~~~30~~~~~&0.050-0.100&0.32\pm0.04^{ +0.11}_{-0.02}\\
            &0.100-0.350&0.52\pm0.05^{ +0.06}_{-0.13}\\
            &0.350-0.600&0.41\pm 0.09^{ +0.20}_{-0.05}\\
            &0.600-0.800&0.46\pm 0.15^{ +0.39}_{-0.14}\\
\hline
~~~~59~~~~~&0.050-0.100&0.37\pm0.06^{ +0.28}_{-0.07}\\
            &0.100-0.350&0.44\pm0.07^{ +0.08}_{-0.07}\\
            &0.350-0.600&0.48\pm 0.09^{ +0.16}_{-0.10}\\
            &0.600-0.800&0.51\pm 0.14^{ +0.48}_{-0.02}\\
\hline
\end{array} 
$$
\end{table}
\begin{table}[ht]
\caption{}
\label{table14}
$$
\begin{array}{|r|c|}
\hline
\,\,<Q^2>~~\!\!&<F_2^{\gamma}/{\alpha}>\\ 
~[GeV^2]~~~&(stat. + syst.)\\
\hline
~~~~9~~~~~ &0.36\pm0.05^{+0.08}_{-0.06}\\
~~~14.5~~~~~&0.41\pm0.04^{+0.04}_{-0.11}\\
~~~30~~~~~&0.48\pm0.05^{+0.06}_{-0.07}\\
~~~59~~~~~&0.46\pm0.06^{+0.07}_{-0.04}\\
\hline
\end{array} 
$$
\end{table}

\vspace*{3cm}

\newpage

\vspace*{11.cm}
\vspace*{4.1cm}

A fit to the new $F_2^{\gamma}$ data at the energies 161-172 GeV and the 
previous {\bf OPAL 97a} set at 91 GeV for $Q^2$ from 7.5 to 135
GeV$^2$, averaged over the $x_{Bj}$ range of 0.1-0.6
(see fig.~\ref{fig:nisius2}), has the form:
\begin{eqnarray}
\nonumber
F_2^{\gamma}(Q^2)/\alpha =(0.16\pm 0.05^{+0.17}_{-0.16})
+(0.10\pm 0.02^{+0.05}_{-0.02})\ln (Q^2/\rm {GeV^2}).
\end{eqnarray}

\begin{figure}[ht]
\vskip 0.cm\relax\noindent\hskip 1.cm
       \relax{\includegraphics{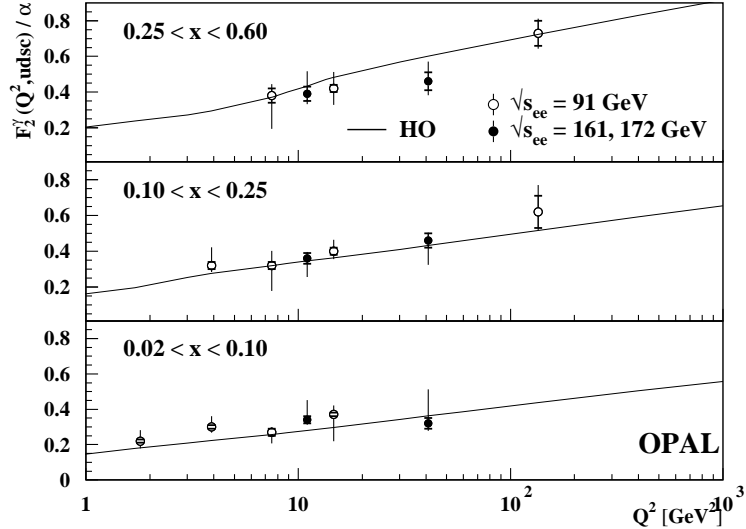}}
\vspace{0.1cm}
\caption{\small\sl The $Q^2$ dependence of $F_2^{\gamma}/{\alpha}$ 
in three $x_{Bj}$ bins. The data  are 
compared to HO predictions based on the GRV HO parametrization 
(from \cite{LP291}).}
\label{fig:LP2914b}
\end{figure}

A special study of the $Q^2$ dependence of $F_2^{\gamma}$
is performed for different $x_{Bj}$ ranges, see 
fig.~\ref{fig:LP2914b}.
\newline

\vspace*{6.1cm}
~\newline
Comment: {\it No correction for P$^2\neq$0 was made. Discrepancies
are observed in the hadronic energy flow between the data and
the HERWIG and PYTHIA simulations, especially
at x$_{vis}<$0.1. Accuracy of the data does not allow to see
the expected  different slope of $F_2^{\gamma}$
versus $ln$Q$^2$ for different x$_{Bj}$ ranges.}

~\newline
$\bullet${\bf{OPAL 2000 \cite{OPAL_PN389} (LEP 1, LEP 2)}}
\newline
The low $x_{Bj}$ data of $F_2^{\gamma}$ obtained at the CM 
energies 91 GeV, 183 and 189 GeV (in 1993-5 and 1997-8)
were collected in the partly overlapping $Q^2$ ranges: 1.5-30 GeV$^2$
at LEP1 and 7-30 GeV$^2$ at LEP2. Three samples:
LEP1 SW ($<Q^2>$ = 1.9 and 3.7 GeV$^2$), 
LEP1 FD ($<Q^2>$ = 8.9 and 17.5 GeV$^2$)
and LEP2 SW ($<Q^2>$ = 10.7 and 17.8 GeV$^2$) were studied.
In the analysis new Monte Carlo programs and improved unfolding methods
were introduced. The final state was studied
using PHOJET 1.05, HERWIG 5.9, ``HERWIG 5.9 + power law $p_t$ (dyn)''
and F2GEN, with the GRV LO parametrization as an input function,
see next section for details.

``HERWIG 5.9 + power law $p_t$ (dyn)'' and  PHOJET 1.05,
giving better description of OPAL data, were applied to obtain
the $F_2^{\gamma}$ structure function.
To reduce dependence of results on the implemented MC model, two-dimensional
unfolding was performed with $x_{cor}$ used as a first variable
and $E_T^{out}/E_{tot}$ as a second one.  
To investigate the low-$x_{Bj}$ range, the unfolding of the
$x_{Bj}$ variable was done on a logarithmic scale. 

The values of $F_2^{\gamma}/{\alpha}$ were obtained as an average
of results of the unfolding with 
``HERWIG 5.9 + power law $p_t$ (dyn)'' and PHOJET 1.05c,
see table \ref{table_OPAL_PN389} ($x_{Bj}^*$ corresponds to the log
centre of the bin for $x_{Bj}$ below the charm threshold.
The $F_2^{\gamma}/{\alpha}$ was unfolded in the quoted bins and 
corrected to the $x_{Bj}^*$ values).

\begin{figure}[ht]
\vskip 0.in\relax\noindent\hskip 2.5cm
       \relax{\includegraphics{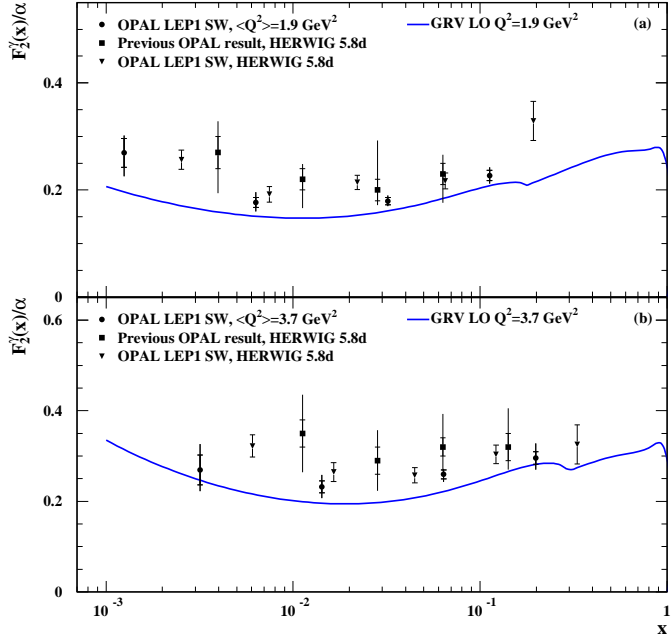}}
\vspace{0.ex}
\caption{\small\sl The $x_{Bj}$-dependence
of $F_2^{\gamma}/{\alpha}$ measured by OPAL for the LEP1 SW sample with 
$<Q^2>$ equal to a) 1.9 GeV$^2$ and b) 3.7 GeV$^2$. For comparison
the previous OPAL data ({\bf OPAL 97b}) and the present sample, both
obtained using HERWIG 5.8d, are  presented.
The GRV LO predictions are also shown
(from \cite{OPAL_PN389}).}
\label{fig:OPAL_PN389_13}
\end{figure}

\begin{figure}[hb]
\vskip 0.cm\relax\noindent\hskip 2.5cm
       \relax{\includegraphics{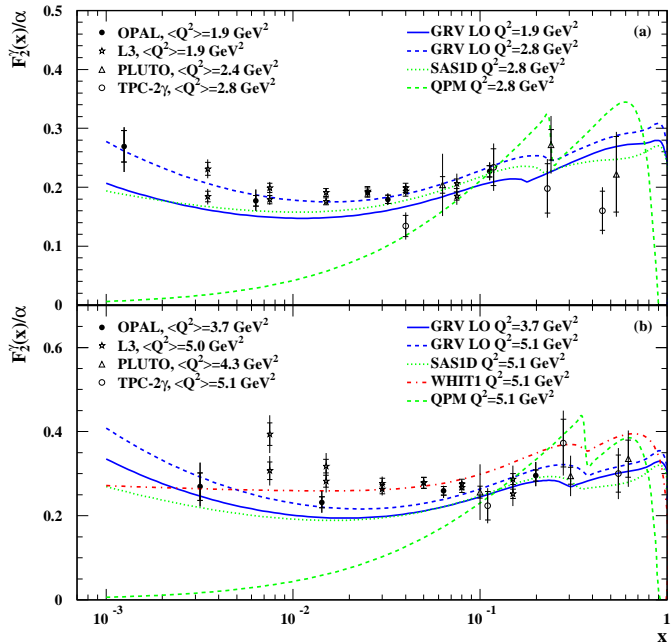}}
\vspace{0cm}
\caption{\small\sl The $x_{Bj}$-dependence
of $F_2^{\gamma}/{\alpha}$ measured by OPAL for the LEP1 SW sample with
$<Q^2>$ equal to a) 1.9 and b) 3.7  GeV$^2$. For comparison
the ({\bf L3 98a}), PLUTO \cite{early} and TPC/2$\gamma$ \cite{early} 
data are shown together with GRV LO, SaS1D and QPM 
parton parametrizations predictions (from \cite{OPAL_PN389}).} 
\label{fig:OPAL_PN389_15}
\end{figure}

The unfolded $F_2^{\gamma}/{\alpha}$ as a function of $x_{Bj}$ 
for $Q^2$=1.9 and 3.7 GeV$^2$ is slightly lower than the previous 
OPAL results obtained using HERWIG 5.8d
({\bf OPAL 97b}), see fig.~\ref{fig:OPAL_PN389_13}.
In fig. \ref{fig:OPAL_PN389_15} the comparison with other measurements 
and parton parametrizations predictions for $Q^2$ range 1.9-5 GeV$^2$ is shown.

\begin{figure}[hb]
\vskip 0.cm\relax\noindent\hskip 2.5cm
       \relax{\includegraphics{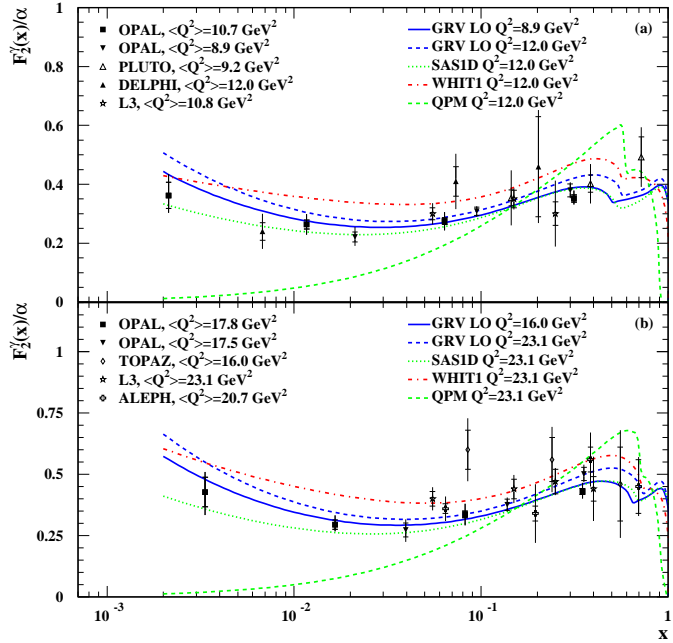}}
\vspace{0cm}
\caption{\small\sl The $x_{Bj}$-dependence
of $F_2^{\gamma}/{\alpha}$ measured by OPAL for 
$<Q^2>$ equal to a) 8.9 (LEP1 FD) and 10.7 GeV$^2$ (LEP2 SW), 
and b) 17.5 (LEP1 FD) and 17.8 (LEP2 SW) GeV$^2$. 
For comparison the {\bf ALEPH 99a}, {\bf DELPHI 96a}, {\bf L3 99a}, 
PLUTO \cite{early} and {\bf TOPAZ 94} data are shown with
predictions of GRV LO, SaS1D, WHIT1 and QPM (from \cite{OPAL_PN389}).}
\label{fig:OPAL_PN389_16}
\end{figure}

The results for the unfolded $F_2^{\gamma}$ at 
$<Q^2>$ = 8.9, 10.7, 17.5 and 17.8 GeV$^2$ (similar to the previous
ones, {\bf OPAL 97b}; comparison not shown) are compared in fig. 
\ref{fig:OPAL_PN389_16} with 
results of other experiments and predictions of various parametrizations,
in similar $Q^2$ range.

\begin{table}[ht]
\caption{}
\label{table_OPAL_PN389}
$$
\begin{array}{|c|c|c|c|}
\hline
<Q^2>&x_{Bj}^{}&x_{Bj}^*&F_2^{\gamma}/{\alpha}\\  
~[GeV^2]& &(log~centre &(stat. + syst.)\\
        & &of~bin) &\\
\hline
1.9&0.0006-0.0028&0.0012&0.269\pm 0.027^{+0.018}_{-0.034}\\
   &0.0028-0.0143&0.0063&0.177\pm 0.009^{+0.017}_{-0.014}\\
   &0.0143-0.0724&0.0322&0.179\pm 0.007^{+0.007}_{-0.006}\\
   &0.0724-0.3679&0.1124&0.227\pm 0.010^{+0.012}_{-0.012}\\
\hline
3.7&0.0015-0.0067&0.0032&0.269\pm 0.033^{+0.047}_{-0.033}\\
   &0.0067-0.0302&0.0143&0.232\pm 0.013^{+0.023}_{-0.021}\\
   &0.0302-0.1353&0.0639&0.259\pm 0.010^{+0.006}_{-0.013}\\
   &0.1353-0.6065&0.1986&0.296\pm 0.014^{+0.029}_{-0.022}\\
\hline
8.9&0.0111-0.0498&0.0235&0.221\pm 0.017^{+0.030}_{-0.026}\\
   &0.0498-0.2231&0.1054&0.308\pm 0.014^{+0.011}_{-0.012}\\
   &0.2231-0.8187&0.3331&0.379\pm 0.022^{+0.017}_{-0.015}\\
\hline
10.7&0.0009-0.0050&0.0021&0.362\pm 0.045^{+0.058}_{-0.039}\\
    &0.0050-0.0273&0.0117&0.263\pm 0.015^{+0.032}_{-0.030}\\
    &0.0273-0.1496&0.0639&0.275\pm 0.011^{+0.029}_{-0.030}\\
    &0.1496-0.8187&0.3143&0.351\pm 0.012^{+0.025}_{-0.016}\\
\hline
17.5&0.0235-0.0821&0.0439&0.273\pm 0.028^{+0.032}_{-0.039}\\
    &0.0821-0.2865&0.1534&0.375\pm 0.023^{+0.020}_{-0.013}\\
    &0.2865-0.9048&0.3945&0.501\pm 0.027^{+0.027}_{-0.019}\\
\hline
17.8&0.0015-0.0074&0.0033&0.428\pm 0.061^{+0.055}_{-0.071}\\
    &0.0074-0.0369&0.0166&0.295\pm 0.019^{+0.033}_{-0.020}\\
    &0.0369-0.1827&0.0821&0.336\pm 0.013^{+0.041}_{-0.042}\\
    &0.1827-0.9048&0.3483&0.430\pm 0.013^{+0.032}_{-0.025}\\
\hline
\end{array} 
$$
\end{table}
\newpage
\vspace*{8.cm}

\vspace*{8.5cm}
\newpage
\vspace*{8cm}

~\newline
Comment: {\it  No correction for $P^2\neq$0 was made.
The $F_2^{\gamma}$ measurement at ``the lowest attainable $x_{Bj}$ 
values'', uses new MC models and
two-dimensional unfolding, on a logarithmic scale.\\
``The GRV LO and SaS1D parametrizations are generally consistent 
with the OPAL data in all the accessible $x_{Bj}$ and $Q^2$ regions,
with the exception of the measurement at the lowest scale,
$<Q^2>$ = 1.9 GeV$^2$, where GRV is too low.''\\
The precision of the measurement is insufficient to determine 
whether or not there is a rise in $F_2^{\gamma}$ at low $x_{Bj}$.}

\begin{figure}[hb]
\vskip 0.in\relax\noindent\hskip 2cm
       \relax{\includegraphics{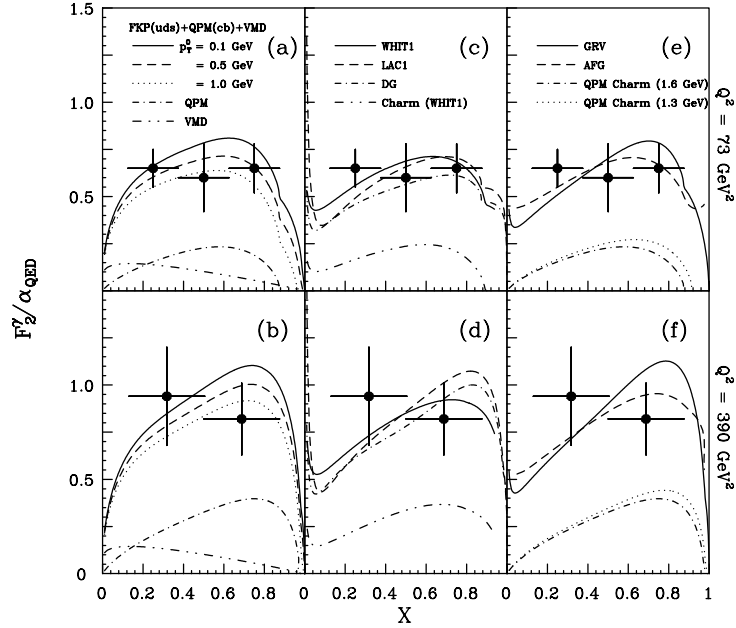}}
\vspace{0.ex}
\caption{\small\sl The $F_2^{\gamma}/{\alpha}$ data from 
the AMY collaboration. Comparison
with the following contributions to the photon structure:
a) and b) FKP (u,d,s) with various $p_T^0$, QPM (c, b) with various 
charm masses, VMD. Comparison with parton parametrizations: c) and d)
WHIT1, LAC1, DG; e) and f) GRV, AFG.
Upper a),c),e) and lower b),d),f) figures correspond to averaged
$Q^2$=73 GeV$^2$ and $Q^2$=390 GeV$^2$, respectively 
(from \cite{amy}). }
\label{fig:scan2}
\end{figure}

~\newline    
$\bullet${\bf{AMY 1.5 95 \cite{amy} (TRISTAN) }}\\
The upgraded AMY 1.5 detector was used to perform
a high $Q^2$ measurement of the photon structure function 
$F_2^{\gamma}$ (with averaged $Q^2$ = 73 and 390 GeV$^2$).
For previous measurement done with AMY 1.0 see \cite{amy2}.
In the Monte Carlo analysis the FKP approach was used
to describe light quarks (with $p_T^0\approx 0.5$ GeV),
the QPM for c and b-quarks, and VMD contribution for soft hadronic
processes were introduced.
Results for unfolded $F_2^{\gamma}$ 
are presented in table \ref{table15} and in fig.~\ref{fig:scan2},
where predictions of various parton parametrizations
as well as individual components are shown.

\begin{table}[ht]
\caption{}
\label{table15}
$$
\begin{array}{|r|c|c|}
\hline
\,\,<Q^2>~\!&<x_{Bj}>&F_2^{\gamma}/{\alpha}\\
~[GeV^2]~~\!&&(stat. + syst.)\\
\hline
~~~~73~~~~~&0.25&0.65\pm0.08\pm0.06\\
&0.50&0.60\pm0.16\pm0.03\\
&0.75&0.65\pm0.11\pm0.08\\
\hline
~~~390~~~~~&0.31&0.94\pm0.23\pm0.10\\
&0.69&0.82\pm0.16\pm0.11\\ 
\hline
\end{array}
$$
\end{table}
\vspace*{7.3cm}

The values of $F_2^{\gamma}$ averaged over 0.3$<x_{Bj}<$0.8
are shown in table \ref{table16}.
\begin{table}[ht]
\caption{}
\label{table16}
$$
\begin{array}{|r|c|}
\hline

\,\,<Q^2>~\!&<F_2^{\gamma}/{\alpha}>\\ 
~[GeV^2]~~\!&\\
\hline
~73~~~~~&0.63\pm0.07\\
390~~~~~&0.85\pm0.18\\
\hline
\end{array} 
$$
\end{table}

~\newline
Comment: {\it The observed 
$x_{Bj}$-behaviour of $F_2^{\gamma}$ is consistent with 
the GRV parametrization and with the FKP one.
The fitted parameter $p_T^0$ is equal to 0.51$\pm$0.39 GeV
if only the AMY data are included, and 0.45$\pm$0.07 GeV if all available, 
at that time, $F_2^{\gamma}$ data are taken.}

\begin{figure}[ht]
\vskip 0.in\relax\noindent\hskip 3.3cm
       \relax{\includegraphics{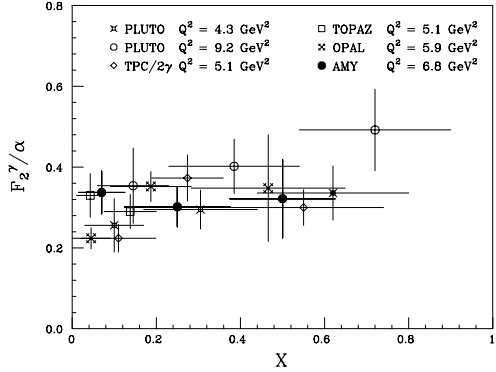}}
\vspace{0.ex}
\caption{\small\sl The $F_2^{\gamma}/{\alpha}$ data from the AMY 
collaboration in comparison with the results
of other experiments at $Q^2\sim$ 4-9 GeV$^2$ 
(from \cite{amy96}).}
\label{fig:amy971}
\end{figure}

\begin{figure}[ht]
\vskip 0.in\relax\noindent\hskip -0.2cm
       \relax{\includegraphics{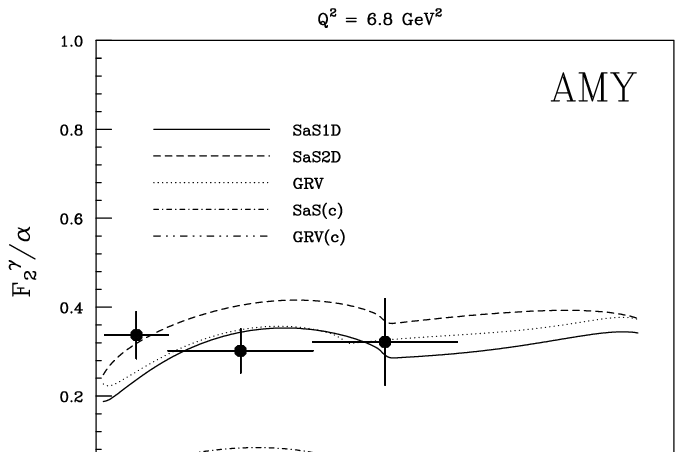}}
\vskip 0.in\relax\noindent\hskip 7.cm
       \relax{\includegraphics{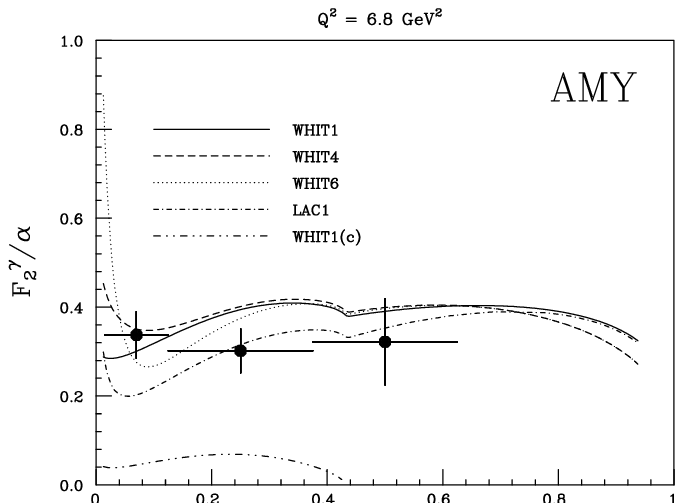}}
\vspace{0.ex}
\caption{\small\sl  The $F_2^{\gamma}/{\alpha}$  as a function of 
$x_{Bj}$. Left: Comparison with the SaS1D, SaS2D and 
GRV LO parton parametrization predictions. 
Also shown are the c-quark contributions 
from the SaS and GRV LO parametrizations. Right: 
Comparison with the WHIT1, WHIT4,
WHIT6 and LAC1 predictions. The c-quark contribution is shown 
for the WHIT1 parametrization (from \cite{amy96}).}
\label{fig:amy972}
\end{figure}

~\newline
$\bullet${\bf {AMY 1.5 97 \cite{amy96} (TRISTAN)}}
\newline
The measurement of $F_2^{\gamma}$ at $<Q^2>$=6.8 GeV$^2$
was performed.
In the Monte Carlo analysis the FKP approach was used
to describe light quarks (with $p_T^0\approx 0.5$ GeV),
the QPM for c-quark, and VMD contribution for soft hadronic
processes were introduced.
The results for 
$F_2^{\gamma}$ versus $x_{Bj}$, together with earlier data,
are presented in fig.~\ref{fig:amy971}. A 
comparison of the data with various parametrizations is given in
fig.~\ref{fig:amy972}.

\vspace*{4.2cm}
\vspace*{4.5cm}

~\newline
Comment: {\it ``The $x_{Bj}$ behaviour of the measured $F_2^{\gamma}$
is consistent with the QCD-based predictions such as SaS, GRV LO and
WHIT models, but inconsistent with the LAC1 prediction for $x_{Bj}$
values around 0.07.''}

~\newline 
$\bullet${\bf {TOPAZ 94 \cite{topaz} (TRISTAN)}}\\
The photon structure function $F_2^{\gamma}$ 
has been measured for averaged $Q^2$ values from 5.1 to 80, and also
338 GeV$^2$.
The Monte Carlo with FKP contribution included for light
quarks (with $p_T^0$ = 0.1 - 1 GeV), the QPM one for charm quark
and VMD contribution for a soft particle production.
The one and two-jet events 
(``hadronic'' events with a stuck quark and the pointlike ones
based on $\gamma^*\gamma\rightarrow q\bar{q}$, respectively)
in this sample are studied, for the first time
in DIS$_{e\gamma}$ process (see also sec. \ref{sec23} and \ref{sec24}).

\begin{table}[ht]
\caption{}
\label{table17}
$$
\begin{array}{|l|r|l|l|c|}
\hline
~\!\!\!<Q^2>&~Q^2\!\!\!~~~~~~&<x_{Bj}>&~~~~~~x_{Bj}&F_2^{\gamma}/{\alpha}\\
~~\!\!\![GeV^2]&[GeV^2]\!\!~~~&&&(stat. + syst.)\\
\hline
~~~~~~5.1&~~3~-~10~&\!~0.043~&0.010-0.076&0.33\pm0.02\pm0.05\\
&&~\!0.138~&0.076-0.20&0.29\pm0.03\pm0.03\\
\hline
~~~~\,16&10~-~30~&~\!0.085~&0.02\,~-0.15&0.60\pm0.08\pm0.06\\        
&&~\!0.24~~&0.15\,~-0.33&0.56\pm0.09\pm0.04\\
&&~\!0.555~&0.33\,~-0.78&0.46\pm0.15\pm0.06\\
\hline
~~~~\,80&45-130~&~\!0.19~~&0.06\,~-0.32&0.68\pm0.26\pm0.05\\
&&~\!0.455~&0.32\,~-0.59&0.83\pm0.22\pm0.05\\
&&~\!0.785~&0.59\,~-0.98&0.53\pm0.21\pm0.05\\
\hline
\end{array}
$$
\end{table}
\vspace*{10cm}
\begin{figure}[ht]
\vskip -4.6cm\relax\noindent\hskip 2.3cm
       \relax{\includegraphics{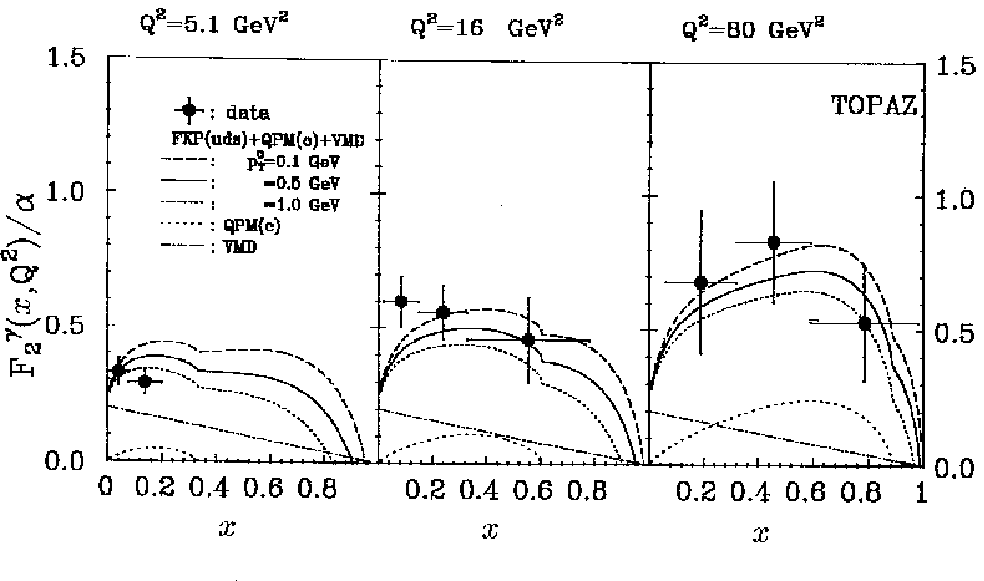}}
\vskip 4.6cm\relax\noindent\hskip 2.3cm
       \relax{\includegraphics{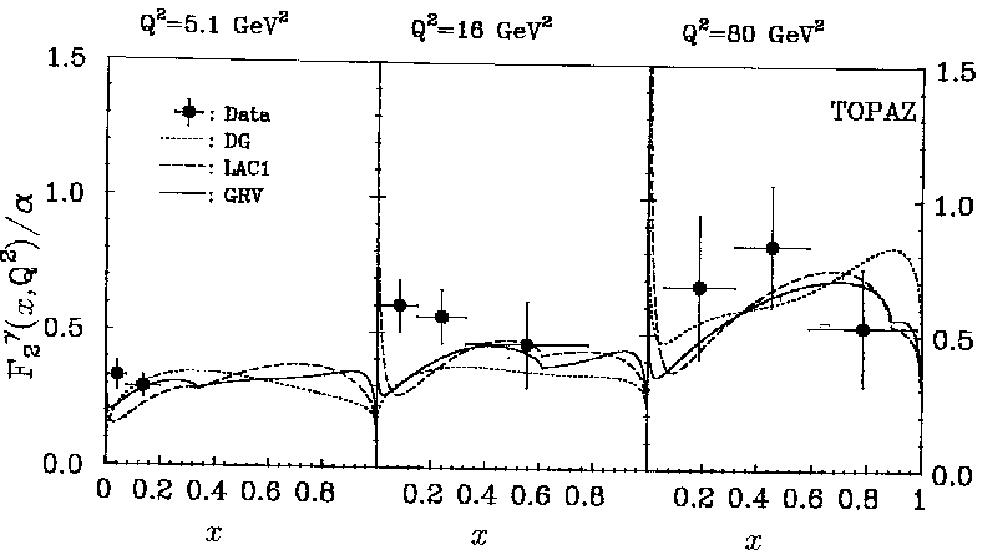}}
\vspace{0ex}
\caption{\small\sl The $F_2^{\gamma}/\alpha$ versus  
$x_{Bj}$ for $<Q^2>$=5.1, 16, 80 GeV$^2$ and a comparison with
the prediction of FKP(u,d,s), QPM (c), VMD contributions
with various $p_T^0$ parameters (upper set)
and with parton parametrizations: DG, LAC1, GRV (lower set) (from \cite{topaz}).}
\label{fig:marysia3}
\end{figure}

\begin{figure}[ht]
\vskip 0.5in\relax\noindent\hskip 0.in
       \relax{\includegraphics{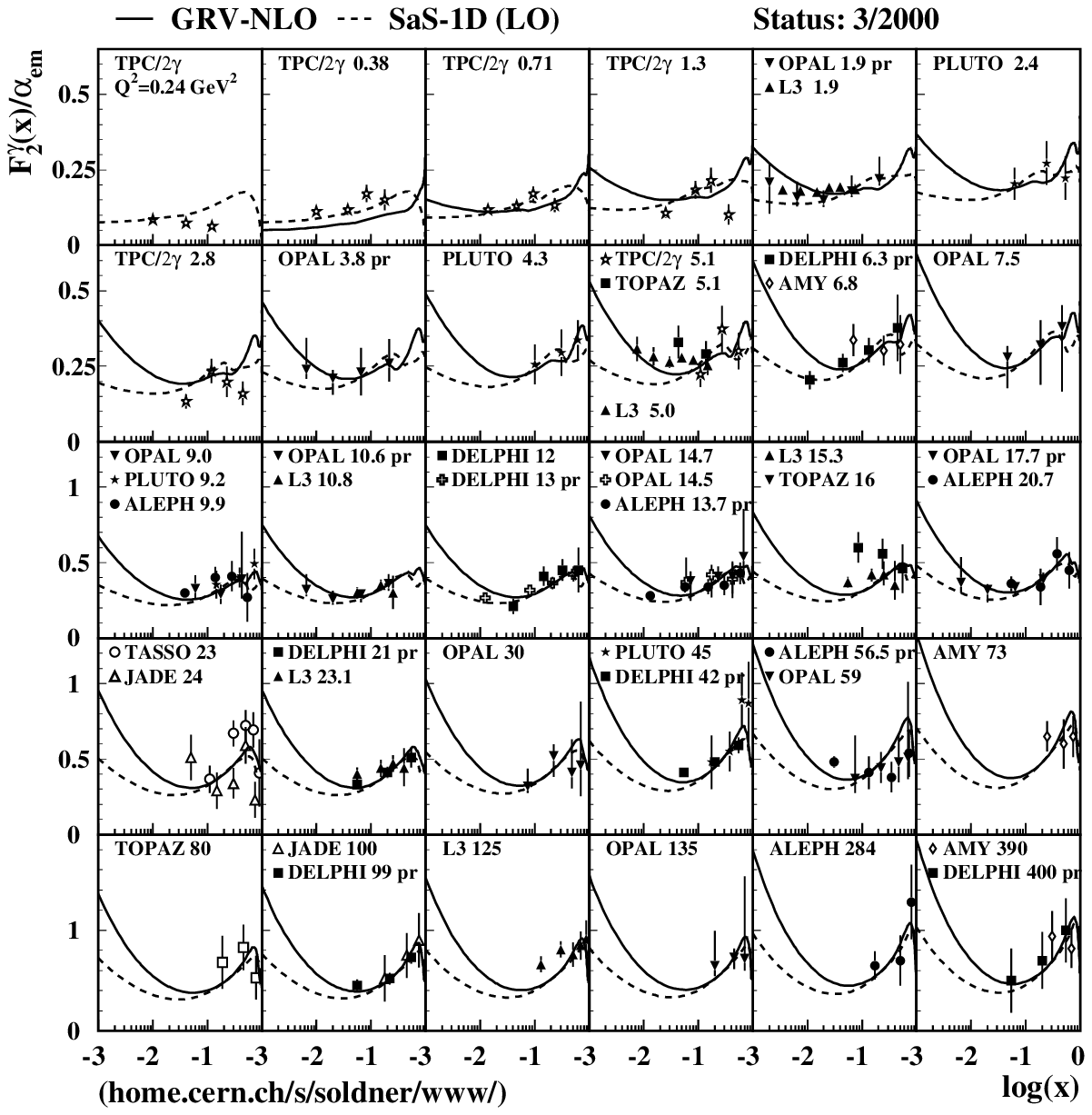}}
\vspace{-27.ex}
\caption{ {\small\sl The photon structure function 
$F_2^{\gamma}/\alpha$ as a function of $x_{Bj}$ in bins of
$Q^2$ compared to the GRV NLO (solid line) and 
SaS1D (LO) (dashed line) parametrizations of parton distributions in
the photon (from \cite{stef}).}}
\label{fig:f2all_bw}
\end{figure}

\begin{figure}[ht]
\vskip 0.cm\relax\noindent\hskip 1.5cm
       \relax{\includegraphics{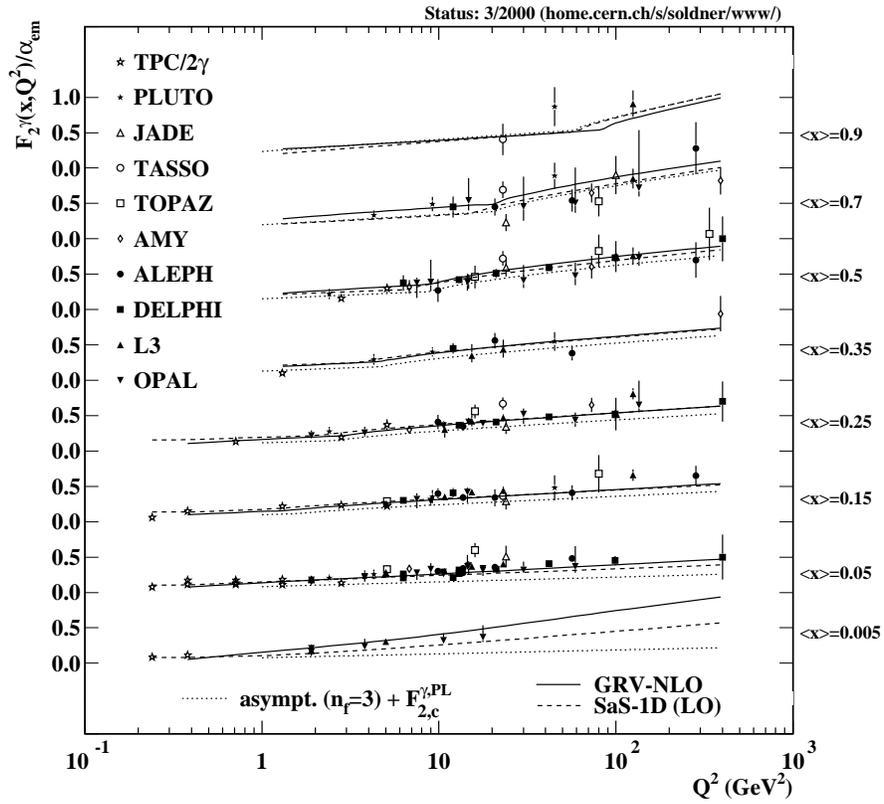}}
\vspace{0ex}
\caption{ {\small\sl The photon structure function
$F_2^{\gamma}/\alpha$ (data from various experiments)
as a function of $Q^2$
(from \cite{stef}).}}
\label{fig:f2q2_bw}
\end{figure}

The results for $<Q^2>$ = 5.1, 16 and 80 GeV$^2$ are presented in 
table~\ref{table17} and in fig.~\ref{fig:marysia3} where the individual
components as well as predictions of various parton parametrizations 
are shown.
To study the $Q^2$ dependence of $F_2^{\gamma}$,
the averaged values of  $F_2^{\gamma}/\alpha$ in the $x_{Bj}$-range
from 0.3 to 0.8 were extracted at $<Q^2>$= 16, 80 
and 338 GeV$^2$ (see table \ref{table18}; the numbers in 
parentheses are  results for the light quarks alone).
\begin{table}[ht]
\caption{}
\label{table18}
$$
\begin{array}{|c|c|}
\hline
\,\,<Q^2>~~~~~\!\!&F_2^{\gamma}/{\alpha}\\ 
~[GeV^2]~~~~~~&\\
\hline
~16&0.47\pm0.08~(0.38\pm0.08)\\
~80&0.70\pm0.15~(0.49\pm0.15)\\
338&1.07\pm0.37~(0.72\pm0.37)\\
\hline
\end{array} 
$$
\end{table}
~\newline
Comment: {\it  The final hadronic state described by QPM(c)+VMD+FKP(u,d,s)
with $p_T^0$=0.1,0.5 and 1  GeV, and also by the GRV, 
DG and LAC1 parton parametrizations, was studied.} 
\vskip 1.cm
\centerline{*****}
\vskip 1.cm
For an overall comparison  
the collective figures of $F_2^{\gamma}/{\alpha}$ versus $x_{Bj}$
(fig.~\ref{fig:f2all_bw}) and $F_2^{\gamma}/{\alpha}$ 
versus $Q^2$ 
(fig.~\ref{fig:f2q2_bw}),
 containing  also earlier data
not discussed here, are presented based on \cite{stef}.
In these figures 
the comparison with theoretical predictions for $x_{Bj}$ 
dependence of $F_2^{\gamma}$ based
on the SaS1D (LO) and GRV NLO parton parametrizations 
(in fig.~\ref{fig:f2q2_bw} in addition the prediction
of the asymptotic solution)  are shown.
For comparison with theoretical predictions of
the $Q^2$ evolution for different $x_{Bj}$ ranges see 
also figs.  
 \ref{fig:delphi9}, 
\ref{fig:tyapkin4}b, \ref{fig:f2q1typ},\ref{fig:f2qtyp},
\ref{fig:l3_f2_98_7}, \ref{fig:l3_q2_98_4},
\ref{fig:l3_highq2_8a}, \ref{fig:scan1}d, 
\ref{fig:nisius2}, \ref{fig:LP2914b}.

\begin{figure}[ht]
\vskip 0.cm\relax\noindent\hskip 1.5cm
       \relax{\includegraphics{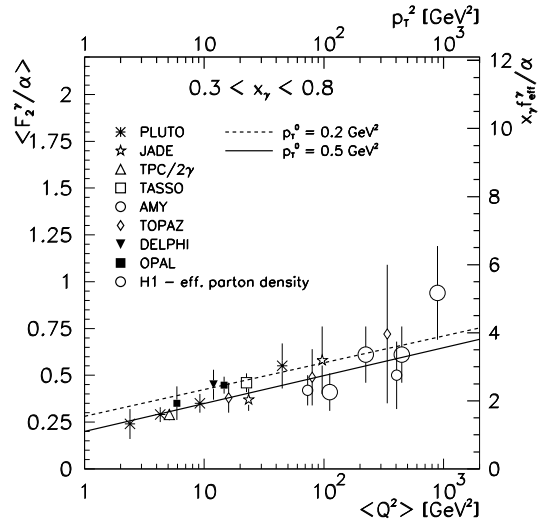}}
\vspace{0.cm}
\caption{\small\sl The scaling violation of the effective parton densities
measured at HERA compared with the $Q^2$ dependence of  $F_2^{\gamma}$
(averaged over the range 0.3$<x_{Bj}<$0.8)
(from \cite{t307}).}
\label{fig:t3074}
\end{figure}

The effective parton density, as measured at HERA collider 
by the H1 group \cite{t307}
in the jet production from resolved photons,
is compared with the $F_2^{\gamma}$ data in 
fig.~\ref{fig:t3074} as well.\\
\vspace*{17.4cm}
\newpage
\vspace*{9.5cm}
\vspace*{6.cm}

\newpage

\subsection{Hadronic final states in the DIS$_{e\gamma}$ experiments 
\label{sec23}}
Although a detailed analysis of hadronic final states, including  
those involving large $p_T$ particles/jets, has not been the main 
aim of DIS$_{e\gamma}$ experiments,  it has proved crucial 
in extracting the unfolded $F_2^{\gamma}$ as mentioned in the previous
section (sec. \ref{sec222}). The obtained experimental 
results, and problems that appear while describing the hadronic energy 
flow, transverse energy out of event plane, 
pseudorapidity and other distributions 
within the existing Monte Carlo models,
 deserve close attention and a separate treatment - this 
section is devoted to this subject. 

The $F_2^{\gamma}$ corresponds to the 
cross section for hadron production in ${\gamma}^*{\gamma}$ 
collisions, eqs. (\ref{si}, \ref{si2}), and obviously the bulk of the 
contribution is due to the soft processes, \ie production of particles
with a relatively small transverse momentum
$p_T$. One should be aware that in the hadronic final state in DIS$_{e\gamma}$
experiments two types of large scale may appear: $Q^2$ and $p_T^2$
\cite{kapusta92}. The bulk of the data, as it was already mentioned,
corresponds to events  with not very large $p_T$;
if for these events the relation $Q^2\gg p^2_T \gg P^2$ holds, the 
interpretation in terms of the  photon interaction between  one 
direct ($\gamma^*$) and  one  resolved real ($\gamma $) photon may 
be introduced. Then it is not clear what scale should be used in the 
parton density for a real photon, $f^{\gamma}(x,\tilde Q^2)$: $\tilde Q^2$
= $p^2_T$, $Q^2$, or ? Moreover the processes 
corresponding to  $Q^2 \ll p^2_T$
should be treated as being resolved from the point of view of both
the real and the virtual photon.

Some of the problems that will appear here are common with those presented 
in sec. \ref{sec33} for the jet production in
$\gamma^*\gamma$ collision and in sec. \ref{sec24}  where dedicated 
measurements of the {\underline {large}} 
$p_T$ jet production in real $\gamma \gamma$ and $\gamma p$ processes
are presented. 

\subsubsection{Modelling of the hadronic final state in $\gamma^*\gamma $
collisions}
In the DIS$_{e\gamma}$ (single-tag) events the whole hadronic 
final state is not 
seen in the detector. A correct unfolding method is necessary 
to reconstruct the true kinematical variables (for definition see beginning
of sec. \ref{sec222}). At this stage  models to generate
 separate samples for the different
components of the photon are  needed, 
 as we have already mentioned in sec. \ref{sec222}.
 The modelling is performed using different MC programs
and it is still a large source of systematic errors in  $F_2^{\gamma}$.

We start with a short description of the FKP approach used in the early 
analyses which bases on the QPM, (soft) VDM and RPC contributions. 
The QPM contribution is generated using the  $e^+e^-\ra e^+e^- q \bar q$
matrix element for massive quarks,  for an arbitrary value of $p_T$ of
produced quarks.  The final state transverse momentum distribution 
is here $d\sigma/dp_T^2\sim p_T^{-4}$.
The QPM  was  applied  mainly  to the  $c$ and $b$ quark 
production.  
The (soft) VMD (called sometimes hadronic) contribution, 
in which the initial real photon is treated as the vector meson(s), 
say $\rho$, with the typical soft interaction with a photon probe,
leads to the production of the final quarks (then hadrons)
 with the $d\sigma/dp_T^2 \sim e^{-b p_T^2}$ distribution 
(where $b\sim$ 3 - 5 GeV$^{-2}$). 
This contribution  should be
negligible for the production of particles with large $p_T$ 
where, on the other hand, other  contributions should be included.
In the FKP approach \cite{fkp} the  cutoff parameter $p_T^0$ is built in
to separate the mentioned above VMD component of the initial photon target
 and the resolved photon 
(QCD) contribution, which develops from 
 the pointlike coupling of the $\gamma$ to the $q\bar q$ pair. This parameter,
$p_T^0$, corresponds therefore to the intrinsic 
transverse momentum of the partons 
in the photon. The additional  cutoff $p_T^{min}$ is introduced in 
the analysis, to define hard ($2\rightarrow 2$)
subprocesses where the pQCD can be used.
 In the description of  the particle production with $p_T>p_T^{min}$  
the resolved photon contribution is introduced (see also  sec. 2.4).
In the FKP approach, where the (soft) VMD contribution was introduced without
 the $Q^2$  dependence, both cutoff parameters are of the same order.

In the DELPHI analysis, 
the TWOGAM generator (``fixed'' in 1993) was used, 
with the QPM, VMD for soft production and RPC contributions 
included within the FKP approach \cite{delphi, tyapkin}. 
This generator treats exactly the kinematics of scattered electron 
and positron, uses exact expression for the two photon luminosity
function, and also in the (single and double)
resolved process the kinematics of the 
partonic system is exact for any photon virtuality. The TWOGAM 
generator was used  also in the last L3 measurements.

In many OPAL analysis the generator F2GEN (developed from
TWOGEN used in older analyses) was used, which can generate events 
according to ''any selected formula for $F_2^{\gamma}$'' 
\cite{miller}a.
In this generator one can easily put a specific assumption
about the final state. For example different angular distributions may 
be assumed in the $\gamma^* \gamma$ CM system for the produced pair of quarks:
 ``pointlike'', \ie 
as for lepton pair from two real photons, ``peripheral'' with an 
exponential distribution of the transverse momentum, or the 
``perimiss'' combination \cite{opal2}.

In recent analyses  general purpose Monte Carlo 
generators are being used in analysis of hadronic final state. 
HERWIG and  PYTHIA (the implementation of the SaS approach)
 apply  chosen parton 
parametrizations. In the PHOJET generator, where hard interaction is described 
using QCD, the description of soft interaction bases on the Regge theory.
The very recent ALEPH, L3, OPAL and TRISTAN $F_2$ data have been 
obtained using 
the modified HERWIG (PYTHIA)
program, where the additional power distribution of the transverse momentum
of partons in the photon, 
of the form $dp_t^2/(p^2_t+\tilde{p}^2_{t0})$, is included,
see {\cite{lauber}} and {\bf ALEPH 99a, ALEPH 99b,conf} or {\bf OPAL 2000}.

As far as the  global hadronic variables like 
$W_{vis},\ Q^2,\ M_{jj}$ 
(invariant mass of two jets)
  and other  distributions  
are concerned,  there is a fairly good description of the data 
by existing Monte Carlo generators.
The problems arose for the 
transverse energy out of the  tag plane 
(the plane defined by the initial and tagged electrons)  or for the
energy flow per event as a function of
 pseudorapidity, $\eta=-\ln(\tan(\theta/2))$ 
\footnote{The polar angle is measured  from the $z-axis$, and 
 the tagged electron(positron) is by definition at the negative rapidity.}. 
The discrepancies were very pronounced especially for small $x_{Bj}$.
 
The first observation of the disagreement between the data and Monte Carlo
models was made by the OPAL collaboration \cite{opal2}. 
``The serious discrepancies between the 
data and any of the available Monte Carlo
models are seen both within the central region of the detector
($|\eta|<2.3$), where the energy flow is well measured, 
and in the forward region, where the energy
can only be sampled.''
It is clear that the unfolding of $F_2^{\gamma}$
``will have large errors as long as the energy 
flow from different models remains in clear disagreement with the energy
flow in the data, in particular in the region of $x_{vis}<$0.1
and $Q^2<$30 GeV$^2$''
(from \cite{opal2}).
A similar effect has been seen by now by other groups, see below.

The  improved description of the mentioned data
is obtained in the ``HERWIG (PYTHIA) + power law $p_t$'' generators.
The PHOJET 1.05 and TWOGAM generators can also satisfactorily describe
the data (DELPHI, L3). Note however that the values of $p_T^{min}$ fitted in 
these analyses are high (2.5 - 3.5 GeV). Another approach which usually
improves the agreement with the data is the inclusion of {\sl multiple
interaction}. The effect is similar in all these approaches and
consists in increasing $p_T$ in processes involving photons.

The recent progress in describing the  hadronic final state in 
single-tag events
at LEP by using the two-dimensional unfolding led to a reduction of the Monte Carlo
modelling error ({\bf OPAL 2000}).  As follows from  first 
 combined results of the newly formed
{\bf LEP-wide Two Photon Physics group}, which used the 
 ``HERWIG 5.9 + power law $p_t$'' and PHOJET 1.05, a good agreement 
is found between experiments in the central regions. 
``Unresolved problems remain in the small angle regions where the results 
of the different experiments showed greater discrepancies than could be 
accounted for by statistical fluctuations...'' \cite{finch-lep}.
It was also found that to describe  low $Q^2$ data an improved, modified
model ``HERWIG 5.9 + power law $p_t$ (dyn)'',
with a dynamical limit $\tilde p^2_{t max}=Q^2 {\rm ~{or}}~Q^2+p_t^2$
 has to be introduced.  

For the general discussion and data - Monte Carlo comparison in the context
of modelling the hadronic final state see \cite{miller,cart},
see also \cite{mariak,finch-lep}
and \cite{butter-jets}, where the jet production in the 
photon-induced processes is studied.

\subsubsection{Measurements of the hadronic final state}
The analyses of the hadronic final state accompanying the 
measurements of $F_2^{\gamma}$ discussed in sec. \ref{sec222}, as well as the 
results from independent analyses of large $p_T$ hadron production 
in {\sl single tagged} events  are presented below.
Variables used in the unfolding procedure are defined and discussed in 
sec. \ref{sec222}.
Note that in some analyses the 
jet topology of  
the final state was studied (see \eg {\bf TOPAZ 94}, {\bf OPAL 97d,conf}).

For the two  photons involved, we  introduce the following notation
for their squared (positive) virtualities:  $P^2_1$ ($=Q^2$)
and $P^2_2$ ($=P^2$), with the relation to variables
used in the discussed DIS$_{e\gamma}$ scattering in parentheses, \ie with
the inequality $P_1^2 \ge P_2^2$ understood.

\newpage
\centerline{\huge DATA} 

~\newline\newline
$\bullet${\bf{ALEPH 97a,conf \cite{LP253} (LEP 1) }}\\
Single tagged events were collected in  years 1992-1994 
at the average $Q^2$=14.2 GeV$^2$.
A dedicated study of the  hadronic final state assuming  the
QPM (for $u$, $d$, $s$ and $c$ quarks) + VMD 
(for the target photon) and using the  HERWIG 5.9 
generator, with the GRV parametrizations, was performed. Also  the modified 
HERWIG generator,
``HERWIG 5.9  + power law $p_t$'', with $\tilde{p}_{t0}=0.66$ GeV 
was used. A cone jet-finding algorithm was used in the
analysis.

\begin{figure}[ht]
\vskip 0.cm\relax\noindent\hskip 0.5cm
       \relax{\includegraphics{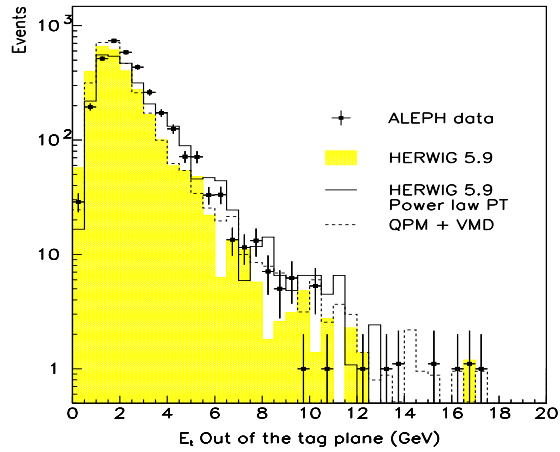}}
\vspace{0.cm}
\caption{\small\sl The energy (out of the plane of the tag 
and the beam) distribution. Comparison with different Monte Carlo
models: HERWIG 5.9 (shaded), ``HERWIG 5.9 + power law $p_t$'' 
(solid line) and QPM + VMD (dashed line) (from \cite{LP253}).}
\label{fig:LP2532b}
\end{figure}

\begin{figure}[ht]
\vskip 0.cm\relax\noindent\hskip 0.5cm
       \relax{\includegraphics{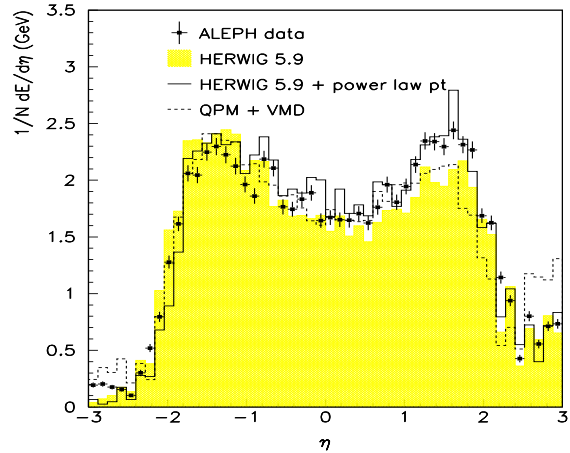}}
\vspace{0.cm}
\caption{\small\sl The energy flow versus 
pseudorapidity of the final hadrons. Comparison with different Monte Carlo
models as in fig. \ref{fig:LP2532b} (from \cite{LP253}).}
\label{fig:LP2532a}
\end{figure}

Results  for energy flow are presented in  figs.~\ref{fig:LP2532b},
\ref{fig:LP2532a}. The discrepancy is observed in the energy 
flow in the positive pseudorapidity region.
The approach ``HERWIG 5.9 + power law $p_t$'' 
leads everywhere to a better description of the data.

\vspace*{5.8cm}
\vspace*{5.1cm}

\begin{figure}[ht]
\vskip 0.cm\relax\noindent\hskip 0.5cm
       \relax{\includegraphics{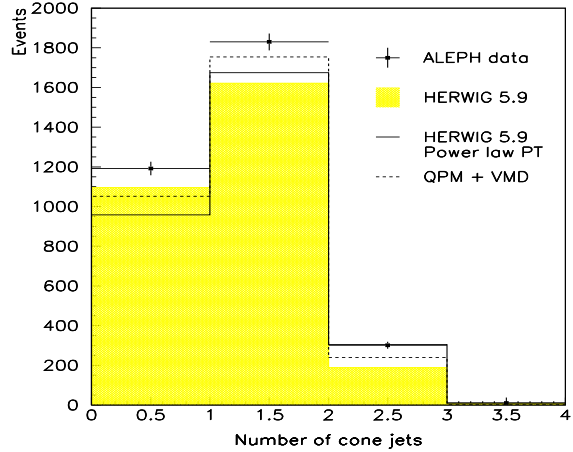}}
\vspace{0.cm}
\caption{\small\sl The number of events classified in four bins
referring to various numbers of jets in an event. 
Comparison with different Monte Carlo
models as in fig. \ref{fig:LP2532b} (from \cite{LP253}).}
\label{fig:LP2533}
\end{figure}

In order to pin down the source of the observed discrepancy 
the analysis of the number of cone jets was performed 
in the  hadronic final state (see fig.\ref{fig:LP2533} for results).
The standard HERWIG model, having additional production mechanisms with
a resolved photon,  should in principle give larger number of two-jet events 
than the QPM+VMD model. On the contrary, it leads to a lower number.

To check the presence of the single 
and double  resolved photon events, the $x_{\gamma}$ and $x_{tag}$
distributions were measured (not shown).
Note, that double resolved photon events contain 
the resolved virtual photon contribution (see also sec. \ref{sec3}).
Also the energy not assigned to jets in two-jet events, 
$E_2^{non-jet}$, was studied (not shown).  The ``HERWIG + power law $p_t$'' 
approach was used successfully to describe these data.

\vspace*{6cm}

~\newline
Comment: {\it ``The ALEPH data confirms recent problems reported by 
OPAL in modelling the hadronic final state.''  \newline
`` 'HERWIG + power law $p_t$' is better at modelling the region
of large $E_{t, out}$'' and the peak in the energy flow 
at positive rapidity.
\newline
The ``double resolved processes are not present to an large extent 
in the data at this $Q^2$''.}

\begin{figure}[ht]
\vskip 0.cm\relax\noindent\hskip 0cm
       \relax{\includegraphics{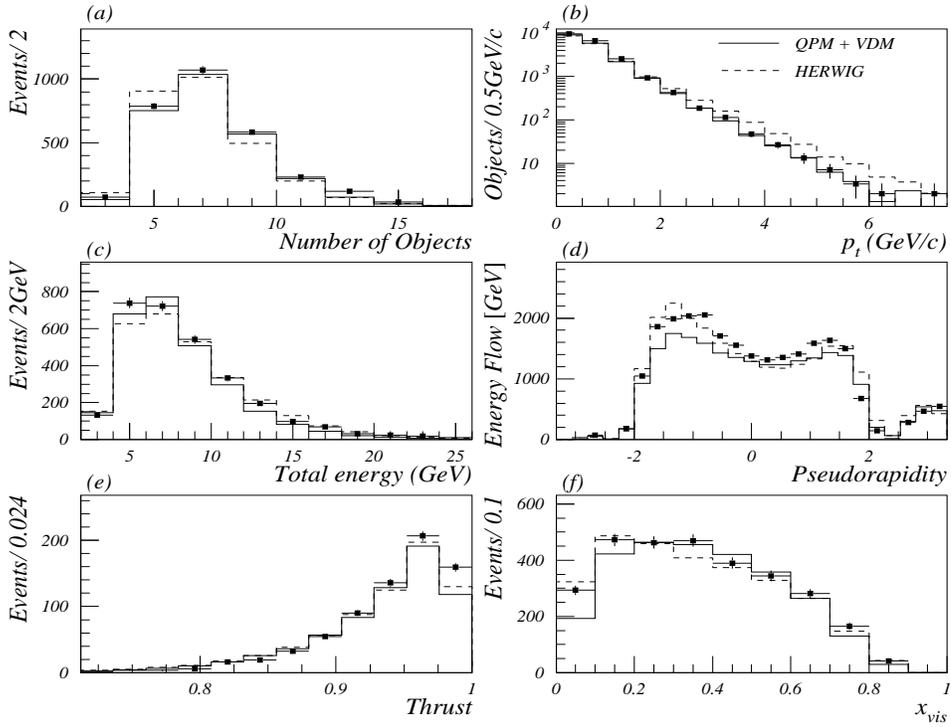}}
\vspace{0.cm}
\caption{\small\sl The data in which the scattered electron
is detected (tagged) in luminosity calorimeter LCAL,
compared with the QPM + VMD model (solid histogram)
and the ``HERWIG 5.9 + power law $p_t$''
model (dashed histogram) (from \cite{finch99}).}
\label{fig:finch_1}
\end{figure}

\begin{figure}[ht]
\vskip 0.cm\relax\noindent\hskip 0cm
       \relax{\includegraphics{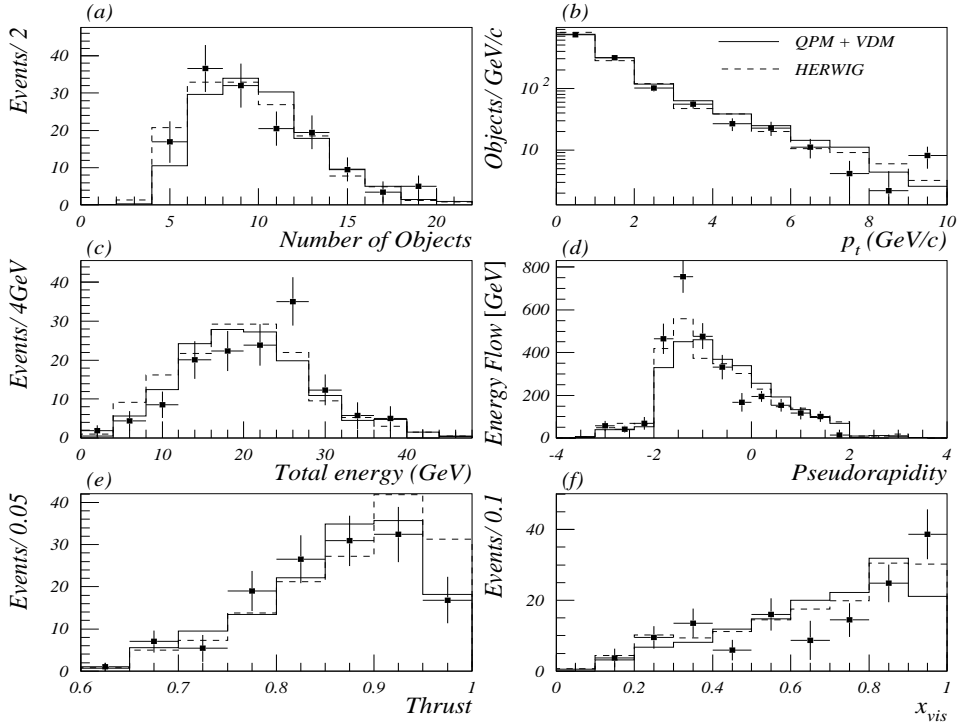}}
\vspace{0.cm}
\caption{\small\sl The data in which the scattered electron
is detected (tagged) in electromagnetic calorimeter ECAL,
compared with the QPM + VMD model (solid histogram)
and the ``HERWIG 5.9 + power law $p_t$'' 
model (dashed histogram) (from \cite{finch99}).}
\label{fig:finch_2}
\end{figure}

~\newline\newline
$\bullet${\bf{ALEPH 99a \cite{finch99} (LEP 1) }}\\
Analysis of the hadronic final state during the $F_2^{\gamma}$
measurement is reported.  
Monte Carlo models: QPM (for all quarks) + VMD (for a target
photon) and ``HERWIG 5.9 + power law $p_t$'', 
with GRV LO parametrization, were studied. 
Results are 
presented in figs. \ref{fig:finch_1} and \ref{fig:finch_2}.
\newline
\vspace*{8cm}
\vspace*{9cm}

~\newline
Comment: {\it The HERWIG program with modified $p_t$ distribution and 
the QPM + VMD
 model give ``an acceptable description of the data''.}

~\newline\newline
$\bullet${\bf {ALEPH 99b,conf \cite{aleph99conf} (LEP 2)}}\\
A new unfolding method was used in  the measurement of $F_2^{\gamma}$ 
at $\sqrt s$=183 GeV based on 1997 data (see also previous section).
Two samples of single-tag events with the tag defined as a cluster
 in one of the small angle luminosity calorimeters SICAL or LCAL,
 for  $Q^2$ ranges: 7-24 GeV$^2$ and 17-200 GeV$^2$, were studied. 

\begin{figure}[ht]
\vskip 0.2cm\relax\noindent\hskip 2.7cm
       \relax{\includegraphics{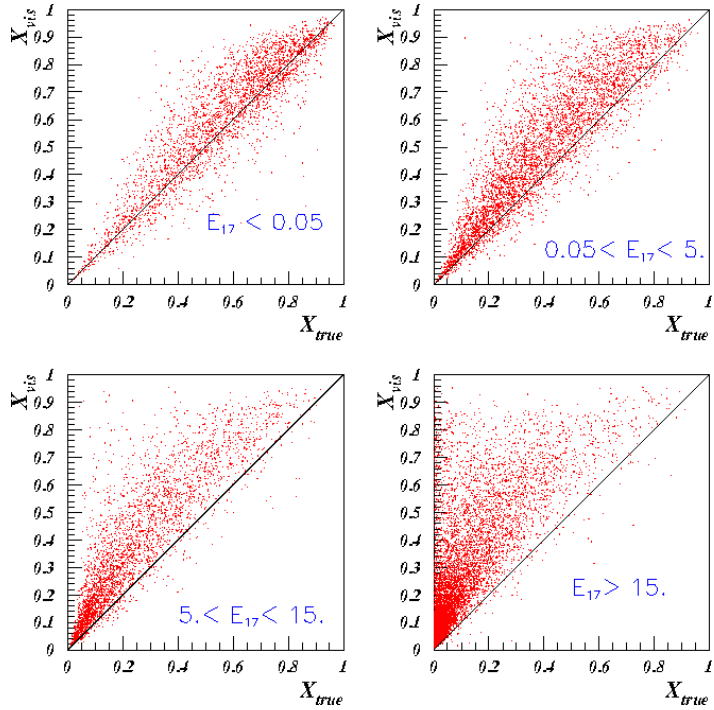}}
\vspace{-0.2cm}
\caption{\small\sl The relationship between reconstructed and
true values for $x_{Bj}$ for four bins of $E_{17}$
(from \cite{aleph99conf}).}
\label{fig:aleph99conf_2}
\end{figure}

The new unfolding procedure, based on the principle of maximum entropy,
was used to correct the distributions for finite resolution and 
acceptance. 
The unfolding was done for $x_{Bj}$ and in additional
for a second kinematical variable, $E_{17}$, defined as the total
energy of particles produced at an angle with respect to the beam
direction smaller  than 17$^o$. In  fig.~\ref{fig:aleph99conf_2}
results of unfolding of the $x$ variable
 for different values of $E_{17}$ obtained using 
``HERWIG 5.9 + power law $p_t$'' (SaS1D)
are presented (LCAL data). One can see that the small $E_{17}$ gives better result.

\vspace*{9.1cm}

\begin{figure}[ht]
\vskip 0.in\relax\noindent\hskip 1.6cm
       \relax{\includegraphics{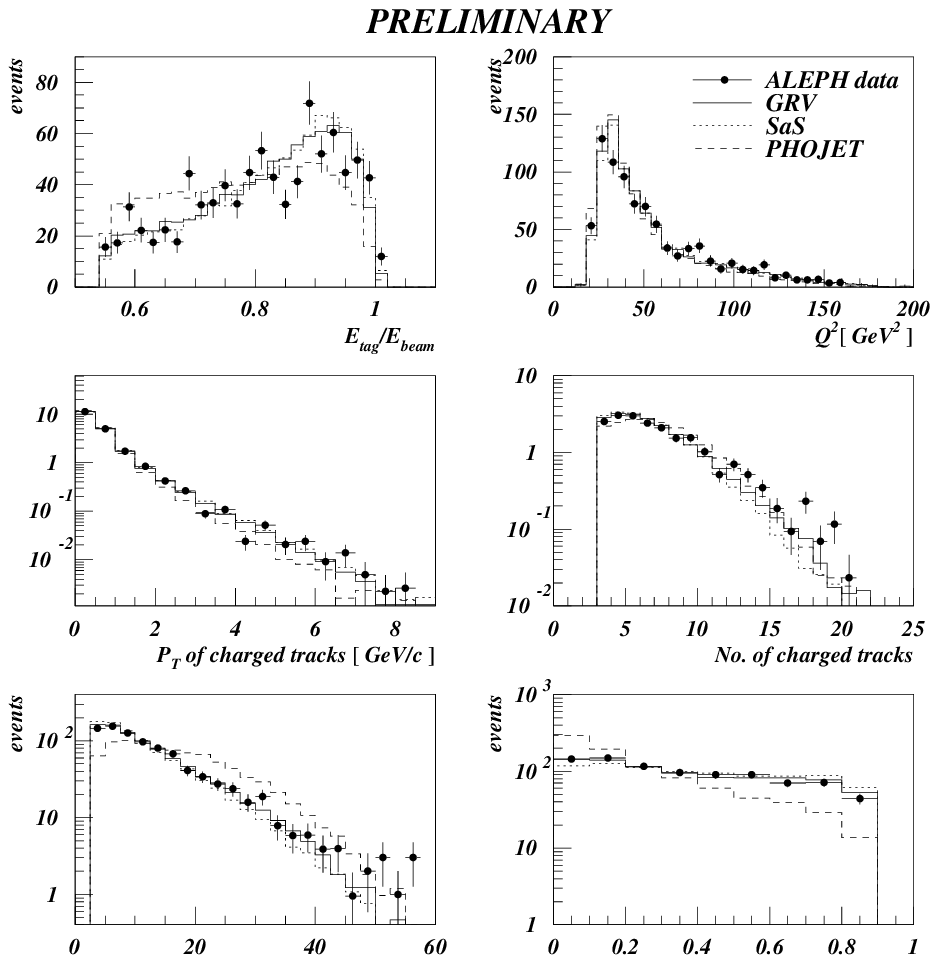}}
\vspace{0.cm}
\caption{\small\sl Comparison of data and simulations 
for the LCAL tagged data.
Predictions of the ``HERWIG 5.9 + power law $p_t$'' model with two parton 
parametrizations: GRV (solid line), SaS1D (dotted line), and
of PHOJET (dashed line) are shown
(from \cite{aleph99conf}).}
\label{fig:aleph99conf_4}
\end{figure}

Although the two-variable unfolding reduces  the dependence on
the Monte Carlo,  some model dependence still remains. This point was studied 
by comparing  several kinematic distributions obtained  using 
``HERWIG 5.9 + power law 
$p_t$'' with the GRV LO and SaS1D parton parametrizations and in addition
using the  PHOJET model. The SICAL-tag data (not shown) are  
described  reasonably by the used models, while for LCAL data
(fig.~\ref{fig:aleph99conf_4}) "PHOJET is in poor agreement with the data".
\newline

\vspace*{9cm}
~\newline
Comment: {\it This unfolding method ``leads to smaller statistical 
errors and a reduced model dependence compared to one-dimensional procedure''.} 

\begin{figure}[hb]
\vskip 0.in\relax\noindent\hskip -0.6cm
       \relax{\includegraphics{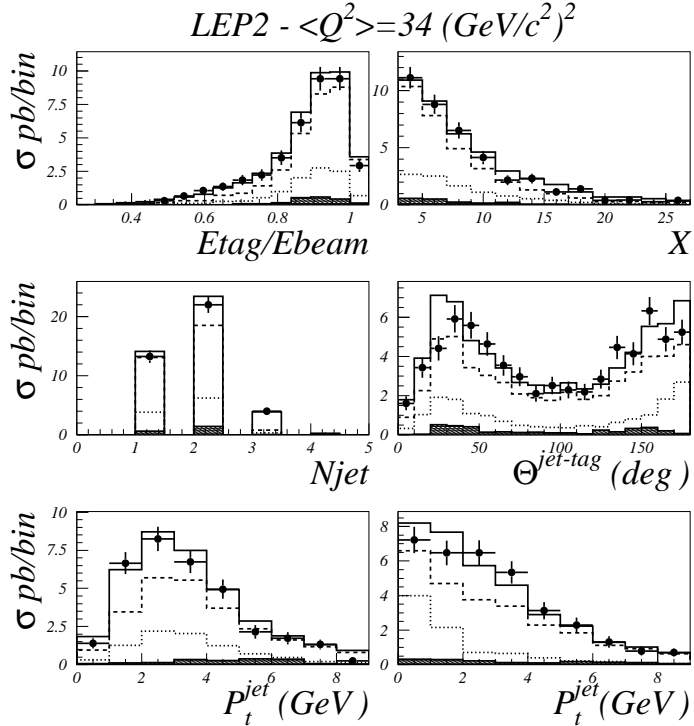}}
\vspace{0ex}
\caption{\small\sl Comparison of data and the Monte Carlo prediction for 
$<Q^2>$ = 34 GeV$^2$ (LEP2) for the variables
: (shown first left, then right in each row) Energy of tagged particle;  The 
invariant mass;  The number of jets;  The jet angle with respect 
to the tagged particle;  $p_T$ for jets in the same
hemisphere as the tagged particle;  $p_T$ for jets
in the opposite hemisphere. Curves show the Monte Carlo predictions:
GVMD+QPM+RPC (GS2) (solid line), 
GVMD+QPM (dots), GVMD (dashes)
(from \cite{tyapkin}).}
\label{fig:had2}
\end{figure}

\begin{figure}[hb]
\vskip 0in\relax\noindent\hskip -0.9cm
       \relax{\includegraphics{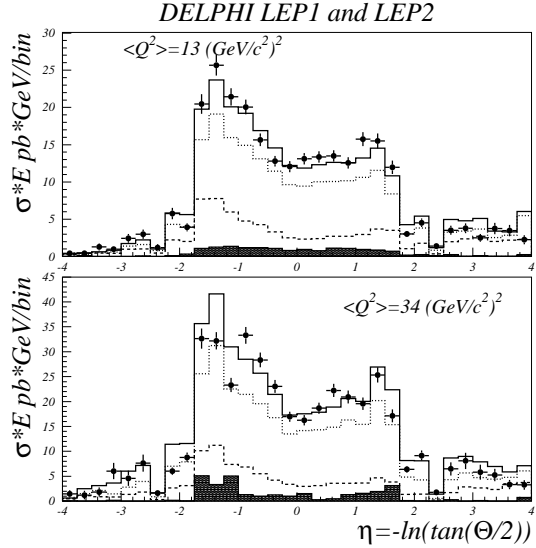}}
\vspace{0ex}
\caption{\small\sl The event energy flow as a function of the pseudorapidity
for $<Q^2>$ = 13 GeV$^2$ (LEP1) and $<Q^2>$ = 34 GeV$^2$ (LEP2).
The notations are as in fig.~\ref{fig:had2} (from \cite{tyapkin}).}
\label{fig:had3}
\end{figure}

~\newline
$\bullet${\bf{DELPHI 95 \cite{delphi3} (LEP 1) }}\\  
First evidence of the hard scattering process in the  single tagged 
events in the data from the 1991-1992 run
is reported.
The values of  $E_T$ of observed jets were larger than  1.5 GeV, 
while the magnitude of the mass of virtual
photon was equal to $<Q^2>\approx$ 0.06 GeV$^2$. This corresponds 
to the standard  resolved (almost real) photon
process, from the point of view of both photons. 
~\newline\newline
Comment: {\it This is not a typical DIS experiment since the 
photon probe
is almost real. The analysis is not typical, either, 
for the standard large $p_T$
jet study. 
``The data are consistent with the predictions 
for quark and gluon density functions 
in the GS parametrization.
The sum of the contributions: VMD + QPM (Quark Parton Model) 
+ [QCD - RPC] is needed in order to describe the data; 
the DO and LAC3 parametrizations
do not adequately describe the data.''}
\newline\newline
$\bullet${\bf{DELPHI 96b,conf \cite{delphi} (LEP 1) }}\\  
The results for the averaged $Q^2$=13 and 106 GeV$^2$
were considered and compared with the TWOGAM generator,
for $F_2^{\gamma}$ results see sec. \ref{sec23}. The QPM, 
GVMD contributions and RPC with the GS2 
parton parametrization were included.
~\newline\newline
Comment: {\it It was shown that the resolved photon contributions are needed 
to obtain the description of the data.}
\newline\newline
$\bullet${\bf{DELPHI 97a,conf \cite{tyapkin} (LEP1, LEP2)}} 
\newline
Analysis of the hadronic final state in the $F_2^{\gamma}$ measurement 
(see sec. \ref{sec222}) was performed for energies around the $Z^0$ mass, and 
for 161-172 GeV (1996 run). The $<Q^2>$= 6.5, 13, 22 and 17, 34, 
63~GeV$^2$ were studied, respectively. The hadronic final state topology 
with events containing jets was studied using the  TWOGAM generator. For 
various  distributions obtained for LEP 2 data,
 see fig.~\ref{fig:had2}. In   fig.~\ref{fig:had3}, the  
 event energy flow as a function of the pseudorapidity
is shown for LEP1 and LEP2 data.

~\newline
Comment: {\it ``All variables were found to be in good agreement with 
TWOGAM prediction.''\\
Note that the $p_T$ range of jets may be of the order of $<Q^2>$
for small $Q^2$ samples.}\\

\vspace*{8.8cm}
\newpage
\vspace*{6.5cm} 
\begin{figure}[ht]
\vskip 0.cm\relax\noindent\hskip 0.0cm
       \relax{\includegraphics{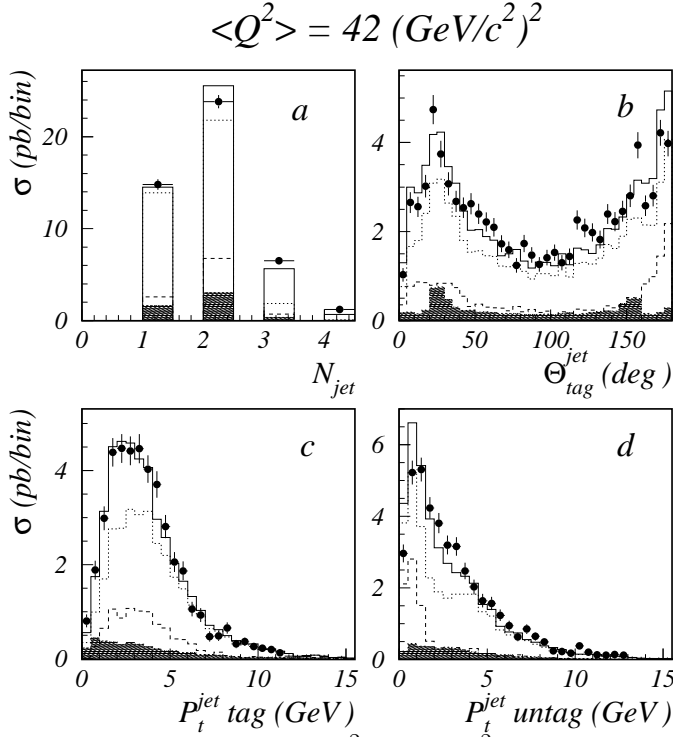}}
\vspace{0.cm}
\caption{\small\sl The cross section for averaged $Q^2$=42 GeV$^2$
as a function of a) The number of jets, b) the jet angle with respect to
the tagged particle and of the transverse momentum of the jet for
c) tagged and d) untagged events. Curves show the Monte Carlo
predictions with QPM + GVMD + RPC (solid line), 
QPM + GVMD (dotted line) and GVMD (dashed line) (from \cite{tyappriv}).}
\label{fig:delphi99}
\end{figure}

\begin{figure}[ht]
\vskip 0.cm\relax\noindent\hskip 0.cm
       \relax{\includegraphics{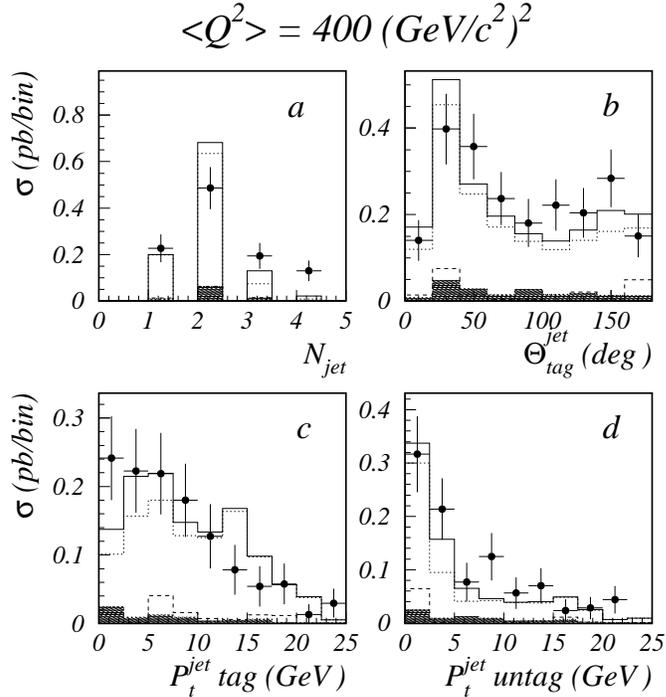}}
\vspace{0.cm}
\caption{\small\sl The cross section for averaged $Q^2$=400 GeV$^2$
as a function of a) number of jets, b) the jet angle with respect to
the tagged particle, and of the transverse momentum of the jet for
c) tagged and d) untagged events. The notations are as in 
fig. \ref{fig:delphi99} (from \cite{tyappriv}).}
\label{fig:400pt}
\end{figure}

~\newline
$\bullet${\bf {DELPHI 98,conf \cite{delphi99,tyappriv} (LEP 2) }}\\
The hadronic final state analysis accompanied the $F_2^{\gamma}$
data taking in the period 1996-98 for $Q^2$ between  
10 and 1000 GeV$^2$ (for the $e^+e^-$ energy of 163-188 GeV), see 
sec. \ref{sec222}.
The three-component description of DIS events was applied:
the TWOGAM generator was used 
for the QPM (all quarks), the soft hadronic GVMD and RPC 
(with the GS2 parton densities in single and double
resolved photon processes) contributions. 
The results are presented in 
figs. ~\ref{fig:delphi99} and  ~\ref{fig:400pt}.
Good description of the data was obtained.
\newpage
\vspace*{8.5cm}
\vspace*{8.5cm}
~\newline
$\bullet${\bf {L3 98a} \cite{l3_f2_98} (LEP 1)}\\
The study of the  hadronic final state was performed while 
the $F_2^{\gamma}$ data were collected in years 1991-95 
for 1.2$\le Q^2\le$9.0 GeV$^2$ (see sec. \ref{sec222}).
Two bins: $<Q^2>$=1.9 GeV$^2$ in the $x_{Bj}$ range 0.002$< x_{Bj}<0.1$  
and $<Q^2>$=5 GeV$^2$ in the $x_{Bj}$ range 0.005$< x_{Bj}<0.2$ were analysed.
A new method for reconstructing kinematical variables, based
on imposing a transverse momentum conservation, was used in the
analysis.

\begin{figure}[hc]
\vskip 0in\relax\noindent\hskip 4.5cm
       \relax{\includegraphics{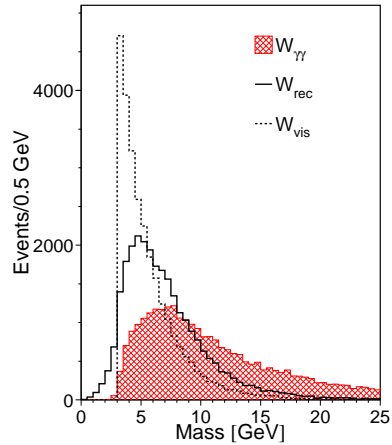}}
\vspace{0cm}
\caption{\small\sl 
Monte Carlo (PHOJET) comparison of the generated two-photon mass $W_{\gamma \gamma}$ distribution with the distributions $W_{vis}$ and $W_{rec}$ 
(from \cite{l3_f2_98}).}
\label{fig:l3_f2_98_2}
\end{figure}

\begin{figure}[hc]
\vskip 0in\relax\noindent\hskip 3.5cm
       \relax{\includegraphics{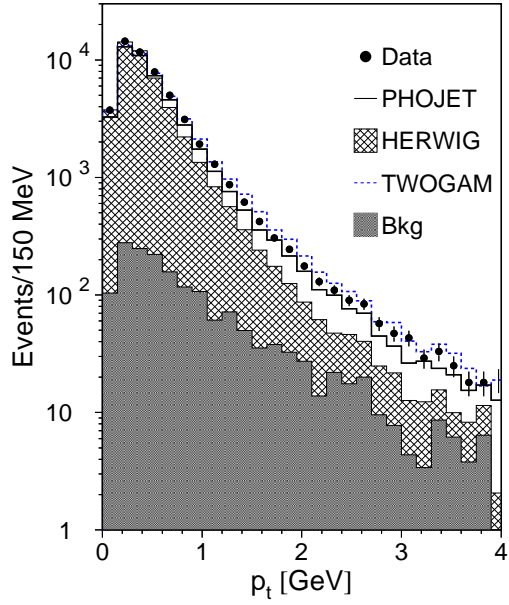}}
\vspace{0cm}
\caption{\small\sl The $p_T$ distribution for charged hadrons: 
data and Monte Carlo predictions (from \cite{l3_f2_98}).}
\label{fig:l3_f2_98_4}
\end{figure}

\begin{figure}[hc]
\vskip 0in\relax\noindent\hskip 3.cm
       \relax{\includegraphics{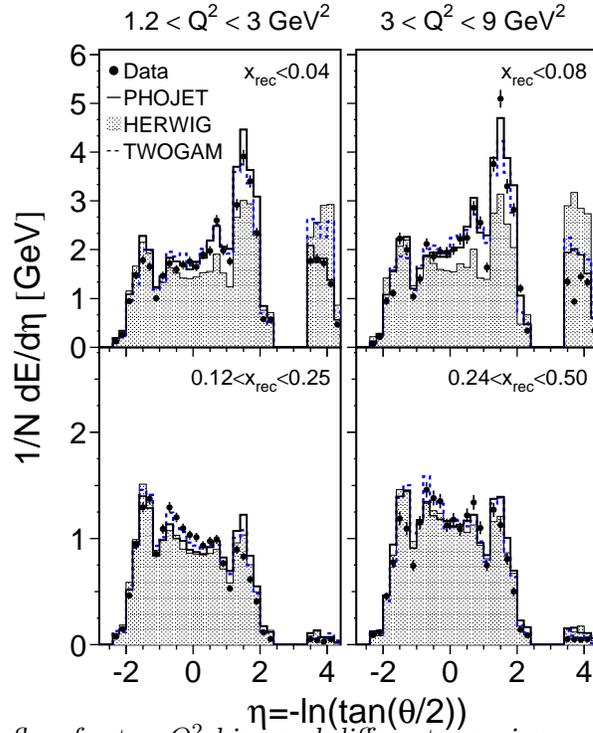}}
\vspace{0cm}
\caption{\small\sl Energy flow for two $Q^2$ bins and different
$x$ regions as a function of the pseudorapidity,
with the polar angle defined with respect to the tagged electron.
Comparison with three Monte Carlo generators is shown
(from \cite{l3_f2_98}).}
\label{fig:l3_f2_98_5}
\end{figure}

The invariant mass of the hadronic system obtained using
PHOJET 1.05 - W$_{\gamma\gamma}$, as well as 
W$_{vis}$ and W$_{rec}$ (from the new method) are compared 
in fig.\ref{fig:l3_f2_98_2}.	
In further analysis of $x_{Bj}$, $p_T$ distributions  
and of the energy flow three Monte Carlo generators : PHOJET 1.05, HERWIG 5.9
and TWOGAM have been used.
The results are shown in figs.~\ref{fig:l3_f2_98_4} and \ref{fig:l3_f2_98_5}.

Two data sets for $F_2^{\gamma}$ based on the PHOJET 
and TWOGAM analyses are presented (they differ up to 28 \%), 
see previous section.

~\newline
Comment: {\it ``A significant improvement is seen in the $W_{rec}$
variable which uses the constraint of transverse momentum conservation.\\
HERWIG disagrees with data in the small $x_{Bj}$ region.'' 
The distribution of the transverse momentum of charged particles 
and energy flow were reproduced by PHOJET and TWOGAM, 
the spectrum of HERWIG was found to be too soft.}
\newline

\vspace*{6cm}
\vspace*{7cm}
\vspace*{8.5cm}
~\newline
$\bullet${\bf {L3 2000  \cite{l3_highq2} (LEP 1)}}\\
The analysis of the hadronic final state 
was performed in the measurement of the structure function for the real 
photon {\underline {and}} virtual photon at $\sqrt s$= 89-92 GeV 
(see also secs. \ref{sec222} and \ref{sec32}).
The large $Q^2$ events , $Q^2$ between 40 to 500 GeV$^2$ 
with $<Q^2>$=120 GeV$^2$, were studied.
The mean virtuality of the target photon is estimated to be 
$<P^2>$=0.014 GeV$^2$ for single-tag events and $<P^2>$=3.7 GeV$^2$
for double-tag events. 
``A new feature in the present work is a fit to
two-photon kinematics imposed on each event.''
 
\begin{figure}[hc]
\vskip 0in\relax\noindent\hskip 3.cm
       \relax{\includegraphics{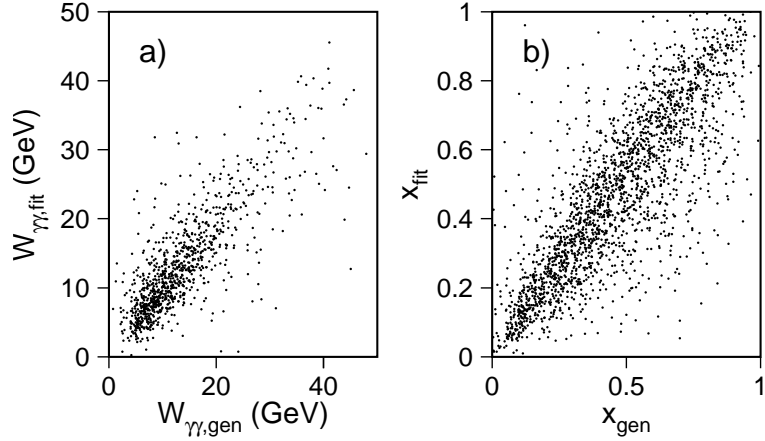}}
\vspace{0cm}
\caption{\small\sl a) Correlation between values of $W$ generated  
and obtained in the unfolding after the fit, b) the same for $x_{Bj}$.
Based on the JAMVG Monte Carlo (from \cite{l3_highq2}).}
\label{fig:l3_highq2_3}
\end{figure}

Single-tag events were analysed using the JAMVG
generator modelling the QPM contribution 
with $N_f$=4, the PHOJET 1.05c with a cutoff 
$p_T^{min}$=2.5 GeV (and charm generated using JAMVG), 
and TWOGAM with $p_T^{min}$=3.5 GeV generating three processes: 
QPM (all quarks), the soft hadronic
VMD and QCD resolved photon contribution (RPC).
Kinematical variables were reconstructed in a fitting procedure
applying constrains from four-momentum conservation.
The correlation between
generated and fitted values for the invariant mass and $x_{Bj}$
is shown in fig.~\ref{fig:l3_highq2_3}.

\begin{figure}[hb]
\vskip 0in\relax\noindent\hskip 3.cm
       \relax{\includegraphics{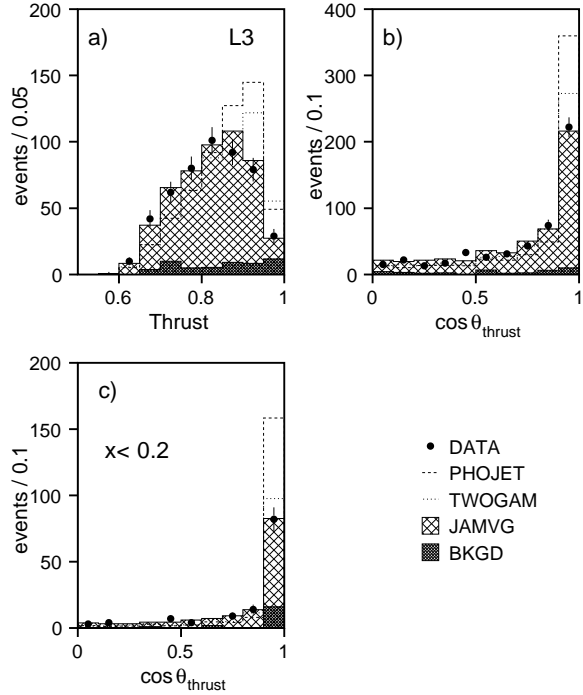}}
\vspace{0cm}
\caption{\small\sl Distributions for the single-tag data. 
Comparison with PHOJET, TWOGAM, JAMVG predictions and with the 
background  estimation is shown. The Monte Carlo distributions 
are normalized to the same number of events as the data (from \cite{l3_highq2}).}
\label{fig:l3_highq2_4}
\end{figure}

\begin{figure}[hc]
\vskip 0in\relax\noindent\hskip 3.cm
       \relax{\includegraphics{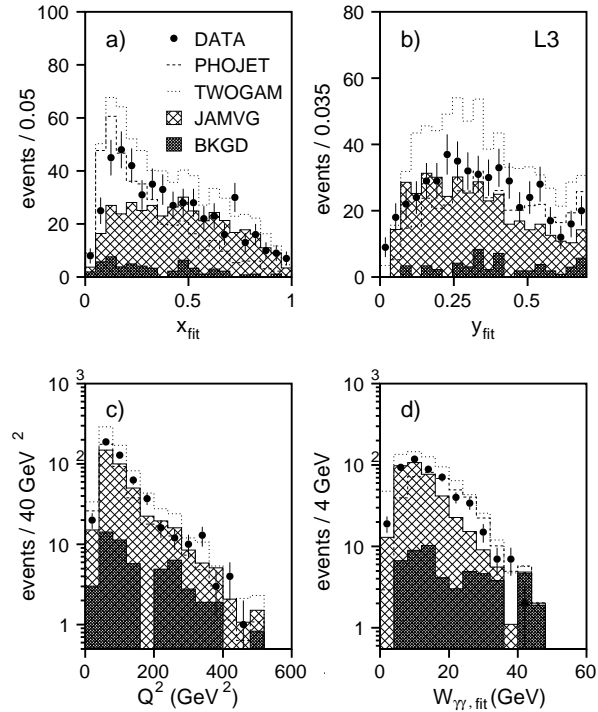}}
\vspace{0cm}
\caption{\small\sl Results from the single-tag events.
a) $x_{fit}$ b) $y_{fit}$ c) $Q^2$ and d) $W_{fit}$ distributions.
Comparison with predictions of the PHOJET, TWOGAM and JAMVG models 
and with an estimation of the background is shown (from \cite{l3_highq2}).}
\label{fig:l3_highq2_5}
\end{figure}

\begin{figure}[hc]
\vskip 0in\relax\noindent\hskip 3.cm
       \relax{\includegraphics{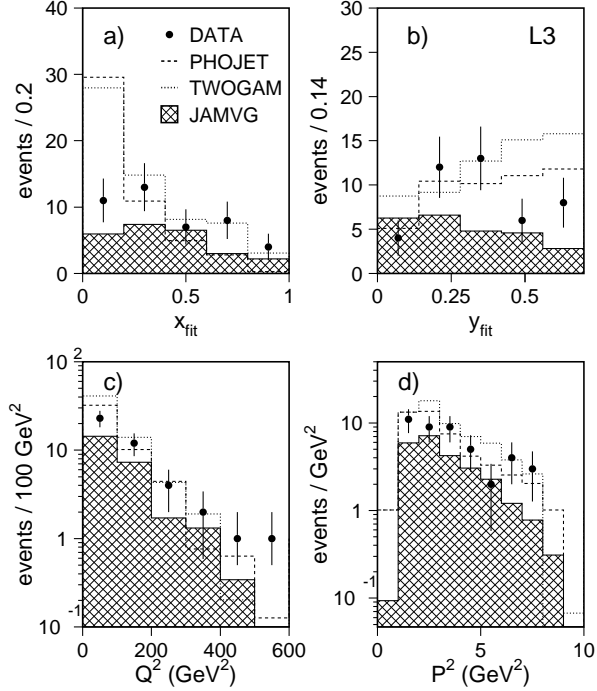}}
\vspace{0cm}
\caption{\small\sl Results from double-tag events.
a), b) and c) as in fig.~\ref{fig:l3_highq2_5}, and d) the  $P^2$ distribution
(from \cite{l3_highq2}).}
\label{fig:l3_highq2_7}
\end{figure}

The Monte Carlo predictions and the data are shown in 
figs.~\ref{fig:l3_highq2_4} and \ref{fig:l3_highq2_5}
for single-tag events. The JAMVG (QPM) is in agreement with the data 
for $x_{Bj}$ above 0.5.  PHOJET agrees with the data at low $x_{Bj}$.
 Similar distributions were studied for double tagged events, see
fig.~\ref{fig:l3_highq2_7}, where some disagreement is seen. 

\vspace*{6cm}
\newpage
\vspace*{8.cm}
\vspace*{8.7cm}
\vspace*{8.1cm}

~\newline
Comment: {\it ``...it is not clear if (this) successful analysis
({\ie unfolding}) can
 be applied also to $Q^2$ and $x_{Bj}$ regions not dominated
by the direct process'' (from \cite{mariak}).
``PHOJET and TWOGAM expectations exceed the data  at low $x_{Bj}$
values'' for double-tag events.}

~\newline
$\bullet${\bf{OPAL 94 \cite{opal94} (LEP 1)}}\\
Data for $F_2^{\gamma}$ 
were collected in 1990-92 for averaged $Q^2$=5.9 and 14.7 GeV$^2$.
 Early analysis of the final hadronic state
(the TWOGEN generator with the contribution based on QPM, 
the soft hadronic VMD process and on the FKP parametrization for
the QCD contribution) was performed.
The parameter $p_T^0 = 0.27 \pm 0.10$ GeV was determined.

~\newline
$\bullet${\bf {OPAL 97a \cite{opal2} (LEP 1)}}\\    
The hadronic final state was analysed in details in the measurement of 
 $F_2^{\gamma}$ with one photon highly virtual ($Q^2$ between 6-30 \g2  - 
{\it low $Q^2$ sample}, and between 60-400 \g2  - {\it high $Q^2$ sample}), 
the other being almost real. 
For generating the hadronic final state the Monte Carlo programs HERWIG 5.8d,
PYTHIA, and for comparison F2GEN were used  with the GRV and SaS1D parton 
parametrizations.

\begin{figure}[hb]
\vskip 3.7cm\relax\noindent\hskip 1.8cm
       \relax{\includegraphics{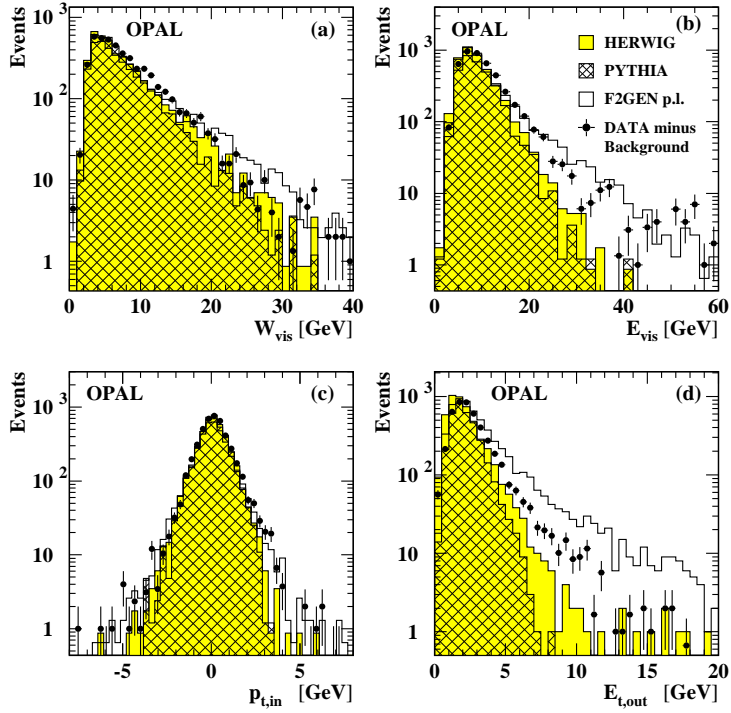}}
\vspace{-3.7cm}
\caption{\small\sl Comparison of the data event quantities in the 
low-$Q^2$ sample with HERWIG, PYTHIA and F2GEN (PL)
Monte Carlo simulations. a) the distribution of the visible
invariant mass; b) the total visible energy of the event;
c) the transverse momentum of the event in the tag plane;
d) the energy out of the tag plane (from \cite{opal2}).}
\label{fig:had4}
\end{figure}

The energy ($E_{t,out}$)
transverse to the plane defined  by the beam axis and the 
tag direction, and other quantities
for the {\it low $Q^2$ sample} are presented in 
fig.~\ref{fig:had4}.
The discrepancy found for the $E_{vis}$ and $E_{t,out}$
distributions in this sample 
(figs.~\ref{fig:had4} b and d, respectively)
is absent in the {\it high $Q^2$ events} (not presented).

\begin{figure}[ht]
\vskip 0.cm\relax\noindent\hskip 2cm
       \relax{\includegraphics{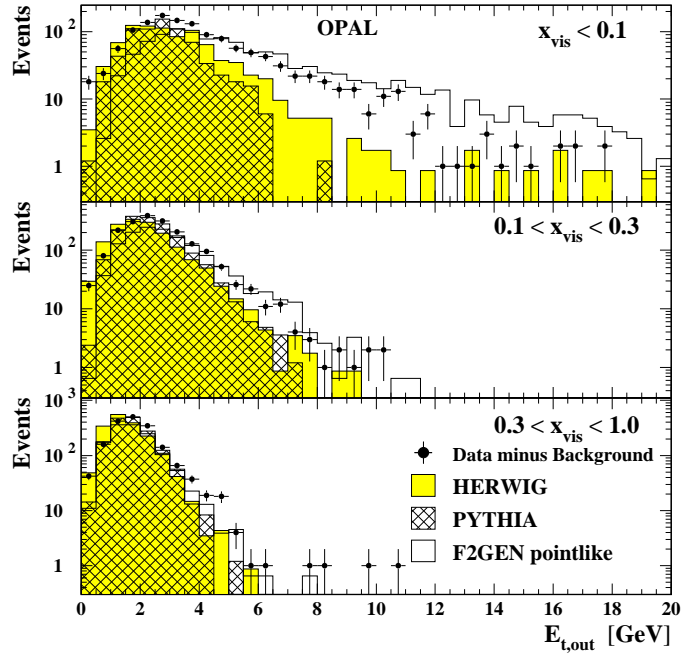}}
\vspace{0cm}
\caption{\small\sl The energy transverse to the tag plane for 
three $x_{vis}$ bins for the {\it low $Q^2$ sample}. 
Monte Carlo generators as in fig.~\ref{fig:had4} (from \cite{opal2}).}
\label{fig:had5}
\end{figure}

The hadronic energy $E_{t,out}$ distributions 
in the three $x_{Bj}$ bins are shown in fig.~\ref{fig:had5}.
The failure of the models in the low $Q^2$ region is 
most visible at low $x_{Bj}$.

\begin{figure}[ht]
\vskip 0.cm\relax\noindent\hskip 2.cm
       \relax{\includegraphics{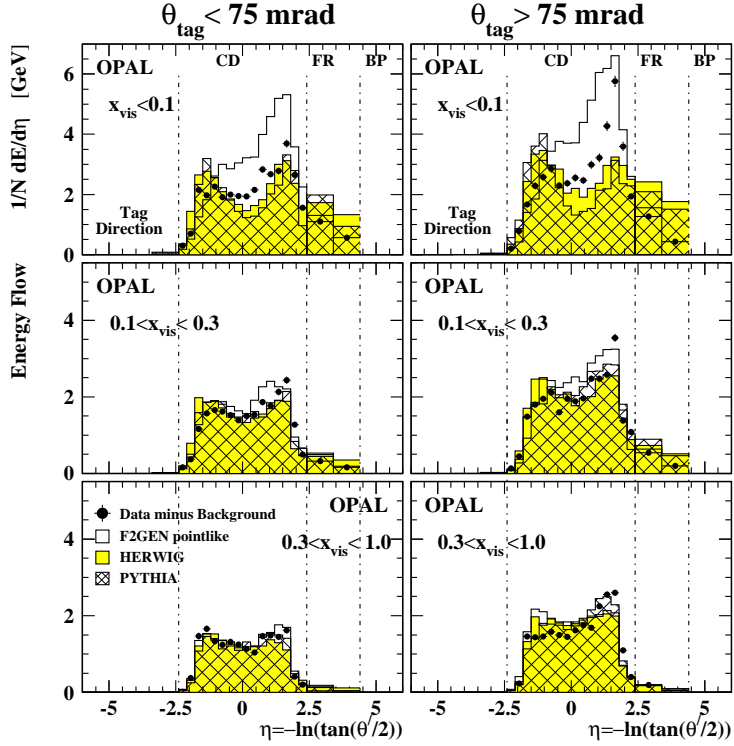}}
\vspace{0.cm}
\caption{\small\sl The hadronic energy flow per event as a function 
of the pseudorapidity $\eta$ for the data and various Monte Carlo simulations,
in various ranges of $x_{vis}$ and $\theta_{tag}$ for the 
{\it low $Q^2$ sample} (from \cite{opal2}.}
\label{fig:had6}
\end{figure}

\begin{figure}[ht]
\vskip 3.5cm\relax\noindent\hskip 3.7cm
       \relax{\includegraphics{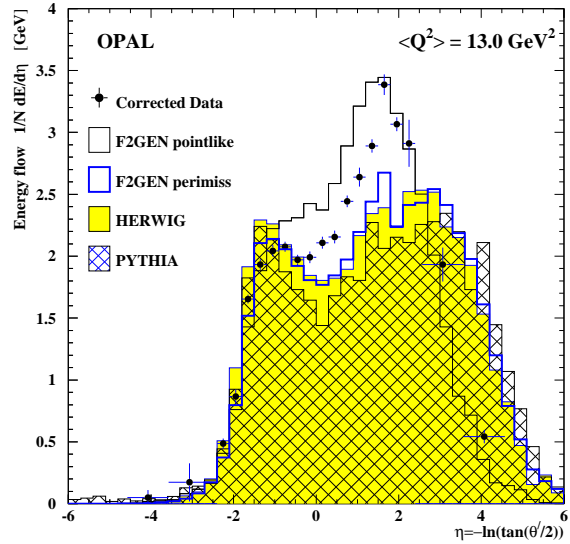}}
\vspace{-3.cm}
\caption{\small\sl The measured energy flow in the low-$Q^2$ sample
corrected for the detector inefficiencies, as a function of the 
pseudorapidity $\eta$, compared to the values generated by various Monte Carlo models
 (from \cite{opal2}). }
\label{fig:had7}
\end{figure}

To establish a source of the discrepancy 
the energy flow per event 
in the {\it low $Q^2$ sample} was also studied as a function of the
pseudorapidity (see fig.~\ref{fig:had6}).
The distribution of pseudorapidity $\eta$ for the 
{\it low $Q^2$ sample}, corrected for the experimental effects,
 is shown in fig.~\ref{fig:had7}.
\newline

\vspace*{8.7cm}
\vspace*{8.2cm}
\vspace*{9.cm}
\vspace*{6.5cm}
\begin{figure}[hb]
\vskip 0.in\relax\noindent\hskip 1.5cm
       \relax{\includegraphics{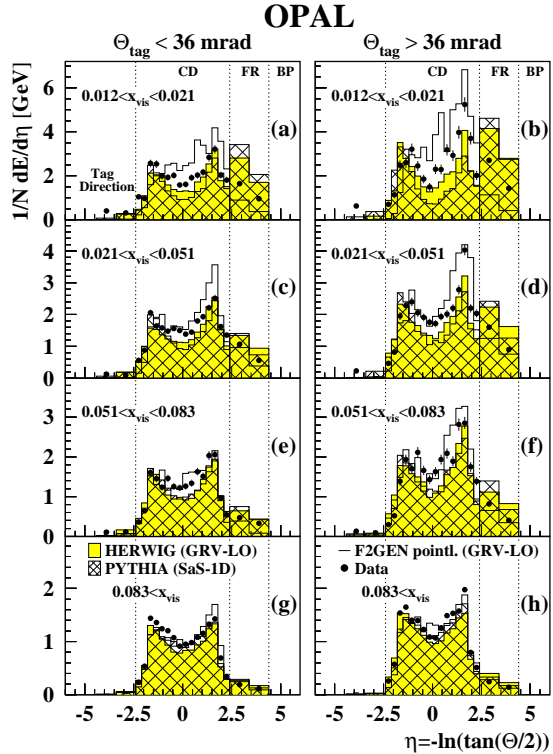}}
\vspace{0ex}
\caption{\small\sl The measured hadronic energy flow as a function of 
the pseudorapidity. Comparison of data with various Monte Carlo 
predictions using the GRV LO and SaS1D parton parametrizations
for three bins in $x_{vis}$ and two bins in $\theta_{tag}$ 
(from \cite{bech}).}
\label{fig:had10}
\end{figure}

~\newline
Comment: {\it 
None of the generators represents the final state accurately; 
$E_{vis}, E_{t,out}$ - distributions as well as the  hadronic flow 
per event show a clear discrepancy.\\
``The failure of the models in the low $Q^2$ region is most marked at
low $x_{Bj}$.'' \\
The differences between the Monte Carlo models and 
the data in the low $Q^2$ region are more pronounced for 
the energy flow per event as a function of the pseudorapidity 
and the azimuthal angle.\\
``Particular attention will need to be given to the 
angular distribution of partons in $\gamma^* \gamma$ system''.

The relation between $p_T^2$ and $Q^2$ in the low $Q^2$ sample may
indicate a need
 to  take into account the structure of the virtual
photon.\\
``The differences between the data and the models contribute significantly 
to the systematic errors on $F_2^{\gamma}$''.} 
~\newline\newline
$\bullet${\bf{OPAL 97b \cite{bech} (LEP 1)}}\\ 
The measurement of $F_2^{\gamma}$ and the modelling of the $\gamma^* \gamma$ 
fragmentation into hadrons at low $Q^2$ region (1.1 to 6.6 GeV$^2$)
and very small $x_{Bj}$ - bins from 0.0025 to 0.2 is reported. 
The hadronic energy flow as a function of $\eta$ for 
three  ${x_{Bj}}$
regions is plotted in fig.~\ref{fig:had10}. 

Differences between the data and Monte Carlo models
(HERWIG, PYTHIA, F2GEN) in the energy flow distributions 
versus the pseudorapidity (fig.~\ref{fig:had10}) and in the summed 
energy transverse to the tag plane (not shown) are found for $x_{vis}<0.05$.\\
\vspace*{8.8cm}
~\newline
Comment: {\it Discrepancy between data and Monte Carlo models in small-$x$
region is found.}
\newpage
~\newline
$\bullet${\bf{OPAL 97c \cite{LP291} (LEP 2)}}
\newline
The data on $F_2^{\gamma}$ at the CM energies 161-172 GeV were
collected in 1996 in two samples: $Q^2$=6-20 GeV$^2$ ($0.02<x_{Bj}<0.6$) 
and  $Q^2$=20-100 GeV$^2$ ($0.05<x_{Bj}<0.8$). The analysis is similar to
that of {\bf OPAL 97a}.  The final hadronic energy flow was studied
using the HERWIG, PYTHIA and F2GEN models.
The LEP 1 data {\bf OPAL 97a, 97b} were used for comparison.
For unfolding the HERWIG program with the GRV parton parametrization was used.
Two types of binning were performed, with $<Q^2>$ = 9, 14.5, 30 and 59 GeV$^2$,
and with $<Q^2>$ = 11 and 41 GeV$^2$.

~\newline
Comment: {\it ``A slightly better agreement between the predictions for 
the hadronic energy flow of the various models and the data in the region 0.1$<x_{vis}<$0.6 is found for the data taken at $\sqrt {s_{ee}}$=161-172 GeV
than for the data collected at $\sqrt {s_{ee}}$=91 GeV. At $x_{vis}<$ 0.1 significant differences persist, as the data prefer a more pointlike hadronic energy flow than assumed in either HERWIG or PYTHIA.''}

~\newline\newline
$\bullet${\bf{OPAL 97d,conf \cite{rooke} (LEP 1)}}\\
An analysis of the hadronic final state was done,
in which the discrepancies between the data and predictions,
reported in {\bf OPAL 97a} \cite{opal2}, were examined from the point
of view of the number of produced jets. The data for
$Q^2\approx 6-30$ GeV$^2$ taken in  years 1994-95 were compared
to the results of the HERWIG, PYTHIA and F2GEN generators.\\

\begin{figure}[ht]
\vskip 0.cm\relax\noindent\hskip 1.5cm
       \relax{\includegraphics{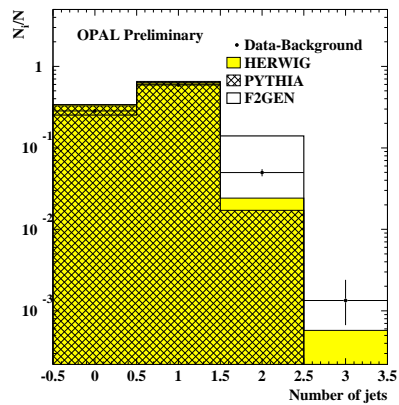}}
\vspace{0.cm}
\caption{\small\sl The fraction of events with 0-3 jets.
The points show the data with the background subtracted, with
statistical errors; histograms were obtained with the HERWIG, PYTHIA
and F2GEN generators
(from \cite{rooke}).}
\label{fig:LP1592}
\end{figure}

The numbers of events for different numbers of jets, divided by
the sum of all events, are
presented in fig.~\ref{fig:LP1592}.\\
\vspace*{4.3cm}

\begin{figure}[hc]
\vskip 0.cm\relax\noindent\hskip 1.5cm
       \relax{\includegraphics{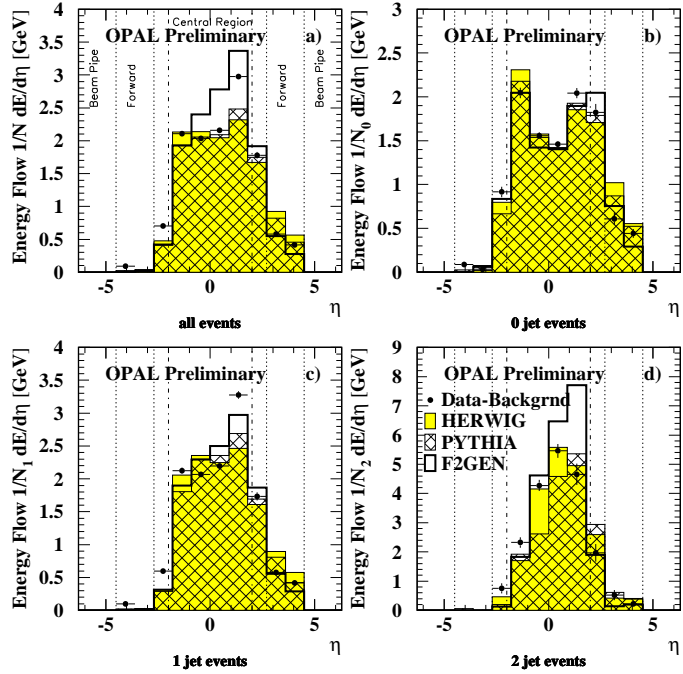}}
\vspace{0.cm}
\caption{\small\sl The hadronic energy flow per event in bins of 
the pseudorapidity $\eta$. Plots a), b), c) and d) show the average 
energy flow per event for summed events, and for events with 0, 1 
and 2 jets, respectively. The samples are represented as in 
fig.~\ref{fig:LP1592} (from \cite{rooke}).}
\label{fig:LP1594}
\end{figure}

The results of further studies of the energy flow versus 
the pseudorapidity $\eta$ for events with different number of jets 
are shown in fig.~\ref{fig:LP1594}.\\

\vspace*{8.cm}

\begin{figure}[ht]
\vskip 0.cm\relax\noindent\hskip 1.5cm
       \relax{\includegraphics{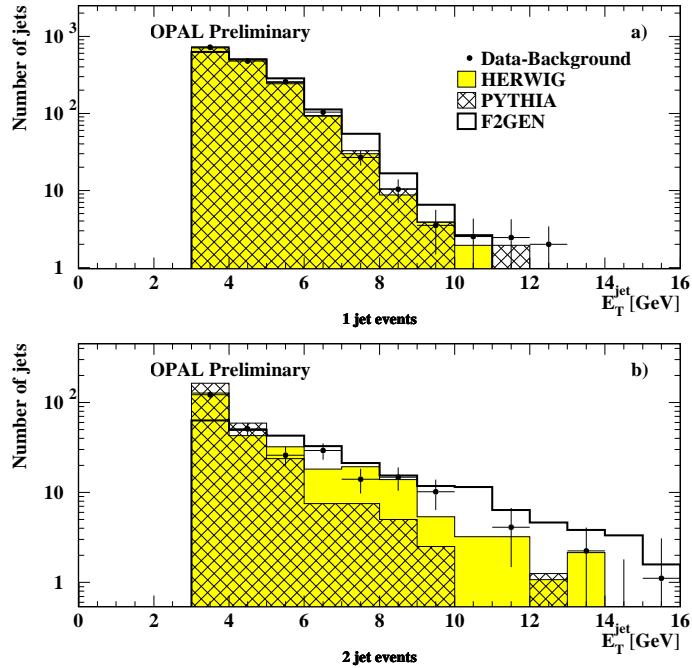}}
\vspace{-0.1cm}
\caption{\small\sl The number of a) one-jet and b) two-jet
events as a function of the jet transverse energy, 
$E_T$. The HERWIG, PYTHIA and F2GEN events are 
normalized to the a) one-jet and b) two-jet data events 
(from \cite{rooke}).}
\label{fig:LP1597}
\end{figure}

The number of events versus the energy transverse to the tag plane, 
obtained by the different generators and observed in the experiment, 
are compared in the fig.~\ref{fig:LP1597}.\\
\vspace*{8.2cm}

~\newline
Comment:{\it ``There is a marked difference between the data
and the Monte Carlo samples in the number of events with 
2 jets''. ``While the F2GEN sample, generated using the 
pointlike approximation, overestimates the 2 jet rate by a
factor of 2.4, HERWIG and PYTHIA are too low by factors of 2.4
and 3.6 respectively.''\\
Disagreement between the data and each of the generators 
for the hadronic energy flow versus the rapidity is seen for all
types of events: 0, 1 and 2-jet.\\   
``All of the Monte Carlo samples model well the $E_{t,out}$
distributions for events with 1 jet, but PYTHIA underestimates 
the $E_{t,out}$ of the events with 2 jets.''}

~\newline
$\bullet${\bf{OPAL 2000 \cite{OPAL_PN389} (LEP 1, LEP 2)}}\\
Hadronic final state analysis accompanying  measurement of $F_2^{\gamma}$
was performed at the CM energies 91, 183 and 189 GeV 
(data from 1993-5 and 1997-8). 
Two detectors were used: the small-angle detector SW allowing 
to access the lowest $x_{Bj}$ region, and the forward detector (FD)
providing the LEP1 data in the same $Q^2$ range as the LEP2 data.
Three samples were studied: 
LEP1 SW with $<Q^2>=$ 1.9, 3.7 GeV$^2$ and $0.0006<x_{Bj}<0.6065$, 
LEP1 FD with $<Q^2>=$ 8.9, 17.5 GeV$^2$ and $0.0111<x_{Bj}<0.9048$
and  LEP2 SW with $<Q^2>=$ 10.7, 17.8 GeV$^2$ and $0.0009<x_{Bj}<0.9048$. 
In the analyses new Monte Carlo programs,
and improved unfolding method - the two-variable unfolding, were used. 
The unfolding of $x_{Bj}$ was done  on a logarithmic scale in order to study 
the low-$x_{Bj}$ region in more detail.

\vspace*{8.8cm}
\begin{figure}[hb]
\vskip 0.in\relax\noindent\hskip -2.3cm
       \relax{\includegraphics{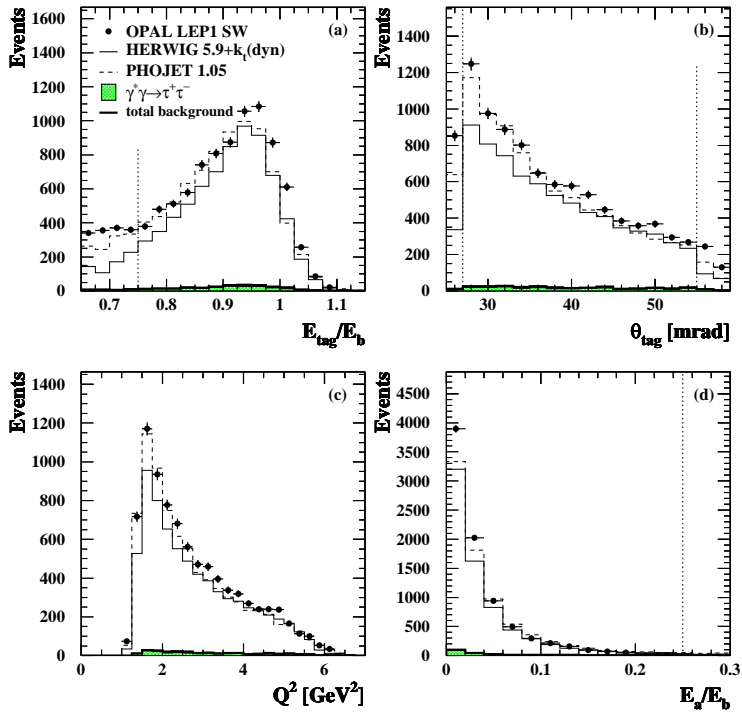}}
\vspace{0.8cm}
\caption{\small\sl 
The comparison of the LEP1 SW data with MC predictions 
(``HERWIG 5.9 + power law $p_t$ (dyn)'' and PHOJET 1.05) for 
a) $E_{tag}/E_b$, where $E_{tag}$ is the energy of tagged electron
and $E_b$ is the energy of the beam, b) electron scattering angle 
$\theta_{tag}$, c) $Q^2$ and d) $E_{a}/E_b$, where $E_a$ is the
highest energy cluster opposite to the tag (from \cite{OPAL_PN389}).}
\label{fig:OPAL_PN389_2}
\end{figure}

The final hadronic state was studied using HERWIG 5.9, 
``HERWIG 5.9 + power law $p_t$ (dyn)'', F2GEN (all with GRV LO 
parametrization as an input function) and PHOJET 1.05 programs.

In fig. \ref{fig:OPAL_PN389_2} the LEP1 SW data are compared 
to the ``HERWIG 5.9 + power law $p_t$ (dyn)'' and PHOJET 1.05 distributions 
for variables related to the scattered electron. Both the 
generators underestimate the data (similar comparisons for LEP1 FD 
and LEP2 SW samples give better agreement - not shown).

\begin{figure}[ht]
\vskip 0.in\relax\noindent\hskip -0.5cm
       \relax{\includegraphics{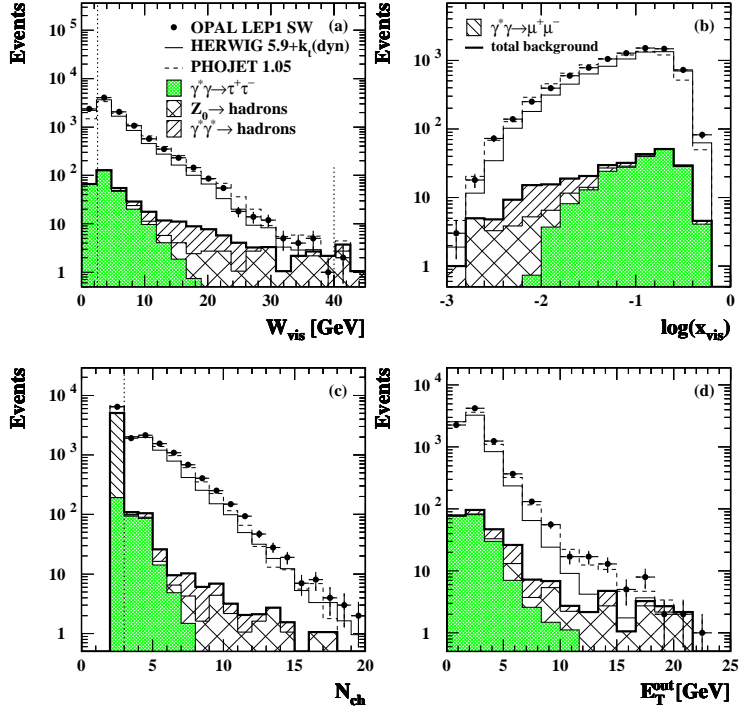}}
\vspace{0ex}
\caption{\small\sl 
The comparison of the data (LEP1 SW) with the MC predictions (``HERWIG 5.9 +
power law $p_t$ (dyn)'' and PHOJET 1.05) for a) invariant mass $W_{vis}$, 
b) $\ln x_{vis}$, c) number of tracks in the event, and d) 
$E^{out}_T$ - the energy out of the tag plane. The dominant background 
contributions are also shown (from \cite{OPAL_PN389}).}
\label{fig:OPAL_PN389_4}
\end{figure}

For the variables related to the hadronic final state there are 
``large discrepancies'' in comparison of data and Monte Carlo,
what can be seen in fig.~\ref{fig:OPAL_PN389_4} where various measured 
distributions for the LEP1 SW sample are shown together with 
``HERWIG 5.9 + power law $p_t$ (dyn)'' and PHOJET 1.05 simulations.

\vspace*{9.5cm}

\begin{figure}[ht]
\vskip 0.in\relax\noindent\hskip -1cm
       \relax{\includegraphics{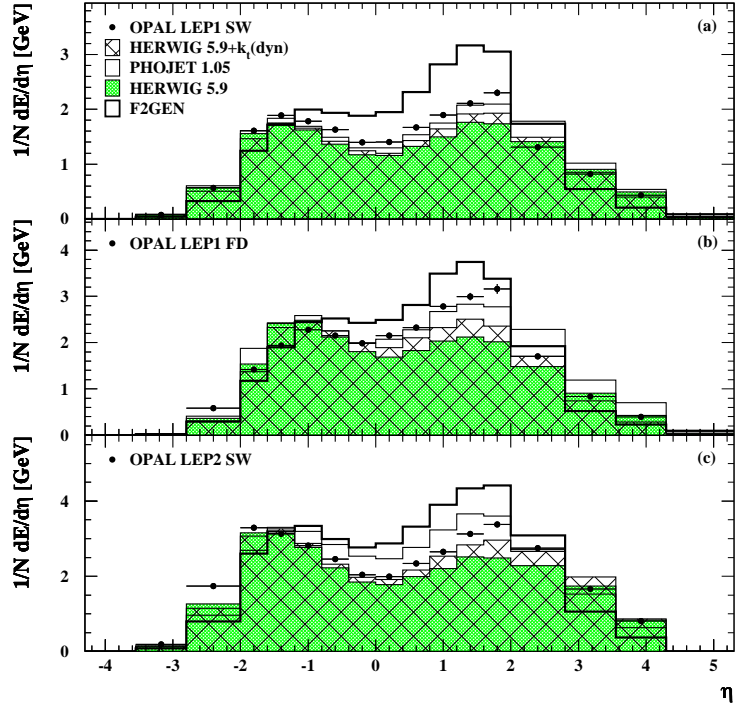}}
\vspace{0ex}
\caption{\small\sl The energy flow per event as a function of the 
pseudorapidity is shown for the a) LEP1 SW, b) LEP1 FD and 
c) LEP2 SW samples in comparison with predictions of F2GEN, 
HERWIG 5.9, ``HERWIG 5.9 + power law $p_t$ (dyn)'' and PHOJET 1.05
 (from \cite{OPAL_PN389}).}
\label{fig:OPAL_PN389_6}
\end{figure}

The hadronic energy flow as a function of pseudorapidity in
various MC models for LEP1 and LEP2 data is presented
in fig. \ref{fig:OPAL_PN389_6}. The HERWIG 5.9 generally
underestimates data, while F2GEN overestimates data
in the rapidity range from -1 to 2.5. The PHOJET 1.05 gives 
the best description of the data for LEP1 SW and LEP1 FD samples.
For LEP2 SW sample ``HERWIG 5.9 + power law $p_t$ (dyn)'' seems to
be in better agreement with data than PHOJET 1.05.

The correlation between the Monte Carlo generated invariant 
mass and its visible, reconstructed and corrected values
was studied (not shown). 

~\newline
\vspace*{8.4cm}

\begin{figure}[hc]
\vskip 0.in\relax\noindent\hskip -1cm
       \relax{\includegraphics{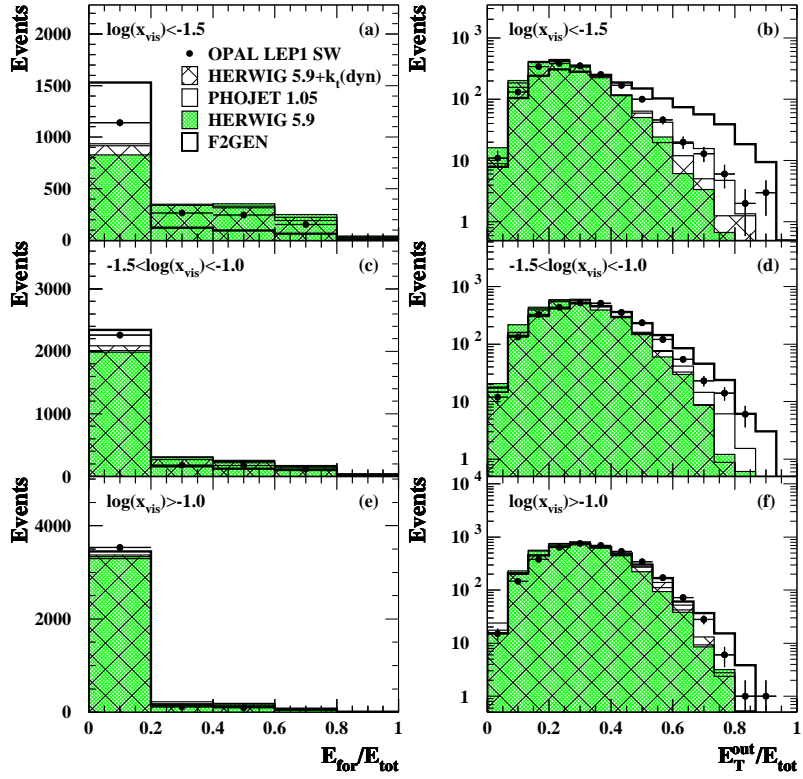}}
\vspace{0.3cm}
\caption{\small\sl  The distribution of the observed energy in
the forward direction divided by the total observed energy,
$E_{for}/E_{tot}$ (a,c,e), and the transverse energy out
of the tag plane divided by the total observed energy,
$E_T^{out}/E_{tot}$ (b,d,f) for three bins of $x_{vis}$.
 The comparison of the data with the predictions of F2GEN, 
HERWIG 5.9, ``HERWIG 5.9 + power law $p_t$ (dyn)'' and PHOJET 1.05 is shown
(from \cite{OPAL_PN389}).}
\label{fig:OPAL_PN389_7}
\end{figure}

In fig.~\ref{fig:OPAL_PN389_7} the usefulness of two choices of the
second, beside $x_{Bj}$, unfolding variable is studied for the LEP1 SW sample.
Both studied variables  $E_{for}/E_{tot}$ and  $E_T^{out}/E_{tot}$
 are sensitive to the angular distribution of the final hadrons.
Similar results for LEP1 FD and LEP2 SW were obtained as well (not shown).

\begin{figure}[ht]
\vskip 0.in\relax\noindent\hskip -0cm
       \relax{\includegraphics{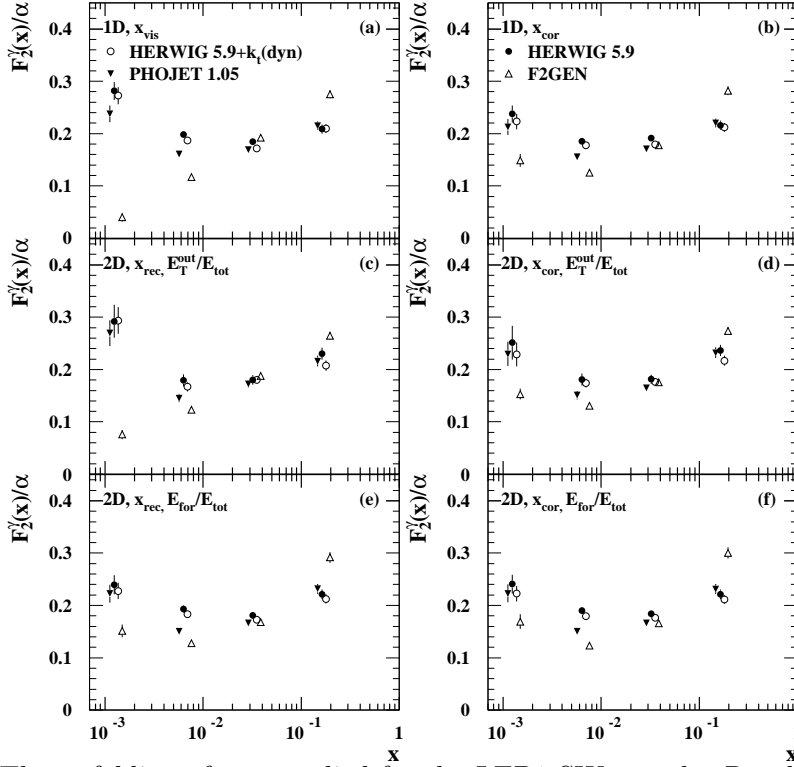}}
\vspace{0ex}
\caption{\small\sl 
The unfolding of $x_{Bj}$ studied for the LEP1 SW sample.
Results based on  one-variable unfolding with a) $x_{vis}$, 
b) $x_{cor}$ and on two-dimensional unfolding with c) $x_{rec}$
and $E_T^{out}/E_{tot}$, d) $x_{cor}$ and $E_T^{out}/E_{tot}$, 
e)  $x_{rec}$ and $E_{for}/E_{tot}$, f) $x_{cor}$ and $E_{for}/E_{tot}$.
The unfolding was done using F2GEN, HERWIG 5.9, ``HERWIG 5.9 + power law 
$p_t$'' (dyn)
and PHOJET 1.05 (from \cite{OPAL_PN389}).}
\label{fig:OPAL_PN389_11}
\end{figure}

\begin{figure}[hc]
\vskip 0.in\relax\noindent\hskip 0cm
       \relax{\includegraphics{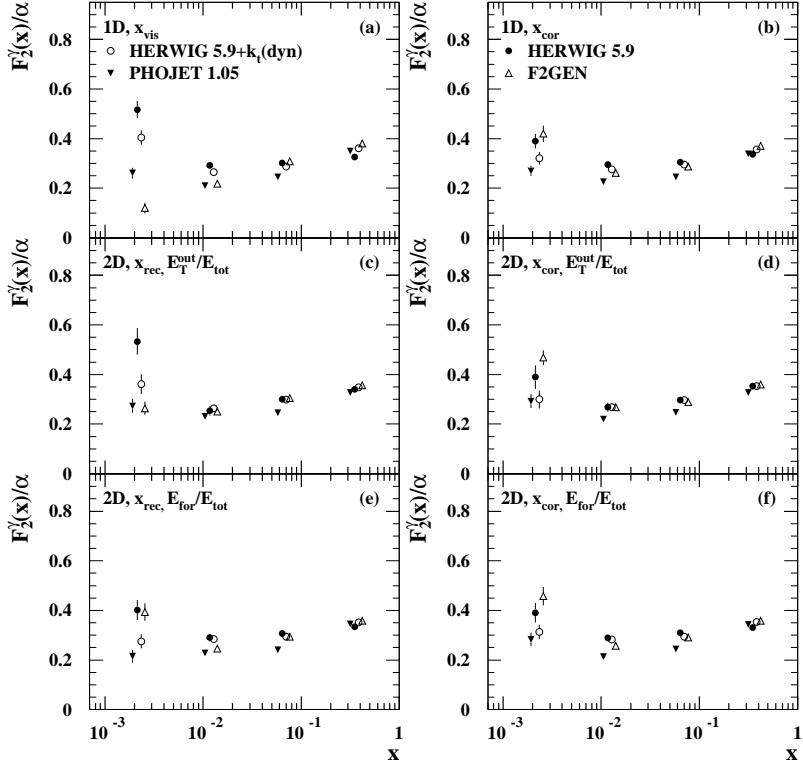}}
\vspace{0.3cm}
\caption{\small\sl As in fig.~\ref{fig:OPAL_PN389_11} for the LEP2 SW sample
(from \cite{OPAL_PN389}).} 
\label{fig:OPAL_PN389_12}
\end{figure}

Test of the unfolding methods was performed, see 
figs.~\ref{fig:OPAL_PN389_11} and \ref{fig:OPAL_PN389_12}, where 
$F_2^{\gamma}$ obtained using different MC models and unfolding 
variables is shown for the LEP1 SW and 
LEP2 SW samples, respectively. Similar test was performed
for LEP1 FD data (not shown). These results allowed to choose
 the most reliable unfolding variables, which are
 $x_{cor}$ and $E_{for}/E_{tot}$. 
\newline
\vspace*{10.cm}
\newpage
\vspace*{10.cm}
\newpage
\vspace*{9.5cm}
~\newline 
Comment: {\it  ``HERWIG 5.9 + power law $p_t$ (dyn)'' 
and PHOJET 1.05 describe the
OPAL data better than the   F2GEN and HERWIG models.\\ 
To reduce the MC ``modelling error,
two-dimensional unfolding has been introduced, using $E_T^{out}/E_{tot}$
as a second unfolding variable.''\\ 
``Monte Carlo modelling of the final state is still a significant source of 
systematic error, but it no longer dominates all other sources.''}

~\newline
$\bullet${\bf {TOPAZ 94 \cite{topaz} (TRISTAN)}}\\ 
The jet production (one and two jets) has been studied for the
first time in  deep inelastic $e{\gamma}$ scattering
(3.0 GeV$^2<Q^2<$30 GeV$^2$), see also sec. \ref{sec222}. Events with 
the transverse momentum  of jets between 2 and 8 GeV and for the $|\eta|<$
0.7 were studied using the jet cone algorithm with R=1 (see next section 
for the definition of the jets).
The Monte Carlo simulation was based on the VMD formula
for the hadronic part. For the perturbative (pointlike) part a QCD
model was used with FKP (uds) and QPM (c) contributions,
with $p_T^0$ =0.5 GeV.

The pointlike
and hadron-like configurations resulting in the different final-state
jet topologies (two jets and single jet, respectively) were studied.
The pointlike part was simulated with the same angular distribution
as for $\gamma\gamma\rightarrow q\bar{q}$.
The hadron-like part involves the struck quark along the $\gamma^*$
direction and the remnant one along the $\gamma$ direction.

~\newline 
Comment: {\it The production of events with  two high-$p_T$ jets is
 consistent with the pointlike perturbative contribution. \\
"...an excess over the pointlike component is observed in the one-jet 
sample, which is direct experimental evidence for the existence of 
the hadron-like component in deep inelastic $e\gamma$ scattering".}

\newpage
\subsection{Large $p_T$  processes in $\gamma \gamma$ 
and $\gamma p$ collisions. Resolved photons \label{sec24}}
The structure function $F_2^{\gamma}$ for the real photon considered 
in sec. \ref{sec22} is sensitive mainly to the combination of quark densities, 
and, moreover, due to its strong  dependence on the charge,
 to the $up$-type quark distributions.
Therefore it is of great importance to have an additional and quite different
from the DIS$_{e\gamma}$-experiments source of information. Hard 
photon(s)-induced processes provide 
such a source, with the cross section given by 
a combination of  different parton densities in the photon(s)  convoluted with
the cross sections for basic partonic subprocesses (see below).
The role of these {\underline{resolved photon}} processes,
\eg  in determining  the  gluon density in the photon,
is unique indeed. They are also crucial in probing the heavy quark content of
the photon, in particular the charm quark density; this topic we discuss
 separately in sec. \ref{sec5}.

As we have already mentioned, the large $p_T$ jet production in $\gamma \gamma$
collision leads to the complementary information on the photon 
structure to that coming from the $F^{\gamma}_2$. 
Moreover, the study of the hadronic final state, with the production of,
predominantly, small $p_T$ particles (jets), is 
a standard part of the structure function analyses 
performed in DIS$_{e\gamma}$ measurements  at LEP and TRISTAN (see discussion in 
secs. \ref{sec22} and \ref{sec23}).   
 
The production cross section of the large $p_T$ jets, which are the hadronic 
representations 
of the hard partons produced in the basic subprocess, does depend 
on the applied jet definition. Although this introduces an additional 
uncertainty in the description of the data, the relatively high rate   makes
large $p_T$ jet production, apart from  $F^{\gamma}_2$, the basic source of 
information on the partonic structure of $\gamma$.
The main goal in such measurements is to extract the individual parton density in the real
photon\footnote{At present most of the data  deal with the almost real photon;
for the jet production in processes with virtual photon, see sec. \ref{sec3}.}.

In this section we focus on the dedicated large $p_T$  
\underline{single} and \underline{double 
jet}\footnote{We count only jets from hard process, not remnant jets.}
production in  the resolved $\gamma \gamma$ and $\gamma p$ collisions, 
where the real photon(s) may interact through its partons\footnote{The 
first evidence for the resolved photon 
contributions  and a need for the gluonic content of photon in $\gamma \gamma$
collision  can be found in 
{\bf AMY 92}. See {\bf H1 92} and {\bf ZEUS 92} for a first evidence for 
 hard processes in the $\gamma p$ collision.}. 
The bulk of data is coming from the HERA collider, where the 
photoproduction of  jets with higher  transverse momentum  
than at LEP and TRISTAN colliders was analysed.
The data  on the three or more jets, the photon remnant or the jet 
shape are  also available.

Beside the jet production also the 
\underline{inclusive one-particle} production
 is  sensitive to the partonic content of the photon,
but it  depends in  addition on  the fragmentation 
functions, or,  as in  the 
\underline{prompt photon} production, on an additional 
assumption (the {isolation criteria}). Nevertheless 
they play a crucial role in establishing 
the structure of the photon, as
 the newest charged particle production measurements at HERA,
 where the gluon content has been derived.
In addition we collect also  the results for
the photoproduction of the prompt photon, which may consist in future 
an additional source of information on the gluon density in the photon.  
The prompt photon production was studied also in $\gamma \gamma$ collision at 
TRISTAN (see {\bf TOPAZ 98,conf}).

In the  previous section we have discussed the single tagged events
with  virtuality of  photons  radiated by the tagged electrons larger 
or much larger than 1 GeV$^2$ (with the exception of one experiment, 
{\bf DELPHI 95}). In this section the initial photons may be considered
\underline{real}, as the median virtuality  of the photon(s) is smaller 
than $\Lambda_{QCD}^2$, while the transverse momenta of the observed jets
are chosen to be much larger than $\Lambda_{QCD}$.

\subsubsection{Theoretical description \label{sec241}}

Below we discuss shortly how the 
{\sl single} and {\sl double} jet production in the photon-induced
processes is described in QCD. 
The specific case of the the charged hadron photoproduction at HERA 
will be described shortly in {\bf H1 99a},
and  the prompt photon photoproduction will be discussed in 
the separate subsection \ref{sec245}.

The production of large $p_T$ jets (or individual particles) may proceed 
directly (the \underline{direct} 
or the (Q)PM contribution), or as a result of partonic
interaction between components of the photon(s) leading to
\underline{resolved} photon contribution. 
In some analyses the soft VMD-type  contributions are included separately.

The resolved photon processes with large 
$p_T$ jets can be characterized by the scale of hardness  $\tilde{Q}^2$,
at which the photon is resolved.
It is usually provided by the $p_T$ of the final jets, \ie $\tilde Q^2 \sim
 p_T^2$. (We introduce here the notation $\tilde{Q}^2$ in order to distinguish 
it from the DIS scale $Q^2$, which is equal to the virtuality of the  photon 
probe.)  
To apply the perturbative QCD
to such  resolved  almost real photons, \ie with $P^2\sim 0$, 
the scale $\tilde{Q}^2$ should be much larger than 
 $\Lambda_{QCD}^2$.   
Events where  virtualities  fulfill the relation  
$\Lambda_{QCD}^2\ll P^2 \ll \tilde{Q^2}$
are discussed in the next section,
where the concept and data on the structure of 
\underbar{virtual} photon will be  introduced. 

\vspace*{3.5cm}
\begin{figure}[ht]
\vskip 0cm\relax\noindent\hskip -0.7cm
       \relax{\includegraphics{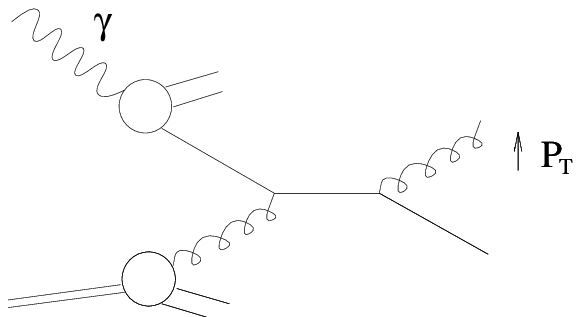}}
\vskip 0.cm\relax\noindent\hskip 6.3cm
       \relax{\includegraphics{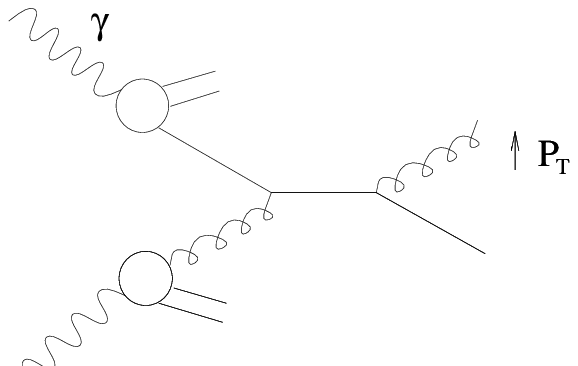}}
\vspace{0ex}
\caption{\small\sl The inclusive large 
$p_T$ jet production in the resolved 
${\gamma}p$ collision (left) and in the (double) resolved
$\gamma\gamma$ collision (right).}
\label{fig:a2}
\end{figure}

Assuming the {\sl factorization} 
between the hard subprocess cross section $\hat{\sigma}$ 
\footnote{The collection of useful formulae for the cross section
$\hat \sigma$  for $2\rightarrow 2$ body partonic processes
 can be found in \cite{combridge}.}
and the parton densities,
the generic LO cross section 
for the two-jet production in  ${\gamma}p$ collision 
(fig.~\ref{fig:a2}, left) is given by 
\be
d\sigma^{{\gamma}p\rightarrow {jet_1}~{jet_2}~X}
=\sum_{i,j}\int \int dx_{\gamma}
dx_{p}f_{i/{\gamma}}(x_{\gamma},\tilde{Q}^2)f_{j/p}(x_p,\tilde{Q}^2) 
\widehat{{\sigma}}^{ij
\rightarrow {jet_1}~{jet_2}} \label{twojets},   
\ee
where $f_{i/{\gamma}}(x_{\gamma},\tilde{Q}^2)$ describes the LL probability 
 of finding a parton of a type ''$i$'' in the photon. 
The corresponding function $f_{j/p}(x_p,\tilde{Q}^2)$ is related to the proton.
The $x_{\gamma}(x_p)$ variable is by definition the part 
of the four-momentum of the photon (proton)
carried by its  parton. The direct interaction of the initial photon
 corresponds to $x_{\gamma}=1$
and $f_{{\gamma}/{\gamma}}=\delta(1-x_{\gamma})$.
 The scale of hardness $\tilde{Q}^2$ 
 plays here a role of the factorization scale between the hard
and soft processes in the $\gamma p$ collision.
Note that the precise definition of the  factorization  scale 
can only be given in the QCD calculation going beyond the LO approach.

For the two-jet production in ${\gamma}{\gamma}$ collision,
${\gamma}{\gamma} \rightarrow {jet_1}~{jet_2}~X$ 
(fig.~\ref{fig:a2}, right) a similar LO formula holds: 
\be
d\sigma^{{\gamma} \gamma \rightarrow {jet_1}~{jet_2}~X}
=\sum_{i,j}\int \int dx_{\gamma}^1
dx_{\gamma}^2f_{i/{\gamma}}(x_{\gamma}^1,\tilde{Q}^2)f_{j/{\gamma}}(x_{\gamma}^2,\tilde{Q}^2) 
\widehat{{\sigma}}^{ij
\rightarrow {jet_1}~{jet_2}} \label{twojet}.   
\ee

If in the final state a specific hadron is measured,
the \underline{fragmentation function} 
will appear in addition in formulae (\ref{twojets})
and (\ref{twojet}), together with the corresponding $z$ variable, 
describing the part of four-momenta of final parton taken by the final hadron.

In the NLO approach, needed to describe the present data, 
the NLL parton densities should be used and the 
$\hat{\sigma}$ should contain  $\alpha_S$ corrections.
Note that the separation between the direct and resolved photon contribution
holds only in the LO approach. 
The theoretical predictions based on the LO \cite{l16} - \cite{l1} or 
NLO (HO) \cite{l2} - \cite{npb507} QCD calculations are available for 
the inclusive one and two-jet
photoproduction in  $e^+e^-$ and $ep$ experiments.

Here we would like to  point out that the way of 
{\sl counting the order} of 
the QCD calculation in the photon-induced processes is still 
a subject of  discussion (see \eg \cite{Chyla}, \cite{NLO-DIC}c 
and earlier papers \cite{DICold}ef). 
The origin of the problem is the presence of the 
definite PM prediction, in form of $\alpha \log (Q^2)$ contribution, 
leading to the inhomogeneous term in the $Q^2$ evolution equations 
for the quark densities in the photon. 
One approach bases on the treatment of the quark density in the photon 
as being of order $q^{\gamma} \sim$ $\alpha/\alpha_s$, 
while in the second one $q^{\gamma} \sim \alpha$.
This difference  leads to different sets of diagrams which 
formally should be included in  the  NLO calculations, see \eg the prompt 
photon production at HERA, where two types of NLO predictions are compared 
to data ({\bf ZEUS 2000a}).
We will not discuss here this issue any further, and in the presentation of 
the results we will use the terminology describing the accuracy of the analysis,
 LO or  NLO,  as given  in the original papers.

The {\sl multiple interaction} between partons
can also be taken into account  in the analysis of the jet production in
$\gamma\gamma$ and $\gamma p$ processes, in analogy to $p \bar p$ processes.
Since the direct events should be free from such multiple
interactions, the $\gamma \gamma$ and $\gamma p$ collisions may
offer a unique laboratory to study this problem.  
The multiple interaction seems to be needed for a good description of the data,
 see below.

\subsubsection{Measurements of the resolved photon processes\label{sec242}} 

At present, the $\gamma \gamma$ collisions arising
in $e^+e^-$ colliders in no-tag\footnote{The name ``untagged'' is also used.
} or antitagged conditions correspond to the
$\gamma \gamma$ events where  both real photons can be described by
 the Weizs\"{a}cker - Williams  energy spectra. In the OPAL experiment 
the maximal  squared target photon masses were equal  $P_{1,2}^2$=
0.8 \g2, while in the ALEPH experiment - around 7 \g2, and in
TOPAZ - 2.6 \g2. Still, the median virtuality in $\gamma \gamma$
collision is small enough (\eg 10$^{-4}$ \g2 in TOPAZ).  
In the $\gamma p$ 
scattering at $ep$ collider HERA a similar Weizs\"{a}cker - Williams  spectrum 
describes the flux of photons coming from the electron
(in year 1996/7 the positron, in 1998 the electron again\footnote{Note 
that starting from 1998 new energies are accessible:
27.5 GeV for $e^-$ and 920 GeV for $p$, see 2.3.4.})- here $P^2<4$ GeV$^2$ 
(but with the median 0.001~GeV$^2$), recently also $P^2<$ 1 GeV$^2$
or below 0.01-0.02 GeV$^2$ if the dedicated  detector is used.

Note that, contrary to the DIS$_{e\gamma}$ experiments, 
in the resolved photon processes there is no one basic observable,
analogous  to  $F^{\gamma}_2$. 
Usually the $p_T$ or $E_T$  and  pseudorapidity $\eta$
distributions for final particle or jet
are compared with the QCD predictions using a specific
parton parametrization.
The agreement with the data justifies {\it a posteriori}
the correctness of the applied  parton parametrization and of 
the QCD calculation.
Only in few cases the 
parton distribution is extracted from data (and only for $\gamma p$
events, see below).

For comparison with  QCD not only the above distributions are important but 
also the study of the structure of the jets (transverse energy
 flow  around the jet axis, \ie jet profile ) and the structure of the 
\underline{underlying event}  ({jet pedestal}) \footnote{\ie the 
transverse energy determined outside of the jet cone},
  corresponding to 
 the hadronic activity outside the jets
where effects due to the  beam {\underline{remnants}}, from spectator partons,
  may be seen. 
Also an additional interaction between these remnants,  
leads to the higher energy level of the underlying event.

For a more  detailed study of properties of the resolved photon
processes the separation between  the direct photon events,
where the photon participates directly in the hard subprocess,
($x_{\gamma}\sim 1$), and the resolved photon events, is
needed.  

In the two-jet events, $x_p$ and $x_{\gamma}$ (for HERA 
measurements)  and  $x_{\gamma}^{1,2}$ (for LEP and TRISTAN data)
 distributions, sensitive to parton densities,
were studied, see \eg {\bf ZEUS 95b, 98a}
and {\bf OPAL 97e}. 
The angular distribution (\eg in the dijet CM system) 
$d\sigma/d\cos\theta^*$, on the other hand, is expected not to be 
sensitive to these ingredients.
So this measurement may help to verify the expectation of  
different angular distributions  due to different subprocesses
which correspond to the direct and resolved photon contributions,
see {\bf ZEUS 96} for the first results, also {\bf OPAL 99a}. 
The first extraction of the gluon density ({\bf H1 95a}) and of 
the \underline{effective parton distribution} 
({\bf H1 98}) was reported by the H1
Collaboration.
The data on prompt photon production have  appeared recently both 
from the  TRISTAN and HERA colliders.

 The main Monte Carlo generators used by the
experimental groups for these kinds of processes are the general purpose 
generators PYTHIA, HERWIG and PHOJET (with or without the multiple parton 
interactions).
The DELPHI collaboration uses the TWOGAM generator to simulate the
QPM, VMD and resolved photon contributions.
All of these MC programs
 use the $p_T^{min}$ cutoffs, separating the soft and hard 
processes in the simulation, being fixed by the requirement to reproduce 
the total cross section. (In case of the multiple interaction $p_T^{mi}$
parameter is used instead of $ p_T^{min}$ ). 
For both  $e^+e^-$ and $ep$ colliders  the additional power
law $p_t$ spectrum has been  included recently in the HERWIG and PYTHIA 
programs. They were used in
\eg {\bf ALEPH 97a,conf, 99a, 99b,conf; OPAL 2000;
TOPAZ 98,conf; ZEUS 95c} (where the spectrum  was first introduced).

The single and double jet samples were studied, reaching 
the highest $p_T$ of jets in $\gamma \gamma$ processes  up to 17 GeV. 
Jets at the HERA collider 
are observed up to 75 GeV ({\bf ZEUS 98c,conf} and {\bf 99b,conf}).   

The jet $E_T$ distributions
are in general in agreement with the expectations for both 
single and double
jet production, both at $e^+e^-$ and $e^{\pm}p$ colliders.
In the pseudorapidity dependence the discrepancy is
observed for the jet rates. The
transverse energy flow around the jet axis, especially for small $E_T$
and small $x_{\gamma}$ is not properly described by existing Monte Carlo generators. Also for the two-jet events it has been found that 
distribution at the small $x_{\gamma}$ lacks proper description.
The agreement with the data
is improved when the  additional power
law $p_t$ spectrum is taken into account  in the analysis,
as in $F_2^{\gamma}$ and hadronic final state analysis.
A need for extra $p_t$ in the simulation of the whole event
 can also be  taken as a
hint that the multiple scattering may be important
in the photon-induced processes.
The multiple interaction included in the experimental analysis
is modelled usually as  in the $p {\bar p}$ experiments 
(see \eg {\bf H1 96a}).
Global study of the photon-induced jet  production with
 the aim of tuning 
the general purpose Monte Carlo models was performed recently 
in \cite{butter-jets}. ``The main ingredients investigated 
in the tuning are the description of the ``underlying event''
and the choice of photon structure''. The conclusion of this analysis 
is that ``the agreement without the multiple interaction is very poor''.

The strong dependence on the 
proper choice of the jet definition and the parameters
like the cone size $R$ (see below),  has also been  observed in the jet 
production
at HERA. There is a  possible relation of this effect
to the problem of describing the
 underlying events observed both in $e^+e^-$  and in $ep$
collisions, that was mentioned above.

The following two \underline{jet definitions} 
are used in the analyses reported below. 
The first is the \underline{jet cone algorithm} 
with the fixed value of the cone variable $R$, 
 defined as $R=\sqrt{(\delta \phi)^2+(\delta\eta)^2}$, with  $\delta\phi$ and 
$\delta\eta$ describing the differences between the cone (jet) 
axis and the particle direction in the azimuthal angle and  
pseudorapidity, respectively. It is  used in PUCELL and EUCELL algorithms.
The second one corresponds to the 
\underline{$k_T$ - cluster algorithm} on which the 
KTCLUS approach is built up. Beside  $R$, also $R_{sep}$ is included in some 
analyses using the cone algorithm. It  corresponds to the additional 
separation between partons, and  the $R_{sep}=2 ~R$ means no restriction. 
(The discussion of jet definitions in the NLO calculation can be found 
\eg in \cite{kkk}.)

The difference of the two jets' transverse energy 
$\Delta E_T$ = $E_{T\,1}$-$E_{T\,2}$
and  analogous difference  for the pseudorapidities, 
$\Delta \eta=|\eta_1-\eta_2|$, are introduced in two-jet events.
(For the analysis of the energy of the underlying event the variable
$\delta \eta=\eta_{cell}-\eta$ is used.) In the analysis of 
two-jet events there appear also the average pseudorapidity 
$\bar \eta=(\eta_1+\eta_2)/2$ (denoted also by $\eta_{jet(s)}$)
 and the average 
transverse energy of jets, $\bar {E_T}$ $(\bar E_{jet(s)})$. 

It is worth noticing the difference between the variables 
describing the produced {\sl parton} and the corresponding quantities
for the {\sl jet}, representing the considered  parton
on the hadronic level. The    unfolding procedure is needed in order to
obtain the "true" partonic variables,
\eg the transverse momentum of the parton
($p_t$) from the transverse momentum or energy 
of the jet ($p_T$ or $E_T$)\footnote{We keep in subscripts small 
letters for  quantities related to the partonic level}.

%
%
\newpage
\subsubsection{Jet production in $\gamma\gamma $ collisions\label{sec243}} 

Here we present the results for the jet production in ${\gamma}{\gamma}$ 
collisions, where  {\it one} or  {\it two} resolved photons may interact. 

As in the previous section  we will denote 
the  virtualities of two involved photons by $P^2_1$
and $P^2_2$. 
The antitagged events at LEP 
correspond to $P^2< $ 4 GeV$^2$, with median approximately 10$^{-4}$ GeV$^2$. 
For the antitagging conditions in {\bf OPAL 97e} 
the maximum value  $P^2$ = 0.8 GeV$^2$ was used.
At TRISTAN  the $P^2$ value in antitagging case lies between 10$^{-8}$ and 
2.6 GeV$^2$ with the mean value $\sim$10$^{-4}$ GeV$^2$.

In $e^+e^-$ machines the parton momentum fractions,  variables 
$x_{\gamma}^{1,2}=x_{\gamma}^{\pm}$, are determined from the two final jets 
with the highest $E_T$, according to the formulae: 
\begin{eqnarray}
x_{\gamma}^{\pm}={{{\Sigma}_{jets}(E\pm p_z)}\over{{\Sigma}_{hadrons}
(E\pm p_z)}}.
\label{xpm}
\end{eqnarray}
Variables referring to the dijet centre of mass system are (in {\bf OPAL 99b})
denoted by star, \eg $\eta^*$.\newline
Below we present the LEP and TRISTAN results.

~\newline\newline
\centerline{\huge DATA}
\newline\newline
$\bullet${\bf{ALEPH 93  \cite{al93} (LEP 1) }}\\
The presence of hard scattering processes is demonstrated in the sample 
from 1990-1991 data. The antitagging condition corresponds here to 
$<P^2_{1,2}>$ = 0.23 GeV$^2$. The VMD, QPM and QCD contributions, 
the latter with $p_T^{min}$ cutoff,  were studied using Vermasseren 
program for QPM, and
TOPAZ MC generator for QCD part. The  $p_T^{min}$ was fitted to be 
 2.5$\pm0.1\pm0.3$.\\
~\newline 
Comment:{\sl ``The data are well described by a combination of three models, namely VMD, QPM and QCD''. \\
The fitted $p_T^{min}$ parameter is  ``higher than the values previously obtained at TRISTAN but consistent with recent HERA photoproduction results''.}
\newline\newline
$\bullet${\bf{DELPHI 94 \cite{delphi94} (LEP 1) }}\\ 
Data were collected in the 
period 1990-92 with  $<P^2_1>\sim 0.12 $ 
\g2. Events are groupped into a minimum bias sample (sample I) and a high $p_T$
jet events sample (sample II). Jets, defined according to the Lund cluster algorithm,
with $p_T$ greater than 1.75 GeV, and a polar angle between 40$^0$ and 140$^0$,
 were observed. The three component  model (VMD for a fully nonperturbative contribution, QPM, and QCD with the $p_T^{min}$)
was used in the analysis. 

\begin{figure}[hb]
\vskip 0.in\relax\noindent\hskip 1.5cm
       \relax{\includegraphics{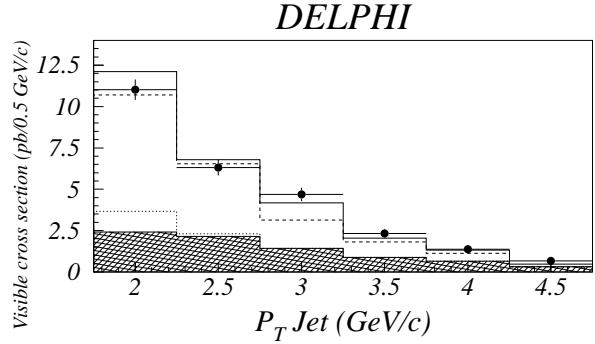}}
\vspace{0ex}
\caption{\small\sl Distribution of jet transverse momentum compared
with the Monte Carlo  predictions (TWOGAM generator). 
Dark area - QPM only; dots - QPM + VMD;
full line - QPM + VMD + DG ($p_T^{min}$ = 1.45 GeV/c);
dashed line - QPM + VMD + DO ($p_T^{min}$ = 1.22 GeV/c)
(from \cite{delphi94}).
}
\label{fig:jet3}
\end{figure}

Results for the jet $p_T$ distributions are shown in  
fig.~\ref{fig:jet3}.  
To model the QCD (resolved photon) contribution 
the FKP approach (see secs. \ref{sec222}, \ref{sec23}) was used with
DG and DO parton parametrizations.
For a comparison also the LAC1 and GS densities were used.
\vspace*{4.2cm}

~\newline
Comment: {\sl `` A three-component model which includes QCD hard scattering subprocesses is needed to describe the data.''\\
Remnant jets, generated along the direction of the initial quasi-real photon,
are studied within the FKP approach.\\
``Each parametrization is associated to a special value of $p_T^{min}$, 
constrained by the description of the visible total cross section''.}

~\newline
$\bullet${\bf{DELPHI 95 \cite{delphi3} (LEP 1) }}\\  
The measurement of  hard processes in $e\gamma$ collision was
performed based on 1991-1992 data, as described in the previous section.
The  average  virtuality of the   
photon from the tagged electron, registered by VSAT, 
was equal to $<Q^2(=P_1^2)>\approx$ 0.06 GeV$^2$ (single-tag events). 
The values of  $E_T$ of observed jets were larger than  1.5 GeV.

Comment: {\sl 
``This study confirms recent results from no-tag experiments, 
indicating that the photon has a significant partonic content.''
}

~\newline\newline
$\bullet${\bf{DELPHI 97b,conf \cite{zimin} (LEP 1, LEP 2) }}\\  
Based on the LEP1 and LEP2  data (1995 and 1996 runs),  with CM energy 172 GeV,
various distributions for the hadronic final state in anti-tag events
were measured. The TWOGAM generator was used in the analysis
with soft VMD, QPM and QCD-RPC (with $p_T^{min}$ cutoff) contributions. 
Different  parton parametrizations for the photon
were used assuming the ``leading order QCD factorization'',
which lead to different values of $p_T^{min}$:
$p_T^{min}(GS2)$ = 1.88$\pm$0.02 GeV and 
$p_T^{min}(GRV)$ = 1.58$\pm$0.018 GeV. 

\begin{figure}[hb]
\vskip 0.in\relax\noindent\hskip 1.5cm
       \relax{\includegraphics{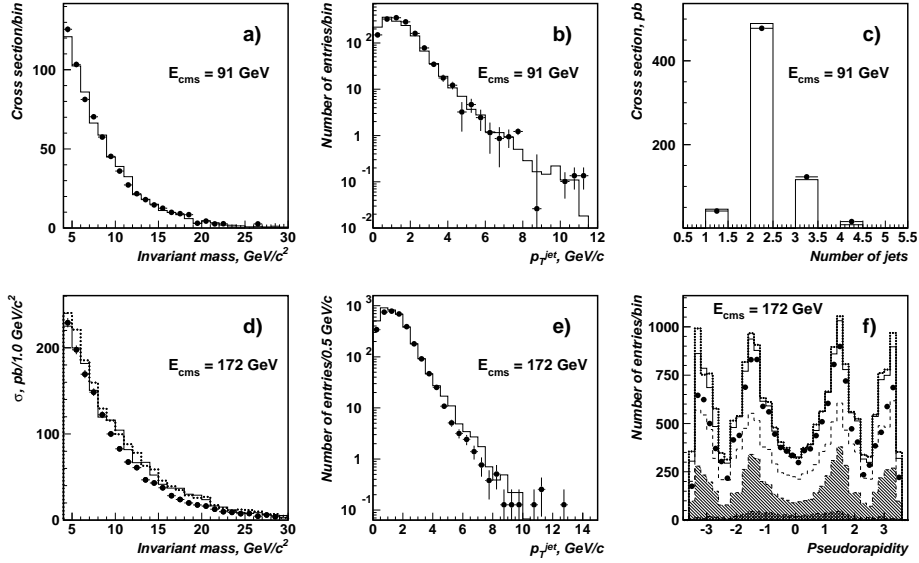}}
\vspace{0ex}
\caption{\small\sl Distributions of: a), d) invariant mass,
b), e) $p_T$, c) number of reconstructed jets,
f) pseudorapidity.
 Solid and dotted lines - predictions based on VMD + QPM + RPC
with the GS2 and GRV parton densities, respectively.
The dashed line - RPC contribution with the GS2 parton densities; the hatched
histogram - the VMD part, the double hatched - the QPM part.
Upper (lower) figures correspond to $\sqrt s$=91 (172) GeV
(from \cite{zimin}).}
\label{fig:jet4}
\end{figure}

Results  are presented in fig.~\ref{fig:jet4}.

\vspace*{7.2cm}

~\newline
Comment:{\it ``Monte Carlo prediction (based on TWOGAM generator)
 gives perfect agreement 
at CM energy = 91 
\gev and slightly exceeds data at   172 \gev''.} 
~\newline\newline
$\bullet${\bf{OPAL 97e \cite{as4} (LEP 1.5)}}\\    
The inclusive one and two-jet 
cross sections have been measured
at $\sqrt{s_{ee}}$ = 130 and 136 GeV (based on the 1995 run).
The antitagging condition was applied 
to  the initial photons corresponding 
to the maximum photon virtuality $P^2_{max}\approx$ 0.8 GeV$^2$. 
The transverse energy of jets $E_T$ is taken to be larger than 3 \gev 
and the \psr lies  within $|\eta |<1.$
 Generators PYTHIA 5.721 (with the SaS1D parton parametrization)
and PHOJET 1.05 (with the GRV LO parton parametrization) were used.
 The cone jet-finding algorithm with R=1 was used (for the first time 
in photon - photon collisions at LEP).

\begin{figure}[ht]
\vskip 0.in\relax\noindent\hskip 1.cm
       \relax{\includegraphics{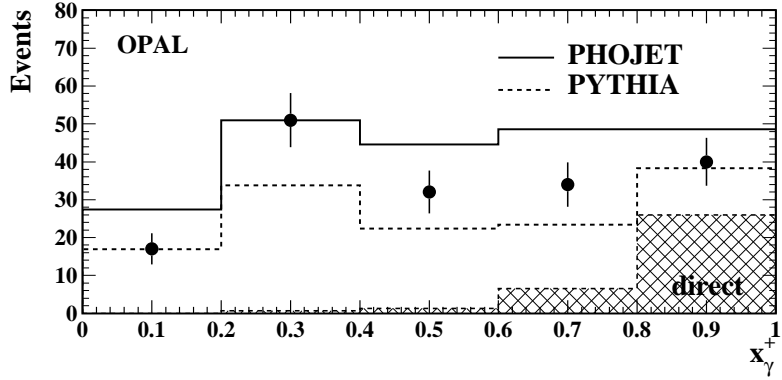}}
\vspace{0ex}
\caption{\small\sl The number of two-jet events as a function of
$x_{\gamma}^+$ compared to the PHOJET (solid line) and PYTHIA
(dashed line) simulations. The hatched histogram is the direct contribution
to PYTHIA events (from \cite{as4}).}
\label{fig:jet5}
\end{figure}

\begin{figure}[ht]
\vskip 0.in\relax\noindent\hskip 1.5cm
       \relax{\includegraphics{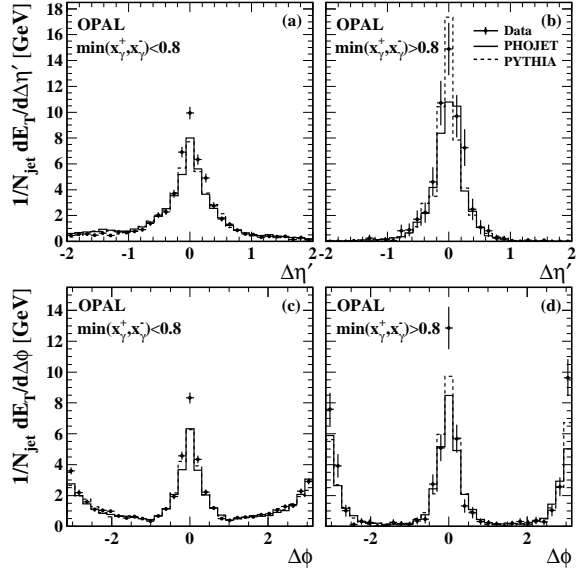}}
\vspace{0ex}
\caption{\small\sl The transverse energy flow around the jets in 
two-jet events. b), d) Direct events 
(min$\{x_{\gamma}^+,x_{\gamma}^-\}>$ 0.8). a),  c) Resolved
events (min$\{x_{\gamma}^+,x_{\gamma}^-\}<$ 0.8).
The data are compared with the PHOJET (solid line)
and PYTHIA (dashed line) simulations. The rapidity $\delta\eta$ distribution 
is shown in a) and b), while the $\delta\phi$ angle distribution
in c) and d) (from \cite{as4}).}
\label{fig:jet6}
\end{figure}

For the two-photon events, the $x_{\gamma}^{\pm}$ distribution 
and the transverse energy flow around the jet, studied separately 
for the direct (min$\{x_{\gamma}^+,x_{\gamma}^-\}>0.8$) and the resolved  
(min$\{x_{\gamma}^+,x_{\gamma}^-\}<0.8$) photon  contributions, were obtained 
(see fig.~\ref{fig:jet5} and fig.~\ref{fig:jet6},  respectively).

\vspace*{4.8cm}
\vspace*{7cm}
\begin{figure}[ht]
\vskip 0.in\relax\noindent\hskip 1.5cm
       \relax{\includegraphics{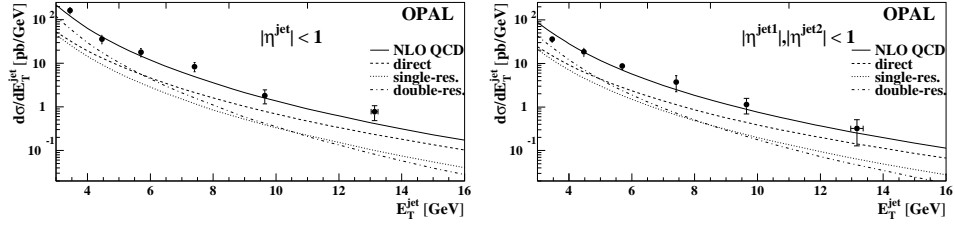}}
\vspace{0ex}
\caption{\small\sl The inclusive one-jet (left) and two-jet 
(right) cross section as a function of $E_T$ for the jets with 
$|\eta| < 1$ compared to the NLO calculations 
\cite{klek}. The solid line is the sum of direct,
single resolved and double resolved contributions that are also shown
separately (from \cite{as4}).}
\label{fig:jet7}
\end{figure}

The one-jet and two-jet cross sections 
${{d{\sigma}}\over {dE_T}}$ 
were measured up to $E_T$= 16 GeV. 
Results are presented in fig.~\ref{fig:jet7}.

\newpage
\vspace*{2.1cm}

\begin{figure}[ht]
\vskip 0.in\relax\noindent\hskip 1.5cm
       \relax{\includegraphics{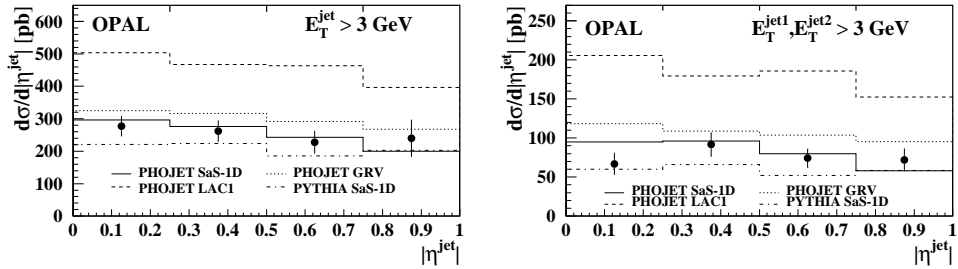}}
\vspace{0ex}
\caption{\sl The inclusive one-jet (left) and two-jet (right)
cross section as a function of $|\eta|$ for the jets with
$E_T\ >$ 3 GeV compared to the LO QCD calculations
of the  PYTHIA (with the SaS1D, LAC1 and GRV parton distributions) and 
PHOJET (with the SaS1D parton distributions) generators (from \cite{as4}).}
\label{fig:jet8}
\end{figure}

The $\eta$ distributions for one and two-jet events were studied 
as well (see fig.~\ref{fig:jet8}).

\vspace*{3.2cm}
~\newline
Comment: {\it "The data on  $\eta$ distributions  agree well with NLO QCD
calculations based on GRV parametrization.
The GRV LO and SaS1D parametrizations describe the data equally well,
the LAC 1, however, gives twice the observed value".
\newline
The large difference between Monte Carlo generators with a fixed
parton parametrization, and for a particular Monte Carlo
generator with different parton parametrizations is observed.}

~\newline
$\bullet${\bf{OPAL 99a  \cite{pr250} (LEP 2)}}\\
The dijet production in two-photon collisions at $e^+e^-$ energies 
161 and 172 \gev in the ``antitagging condition''  was measured.
The transverse energy of the jets is taken to be  $E_T>3$ \gev 
and the \psr lies within $|\eta |<2.$
The Monte Carlo generators PYTHIA 5.722 and PHOJET 1.05c were used with
the  SaS1D
and GRV LO parton parametrizations, respectively (with $p_t^0=1.4$ GeV, as a default value). Both generators use the  multiple 
interaction  to model the underlying events (a default $p_T^{mi}$=1.4 GeV).
The cone jet-finding algorithm with R=1 was used. 

The contributions of direct and  resolved photon subprocesses 
($x_{\gamma}^{\pm}\ ^>_<$ 0.8)
were studied for various distributions.

\begin{figure}[hb]
\vskip 0.in\relax\noindent\hskip 2.2cm
       \relax{\includegraphics{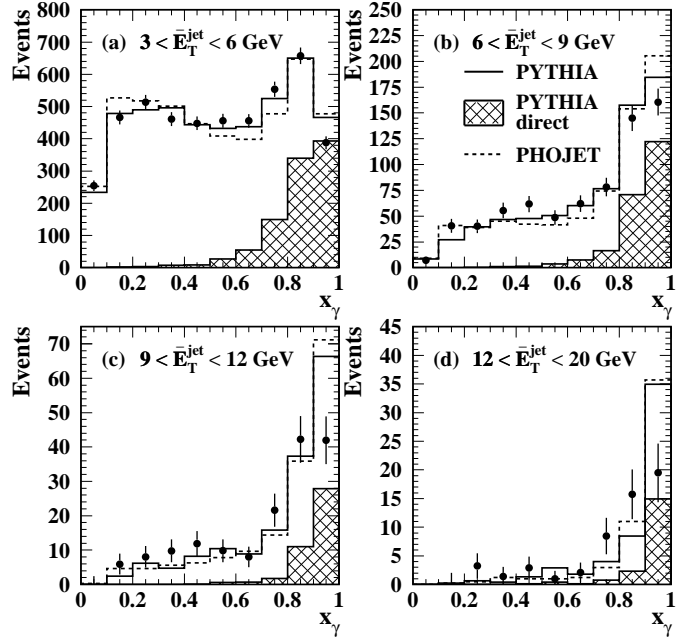}}
\vspace{-1.3cm}
\caption{\small\sl 
The uncorrected $x_{\gamma}$ distributions in bins of $\bar {E}_T$, 
compared to PYTHIA and PHOJET predictions
(from \cite{pr250}).}
\label{fig:jet91}
\end{figure}

The $x_{\gamma}^{\pm}$ distributions were measured for different 
$\bar E_T$ bins, from 3 to 20 GeV, showing the large contribution
of the resolved photon processes (fig.~\ref{fig:jet91}).
\newline
\vspace*{9.7cm}

\begin{figure}[hb]
\vskip 0.in\relax\noindent\hskip 0.cm
       \relax{\includegraphics{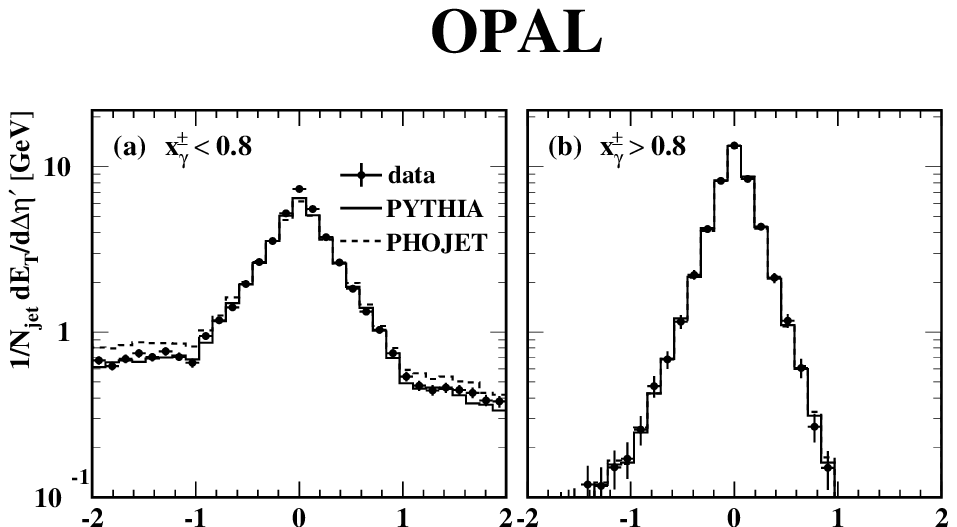}}
\vspace{0ex}
\caption{\small\sl The transverse energy flow around the jet axis
as a function of $\delta\eta$
in two-jet events; a) double resolved events, (b) direct events.
Comparison with PYTHIA (solid line) and PHOJET (dashed line)
(from \cite{pr250}).}
\label{fig:jet9}
\end{figure}

The transverse energy flow around the jet axis is presented in 
fig.~\ref{fig:jet9}
for the double resolved and direct photon processes.
Larger hadronic activity for the resolved photon sample
is observed as expected.

\vspace*{5.5cm}

\begin{figure}[ht]
\vskip 0.in\relax\noindent\hskip 1.5cm
       \relax{\includegraphics{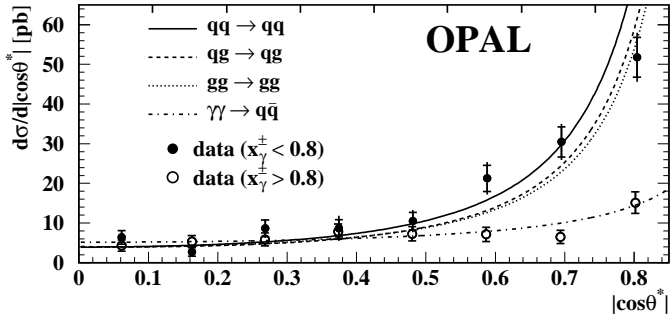}}
\vspace{0ex}
\caption{\small\sl The angular distribution of events with $E_{T1}$
= $E_{T2}$ and collinear in $\phi$, separated into 
the direct and the double resolved contributions, together with the 
LO QCD expectations based on matrix elements \cite{combridge}
(from \cite{pr250}).}
\label{fig:jet10}
\end{figure}

The angular dependence plotted separately for the three basic resolved
photon subprocesses and for the direct events 
is presented in comparison with matrix element predictions \cite{combridge}
in fig.~\ref{fig:jet10}. A comparison with the NLO calculation with 
the GRV HO parton parametrization was also made (not shown).
\newline

\vspace*{3.8cm}

\begin{figure}[ht]
\vskip 0.in\relax\noindent\hskip 1.5cm
       \relax{\includegraphics{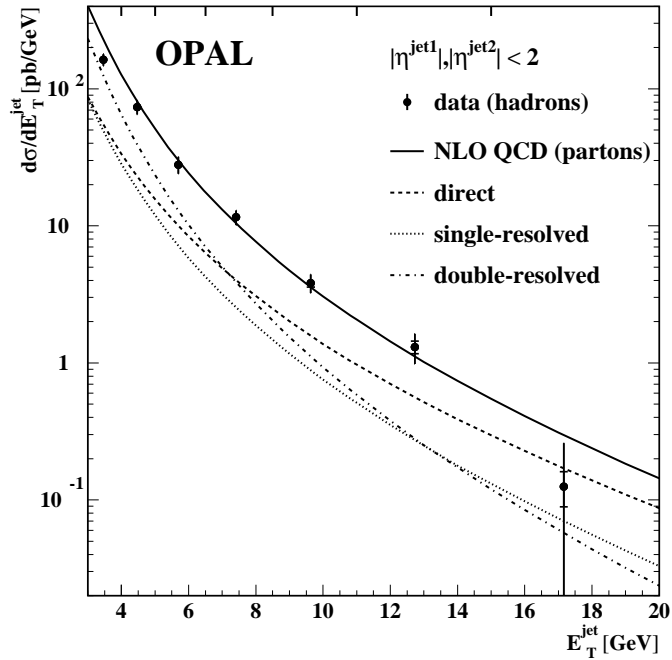}}
\vspace{0ex}
\caption{\small\sl The inclusive two-jet cross section as a function of
$E_T$ for jets with $|\eta|\ <$ 2, compared to the
NLO calculations \cite{klek}. Solid line is the sum of the direct,
single resolved and double resolved cross sections that are also shown
separately (from \cite{pr250}). }
\label{fig:jet11}
\end{figure}

The cross sections for the direct, single resolved and double resolved
two-jet events versus $E_T$,
compared to  the NLO calculation with 
the GRV HO parton parametrization,  are  shown in fig.~\ref{fig:jet11}.\\
\vspace*{8.cm}

\begin{figure}[hb]
\vskip 0.in\relax\noindent\hskip 2.cm
       \relax{\includegraphics{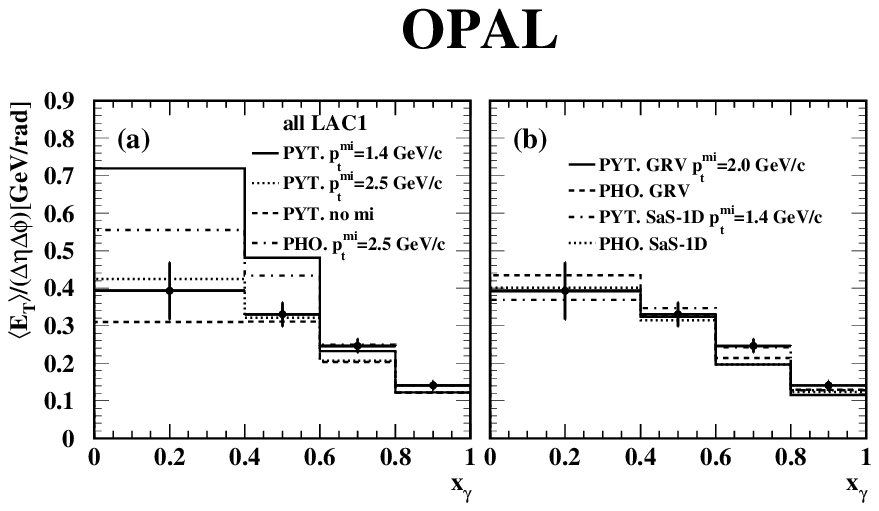}}
\vspace{-0.7cm}
\caption{\small\sl The transverse energy flow outside the jets for 
$|\eta^*|<$1 as a function of $x_{\gamma}$ for various parton parametrizations,
MC models and values of $p_t^{mi}$
(from \cite{pr250}).}
\label{fig:jet111}
\end{figure}

The influence of the ``underlying event'' has been studied.
The transverse energy flow outside the jets was measured as
a function of $x_{\gamma}$, in order to study the 
possible contribution from multiple interactions. The Monte Carlo simulations
(PYTHIA, PHOJET) were performed with various parametrizations
and values of the $p_t^{mi}$ - cutoff parameter for multiple
interactions (fig.~\ref{fig:jet111}).

\begin{figure}[ht]
\vskip 0.in\relax\noindent\hskip 2.4cm
       \relax{\includegraphics{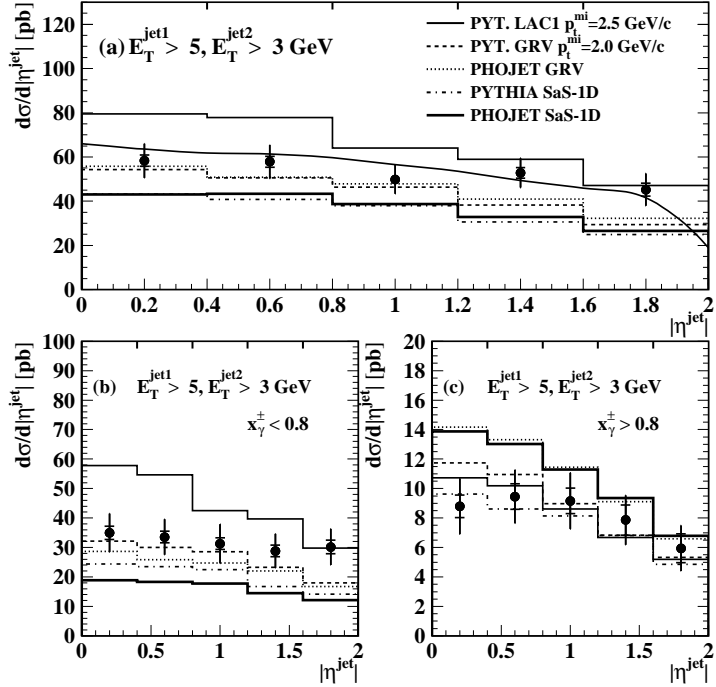}}
\vspace{0cm}
\caption{\small\sl The inclusive two-jet cross sections as a function 
of $|\eta|$ for jets with $E_T^{jet1}\ >$ 5 GeV and $E_T^{jet2}\ >$ 3 GeV:
 a) for all events, b) for direct events, c) for double resolved events
(from \cite{pr250}).}
\label{fig:jet12}
\end{figure}

The two-jet pseudorapidity  distribution  is presented in fig.~\ref{fig:jet12}    
together with the predictions from the PYTHIA and PHOJET generators,
with various parton parametrizations. Here the asymmetric cuts were applied,
to avoid infrared instability in the NLO calculations.
The results are also shown for two samples, corresponding to the direct and double resolved photon processes.

\vspace*{5.3cm}
\vspace*{9cm}
 To reduce the influence from multiple interactions and hadronization effects the two-jet cross section was measured also for  $E_T^{jet1}\ >$ 5 GeV (not shown).

~\newline
Comment: {\it "The $E_T$ dependent two-jet cross section is in good agreement 
with NLO QCD calculation.''\newline
``Using PYTHIA and PHOJET the GRV LO parametrization is able to describe 
the two-jet cross section; prediction based on the SaS1D is too low,
and that for LAC1 - significantly too high.''\newline
``Within the errors of the measurement we are unable to differentiate 
between models with and without multiple interactions."}

\begin{figure}[ht]
\vskip 0.in\relax\noindent\hskip 2.2cm
       \relax{\includegraphics{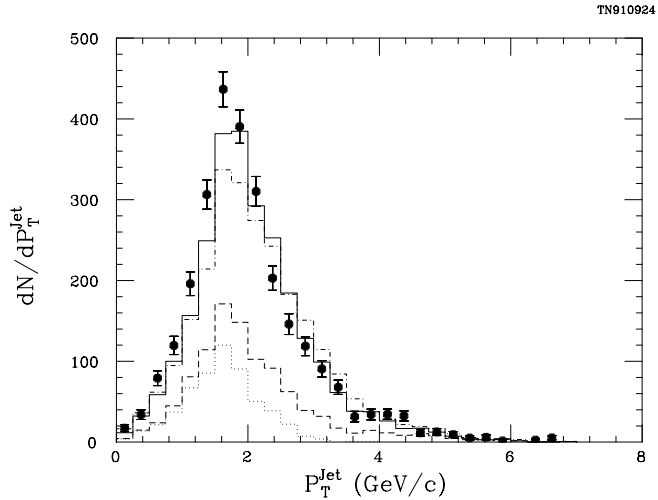}}
\vspace{0ex}
\caption{\small\sl The  $p_T$ distribution compared 
with the predictions of QPM (dotted histogram), QPM + VMD
(dashed histogram) and QPM + VMD + MJET (solid histogram)
with $p_T^{min}$ = 1.6 GeV
(from \cite{amy3}).}
\label{fig:jet13}
\end{figure}

~\newline
$\bullet${\bf{AMY 1.0 92 \cite{amy3} (TRISTAN)}}\\
The measurement of the high $p_T$ hadron production in the quasi-real
${\gamma} {\gamma}$ collision, 
for the $e^+e^-$ collision
energy between 55 and 61.4 GeV, was performed.
In the observed three and four-jet events there are one
or two  spectator jets coming from the
resolved photon(s). 
The QPM+VMD contribution, and the MJET one, which corresponds to 
the RPC contributions, were included in the analysis. 
The DG parton parametrization was used with the $p_T^{min}$ equal to 1.6 GeV.
Various distributions were studied, the results for the $p_T$ distribution
are shown in   
fig.~\ref{fig:jet13}.\\
\vspace*{6.2cm}
The sensitivity of the results to the gluonic content of the photon
was studied.
~\newline
Comment: {\it The ``first clear'' evidence is found for the hard 
scattering of hadronic 
constituents of photons in the photon - photon collisions at TRISTAN.\newline
``...without the gluonic component it is impossible
to reproduce the data''.}
~\newline\newline
$\bullet${\bf {AMY 1.5 94 \cite{as2} (TRISTAN)}}\\
AMY 1.5 (the upgraded AMY detector at the energy 60 GeV) measured
the inclusive single and 
double jet cross section ${{d{\sigma}}\over{ dp_T}}$
in quasi-real $\gamma\gamma$ collisions. 
Jets with  $p_T$ above 2.5 GeV up to ~8 GeV and $|\eta|<$ 1 were studied.
The QPM + GVMD (for the diffractive production of the hadrons) and the improved MJET generator for the resolved photon processes (using  DG, LAC 1,2,3 and GRV parton parametrizations, and  various $p_T^{min}$ parameters) were applied
in the analysis. The cone algorithm with R=1 was used.

\begin{figure}[ht]
\vskip 0.in\relax\noindent\hskip 0.5cm
       \relax{\includegraphics{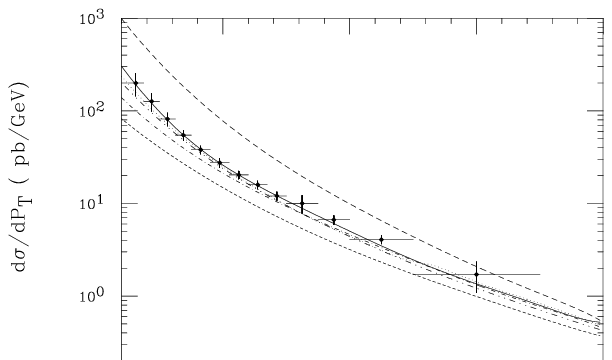}}
\vskip 0.in\relax\noindent\hskip 8.cm
       \relax{\includegraphics{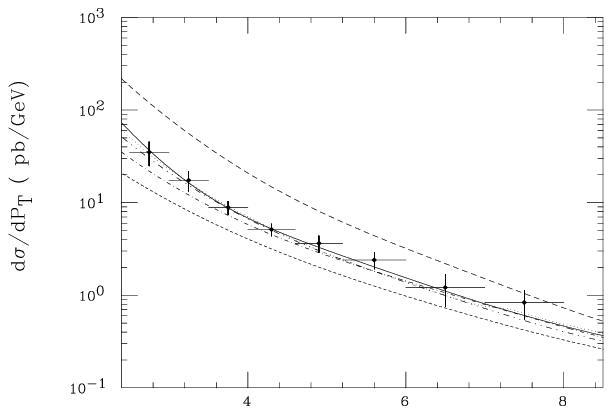}}
\vspace{0ex}
\caption{\small\sl The inclusive jet 
cross section as a function of $p_T$
integrated over $|\eta|\ <$ 1.0 for one-jet (left) and
two-jet events (right). The curves represent the sum of the QPM
(direct) and MJET (resolved) cross sections using the parton 
parametrizations: LAC1
(full line), GRV LO (double dot - dashed), DG (dotted line),
LAC3 (dashed line) and LAC1 without the gluon component
(dot - dashed line). The short - dashed curve corresponds 
to the QPM cross section (from \cite{as2}).}
\label{fig:jet14}
\end{figure}

In fig.~\ref{fig:jet14} the $p_T$ 
distributions for one and two jets, for $|\eta|< 1.0$,
are presented.
\newpage
\vspace*{3cm}

\begin{figure}[ht]
\vskip 0.in\relax\noindent\hskip 2.7cm
       \relax{\includegraphics{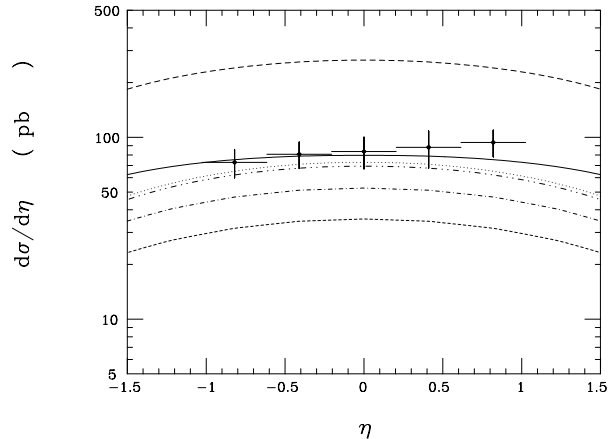}}
\vspace{0ex}
\caption{\small\sl The inclusive jet cross section as a function of
$\eta$ integrated over $|p_T|\geq$ 2.5 GeV. The lines
convention is the same as in fig.~\ref{fig:jet14}
(from \cite{as2}).}
\label{fig:jet15}
\end{figure}

The $\eta$ dependence for the jet with $p_T$ larger than 2.5 
is presented in fig.~\ref{fig:jet15}.
\newline
\vspace*{5cm}

~\newline
Comment: {\it ``The data are in good agreement with LO QCD 
calculations based on either
LAC 1, DG or GRV parametrizations of the parton densities in the photon.
The calculation based on LAC 3 disagrees with the data
and we do not see the deviation that is observed in the 
$\eta$ distribution by the H1 experiment'' (here denoted as {\bf H1 93}).}

~\newline
$\bullet${\bf {TOPAZ 93 \cite{as3} (TRISTAN)}}\\
The inclusive jet cross section ${{d{\sigma}}\over {dp_T}}$ was measured 
at the collision energy $\sqrt{s_{ee}}$=58 GeV and virtuality of photon
below 2.6 GeV$^2$. Jets with
 $p_T$ between  2.5 GeV
and 7.5  GeV and $|\eta|<$ 0.7 were analysed.
 The direct and resolved photon contributions were included, 
as well as the VMD one. The resolved contributions
were calculated with various  parton parametrizations (DO, DG and LAC1,2,3),
using the $p_T^{min}$ parameter. The BASES/SPRING generator was used.
  To describe jets, the cone algorithm with $R=1$ was applied.

The energy flow within the jet (per jet) as a function of
 $\delta \eta$ and $\delta \phi$ was also studied
and found in agreement with the Monte Carlo simulation (not shown). 

\begin{figure}[ht]
\vskip 0.in\relax\noindent\hskip 3.7cm
       \relax{\includegraphics{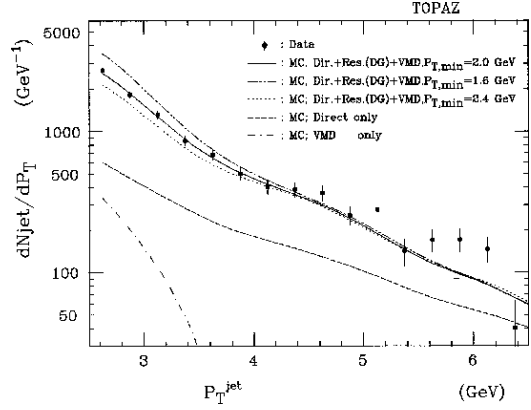}}
\vspace{0ex}
\caption{\small\sl The jet $p_T$ distribution for a no-tag sample.
Predictions are shown for the sum of
direct + resolved photon (with the DG parton parametrization) + VMD 
processes for different $p_T^{min}$
values. Contributions of the direct
(dashed line) and the VMD (dot - dashed line) processes are shown.
Parton parametrization according to DG
(from \cite{as3}).}
\label{fig:jet16}
\end{figure}

\begin{figure}[ht]
\vskip 0.in\relax\noindent\hskip 4.3cm
       \relax{\includegraphics{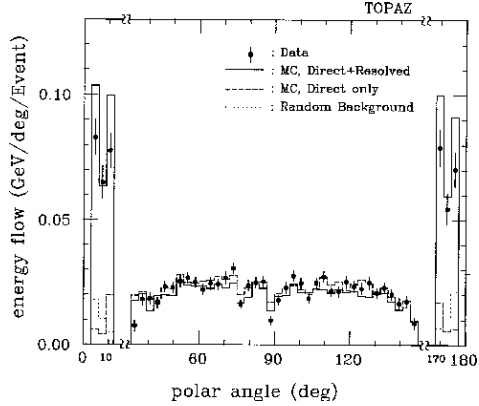}}
\vspace{0ex}
\caption{\small\sl The energy flow versus the polar angle $\theta$ 
for the jet sample. The data are compared with the Monte Carlo predictions 
for the sum of the direct and the resolved processes (solid histogram)
and the direct process only (dashed histogram)
(from \cite{as3}).}
\label{fig:jet17}
\end{figure}

\begin{figure}[ht]
\vskip -5cm\relax\noindent\hskip 3.7cm
       \relax{\includegraphics{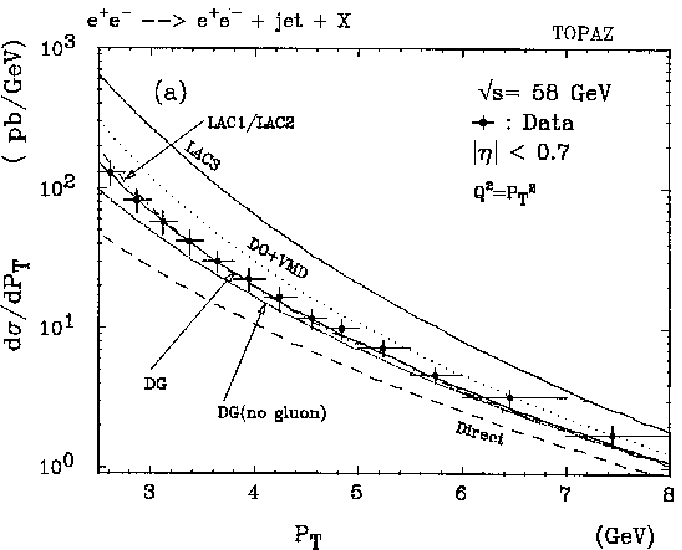}}
\vskip 5cm\relax\noindent\hskip 3.7cm
       \relax{\includegraphics{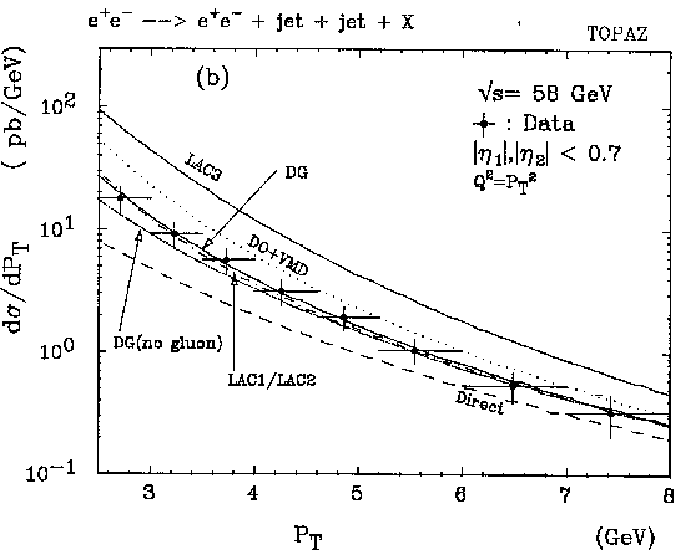}}
\vspace{0.ex}
\caption{ {\small\sl 
The inclusive a) jet and b) two-jet cross sections as 
a function of
 $p_T$ for the pseudorapidity $\mid \eta \mid \leq0.7$.
TOPAZ Collaboration results are compared with the predictions of
different parton parametrizations: LAC3, DO + VMD, DG, DG 
(without gluon), LAC1 and LAC2. Also a direct contribution is shown
(from \cite{as3}).}}
\label{fig:rys}
\end{figure}

In figs.~\ref{fig:jet16} and  \ref{fig:jet17}  
the $p_T$ distribution of the jet and the
energy flow per event
as a function of the polar angle of the jet ($\theta$) are presented, 
respectively.
Fig.~\ref{fig:rys} shows the inclusive one-jet and 
two-jet cross sections.
\newline
\vspace*{4.7cm}
\vspace*{5.cm}
\vspace*{10cm}
~\newline
Comment: {\it The polar angle distribution gives ..''direct evidence
of the presence of the resolved processes.
The data exclude the parametrizations of LAC 3 and DO + VMD,
which predict a very large gluon content even at large 
$x_{\gamma}$.''}

~\newline
$\bullet$ {\bf TOPAZ 95,conf \cite{topaz95} (TRISTAN)}
\newline
The updated results of the jet study 
in (quasi) real $\gamma\gamma$ collisions are given, for a larger 
data sample than in \cite{as3}. Anti-tag condition corresponds
to the maximum photon virtuality of $P^2_{max}$ = 2.6 GeV$^2$.
Production of jets with $p_T$ between 1 (2) and 8 GeV, and $|\eta|<$ 0.7
was measured.
The direct (QPM) + VMD + resolved (RPC) contributions were included
in the analysis of the data. The parton density parametrization
used is that of DG, and LAC1,
with  optimalized cutoff parameters $p_T^{min}$.

\begin{figure}[ht]
\vskip 0.in\relax\noindent\hskip 2.cm
       \relax{\includegraphics{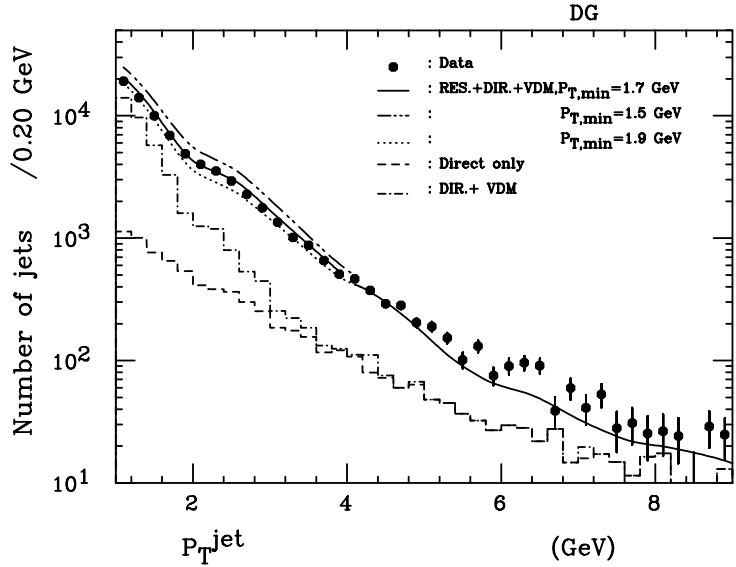}}
\vspace{-1cm}
\caption{ {\small\sl The inclusive jet distribution as a function of
$p_T$ for $|\eta |\leq 0.7$. The solid line is a prediction
of the direct + resolved (with DG parton parametrization) 
+ VMD processes ($p_T^{min}=1.7$ GeV). The
contributions of the direct and of direct + VMD processes are shown
as dashed and dot-dashed histograms, respectively (from \cite{topaz95}).}}
\label{fig:topaz95_2}
\end{figure}

The $p_T$ distribution for the inclusive jet production  has
been studied down to $1$ GeV,
and the data show clearly the evidence 
for the resolved processes (see fig.~\ref{fig:topaz95_2}). 

~\newline
\vspace*{7cm}

\begin{figure}[ht]
\vskip 0.in\relax\noindent\hskip 2.5cm
       \relax{\includegraphics{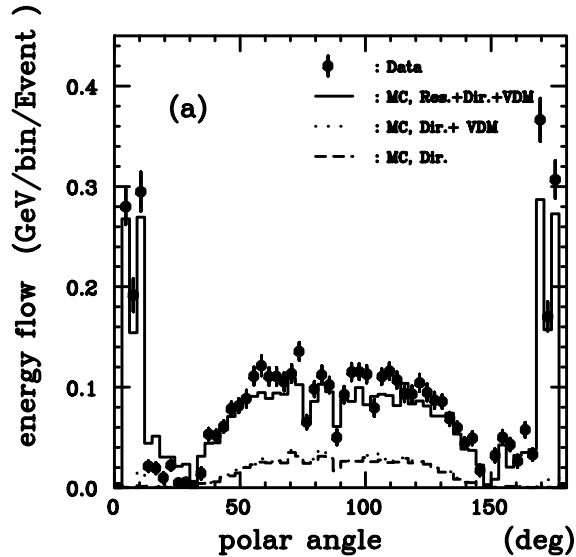}}
\vspace{-3cm}
\caption{ {\small\sl The energy flow distribution for the central
two-jet sample with $|\eta |\leq 0.7$ and $p_T\geq 2$ GeV.
Comparison with prediction of Monte Carlo models as in fig. 
\ref{fig:topaz95_2} (from \cite{topaz95}). }}
\label{fig:topaz95_3a}
\end{figure}

The evidence for the remnant-jet activity in small-angle region
was observed. Fig.~\ref{fig:topaz95_3a} shows the particle energy flow for the 
sample with two high - $p_T$ jets in the central region, where the
energy deposit from direct (or VMD) processes is rare.

\vspace*{10cm}
~\newline
Comment: {\it The existence of the resolved processes in $\gamma\gamma$
collisions is confirmed by the jet rate and the direct detection
of the energy clusters coming from the remnant  jet. The rate 
of the remnant jet is also consistent with the Monte Carlo
expectations.}

\newpage
~\newline
$\bullet$ {\bf TOPAZ 98,conf \cite{topaz98} (TRISTAN)}
\newline
A study of dijet production in (quasi) real $\gamma\gamma$
collisions (anti-tag conditions: $P^2$ from 10$^{-8}$ to 2.6 GeV$^2$
with the mean value $\sim$ 10$^{-4}$ GeV$^2$) at $\sqrt{s_{ee}}=58$ GeV based on 
the full data taken in the years 1990-95
was performed. (Also a prompt photon production cross section was measured.) 
Limits: $E_T\geq 3$ GeV and $E_T\geq 2$ GeV for
two jets with the highest transverse energy were applied.
 The Monte Carlo programs PYTHIA 
(also with additional power law $p_t$ distribution) and the PHOJET with
GRV LO and SaS1D parton distributions were used.
The jets were reconstructed using a cone jet-finding algorithm
with fixed value of cone radius $R = 1.0$.

\begin{figure}[ht]
\vskip 0.in\relax\noindent\hskip 0.cm
       \relax{\includegraphics{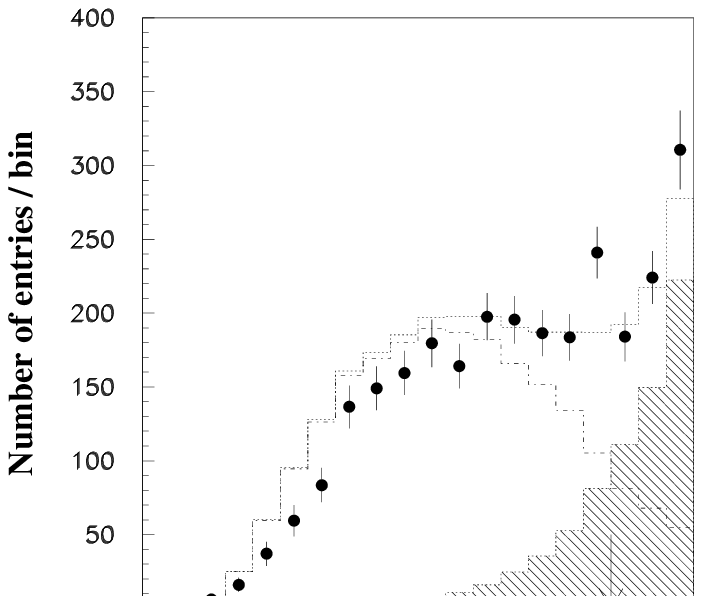}}
\vskip 0.in\relax\noindent\hskip 7.5cm
       \relax{\includegraphics{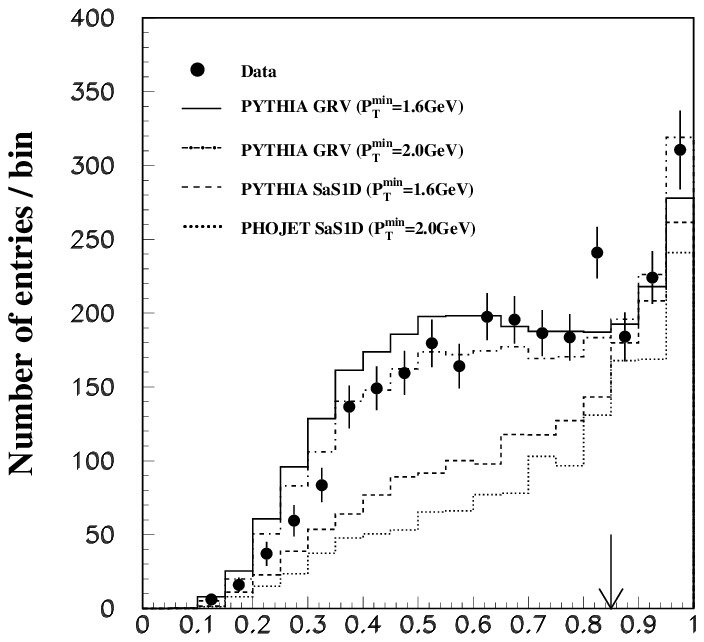}}
\vspace{0.ex}
\caption{ {\small\sl The $x_{\gamma}^{min}$ distribution of the
dijet sample. Left: Comparison with the PYTHIA predictions with the GRV LO
photon structure function; the shaded histogram - contribution
from the direct process, the dashed one - from the resolved process, 
the dotted one - sum of both. Right: The same distribution compared 
with the various Monte Carlo predictions (with $p_T^0$ = 1.6 GeV
or 2 GeV) (from \cite{topaz98}).}}
\label{fig:topaz98_2}
\end{figure}

The study of the hadronic final state (the 
transverse energy distribution, the jet energy profile and the
azimuthal angle dependence of the number of dijet events) was carried out by 
separating the direct and resolved sample by a $x_{\gamma}$ 
reconstruction method developed at HERA $\gamma p$ experiments 
(see \eg sec. \ref{sec22} and \ref{sec244}). The number of dijets as a function of 
$x_{\gamma}$ is shown in fig.~\ref{fig:topaz98_2}.
The predictions of the PYTHIA generator with the GRV LO 
structure function agree with the data, but that with SaS1D
are too low by a factor 2 for $x_{\gamma}^{min}$
(defined as min($x_{\gamma}^+,x_{\gamma}^-$)) $\sim 0.5$.
This observation appears for both PYTHIA and PHOJET programs, and
also for different values of  parameter $p_T^{min}$,
 varying from 1.6 to 2 GeV.\newline

\vspace*{6cm}

\begin{figure}[ht]
\vskip 0.in\relax\noindent\hskip 0.cm
       \relax{\includegraphics{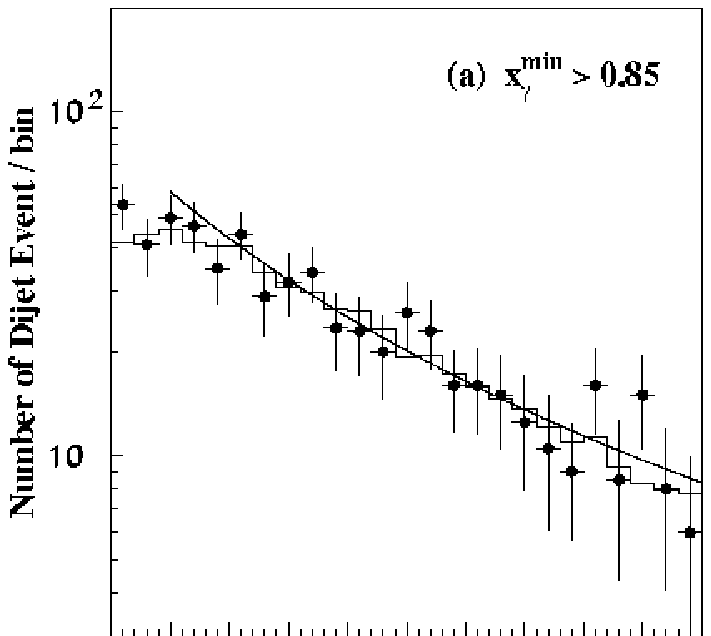}}
\vskip 0.in\relax\noindent\hskip 7.5cm
       \relax{\includegraphics{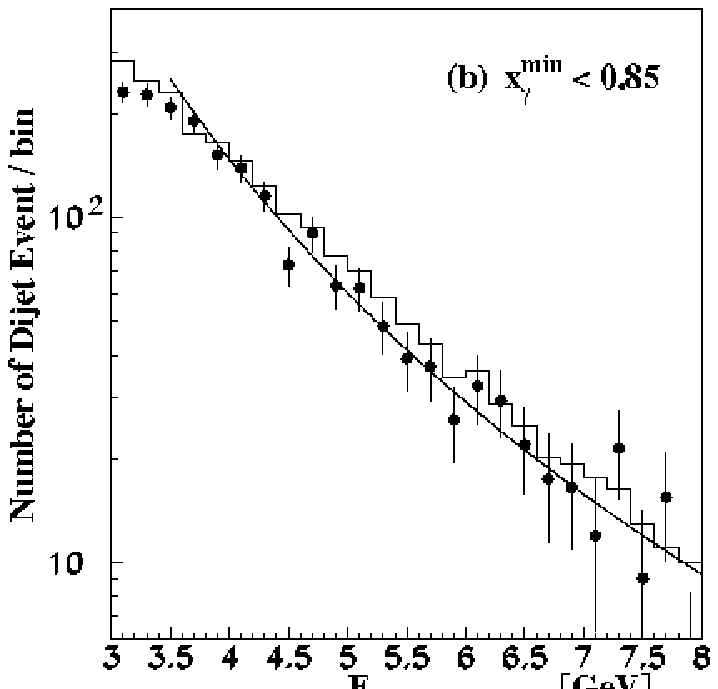}}
\vspace{0.ex}
\caption{ {\small\sl The  $E_T$ distributions for (a) the direct -
enriched sample and (b) the resolved - enriched sample. The histograms
are the PYTHIA predictions with the GRV LO structure function
(from \cite{topaz98}). }}
\label{fig:topaz98_3}
\end{figure}

The measured $E_T$ distributions for the resolved -
enriched sample ($x_{\gamma}^{min} < 0.8$) 
and for the direct - enriched sample ($x_{\gamma}^{min} > 0.8$)
are shown in fig.~\ref{fig:topaz98_3}. 
\newpage

\vspace*{6.2cm}

\begin{figure}[ht]
\vskip 0.in\relax\noindent\hskip 0.cm
       \relax{\includegraphics{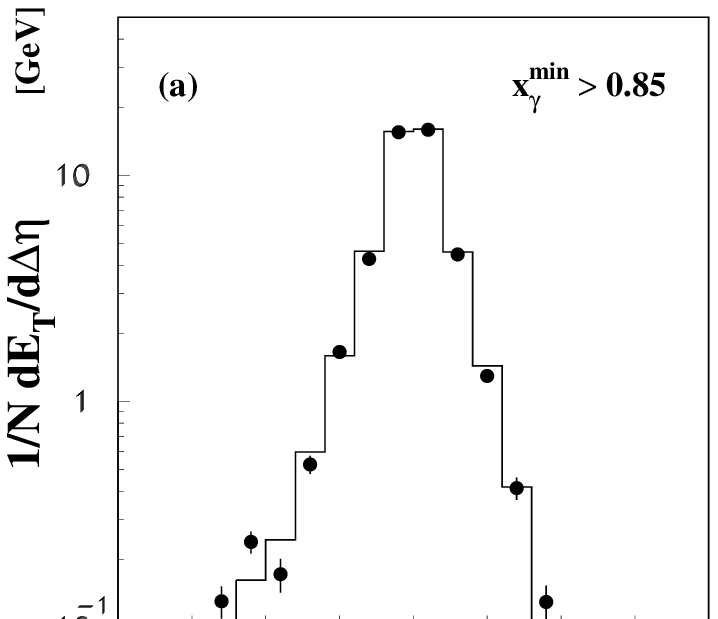}}
\vskip 0.in\relax\noindent\hskip 7.5cm
       \relax{\includegraphics{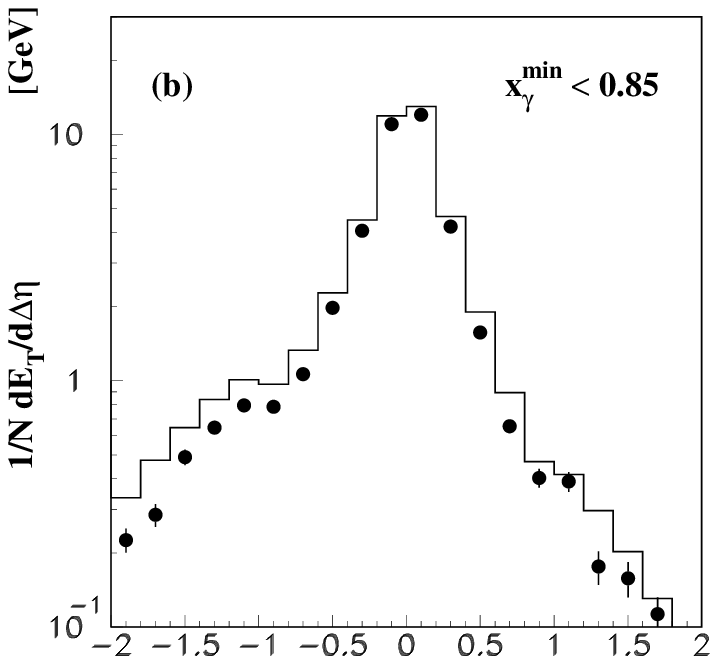}}
\vspace{0.ex}
\caption{ {\small\sl Transverse energy flow relative to the direction
of each jet in dijet events as a function of $\delta\eta$ for (a)
direct - enriched and (b) resolved - enriched sample. The histogram
is the result of the PYTHIA simulation (from \cite{topaz98}). }}
\label{fig:topaz98_4}
\end{figure}

The transverse energy flow as a function of the pseudorapidity 
$\delta\eta$ was measured 
with respect to the jet direction for different dijet samples,
see fig.~\ref{fig:topaz98_4}. The direct - enriched sample is nearly 
symmetric and there is almost no activity outside the jet, while 
the resolved - enriched sample shows considerable activity 
outside the jet cone for $|\delta\eta| > 1$. The energy flow 
outside the jets is well modelled by the PYTHIA generator. 
\newline

\vspace*{5.8cm}

\begin{figure}[ht]
\vskip 0.in\relax\noindent\hskip 0.cm
       \relax{\includegraphics{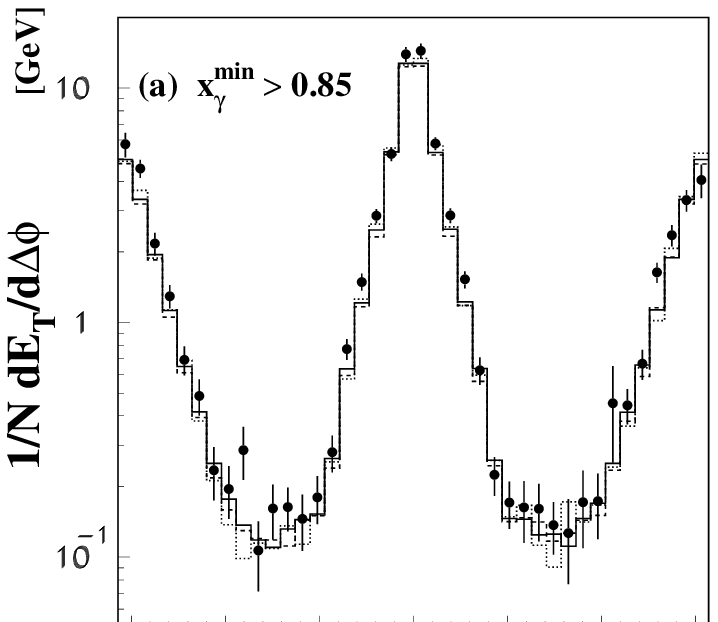}}
\vskip 0.in\relax\noindent\hskip 7.5cm
       \relax{\includegraphics{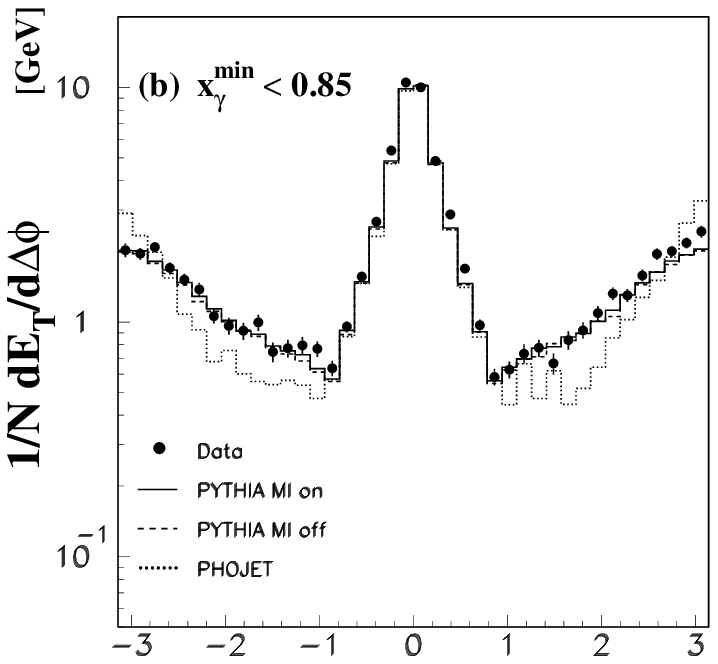}}
\vspace{0.ex}
\caption{ {\small\sl The transverse energy flow relative to each jet
direction as a function of the azimuthal angle $\delta\phi$ for
(a) direct - enriched and (b) resolved - enriched sample. The PYTHIA
(with and without multiple interaction, solid and dashed line,
respectively) and PHOJET (dotted line) predictions are shown 
(from \cite{topaz98}). }}
\label{fig:topaz98_5}
\end{figure}

The transverse energy flow versus the azimuthal angle around the jet 
direction is shown in fig.~\ref{fig:topaz98_5} for the two above
samples together with the Monte Carlo predictions with and without
the multiple interactions (MI).
The activity observed for the resolved - enriched
sample is much larger than for the direct - enriched; at 
$|\delta\phi| \sim 1.5$ the difference is by a factor 8. 
\newline
\vspace*{6cm}

The dependence of the azimuthal angle difference between two jets with the 
highest $E_T$ was studied. Here the effect of the primordial
$p_t$ distribution of partons -- gaussian and the power-law type --  was
analysed using the PYTHIA program (with the GRV parton parametrization).
These data (not shown) are reproduced very well when the initial
state radiation (IR) is included and the power-law $p_t$ distribution
is assumed.
 
~\newline
Comment: {\it Too low prediction for $x_{\gamma}$ distribution
from the PYTHIA and PHOJET programs if based on the SaS1D 
parton parametrizations.

``The PYTHIA predictions on the direct and resolved 
processes reproduce the transverse energy flow reasonably well. 
While the PHOJET prediction for the resolved sample shows too small 
activity in the middle region between two jets, the direct 
sample is reproduced well.''

The transverse energy flow around the jet shows that... ``In 
$\gamma\gamma$ collisions at TRISTAN, the PYTHIA predictions 
with and without multiple interactions are almost identical, 
indicating that there are no effects of the multiple interaction 
at the TRISTAN energy region'', although they have been found at HERA 
in $\gamma p$ reaction.

The data on the azimuthal angle between jets are very well described when
the power law $p_t$ distribution in PYTHIA is assumed.}

%
%
\newpage
\subsubsection{Jet production in $\gamma p$ collisions\label{sec244}}

Here we present the data for the jet production in $\gamma p$ scattering
 taken at HERA collider at $\sqrt {s_{ep}} \sim 300$ GeV 
\footnote{The energy of the electron
at the beginning of running of the HERA collider was equal to 26.7 GeV;
starting from 1994 the positron replaced the electron and its
energy increased to 27.5 GeV, with the energy of the proton 
820 GeV. Starting from 1998 the  electron beam with energy 27.5 GeV is used
again while the 
proton energy is higher than before, 920 GeV.}, 
where only   {\it one}  photon may be resolved\footnote{The 
$\gamma \gamma $ events leading to the large $p_T$
jets are rare at HERA.}.
The photoproduction events correspond here to the limit
of virtuality of the initial photon: $P^2 \sim 4$ GeV$^2$ (with median 0.001 
 GeV$^2$), at present also $\sim$ 1 GeV$^2$
or (with a special condition) $\sim$ 0.01-0.02 \g2 (see below).
We will concentrate below on the single and double jet production
and in addition we will include in the presentation
some  results for the charged {\sl particle}  production in the
context of  extraction of the gluonic content of the photon.

The partonic variables,  related both  
to the initial and final states in the hard partonic subprocesses,
are reconstructed from the corresponding quantities 
for the final state hadronic jets.
The relation between variables corresponding to  these two levels
depends on the order of the perturbative QCD calculation,
as  was mentioned before.  
The  parton momentum fraction $x^{jet}_{\gamma}$
 (called also $x_{\gamma}^{vis}$ or 
$x_{\gamma}^{obs}$) is in the LO case  equal to $x_{\gamma}$.
 In practice
it is reconstructed 
 using  two jets  with the highest transverse energy $E_T$ (or $E_T^{jet}$) 
in the event, using the following relation:
\be
x^{jet}_{\gamma}=
{E_{T1}e^{-\eta_1}+E_{T2}e^{-\eta_2}\over
2E_{\gamma}},
\ee
where also jet pseudorapidities,
${\eta}=-\ln (\tan(\theta/2))$, and
the energy of the photon $E_{\gamma}$ (= $yE_e$) enter. 
(The positive pseudorapidity
corresponds to the proton direction.)
The scaled energy $y$ of the initial photon (eq.~\ref{xy}) 
is measured from the transverse energy $E_T^h$ and pseudorapidity  
of hadrons $\eta^h$ according to the 
 formula
\be
y={{1}\over{2E_e}}\sum_h E^h_Te^{-\eta^h},
\ee
where the sum is over the produced hadrons.\newline

The rapidity in the $\gamma p$ CM system is usually denoted by $\eta^*$
(typically  $\eta-\eta^{*}  \sim 2$, where $\eta$
is the value in the laboratory frame).
In the angular distribution analysis ({\bf ZEUS 96, 98d,conf}) the polar 
angle in the dijet CM system, $\theta^*$, is introduced.

~\newline\newline
\centerline{\bf \huge DATA}
\newline\newline
$\bullet${\bf {H1 92 \cite{h192} (HERA)  }}\\
The evidence for the   hard photoproduction of a jet 
and single particle with   
$E_T > 10 $ \gev is reported. 
The ``soft''  contribution and the hard processes were simulated,
the latter ones  using
the PYTHIA program
with the MT B1 parton parametrization for the proton and DG for the photon,
and  $p_T^{min}$ = 2.5 GeV.
The jet cone algorithm with $R<1$ was used in the analysis.
\newpage
~\newline
$\bullet${\bf {H1 93 \cite{h193} (HERA)  }}\\
First measurement  of the inclusive jet cross section 
at the $ep$ collider HERA (based on the 1992 data) 
is reported for  the 
photon virtuality $P^2$   smaller than 0.01 GeV$^2$. 
Events with   the scaled energy $y$ of the initial photon
between 0.25 and 0.7 were collected.
The photoproduction of jet was studied   
for $E_T$   from 7 to 20 GeV and the \psr 
 ~interval $-1 <{\eta} < 1.5$. The PYTHIA 5.6 program (with the GRV LO 
parton parametrization for the proton and GRV LO for the photon) with the 
jet cone algorithm assuming  R=1 was used in the analysis. 

\begin{figure}[ht] 
\vskip 0.in\relax\noindent\hskip 1.5cm
       \relax{\includegraphics{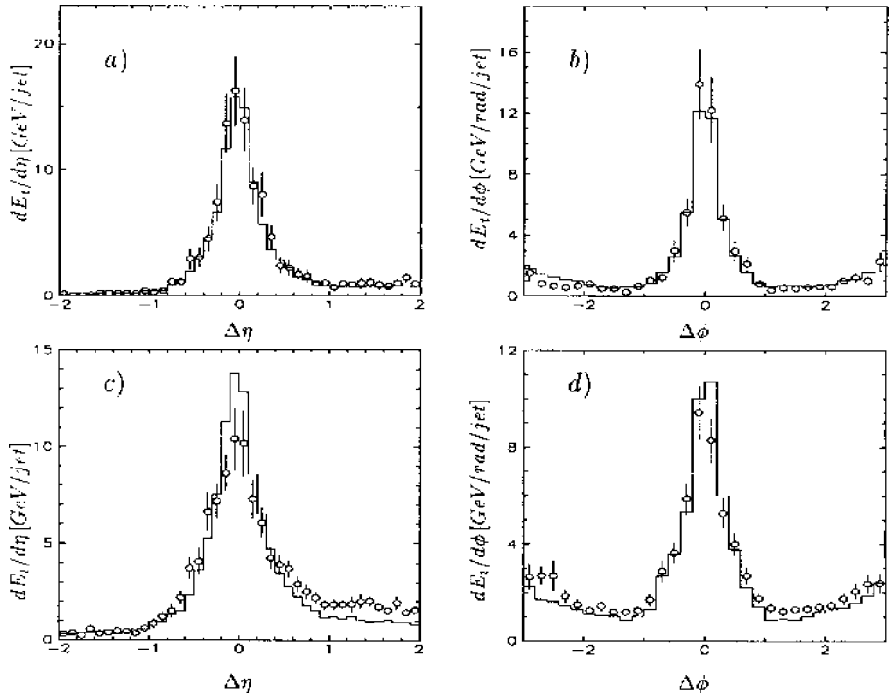}}
\vspace{0ex}
\caption{\small\sl The transverse energy flow as a function of 
$\delta\eta$ (integrated over $|\delta\phi |\ <$ 1.0) (a, c)
and as a function of $\delta\phi$ (integrated over 
$|\delta\eta |\ <$ 1.0) (b, d). Figs. a) and b) correspond to
-1.0 $<\ \eta\ <$ 0.5, c) and d) to 0.5 $<\ \eta\ <$ 1.5
(from \cite{h193}).}
\label{fig:jet20}
\end{figure}

The transverse energy flow around the jet axis was studied 
 and the discrepancy was found
in a form of  too large transverse energy flow on the forward side of the jet
(see fig.~\ref{fig:jet20}). 
Among others, multiple parton interactions were
mentioned as a possible  explanation of this effect.\\
\vspace*{8.3cm}

\begin{figure}[hb]
\vskip 0.in\relax\noindent\hskip 4.cm
       \relax{\includegraphics{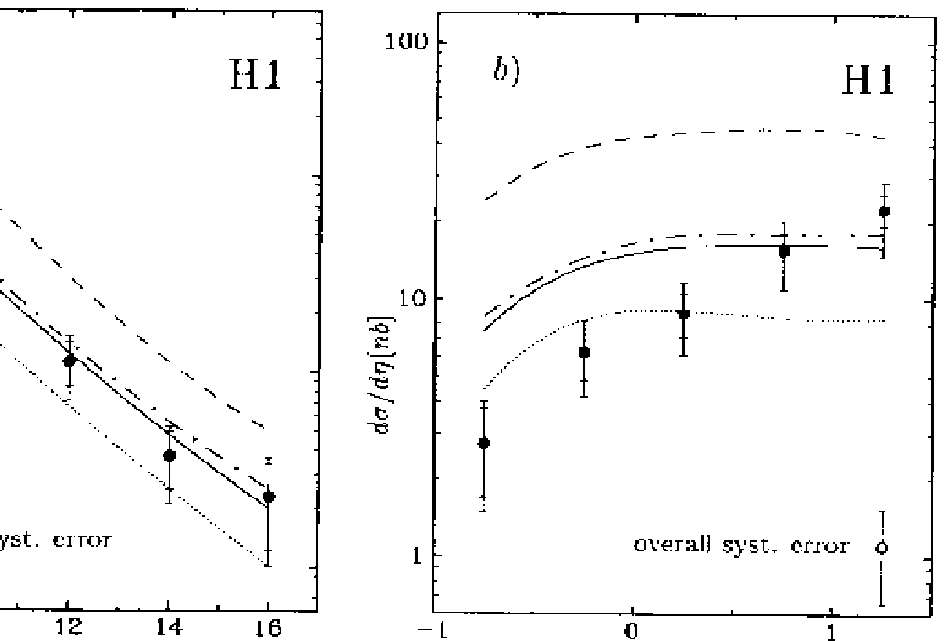}}
\vspace{0ex}
\caption{\small\sl The inclusive jet $E_T$ spectrum integrated over the $\eta$
interval -1.0 $<\ \eta\ <$ 1.5 (a) and inclusive $\eta$
spectrum (b) for jets with $E_T\ >$ 7 GeV. The LO QCD predictions from
the  PYTHIA generator using parametrizations LAC3 (dashed line),
LAC2 (dash-dotted line), GRV LO (full line) and GRV LO (without
gluons, dotted line) (from \cite{h193}).}
\label{fig:jet21}
\end{figure}

The inclusive jet cross sections 
$d\sigma/dE_T$ versus $E_T$ and $d\sigma/d\eta$ as a function of 
$\eta$ integrated over the corresponding range of the $\eta$ and $E_T$, 
respectively, were measured  and compared with the LO calculation
using the following parton parametrizations: LAC2, LAC3 and GRV LO 
for the photon, and GRV LO for the proton (see fig.~\ref{fig:jet21}).\\
\vspace*{5.7cm}

The shape of the $d\sigma/dE_T$ is well described in the range 
of $\eta$ between -1 and 1.5. It is a problem to describe
the $\eta$ distribution of jets.

~\newline
Comment: {\it 
"In the (pseudorapidity) range 0.5 to 1.5  the data show 
larger average values of the transverse energy flow
outside the jet cone on the forward side of the 
jet than predicted by the Monte Carlo."\\
LAC3 gives cross section higher by factor 3 
than data for the  $E_T$ distribution of the jets.\\
"None of the models describe well the measured 
$\eta$ dependence (for jets)." \\
This discrepancy is absent in
{\bf H1 96a} analysis, where multiple interactions are added in
the Monte Carlo programs.}
~\newline\newline
$\bullet${\bf {H1 95a \cite{h1} (HERA)  }}\\
The photoproduction of two-jet events (the 1993 data) was studied 
for the photon virtuality $P^2$   smaller than 0.01 GeV$^2$.
The scaled  energy $y$ of the initial photon was between 0.25 and 0.7.
The jet $E_T$  range from 7 to 20 GeV and the \psr 
interval $0\le{\eta}\le2.5$  with $|\Delta \eta|\le1.2$
(between the most energetic jets) (see also {\bf H1 96a}) were investigated. 

The cone jet-finding algorithm with R=1 
(and 0.7 for cross checks) was used.
The 
PYTHIA 5.6 generator with the GRV LO parton parametrizations
 for the proton and the photon, with $p_T^{min}$=2 GeV  was used 
in the analysis.

For the first time the inclusive (LO) cross sections 
were derived for the  parton level, and the gluon density 
in the photon was measured.

The transverse energy flow around the jet direction per event
versus the rapidity distance from the jet direction
was studied for 7 $\le E_T \le $ 8  GeV and  $ 0\le {\eta} \le 1$
and found to be asymmetric and different for  samples with 
$x_{\gamma}> 0.4$ and  $x_{\gamma} <0.4$.
The data were compared with the PYTHIA (with MI) predictions (not shown).

\begin{figure}[ht]
\vskip 0.in\relax\noindent\hskip 3.5cm
       \relax{\includegraphics{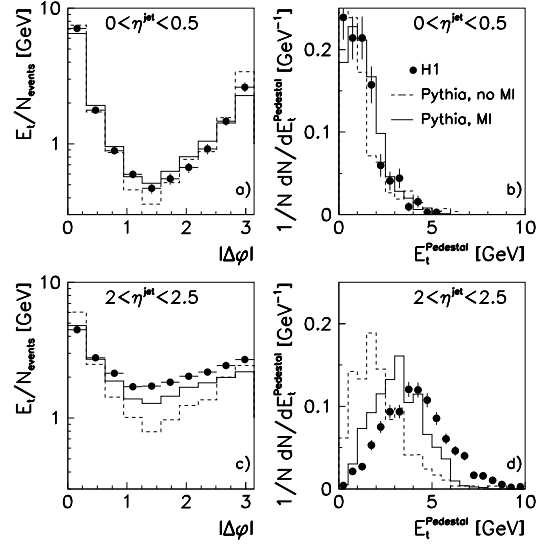}}
\vspace{-.4cm}
\caption{\small\sl (a, c) The transverse energy flow versus the azimuthal
angle with respect to the jet direction in two rapidity bins.
(b, d) Distributions of the transverse energy measured outside of 
jets. Histograms show the PYTHIA simulations with (full line)
and without (dashed line) the multiple interactions
(from \cite{h1}).}
\label{fig:jet22}
\end{figure}

The transverse  energy flow versus the azimuthal angle around 
the jet direction
 and the transverse energy of the underlying events 
outside the jets (here named $E_t^{pedestal}$)  was studied for $7\le E_T \le 8$ GeV in a cone size $R=1$.
Results 
(see fig.~\ref{fig:jet22}) compared to the prediction of the PYTHIA program 
(with and without MI) lead to  the conclusion:
``the multiple interaction gives an improved description''.\\
\vspace*{7.cm}

\begin{figure}[ht]
\vskip 0.in\relax\noindent\hskip 4.3cm
       \relax{\includegraphics{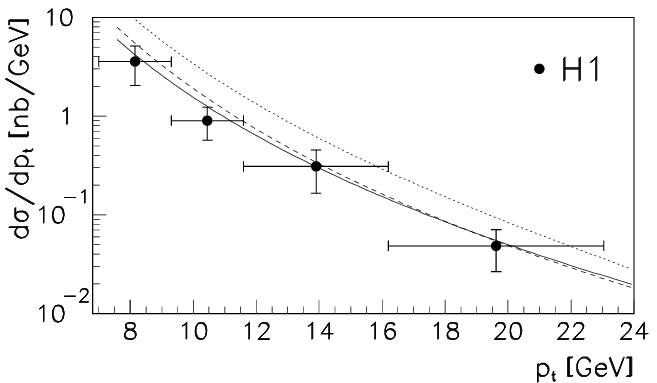}}
\vskip 3.5cm\relax\noindent\hskip 4.3cm
       \relax{\includegraphics{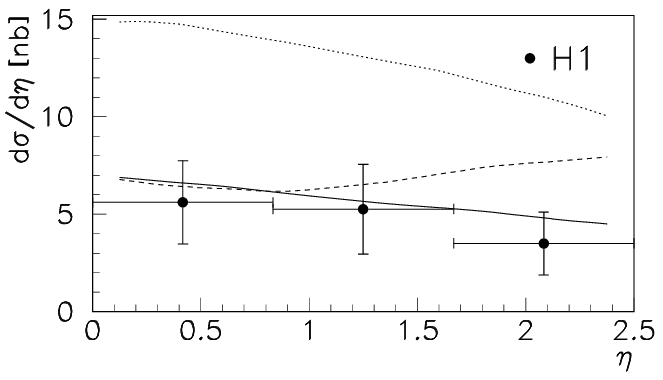}}
\vspace{-0.5cm}
\caption{\small\sl The single parton cross sections: (top) $d\sigma$/$dp_t$
integrated over the pseudorapidity range 0 $<\ \eta\ <$ 2.5,
(bottom) $d\sigma$/$d\eta$ for $p_t\ >$ 7 GeV. The solid line - the LO QCD
calculation with the GRV LO parametrization for partons in the proton
and the photon. The dashed (dotted) line - the same for the LAC1 (LAC3)
parametrization for the photon (GRV LO for the   proton)
(from \cite{h1}).} 
\label{fig:jet23}
\end{figure}

To achieve the goal which was here the extraction of the 
gluon distribution,
the single {\sl parton}  cross section $d\sigma/dp_t$ 
integrated over the parton rapidity range, as well as
the single {\sl parton} cross section 
 $d\sigma/d\eta$ were studied and compared with the prediction based on 
 the LO
parametrizations for the photon: GRV, LAC 1 and LAC 3
(see fig.~\ref{fig:jet23}).
\newpage
\vspace*{3.3cm}

\begin{figure}[ht]
\vskip 0.in\relax\noindent\hskip 4.2cm
       \relax{\includegraphics{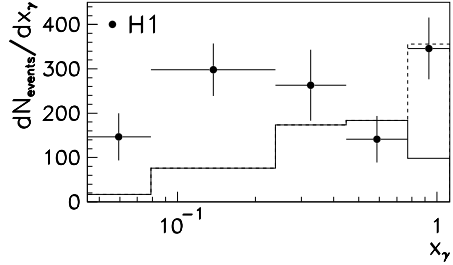}}
\vspace{-0.5cm}
\caption{\small\sl The distribution of $x_{\gamma}$
in the photon. The solid line - the contribution from 
the quark-resolved photon processes; the dashed line - the direct 
photon contribution, from the PYTHIA Monte Carlo (from \cite{h1}).}
\label{fig:jet24}
\end{figure}

To extract information on the subprocesses, 
the full two-jet kinematics was used and 
the distributions of  $\Delta \eta$, 
$\Delta E_T$, and of  $x_{\gamma}$ and $x_{p}$ were  studied. 
In fig.~\ref{fig:jet24} we present the distribution of 
 $x_{\gamma}$.\\
\vspace*{3.5cm}

\begin{figure}[ht]
\vskip -0.8in\relax\noindent\hskip 6.3cm
       \relax{\includegraphics{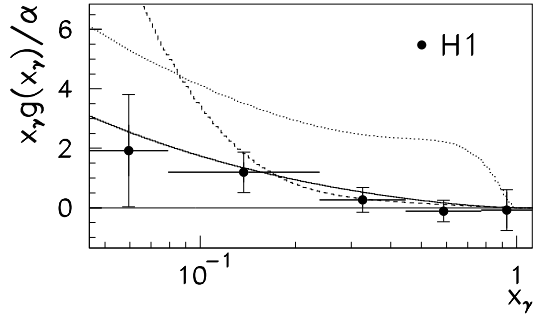}}
\vspace{0.ex}
\caption{\small\sl The gluon distribution 
extracted from the resolved photon processes at
$<$~$\tilde{Q}^2>=<p_t^2>=75$ GeV$^2$. 
For comparison the GRV LO (full line), LAC1 (dashed) and LAC3 (dotted)
gluon parametrizations are shown (from \cite{h1}).}
\label{fig:mary13}
\end{figure}

The (LO) gluon distribution in the photon was derived,    
at the average (factorization and renormalization) scale  
$<\tilde{Q}^2>$=$<p_t^2>$=75 GeV$^2$ 
for 0.04$\leq x_{\gamma} \leq$1, see table \ref{table19} and 
fig.~\ref{fig:mary13} for results.
\begin{table}[ht]
\caption{}
\label{table19} 
$$
\begin{array}{|c|c|c|}
\hline
<\tilde{Q}^2>&<x_{\gamma}>&x_{\gamma}G(x_{\gamma})/{\alpha}\\
~[GeV^2]~&&(stat. + syst.) \\
\hline
75&~\,0.059&~~1.92\pm0.87\pm1.68\\
&0.14&~~1.19\pm0.34\pm0.59\\
&0.33&~~0.26\pm0.24\pm0.33\\
&0.59&-0.12\pm0.15\pm0.33\\
&0.93&-0.08\pm0.61\pm0.30\\ 
\hline
\end{array}
$$
\end{table}

\vspace*{4.9cm}

~\newline
Comment: {\it
 "The multiple interaction option gives an improved description"
 of the jet profiles and pedestal distributions 
(still some "deviations from data at large jet rapidities 2 $< \eta < $2.5"
remain). \\
In the extracting the gluon density
$q^{\gamma}$ was taken as 
determined by the two-photon experiments at LEP and TRISTAN, in 
the form given by the 
GRV LO parametrization.\\
"A high gluon density at large parton momenta as suggested 
by the LAC 3 parametrization is clearly
excluded. The strong rise of the LAC1 parametrization 
below $x_{\gamma}\le 0.08$ is not supported."}

~\newline
$\bullet${\bf{H1 96a \cite{h196} (HERA)}}\\
The jet production with $E_T\geq 7$ \gev (and -1 $<\eta<$ 2.5)
was  measured   in $ep$ collisions (data from 1994) 
with the scaled photon energy $0.25 < y<0.7$ and $P^2$ below 0.01 GeV$^2$.
The properties of the hadronic final state 
and the distribution of the transverse energy are studied in detail.
The  PYTHIA 5.7 (with $p_T^{min}$ = 2 GeV), HERWIG 5.8 ($p_T^{min}$ = 2 GeV)
and PHOJET 1.0 ($p_T^{min}$ = 3 GeV) generators with 
the GRV LO parton parametrizations for the proton and the photon)
were used in the analysis.
The PYTHIA program were used also with MI, then with $p_T^{mi}$ = 1.2 GeV.
The cone algorithm with R=1 was used.

\begin{figure}[ht]
\vskip 0.in\relax\noindent\hskip 3.7cm
       \relax{\includegraphics{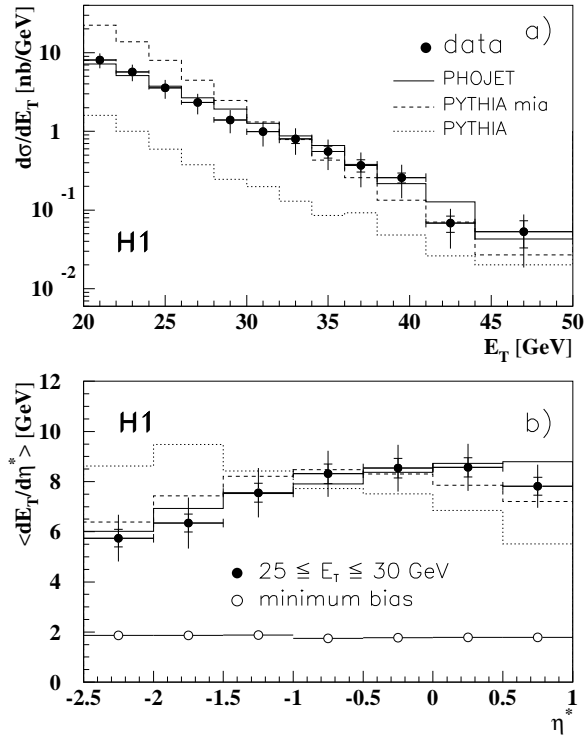}}
\vspace{-0.6cm}
\caption{\small\sl {a) The differential transverse energy cross 
section integrated over the pseudorapidity 
(-2.5 $\leq\eta^*\leq$ 1). The histograms are 
the results of the simulations with
interactions of the beam remnants (full line - PHOJET,
dashed - PYTHIA) and without them (dotted line - PYTHIA).
b) The corrected transverse energy flow versus
$\eta^*$ ($\eta^*\ >$ 0 corresponds to the proton direction).
The pseudorapidity range and histograms as in a)
(from \cite{h196}).}}
\label{fig:jet26}
\end{figure}

The integration over the $\gamma p$ CM 
system pseudorapidity -2.5 $<$\-$\eta^*<$ 1
leads to the   total transverse event energy  distribution shown in 
fig.~\ref{fig:jet26}a (here $0.3 < y<$ 0.7).
The average transverse energy flow versus $\eta^*$ 
for the total $E_T$ range between 25 and 30 GeV was also measured 
(see  fig.~\ref{fig:jet26}b). The shape of both distributions
 may indicate the need of 
the multiple interactions.\\
\vspace*{9.5cm}

\begin{figure}[ht]
\vskip 0.in\relax\noindent\hskip 4.2cm
       \relax{\includegraphics{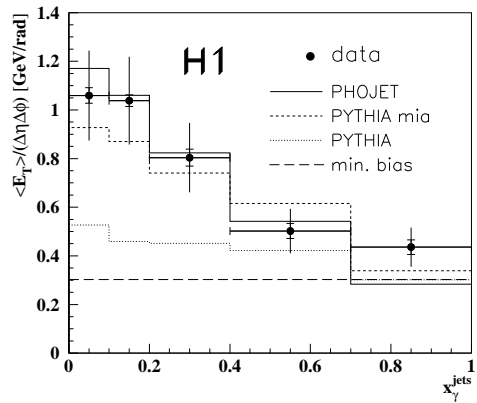}}
\vspace{-0.6cm}
\caption{\small\sl The corrected transverse 
energy density in the region
$|\eta^*|\ <$ 1 outside the jets, as a function of 
$x_{\gamma}$. The histograms are as in fig.~\ref{fig:jet26}
(from \cite{h196}).}
\label{fig:jet28}
\end{figure}

To get an insight into the details of the considered events 
 the transverse energy flow outside of the two jets 
with the highest $E_T$ was studied as a function of  
$x_{\gamma}$  for the $|\eta^*|<1$ and $\Delta \eta <$ 1.2.
Results for the transverse energy density can be found in 
fig.~\ref{fig:jet28}.\\
\vspace*{5cm}

The distribution of the transverse energy around the jet axis was also
 measured as a function of  $\delta \phi$ (not shown).
 The jet width obtained in this analysis is
similar to the corresponding quantity in the $p {\bar p}$ collision.

\begin{figure}[ht]
\vskip 0.in\relax\noindent\hskip 2.5cm
       \relax{\includegraphics{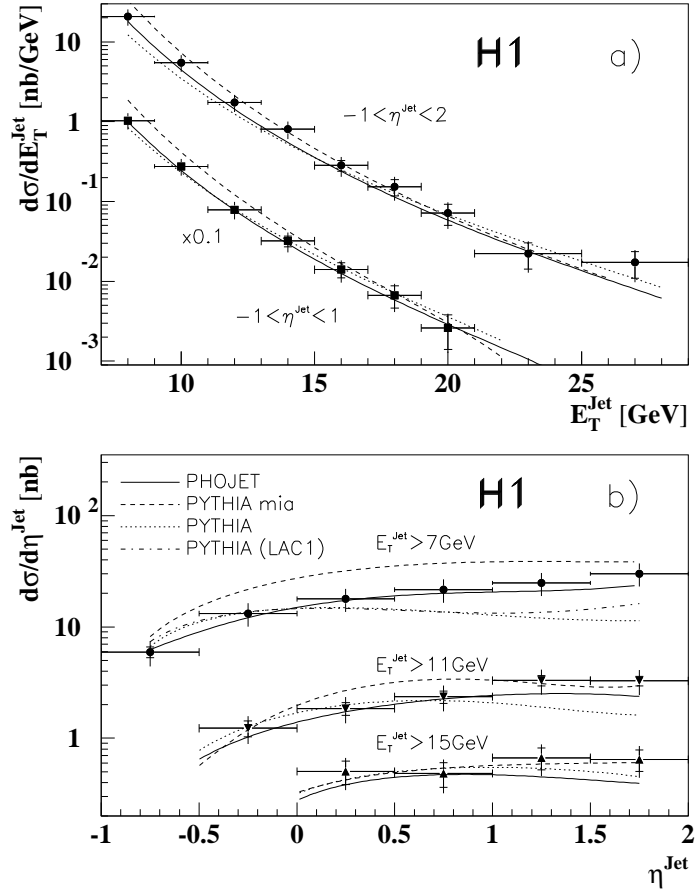}}
\vspace{-0.5cm}
\caption{\small\sl a) The cross section $d\sigma$/$dE_T$ for the jet production
for $E_T\ >$ 7 GeV, in two $\eta$ ranges:
-1 $<\ \eta\ <$ 2 
and -1 $<\ \eta\ <$ 1. The curves 
show the Monte Carlo simulations with interactions of the beam 
remnants (full line - PHOJET, dashed - PYTHIA) and
without them (dotted - PYTHIA).
b) Cross section $d\sigma$/$d\eta$ versus $\eta$
for different thresholds in $E_T$: 7, 11 and 15 GeV. The curves
are as in a); additional dash-dotted curve
- PYTHIA with the LAC1 parametrization
(from \cite{h196}).}
\label{fig:jet29}
\end{figure}

Results 
obtained for the jet cross section (here 0.25$ <y<$0.7)
are presented in  fig.~\ref{fig:jet29}a, 
where  $d\sigma/dE_T$ for the jet production in two $\eta$
regions is shown, and in fig.~\ref{fig:jet29}b, where 
the distributions $d\sigma/d\eta$ 
for the events with the transverse jet energy 
$E_T>$ 7, 11, 15 \gev are  presented.
The  comparison with the PHOJET and PYTHIA simulations, with and without 
multiple interactions, was done for both kinds of distributions.
Note that the rapidity distribution is 
   sensitive rather to 
the photon structure functions, while the $E_T$ cross section to 
the matrix elements for the hard processes.
Note also that  "for $E_T$ bigger than 7 GeV 
previous measurements ({\bf H1 93}) suffered from a defect
 and are superceded by this new measurement".\\

\vspace*{11.7cm}

~\newline 
Comment: {\it In addition to the primary hard scattering 
process, the interaction between the two beam remnants is included
in the analysis (\eg using PYTHIA with
$p_T^{mi} \geq 1.2$ (for GRV LO) or 2.0 GeV (LAC1)). 
It  gives ``adequate description of data''
for the transverse energy versus pseudorapidity and the average 
energy flow obtained in this analysis.\\
``For the first time  the underlying event energy has been measured 
in jet events using direct and resolved photon probes.''\\
``The strong influence of the underlying event energy on the measured cross section is demonstrated.''

The multiple interaction seems to improve also $d\sigma/dE_T$
and $d\sigma/d\eta$ distributions for jets;
within this approach the low $E_T$ and positive 
$\eta$ range is still not properly described by the two considered 
LO parton parametrizations:
LAC1 and  GRV within the PYTHIA program. PHOJET describes these data.}

~\newline
$\bullet${\bf{H1 98 \cite{h198} (HERA)}}\\
The  two-jet events 
(the 1994 data) corresponding to  to $P^2$ lower than 4 GeV$^2$
and the scaled energy $y$ between 0.2 and 0.83 were measured.
All jets have $E_T >$ 7.5 GeV, with $\bar{E}_T$ ranges above 10 \gev 
and $\Delta E_T<0.25 {\bar E_T}$, and pseudorapidity 
0 $< {\bar {\eta}} <$ 2, with $\Delta \eta <$ 1.
The cone algorithm with R=0.7 was used. 
The PYTHIA 5.7 generator with the multiple interaction using the GRV LO
parton parametrizations for the proton and the photon 
 was used in the analysis ($p_T^{mi}$=1.2 GeV). The PHOJET 1.06 was 
also used with the same parton parametrizations 
(with the $p_T$ cutoff 2.5 GeV).

\begin{figure}[ht]
\vskip 0.cm\relax\noindent\hskip 1.5cm
       \relax{\includegraphics{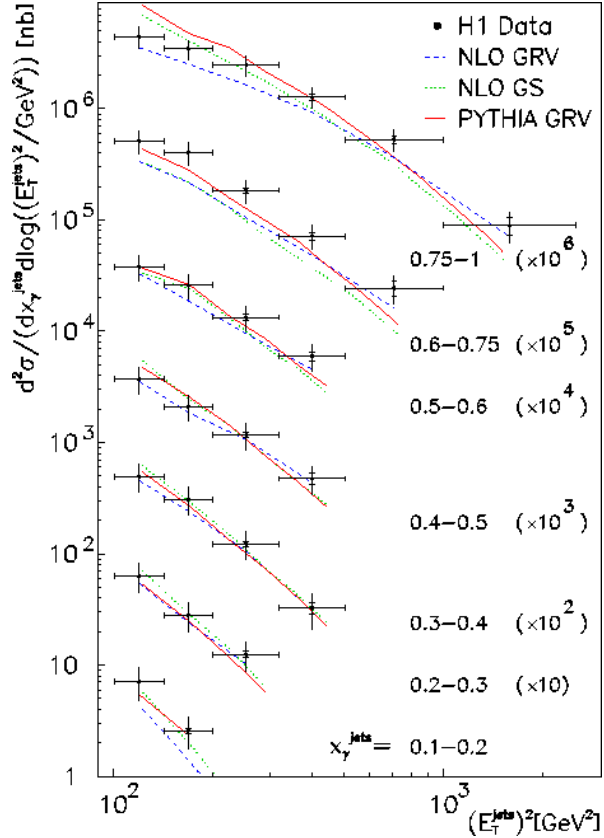}}
\vspace{0.cm}
\caption{\small\sl The double differential cross section 
as a function of the square of the averaged jet transverse energy 
 for the different $x_{\gamma}$ bins, compared to the PYTHIA (GRV)
 and the NLO calculation (GRV and GS) (from \cite{t307}).}
\label{fig:t3072}
\end{figure}

The double differential dijet cross sections,
$d{\sigma}/dx_{\gamma}/d\log(E_T^2/E_0^2)$, as a function of 
$\bar E_T^2$ for few ranges of $x_{\gamma}$ are compared with the 
NLO QCD calculation \cite{l10} and PYTHIA (GRV) simulation, 
see fig.~\ref{fig:t3072}.\\

\vspace*{10.5cm}
\vspace*{4.2cm}
\begin{figure}[ht]
\vskip 0.cm\relax\noindent\hskip 1.5cm
       \relax{\includegraphics{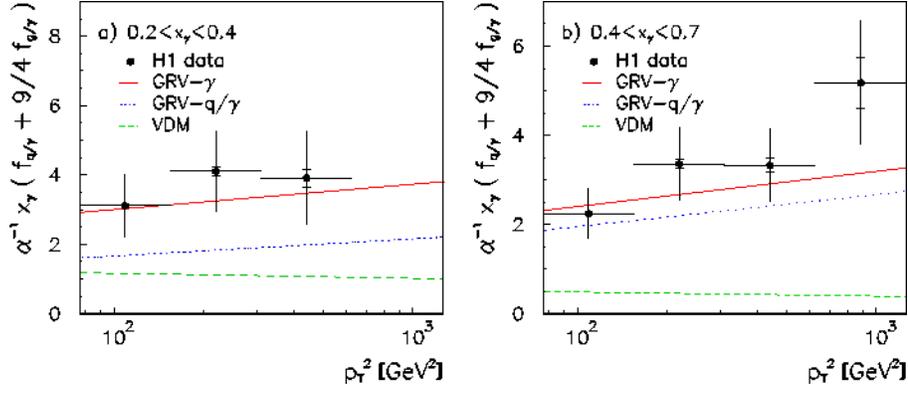}}
\vspace{0.cm}
\caption{\small\sl The effective parton density in the photon
(from \cite{t307}).}
\label{fig:t3073}
\end{figure}

The method based on the Single Effective Subprocess Approximation \cite{com-max}
was used to extract, for the first time,
 the effective (LO) parton density of the photon:
\be
{\alpha}^{-1}x_{\gamma}(\tilde q^{\gamma}+{9\over 4}G^{\gamma}),
\ee 
with $\tilde q^{\gamma} = \sum (q^{\gamma} + {\bar q^{\gamma}}$).

This was done  for 
63 GeV$^2<p_t^2<$1000 GeV$^2$,
for 
0.2 $<x_{\gamma}<$ 0.4 and 0.4 $<x_{\gamma}<$ 0.7.
The dependence of the effective parton density 
on the $\tilde Q^2$ (= $p_t^2$) scale is shown in 
fig.~\ref{fig:t3073} and in table \ref{table20} (where the statistical
and systematical errors are added in quadrature).
\begin{table}[ht]
\caption{}
\label{table20} 
$$
\begin{array}{|c|c|c|}
\hline
<x_{\gamma}>&\tilde{Q}^2 [GeV^2]&x_{\gamma}\tilde{f}(x_{\gamma})/{\alpha}\\
\hline
0.3& 112 & 3.11 \pm 0.88\\
   & 224 & 4.10 \pm 1.17\\
   & 447 & 3.91 \pm 1.34\\
\hline
0.55& 112 & 2.25 \pm 0.55\\
    & 224 & 3.36 \pm 0.82\\
    & 447 & 3.33 \pm 0.82\\
    & 891 & 5.18 \pm 1.38\\
\hline
\end{array}
$$
\end{table}

~\newline
Comment: {\it "Satisfactory overall description 
(of the double differential cross section for jet) 
except for $x_{\gamma}>0.6$".\newline
 "The effective parton distribution grows with 
the scale $p_t^2$, although the increase appears 
slightly steeper than expected from the GRV LO parametrization."}

\begin{figure}[ht]
\vskip -1.cm\relax\noindent\hskip -0.2cm
       \relax{\includegraphics{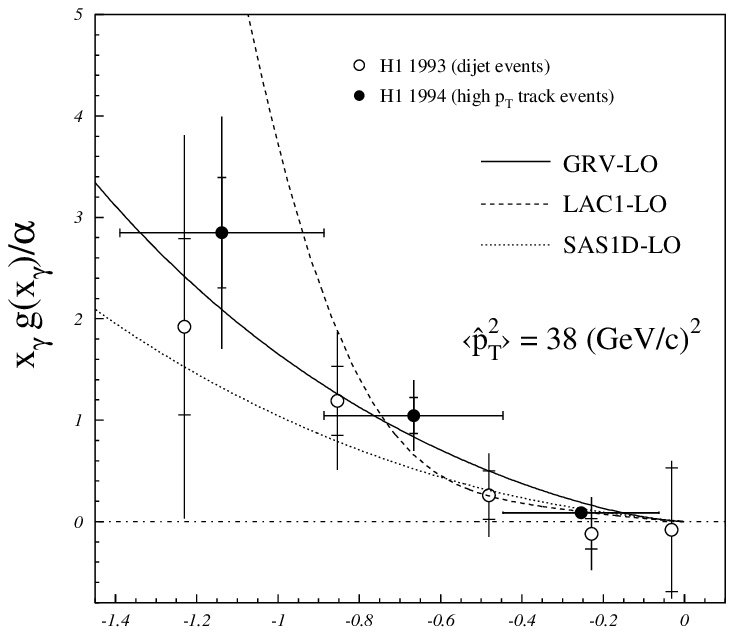}}
\vspace{0.2cm}
\caption{\small\sl The LO gluon density data extracted from 
charged particles events at $<p_T^2>$=38 GeV$^2$ ($\bullet$)
together with predictions of the GRV LO, LAC1 LO and SaS1D LO
parametrizations. Also the {\bf H1 95a} \cite{h1} data extracted 
from dijet events at  $<p_t^2>$=75 GeV$^2$ are shown
($\circ$) (from \cite{char98}).}
\label{fig:char}
\end{figure}

~\newline
$\bullet${\bf {H1 99a \cite{char98} (HERA) }}\\
The new method of extracting the gluon density in the (real) photon
from the processes with {\sl charged particles}
is introduced. Events collected in 1994 with  
$0.3<y<0.7$, $P^2<0.01$ GeV$^2$, $|\eta|<1$ and $p_T>$ 2,3 GeV were  used.
The reconstruction of the $x_{\gamma}$ variables from the highest 
$p_T$ charged 
tracks follows closely the analysis {\bf H1 95a}.
The PYTHIA 5.7 (with and without MI) 
using  the GRV LO parton parametrization for the proton,
and for the photon : GRV LO, SaS1D, LAC1, with the corresponding
$p_T^{min}$ cutoff 1.2 GeV, 1.0 GeV and 2 GeV. 
The $p_T$ and $\eta$ distributions were measured.
The  result on the LO gluon density at $\tilde Q^2=<p_t^2>$=38 GeV$^2$
 is presented in fig.~\ref{fig:char}  and compared with the 
jet data based on the 1993 runs at $<p_t^2>$=75 GeV$^2$ 
(see also fig.~\ref{fig:mary13}).\\
\vspace*{6.8cm}

~\newline
$\bullet${\bf {H1 2000a \cite{grin} (HERA) }}\\
The more precise dijet data taken in 1996 are analysed in regions
$0.5<y<0.7$, $P^2<0.01$ GeV$^2$ and $E_T>$ 4 GeV or 
$E_T>$ 6 GeV.
The rapidities lie in the range $-0.5<\bar\eta <2.5$ and
$\Delta\eta <1$.
In the cross section for $E_T>$ 4 GeV the cut on 
the invariant mass of the jets was introduced: $M_{jj}>$ 12 GeV.
The two Monte Carlo models were used : PHOJET and ``PYTHIA 5.7 + $power$
$law$ $p_T$'' (with $\tilde{p}_{t_0}$ = 1.55 GeV). 
Multiple interaction was included in both generators. For 
PYTHIA $p_T^{min}$ = $p_T^{mi}$ = 1.2 GeV, while for PHOJET
$p_T$ cutoff equals 2.5 GeV.
The GRV LO parton parametrizations in the photon and the proton were used.
The CDFCONE algorithm with R=0.7 was assumed for jet reconstruction. 

\begin{figure}[ht]
\vskip 0.cm\relax\noindent\hskip 4cm
       \relax{\includegraphics{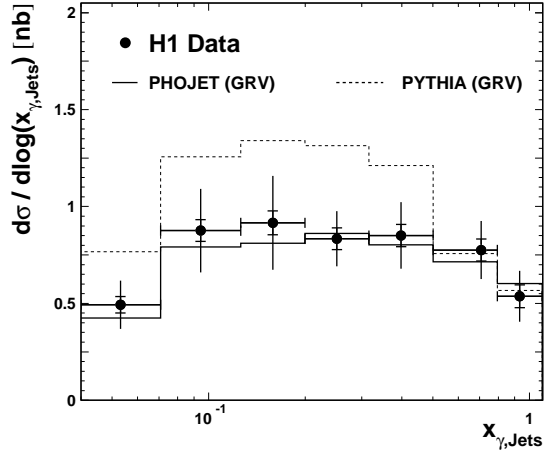}}
\vspace{-0.8cm}
\caption{\small\sl The $x_{\gamma}$ distribution of dijet events
with $E_T>$ 4 GeV (from \cite{grin}).}
\label{fig:p99-kaufmann-gpgluon-f1}
\end{figure}

\begin{figure}[ht]
\vskip 0.cm\relax\noindent\hskip 1.9cm
       \relax{\includegraphics{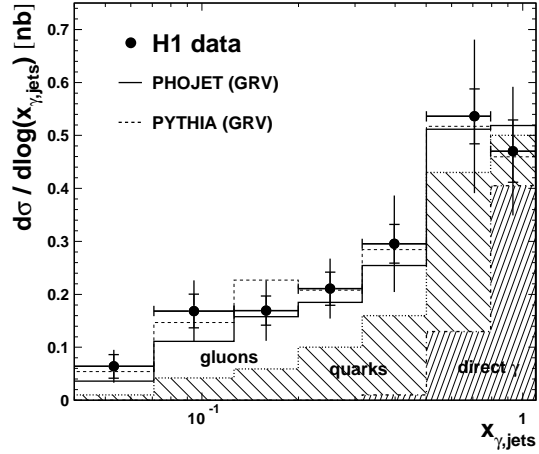}}
\vspace{-1.2cm}
\caption{\small\sl The $x_{\gamma}$ distribution of dijet events
with $E_T>$ 6 GeV (from \cite{grin}).}
\label{fig:p99-kaufmann-gpgluon-f2}
\end{figure}

The measured cross section as a function of $x_{\gamma}$ for 
$E_T>$ 4 GeV is presented in fig.~\ref{fig:p99-kaufmann-gpgluon-f1}.
Similar distribution for $E_T>$ 6 GeV (obtained after correcting 
for pedestal energy due to the underlying event) is shown in
fig.~\ref{fig:p99-kaufmann-gpgluon-f2}, where the contribution
from direct, and resolved - quark and gluon initiated - processes are
indicated separately.\\

\vspace*{5.7cm}
\vspace*{6.9cm}

\begin{figure}[ht]
\vskip 0.cm\relax\noindent\hskip 0.8cm
       \relax{\includegraphics{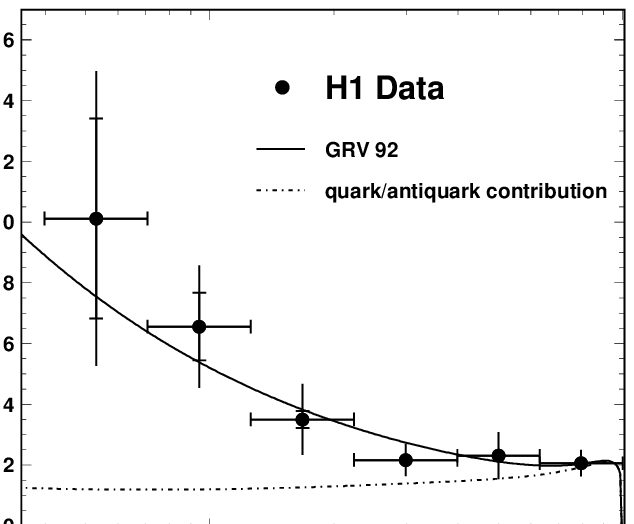}}
\vskip 0.cm\relax\noindent\hskip 6.5cm
       \relax{\includegraphics{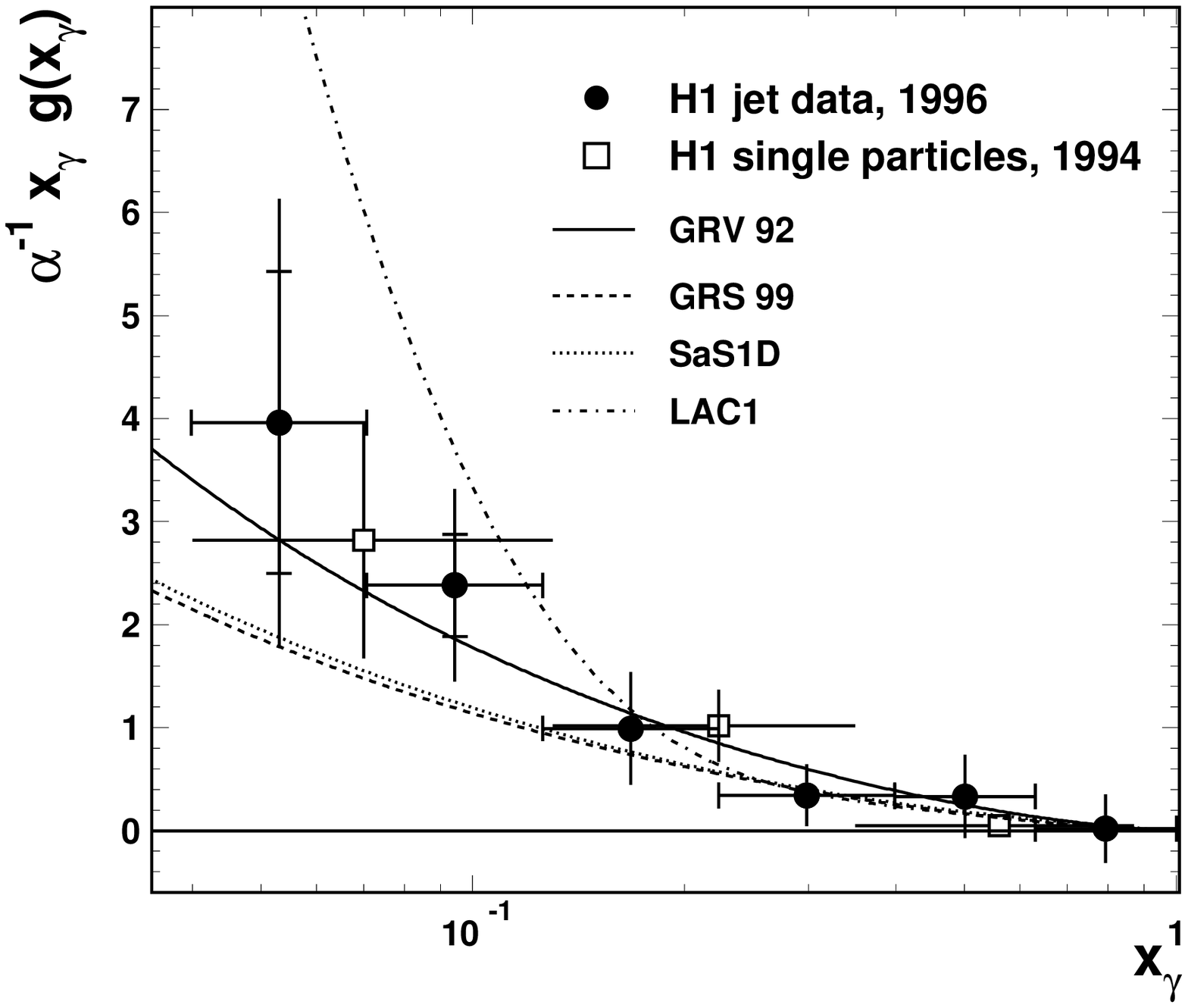}}
\vspace{0.cm}
\caption{\small\sl The $x_{\gamma}$ distributions at $<
\tilde Q^2>=<p_t^2>$=74 GeV$^2$ of the effective 
parton density (left) and the gluon density (right).
The comparison with GRV LO (left) and with GRV LO, GRS LO, SaS1D, LAC1 
predictions and the {\bf H1 99a} data (right) is shown (from \cite{grin}).}
\label{fig:p99-kaufmann-gpgluon-f4}
\end{figure}

Using the Single Effective Subprocess Approximation \cite{com-max}
the LO effective parton densities of the photon 
were determined as a function of $x_{\gamma}$
for $< \tilde Q^2>(=<p_t^2>)$=74 GeV$^2$.
The result is presented in fig.~\ref{fig:p99-kaufmann-gpgluon-f4} 
(left) and in table \ref{tableH1}. 
The gluon density is determined from the effective parton density by 
subtracting the quark densities (as given by GRV LO in agreement with
$F_2^{\gamma}$ data). Results of this analysis are presented in 
fig.~\ref{fig:p99-kaufmann-gpgluon-f4} (right)
and in table \ref{tableH1}. 
The obtained gluon density is  in agreement with that from
the single particle data presented in {\bf H1 99a}.

\newpage
\vspace*{4.5cm}

\begin{table}[ht]
\caption{}
\label{tableH1} 
$$
\begin{array}{|c|c|c|c|}
\hline
<\tilde{Q}^2>&<x_{\gamma}>&x_{\gamma}f_{\gamma ,eff}(x_{\gamma})/{\alpha}
&x_{\gamma}G(x_{\gamma})/{\alpha}\\
~[GeV^2]~&&total (stat.)&total (stat.)\\
\hline
74&0.053&10.1\pm 4.9 (3.3)&4.0\pm 2.1 (1.4)\\
&0.094&6.6\pm 2.0 (1.1)&2.4\pm 0.9 (0.5)\\
&0.17&3.5\pm 1.2 (0.3)&0.99\pm 0.55 (0.12)\\
&0.30&2.2\pm 0.5 (0.1)&0.34\pm 0.30 (0.03)\\
&0.50&2.3\pm 0.8 (0.2)&0.33\pm 0.40 (0.08)\\
&0.79&2.1\pm 0.4 (0.1)&0.02\pm 0.33 (0.03)\\
\hline
\end{array}
$$
\end{table}

~\newline
Comment: {\it The data reach parton fractional energies down
to $x_{\gamma}$ = 0.05. Leading order QCD gives a good 
description of the $E_T > 6$ GeV data (after subtraction
of the underlying event energy) which makes possible a 
determination of the effective parton density in the photon.
``This quantity is dominated by the gluon density for
$x_{\gamma} < 0.2$ which is found to rise strongly towards small
$x_{\gamma}$.''}

\newpage
~\newline
$\bullet${\bf {ZEUS 92 \cite{zeus92} (HERA) }}\\
The evidence for the hard scattering (jet production and resolved
photon processes) in the photoproduction
has been found.
The cross section for jets with $E_T\ge $ 10 GeV at HERA
was obtained.  The PYTHIA and HERWIG generators, with GS and MT B1 
parton parametrization for the photon and the proton, respectively,
were used (with $p_T^{min}$=1.5 GeV).  In the analysis the  
cone algorithm with $R=1$ was applied.

~\newline
Comment: {\sl "Evidence for the photon remnant jets was found.''}

~\newline
$\bullet${\bf {ZEUS 94 \cite{zeus94} (HERA) }}\\
The measurement was based on the 1992 data
for the single and double jet photoproduction,
with  $y$ between 0.2 and 0.7.
For tagged events  $P^2$ was below 0.02 GeV$^2$,
 otherwise below 4  GeV$^2$.
The analysis of the direct and resolved photon
processes was made using the HERWIG 5.7 and PYTHIA 5.6 generators.
In generation of events the parton parametrization GRV for the 
photon and MRSD0 for the proton
(in addition also DG, LAC and MRSD$_-$) were used, with 
$p_T^{min}$ =2.5 GeV. The jet-finding cone algorithm with $R$=1 was applied.
\newline
\vspace*{10.3cm}
\begin{figure}[hb]
\vskip 0.2cm\relax\noindent\hskip 3.5cm
       \relax{\includegraphics{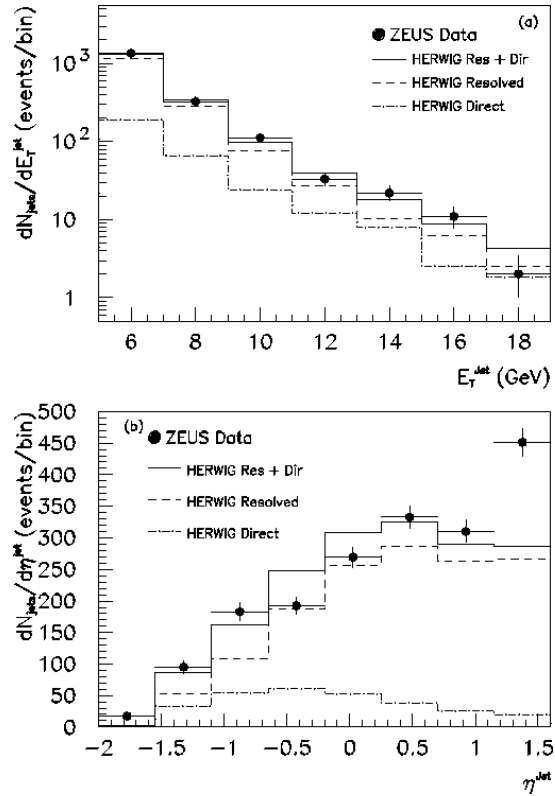}}
\vspace{-0.4cm}
\caption{\small\sl Inclusive jet distributions for (a) transverse energy
of jets, (b) pseudorapidity of jets. Comparison with the HERWIG  
prediction for the direct and resolved contributions is shown
(from \cite{zeus94}).}
\label{fig:jet34}
\end{figure}

The results for the single jets' concerning  the  $E_T$ distribution   
up to $E_T$ = 18 GeV,
integrated over rapidity $ \eta $ below 1.6, are presented in 
fig.~\ref{fig:jet34}a.
Fig.~\ref{fig:jet34}b shows 
 the $ d\sigma/d\eta$ 
data where the disagreement with the Monte Carlo prediction occurs 
for the positive $\eta$.

\begin{figure}[ht]
\vskip -0.3cm\relax\noindent\hskip 2.cm
       \relax{\includegraphics{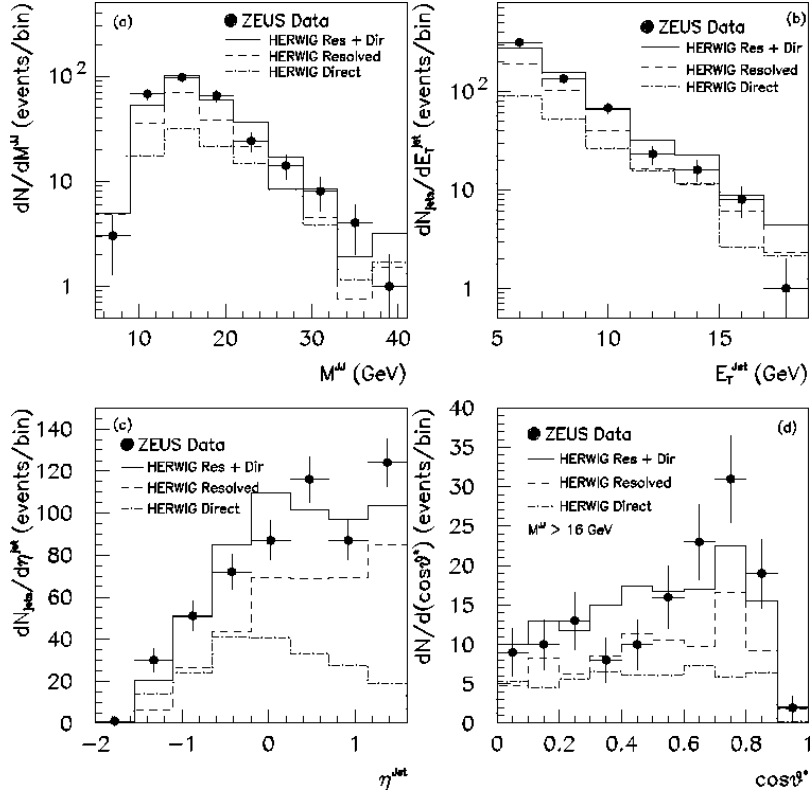}}
\vspace{-0.4cm}
\caption{\small\sl Kinematic distributions for events with two or
more jets: (a) the jet pair invariant mass, (b) the transverse
energy of jets, (c) the pseudorapidity, (d) $\cos\theta^*$
of jet angles in jet-jet CM with respect to the proton momentum
for events with $M_{ij} >$ 16 GeV. The comparison with the 
Monte Carlo simulations 
is shown (from \cite{zeus94}).}
\label{fig:jet35}
\end{figure}

Dijet production has been studied by selecting events 
with two or more jets with $E_T> 5$ GeV, for $\eta $ smaller than 1.6.
(fig.~\ref{fig:jet35}).

The $x_{\gamma}$ and $x_{proton}$ distributions were studied as well 
for events with $| \Delta \eta |< $1.5, $|\Delta \phi|>$120$^o$ and 
the invariant mass of two jets $M_{ij}$ larger than 16 \gev (not shown).\\

\vspace*{10.2cm}
~\newline
$\bullet${\bf {ZEUS 95a \cite{zeus95} (HERA) }}\\
The 1993 data for the production of at least  one jet  with $E_T>$ 
6 \gev (up to 41 \gev) are presented. Events correspond to:  
$P^2$ below 4 \g2, $y$ 
between 0.2 and 0.85 and the jet pseudorapidity between -1 and 2.
The PYTHIA 5.6 and HERWIG 5.7 generators with 
the GRV parametrization for the photon and MRSD\_ for the proton
(and in addition LAC1  and MRSD$_0$)
were used.  The diffraction contribution was modelled by the
POMPYT program ($p_T^{min}$=3 GeV).
\\
The cone algorithm with R=1 was assumed for jets. 

\begin{figure}[ht]
\vskip -0.4cm\relax\noindent\hskip 1.4cm
       \relax{\includegraphics{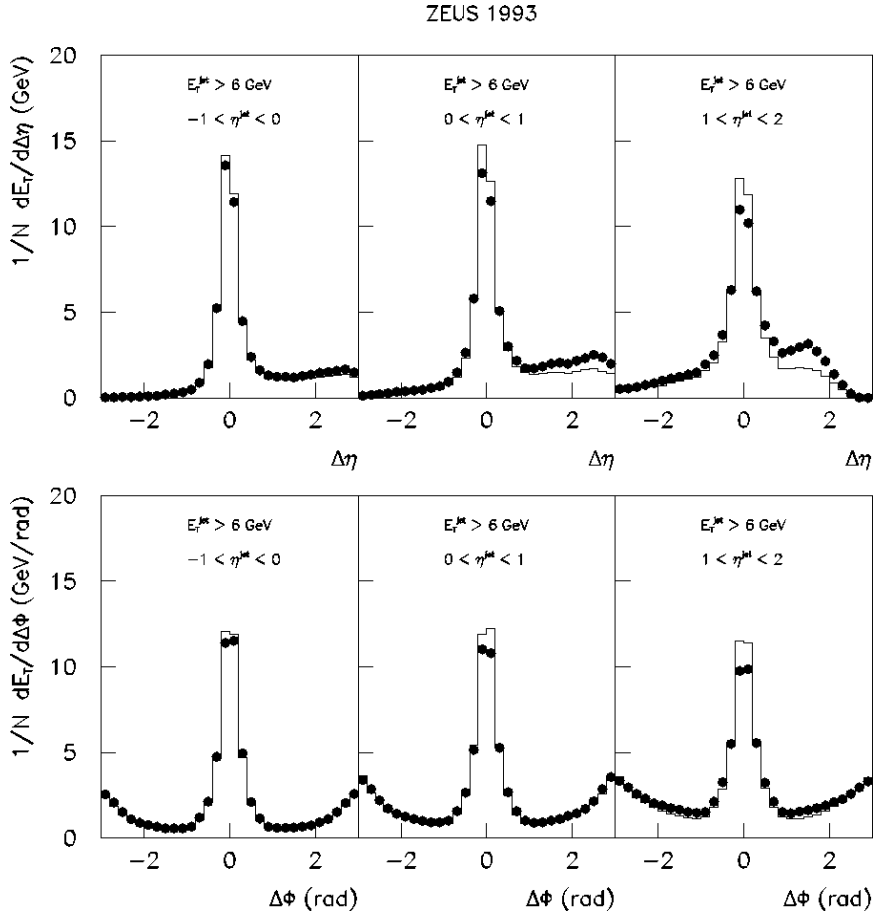}}
\vspace{-0.3cm}
\caption{\small\sl Transverse energy profiles as functions of
$\delta\eta$ integrated over $|{\delta \phi}|<\pi/2$ 
(top row) and $\delta\phi$ (bottom row) (see the text). 
Results from the PYTHIA simulation (with both resolved and direct 
processes) are shown (from \cite{zeus95}).}
\label{fig:jet36}
\end{figure}

The transverse energy flow  around jet axis was studied. Results in three
$\eta$ ranges are 
presented in fig.~\ref{fig:jet36}, where ``there is some discrepancy 
for the forward-going jets in the $\delta \eta>$ 1''.

\begin{figure}[ht]
\vskip 0.in\relax\noindent\hskip 2.8cm
       \relax{\includegraphics{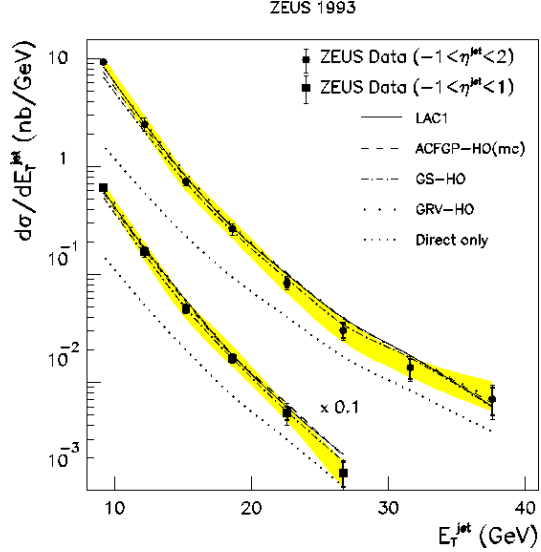}}
\vspace{-1.5cm}
\caption{\small\sl The $E_T$ distributions for jets (see text). 
Results of the PYTHIA simulations with the LAC1, ACFGP HO, GS HO and 
GRV HO parametrizations of the parton densities in the photon are 
shown (from \cite{zeus95}).}
\label{fig:jet37}
\end{figure}

\begin{figure}[ht]
\vskip 0.in\relax\noindent\hskip -0.5cm
       \relax{\includegraphics{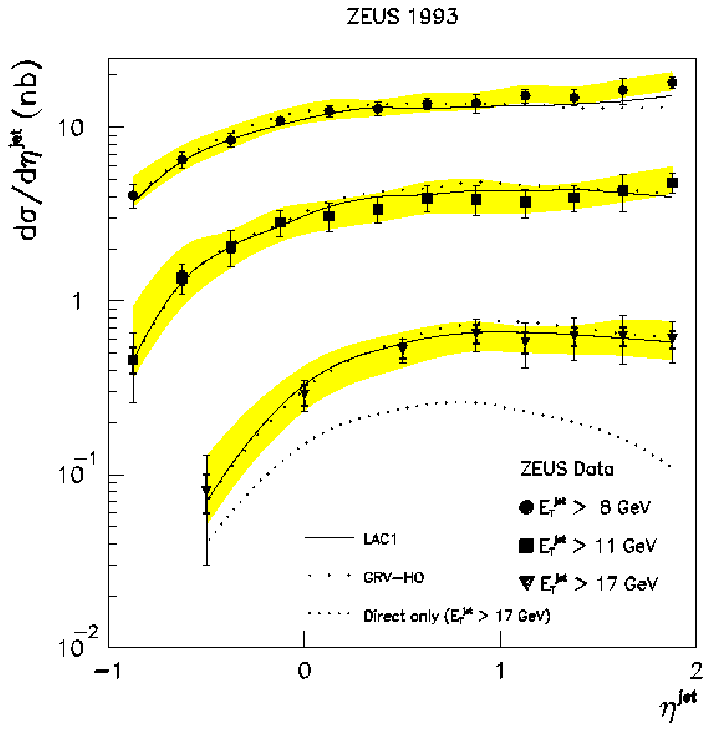}}
\vskip -0.45cm\relax\noindent\hskip 6.6cm
       \relax{\includegraphics{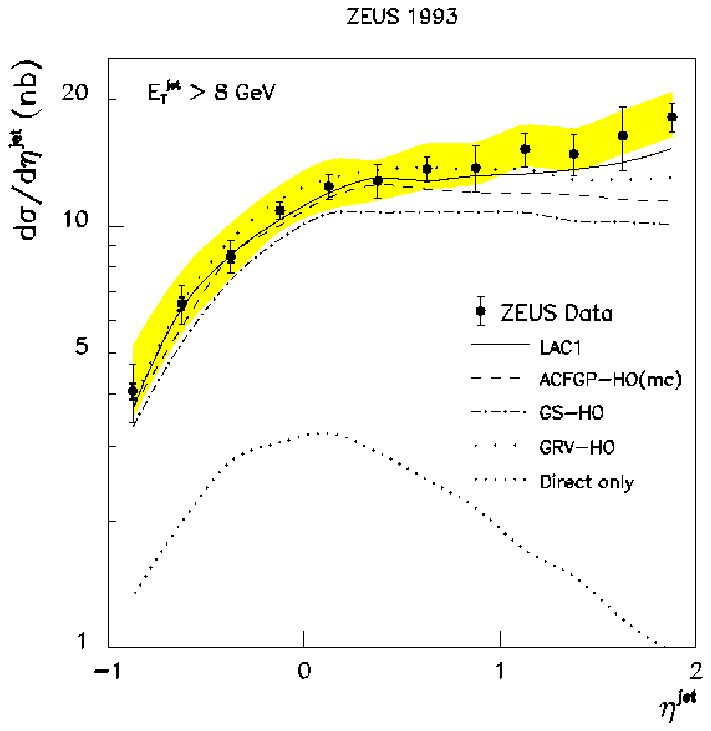}}
\vspace{-1.7cm}
\caption{\small\sl The jet pseudorapidity distributions: (left) 
integrated above three energy thresholds, $E_T\ >$ 8, 11 and 17 GeV 
(see text); the comparison with the PYTHIA simulations using the LAC1 
and GRV HO parametrizations for the photon is shown; (right)
 the $E_T\ >$ 8 sample compared in 
addition with the ACFGP HO and GS HO parametrizations for the photon 
(from \cite{zeus95}).}
\label{fig:jet38}
\end{figure}

The $E_T$ distributions integrated over two different pseudorapidity 
ranges and $d\sigma/d\eta$ distribution integrated above three $E_T$  
thresholds: 8, 11 and 17 GeV were measured. The results 
are presented in  figs.~\ref{fig:jet37} and
\ref{fig:jet38}, respectively, and compared to the PYTHIA prediction 
(for LAC1, GRV HO, ACFGP HO, GS parton parametrizations), 
with $p_T^{min}$ =5 GeV. The discrepancies between the measurement and the 
LO QCD prediction of PYTHIA are restricted to low $E_T$, very forward jet.
\newline

\vspace*{12.cm}
\newpage
\vspace*{8cm}
\vspace*{8.1cm}
~\newline
Comment: {\it "In the jet profiles, there is a significant 
excess of the transverse energy density in the data with 
respect to the Monte Carlo expectations
for jets in the region $1<\eta<2$. This excess is located outside 
of the jet in the forward direction, \ie $\Delta \eta>1$."\\
``Except for the region of very forward, low $E_T$ jets,
these measurements  are fully consistent with LO QCD 
in this new kinematic regime of the structure of the photon'' . 
 The result (for $d\sigma/d\eta$ with $E_T>$ 8 GeV$^2$
 and $-1<\eta<1$) ``does not support the discrepancy of
$d\sigma/d\eta$ with respect to LO QCD calculations 
observed by the H1 Collaboration [{\bf{H1 93}}]".}

~\newline
$\bullet${\bf {ZEUS 95b \cite{zeus95b} (HERA) }}\\
The photoproduction of dijets, with at least two jets of $E_T$ larger 
than 6 \gev, is considered in the 1993 data. Events corresponding to 
the scaled energy $y$ between 0.2 and 0.8 and $P^2$ lower than 4 \g2, 
with median $\sim 10^{-3}$ \g2
(for $|\Delta\eta |<0.5$) were grouped in  the resolved and direct 
processes samples. The cone algorithm with R=1 was used within the
HERWIG 5.7 and PYTHIA 5.6 generators with the GRV LO parametrization
for the photon and the MRSD\_ for the proton (with $p_T^{min}$=2.5 GeV).

The $x_{\gamma}$ distribution was studied,
for more recent results see 
{\bf ZEUS 96, 98a}. The cut on the $x_{\gamma}$, equal to
0.75, was introduced later to enhance the resolved or the direct
photon contributions, and a few distributions were studied 
separately for these samples.

\begin{figure}[ht]
\vskip 0.in\relax\noindent\hskip 1.5cm
       \relax{\includegraphics{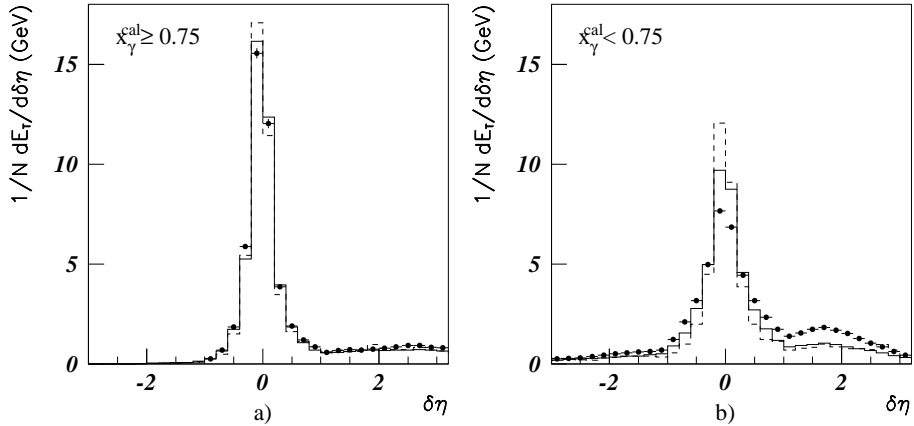}}
\vspace{0ex}
\caption{\small\sl The transverse energy flow around the jet axis versus
$\delta\eta$: (left) $x_{\gamma}\ >$ 0.75 (direct events);  
(right) $x_{\gamma}\ <$ 0.75 (resolved events). The solid (dashed) line - 
the PYTHIA (HERWIG) simulation 
(from \cite{zeus95b}).}
\label{fig:jet39}
\end{figure}

In fig.~\ref{fig:jet39} the transverse energy flow around the jet 
axis versus $\delta \eta$ is shown, for the first time separately 
for the resolved photon and direct photon contributions (with 
the failure to describe low $x_{\gamma}$ data).\\
\vspace*{4.5cm}

The $d\sigma/d{\bar {\eta}}$ was also measured 
for the direct and resolved photon events (not shown, 
see below for comments and new data in {\bf ZEUS 96}).
~\newline\newline
Comment: {\it 
"Both simulations fail to describe the transverse energy flow 
in the forward region (see also {\bf ZEUS 95a} and {\bf H1 93})."

"The  LO QCD predictions (with DG, GRV and GS2
parton parametrizations) 
 lie below  the dijet cross section 
$d\sigma/d{\bar {\eta}}$  data by factor 
1.5-2".

 The importance of the NLO calculation is stressed.}  

~\newline
$\bullet${\bf {ZEUS 95c  \cite{plb354} (HERA) }}\\
Photoproduction of three jets is studied for $\gamma p$ 
centre of mass energies in the range 130-270 GeV (1993 data). 
Events with two high-$p_T$ jets ($p_T  > 6$ GeV, $\eta < 1.6$) 
were selected. A third cluster in the 
approximate direction of the electron beam is isolated using a 
$k_T$ clustering algorithm and identified as the
photon remnant. Its properties (\eg transverse and 
longitudinal energy flows around the axis of the cluster) 
were studied. 
The Monte Carlo PYTHIA both with default and with harder $p_t$ spectrum was 
applied (the power law with $\tilde{p}_{t0} = 0.66$ GeV).
For comparison also HERWIG was used.
The GRV LO (also LAC1) and MRSD$_-$ parton parametrizations were used
for the photon and proton, respectively. The photon remnant jet was 
isolated for the first time.

~\newline
Comment: {\it Properties of the photon remnant jet isolated for
the first time ``are consistent with those commonly attributed
to jets, and in particular with those found for the two jets in 
these events. The mean value of the photon remnant $p_T$ with
respect to the beam axis is measured to be $2.1 \pm 0.2$ GeV,
which demonstrates substantial mean transverse momenta for the 
photon remnant.'' \newline
``The leading order QCD Monte Carlo simulation, with default 
parameters, does not reproduce the pseudorapidity distribution
or the transverse momentum distribution (with respect to the 
incident photon) of the photon remnant. The mean value of $p_T$
for the photon remnant, $2.1 \pm 0.2$ GeV, is substantially larger 
than the Monte Carlo expectation. Better agreement can be obtained
by increasing the average intrinsic transverse momenta of the
partons in the photon to about 1.7 GeV.''}

~\newline
$\bullet${\bf {ZEUS 96  \cite{zeus96} (HERA) }}\\
Analysis of the 1994 data for dijets (in events with two or more jets) 
for $E_T$ above 6 GeV and
with the jet pair invariant  mass above 23 GeV was performed.
Events correspond to the range of  
$y$ between 0.25 and 0.8 and $P^2$  below 4 \g2.
The cone algorithm with R=1 was used.  
The PYTHIA 5.7 and HERWIG 5.8 (with the multiple interaction) generators
(with the MRSA parton parametrization for the proton and the GRV LO for 
the photon) were applied with the $p_T$ cutoff 2.5 GeV.

To obtain the scattering angle $\cos \theta^*$ 
distribution, sensitive to the parton dynamics and 
not parton densities as in analysis above, the cut 
was introduced not on 
$\Delta \eta $ (as in previous analysis) but on $\bar {\eta}$.

\begin{figure}[ht]
\vskip 0.in\relax\noindent\hskip 0.cm
       \relax{\includegraphics{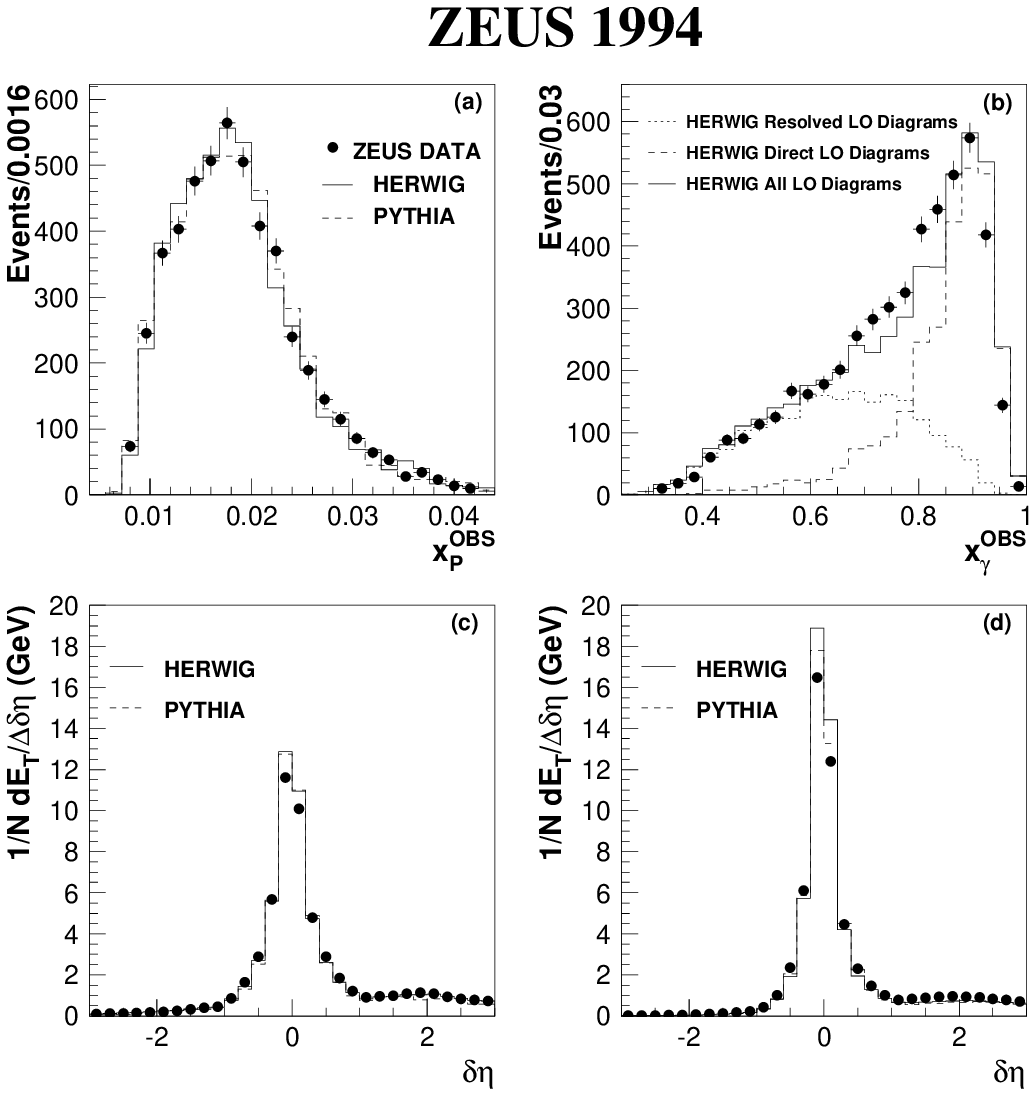}}
\vspace{0.4cm}
\caption{\small\sl The uncorrected distributions  in $x_p$ (a),
$x_{\gamma}$ (b) and of the transverse energy flow (c,d). The resolved
($x_{\gamma} <$ 0.75) and the direct ($x_{\gamma} >$ 0.75)
events as a function of $\delta\eta$ are presented separately 
(c and d, respectively).
The Monte Carlo (HERWIG) results are also shown
(from \cite{zeus96}).}
\label{fig:jet40}
\end{figure}

The results for the uncorrected $x_{\gamma}$, $x_p$ distributions 
and for the transverse
energy flow as a function of $\delta \eta$ 
are shown in fig.~\ref{fig:jet40}. Both the HERWIG (with MI) and PYTHIA 
(without MI) describe the jet profiles data.
Due to the  cut on  $\bar {\eta}$ the absolute
value of $\eta$ is restricted to be below 1.8.
Note that the applied cut on the
invariant mass suppresses events with low $x_{\gamma}$.\\ 
\vspace*{10.cm}

\begin{figure}[ht]
\vskip 0.in\relax\noindent\hskip 1.5cm
       \relax{\includegraphics{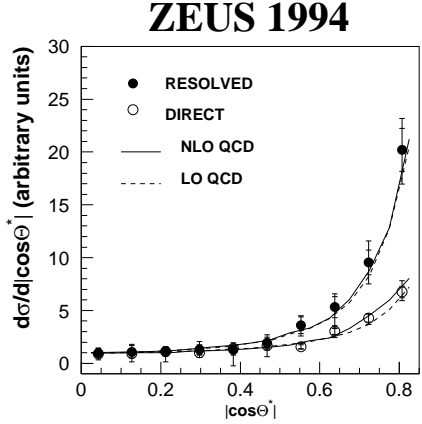}}
\vspace{0ex}
\caption{\small\sl The dijet angular distributions for the resolved
and the direct contributions (see text)
(from \cite{zeus96}).}
\label{fig:jet41}
\end{figure}

The important results concerning the angular distributions
due to various partonic subprocesses were obtained for the first 
time in the large $p_T$ resolved photon processes.
In fig.~\ref{fig:jet41} the angular distributions $d\sigma/ d\cos \theta^*$
for the  resolved and direct processes
together with the LO and NLO calculation based on the CTEQ3M 
parametrization for the proton and the GRV for the photon are presented.
The comparison was also made with the 
HERWIG and PYTHIA simulations (not shown).

~\newline
Comment: {\it The transverse energy flow is described properly,
the "requirements of high mass and small boost remove the 
disagreement in the forward flow between data and the simulations 
which has been reported elsewhere in hard photoproduction at HERA."

The dijet angular dependence  is well described by the  LO and NLO
QCD calculations, and also by the HERWIG and PYTHIA models.}\\
\vspace*{4.7cm}

~\newline
$\bullet${\bf {ZEUS 98a \cite{zeus18v2} (HERA) }}\\
This is the extension of previous analyses ({\bf {ZEUS 95b, 96}}) based 
also on the 1994 data 
on the dijet production with $P^2$ lower than 4 \gev 
and  $y$  between 0.2 and 0.8. 
Results for the production of at least two jets with the \psr 
between -1.375 and 1.875
and for $E_T^{min}$ = 6, 8, 11 and 15 \gev are presented,  
assuming $|\Delta \eta|<0.5$.
In the data analysis different jet-finding algorithms were
applied: the
cone algorithms EUCELL  and PUCELL (both with R=1) and the $k_T$ - cluster 
algorithm        
KTCLUS. 
The Monte Carlo simulation of HERWIG 5.8 and PYTHIA 5.7
with or without the multiple interaction was performed 
using MRSA and GRV LO  parton parametrizations for the proton and 
the photon, respectively. The cutoff  for no MI option 
was $p_T^{min}$=2.5 GeV, while
in the MI approach: $p_T^{mi}$ =2.5 GeV (HERWIG) and 1.4 GeV (PYTHIA). 
The data  were compared to the predictions from a NLO
QCD calculation, with the additional parameter 
describing the separation of jets $R_{sep}=R$ or 2$R$.

The resolved cross section was measured in the range 
$0.3<x_{\gamma}<0.75$ and the direct one - for $x_{\gamma}>0.75$.
Analysis of the $x_\gamma$ distribution 
 was performed;
jet profiles in form of the transverse energy flow around the jet axis
and the $d\sigma/d{\bar \eta}$ 
for the various jet definitions and transverse energy thresholds
were studied.

\begin{figure}[ht]
\vskip 0.in\relax\noindent\hskip 1.cm
       \relax{\includegraphics{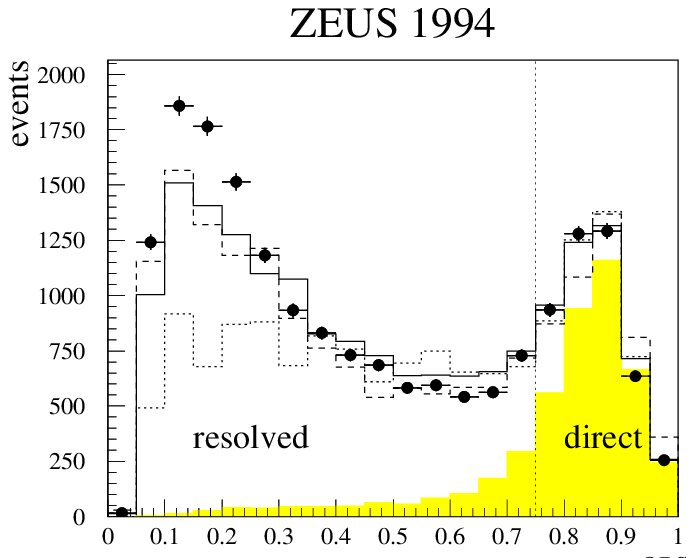}}
\vspace{0.cm}
\caption{\small \sl The corrected $x_{\gamma}^{obs}$ distribution.
Solid line - HERWIG with the multiple interaction, dashed line -
PYTHIA with the multiple interaction, dotted line - HERWIG without 
the multiple interaction (from \cite{zeus18v2}).}
\label{fig:z18v2.1}
\end{figure}

The  corrected $x_{\gamma} $ distribution obtained 
from the two-jet events 
indicates a need for the resolved photon contribution,
 see also {\bf ZEUS 95b} and {\bf 96}.
The present analysis is based on the KTCLUS algorithm, for results
see fig.~\ref{fig:z18v2.1}.
The small $x_{\gamma} $ region is not properly described by the Monte Carlo
simulations both with and without the multiple interaction.

\begin{figure}[ht]
\vskip 0.in\relax\noindent\hskip 1.3cm
       \relax{\includegraphics{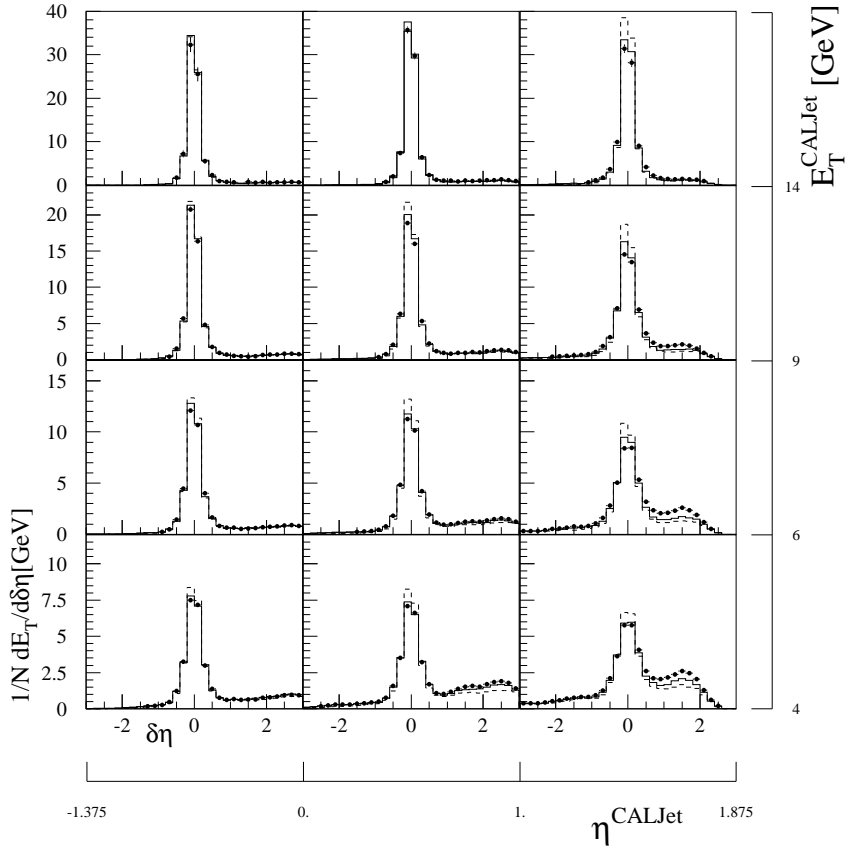}}
\vspace{-0.8cm}
\caption{\small\sl The uncorrected transverse energy flow around the 
jet. Solid line - HERWIG with the multiple interaction, 
dashed line  - HERWIG without the multiple interaction
(from \cite{zeus18v2}).}
\label{fig:z18v2.2}
\end{figure}

The transverse energy flow 
obtained using the KTCLUS algorithm is presented 
for different $E_T$ in fig.~\ref{fig:z18v2.2}, with a similar 
discrepancy in comparison to MC (HERWIG, with and without MI)
seen in the forward direction, as in previous measurements.
\newline

\vspace*{5.cm}

\vspace*{11.7cm}
 
\begin{figure}[ht]
\vskip  0.in\relax\noindent\hskip 0.cm
       \relax{\includegraphics{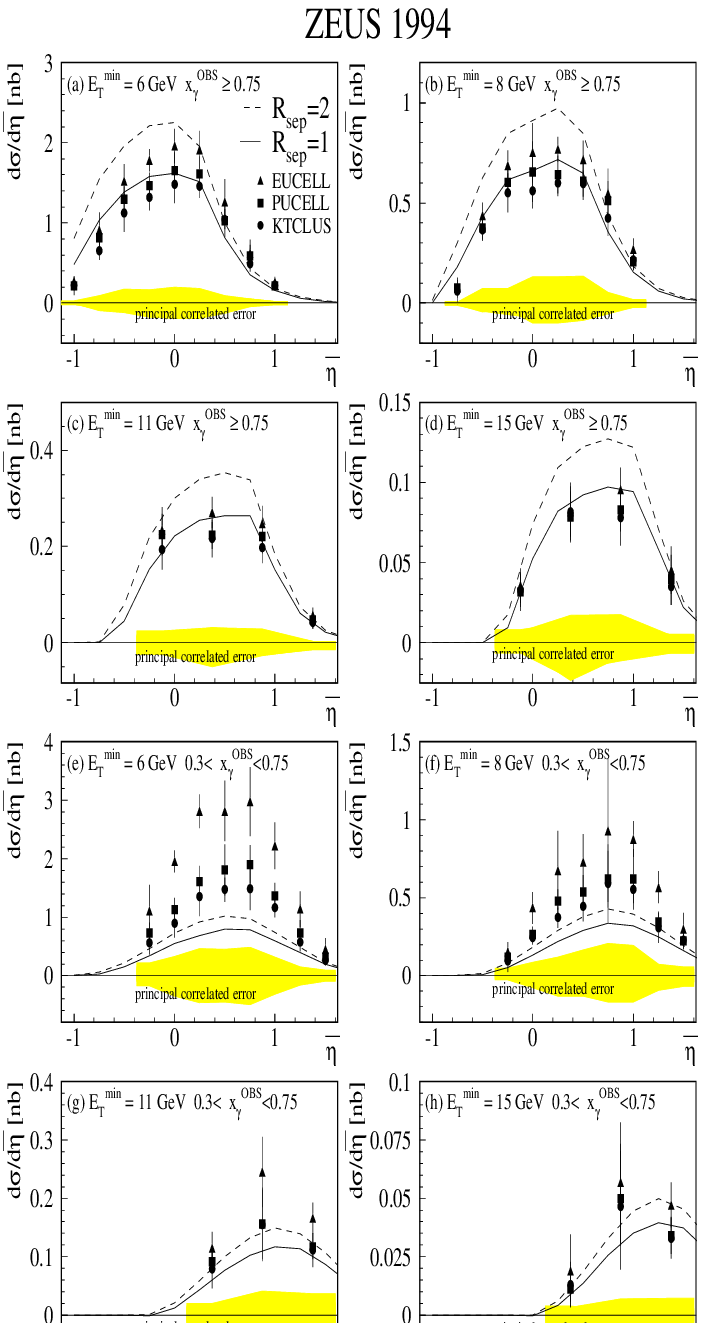}}
\vskip 0.in\relax\noindent\hskip  8.2cm
       \relax{\includegraphics{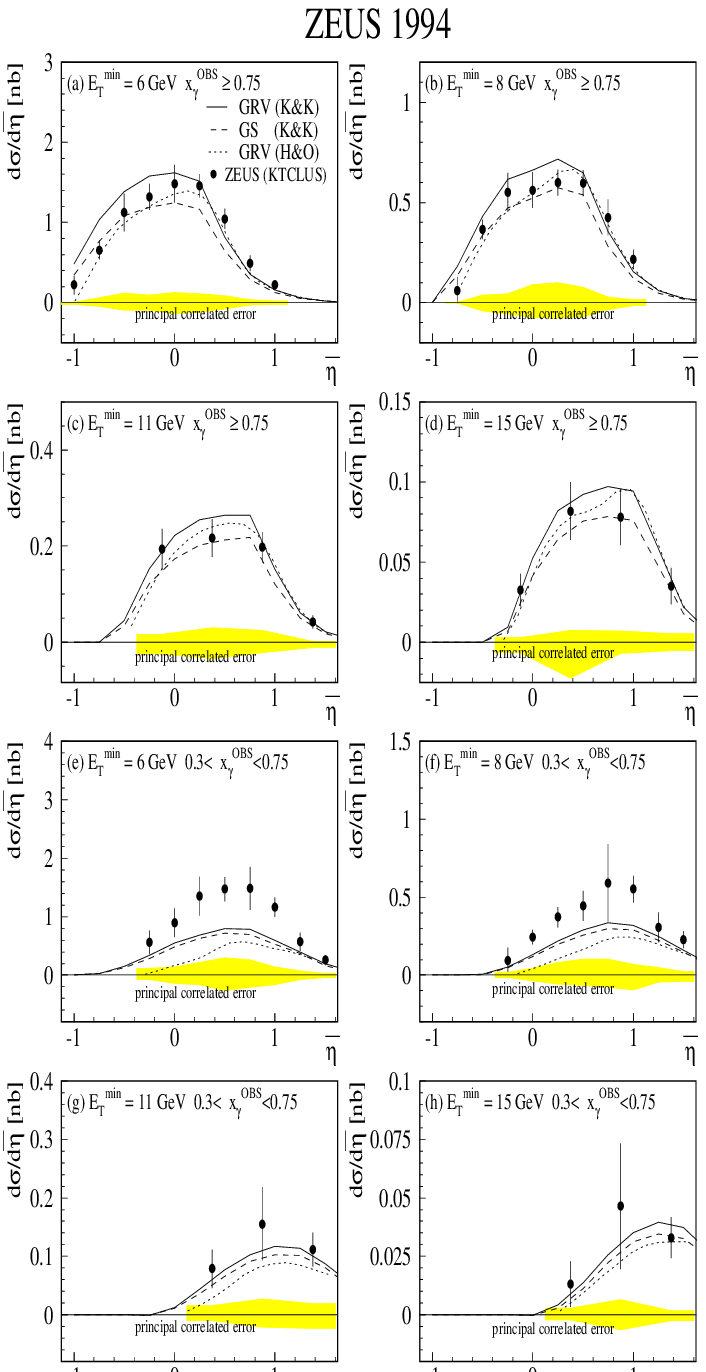}}
\vspace{0cm}
\caption{\small\sl Left: Dijet cross sections obtained using
different jet algorithms: EUCEL, PUCEL and KTCLUS. The curves show 
the results of the NLO calculation \cite{l10} using GS parton 
parametrization with $R_{sep}$ = 1 (solid line) and $R_{sep}$ = 2 (dashed).
Right: Dijet cross sections obtained using the KTCLUS jet algorithm. 
The curves show the results of the NLO calculations K\&K \cite{l10} 
and H\&O \cite{l13} with $R_{sep}$ = 1, using different NLO parton 
parametrizations: GRV (K\&K - solid, H\&O - dotted line) 
and GS (K\&K - dotted line) (from \cite{zeus18v2}).}
\label{fig:z18v2.4}
\end{figure}

The dijet cross section 
$d\sigma/d\bar {\eta}$ obtained under condition  $ | \Delta \eta | < $ 0.5,
for $E_T >$ 6 GeV was measured and the effect of the MI was studied 
(not shown).
This cross section was also studied 
for different $E_T$ thresholds and with  different jet definitions, separately for $x_{\gamma}>$ above and below  0.75. 
The results are  
 plotted in fig.~\ref{fig:z18v2.4} (left) together with the predictions of     
the NLO QCD approach (the CTEQ3M parton parametrization for the proton 
and GRV HO for the photon; R$_{sep}$=2 or 1) \cite{l10}.

The same cross section, 
now with the KTCLUS jet definition,  
is  plotted in fig.~\ref{fig:z18v2.4} (right). The predictions of the    
NLO QCD approach \cite{l10}
with the different parton pa\-ra\-me\-tri\-za\-tions for the photon (GS and GRV HO)
are compared with the data.\\
\vspace*{12.7cm}
~\newline
``The difference between the parton distributions is largest in the 
direct photoproduction region. This is due to differences between the 
quark distributions in the photon for $x_{\gamma}>$ 0.8''.

~\newline
Comment: {\it There is a  lack of a good description of the $x_{\gamma}$ distribution for low
$x_{\gamma}$. The discrepancy in the $\bar {\eta}$ distributions for
resolved photon contributions
was found for events with $E_T^{min}>6 $ GeV.}

~\newline
$\bullet${\bf {ZEUS 98b \cite{zeus650} (HERA) }}\\
The inclusive single jet  
cross section for the $P^2$ below 4 \g2 
with $y$ between 0.2 and 0.8 and for three energy regions for 
$W$ between 134 and 277 GeV was studied. 
The measurement  of the transverse energy of the jets bigger 
than 14 \gev and  the pseudorapidity range from  -1 to 2
was performed. The Monte Carlo generators
HERWIG 5.8  and PYTHIA 5.7 (default) ($p_T^{mi}$=1 GeV, and $p_T^{mi}$ for two partons equal to 8 GeV),
including the multiparton interaction, were introduced
(using  jet cone algorithm with R=1, 0.7 and 0.5)  
in the analysis. Events were generated with 
the MRSA parton parametrization describing  the proton structure and the GRV HO
 and LAC1 parton parametrizations for the photon.

\begin{figure}[ht]
\vskip  0.in\relax\noindent\hskip 0.cm
       \relax{\includegraphics{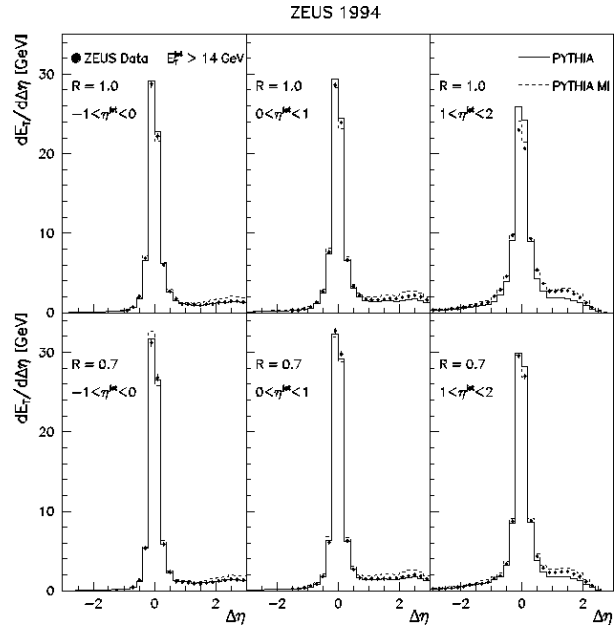}}
\vspace{0cm}
\caption{\small\sl The uncorrected jet profiles as a function of $\delta \eta$
(integrated over $|\delta \phi| < \pi/2$),  for $E_T >$ 14 GeV, 
in three $\eta$ regions, compared with PYTHIA (with and without MI)
for R=1 and 0.7
(from \cite{zeus650}).}
\label{fig:9802012_2}
\end{figure}

Transverse energy profiles for jets with $E_T>14$ GeV
are given for R=1.0 and 0.7 using the PYTHIA (with and without MI), 
see  fig.~\ref{fig:9802012_2}.
The discrepancies are observed for the 
jets with the lowest $E_T$ and for $\eta >1$ for R=1, they are  reduced 
when $R=0.7$ is used. The multiple interaction improves the description of the data for forward jets with low $E_T$, but is worse in describing the smaller
$\eta$ region for R=1, and for whole $\eta$ region when R=0.7 is used. 
For jets with $E_T> 21$ GeV no significant discrepancies for jet profiles 
are observed between data and the Monte Carlo simulations (not shown).

\vspace*{8.cm}

\begin{figure}[ht]
\vskip -9.cm\relax\noindent\hskip 2.7cm
       \relax{\includegraphics{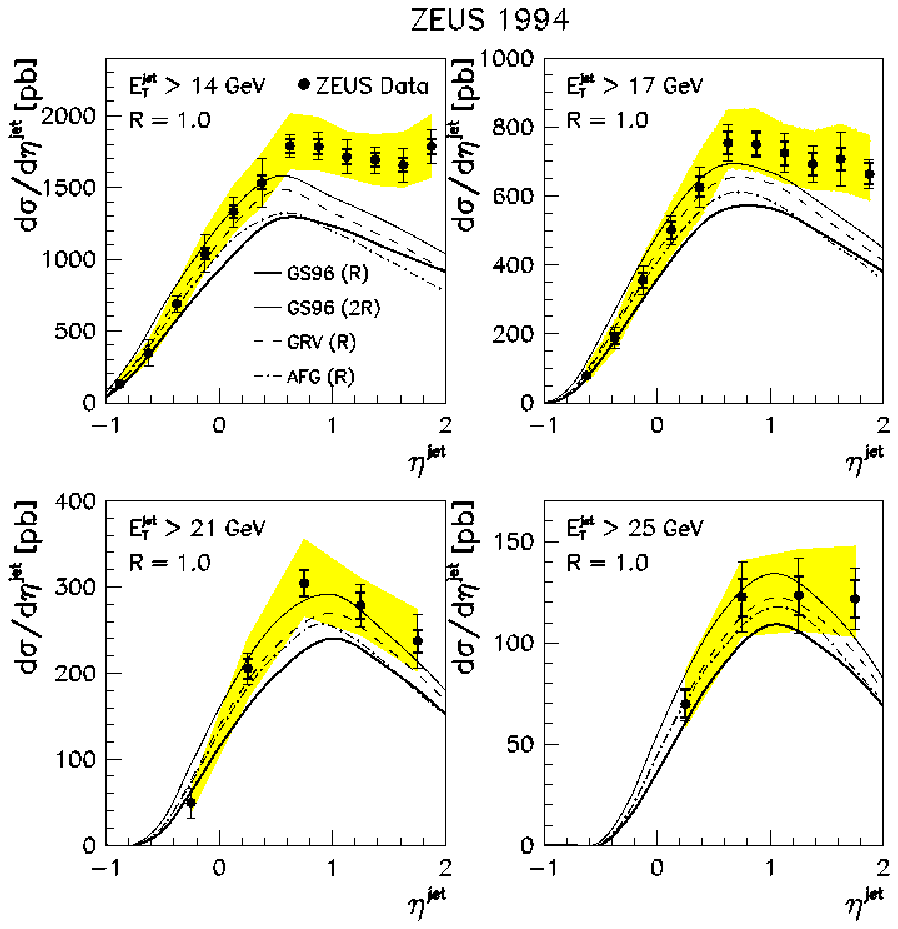}}
\vskip 9.cm\relax\noindent\hskip 2.7cm
       \relax{\includegraphics{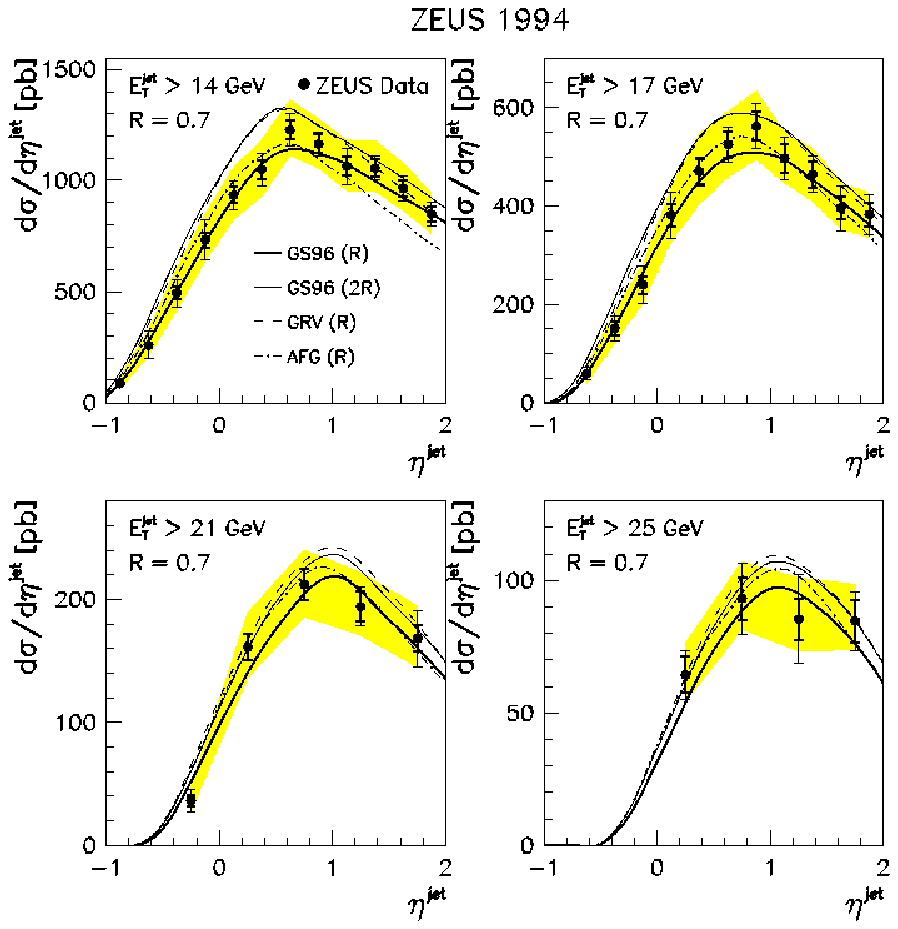}}
\vspace{-0.6cm}
\caption{\small\sl The differential cross section
$d\sigma$/$d\eta$ integrated over $E_T$
for four thresholds: $E_T >$ 14, 17, 21 and 25 GeV.
Curves based on the
NLO calculations \cite{l10} 
 using the GRV HO (R$_{sep}$ = R), GS HO (R$_{sep}$ = R and 2R) 
and AGF (R$_{sep}$ = R) 
 parton parametrizations
for the photon and the CTEQ4M for the proton with  $R$=1 (upper figures),
 $R$=0.7 (lower figures)
are shown (from \cite{zeus650}).}
\label{fig:jet6501}
\end{figure}

 The distributions of jets  
with $E_T^{min}$=14, 17, 21 and 25 GeV as a
function of $\eta$ ($-1<\eta<2$) were measured, see for results 
fig.~\ref{fig:jet6501}. 
They are not properly 
 described in the forward low $E_T$ region by the NLO calculation
\cite{l10} if $R$=1 is used (upper figures),  for $R$=0.7 the agreement
is obtained (lower figures).   
In the analysis  $R_{sep}=R$ or 2$R$ was 
applied.\\

\vspace*{17cm}

\begin{figure}[ht]
\vskip 0.cm\relax\noindent\hskip 1.6cm
       \relax{\includegraphics{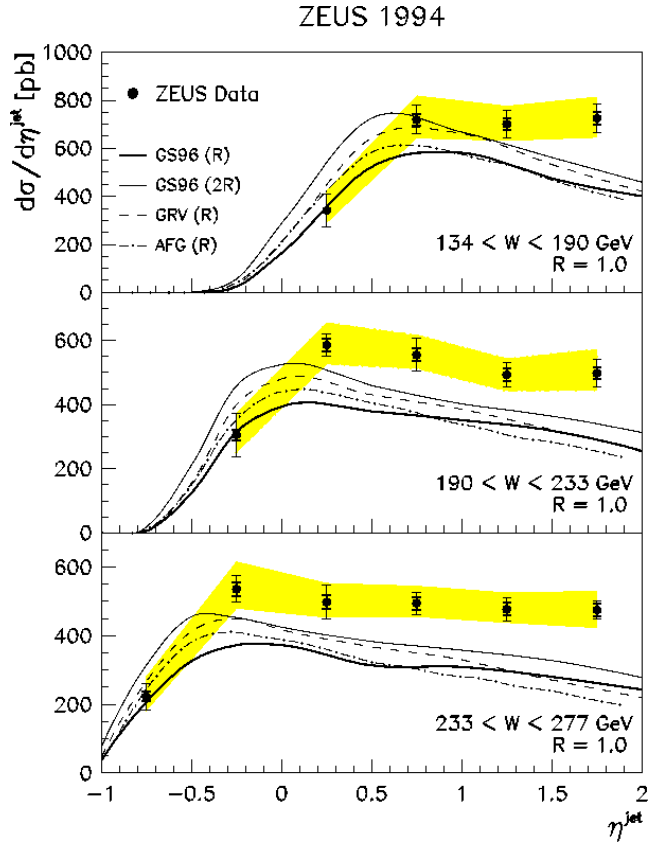}}
\vspace{-0.6cm}
\caption{\small\sl The differential cross 
section $d\sigma$/$d\eta$
for three regions of the energy W. The PYTHIA results with the MRSA 
for the proton and with different
parton distributions in the photon (with R$_{sep}$ = R or 2R) are shown
(from \cite{zeus650}).}
\label{fig:DESY-98-018_6}
\end{figure}

The $\eta$ distributions in three regions of energy W
are not in  agreement with the NLO calculation for the $R$=1,
as can be seen in fig.~\ref{fig:DESY-98-018_6}.
The data for $R=0.7$ (not shown) are  in agreement with a QCD 
calculation, the same as in the previous analysis, {\bf ZEUS 98a}.\\
\vspace*{10.cm}

\begin{figure}[ht]
\vskip 0.cm\relax\noindent\hskip 2.7cm
       \relax{\includegraphics{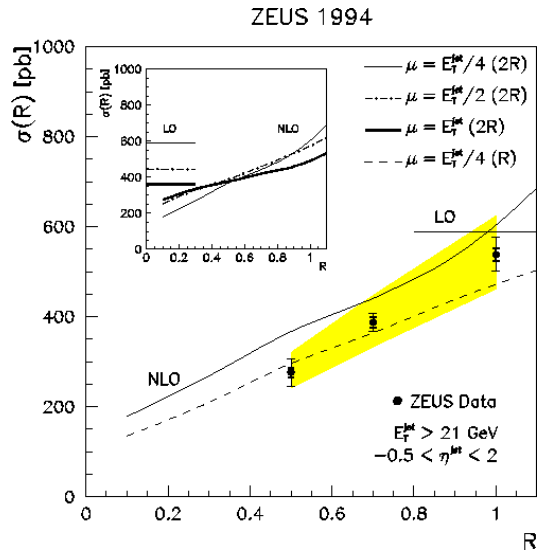}}
\vspace{0cm}
\caption{\small\sl 
The cross section dependence on the R parameter  for 
$E_T >$ 21 GeV and -0.5 $< \eta < $ 2 obtained using various 
factorization scale $\tilde Q$ $(= \mu)$ in the LO and NLO calculations 
(with the  CTEQ4 and GS96 parton parametrization for the proton and photon, 
respectively) \cite{l10} 
(from \cite{zeus650}).}
\label{fig:DESY-98-018_8}
\end{figure}

The cone radius dependence of the cross section was studied as well
to elucidate effects of a possible underlying event,
see fig.~\ref{fig:DESY-98-018_8}.  
\vspace*{7.cm}
~\newline
Comment: {\it ``...the uncertainties on the jet measurements due to 
possible underlying event contributions become reduced at high 
$E_T$ ($E_T >$ 21 GeV) or when using a reduced cone radius (R = 0.7).''
  
"The measured cross sections for jets with $R$=0.7 are 
well described by the [NLO QCD] calculations in the entire 
measured range of $\eta$ and $E_T$.}"

~\newline
$\bullet${\bf {ZEUS 98c,conf  \cite{ICHEP98-812} (HERA) }}\\
The $E_T$ measurement for jet photoproduction ($P^2 < 4$ GeV$^2$) at 
134 $<W<277$ GeV
is presented. Data collected in 1995-1997 lie in the large $E_T$ range: 17--74
GeV and -0.75 $< \eta< 2.5$. Two methods of  jet identification were 
used - with the iterative cone algorithm (R=0.7 and 1) and with the $k_T$ cluster algorithm. Comparison was made with the NLO calculation (with R$_{sep}$=R), based on the 
CTEQ4M and the GRV HO, GS 96 and AGF parton parametrization for the proton and photon, respectively.

\begin{figure}[ht]
\vskip 0.in\relax\noindent\hskip -0.3cm
       \relax{\includegraphics{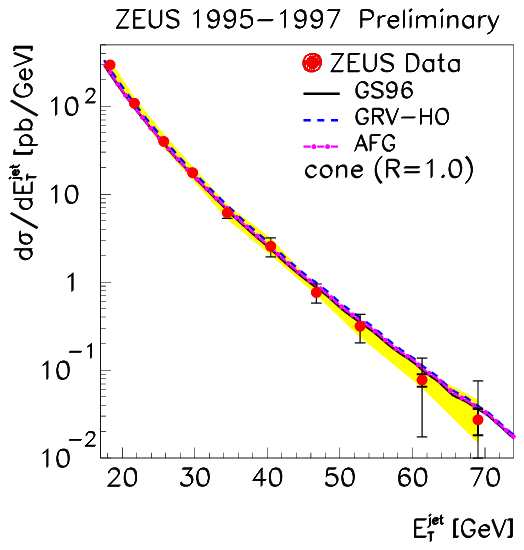}}
\vskip -0.48cm\relax\noindent\hskip 4.65cm
       \relax{\includegraphics{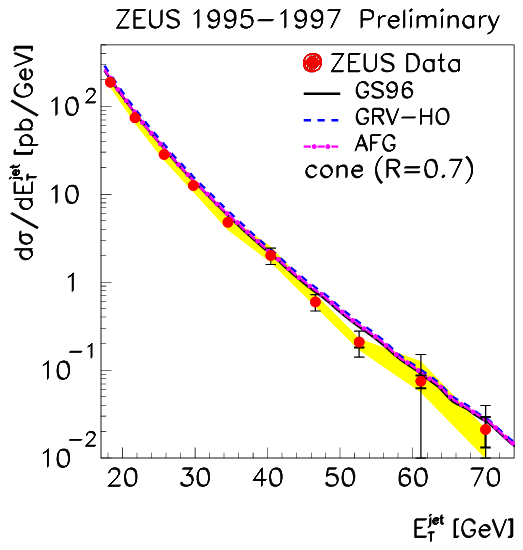}}
\vskip -0.6cm\relax\noindent\hskip 9.6cm
       \relax{\includegraphics{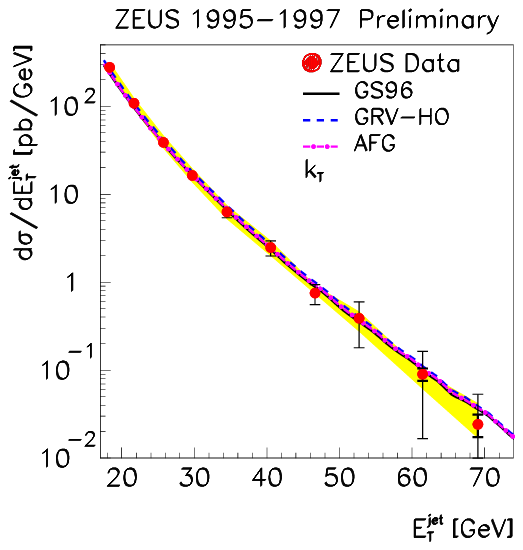}}
\vspace{-0.8cm}
\vspace{0ex}
\caption{\small\sl The $E_T$ distributions obtained using cone algorithm
(left with R=1 and center with R=0.7) and the  $k_T$ cluster method (right),
compared to the NLO calculation \cite{l10}
(from \cite{ICHEP98-812}).}
\label{fig:fig812_1}
\end{figure}

\begin{figure}[ht]
\vskip 0.in\relax\noindent\hskip -0.3cm
       \relax{\includegraphics{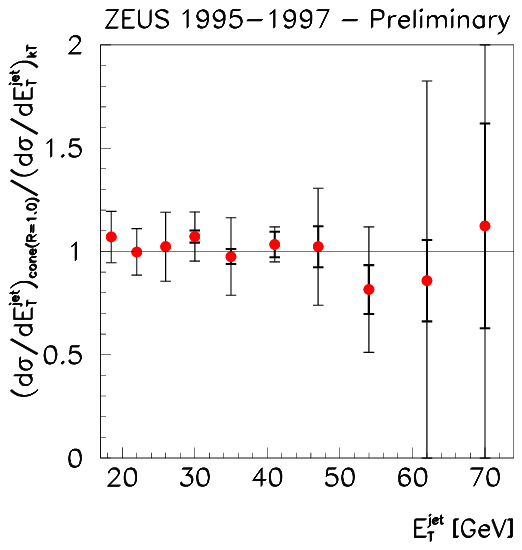}}
\vskip -0.48cm\relax\noindent\hskip 4.65cm
       \relax{\includegraphics{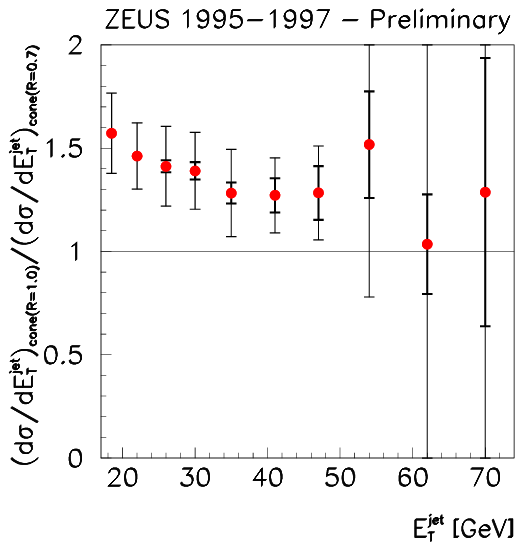}}
\vskip -0.6cm\relax\noindent\hskip 9.6cm
       \relax{\includegraphics{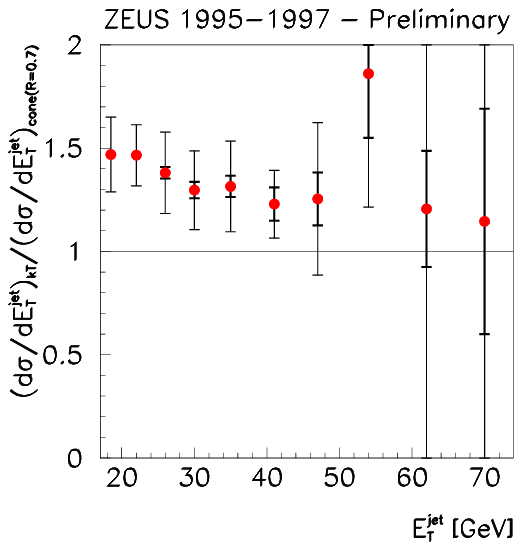}}
\vspace{-0.8cm}
\vspace{0ex}
\caption{\small\sl The ratios of the $E_T$ distributions obtained with the
jet definitions, as used in fig.~\ref{fig:fig812_1}
(from \cite{ICHEP98-812}).}
\label{fig:fig812_4}
\end{figure}

The $E_T$ distributions are presented in fig.~\ref{fig:fig812_1}
for various jet definitions:
cone algorithm with R=1 (left), cone algorithm with R=0.7 (center) 
and $k_T$ cluster method (right). Fig. \ref{fig:fig812_4} shows the 
comparison of different ratios of $d\sigma/dE_T$.
The cone algorithm with R=1 and the  $k_T$ cluster approach 
give similar results, larger than results obtained with R=0.7.
  
\vspace*{6.cm}

\vspace*{5.5cm}

\begin{figure}[ht]
\vskip 0.in\relax\noindent\hskip -0.3cm
       \relax{\includegraphics{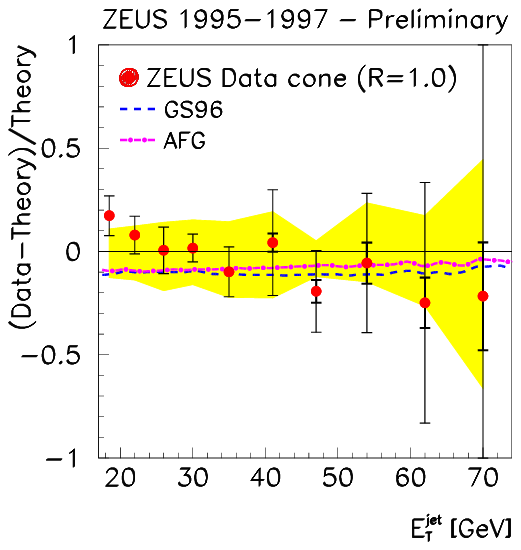}}
\vskip -0.48cm\relax\noindent\hskip 4.65cm
       \relax{\includegraphics{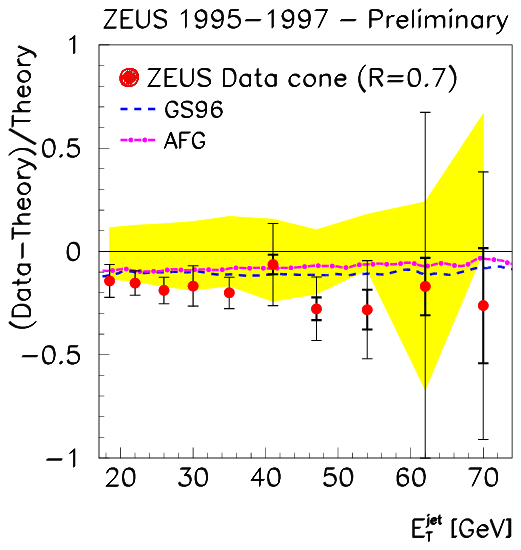}}
\vskip -0.5cm\relax\noindent\hskip 9.6cm
       \relax{\includegraphics{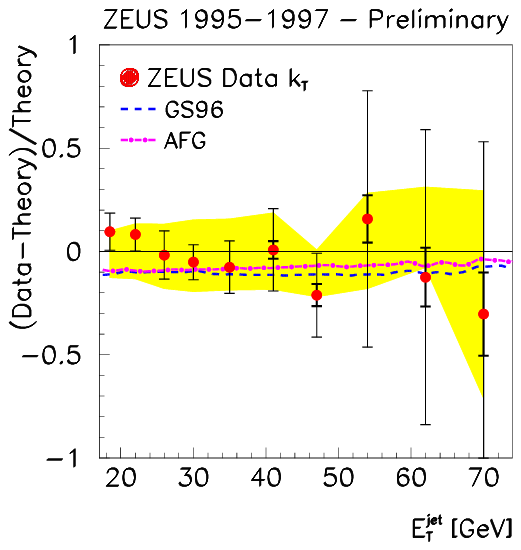}}
\vspace{-0.8cm}
\vspace{0ex}
\caption{\small\sl The fractional differences between the measured 
$d\sigma/dE_T$ for jets and the NLO calculations (using the GRV HO 
parametrizations - black points) with the  cone algorithm
(left with R=1, center with R=0.7) and the  $k_T$ cluster method (right).
The comparison with the NLO calculations with 
different parton parametrization for the photon:
GS96 (dashed lines) and AGF (dash-dotted lines)
is also shown
(from \cite{ICHEP98-805}).}
\label{fig:fig812_5}
\end{figure}

The comparison with the theoretical predictions using the CTEQ4M parton
 parametrization for the proton and the GRV HO, GS96 and AGF densities
for the photon, with $R_{sep}=R$, and different jet definitions
is presented in fig.~\ref{fig:fig812_5}.\\

\vspace*{6.2cm}

~\newline
Comment: {\it ``The NLO calculations give a reasonable description of 
the measured differential cross section in magnitude and shape.''}
~\newline\newline
$\bullet${\bf {ZEUS 99a  \cite{DESY-99-057} (HERA) }}\\
The 1995 measurement  of the photoproduction of dijets 
for $P^2<$ 1 \gev$^2$, and $y$ between 0.2 and 0.85,
  corresponding to 134 $<W<$ 277 GeV was performed, also a narrower $y$ range 
0.50 to 0.85 was studied.
  The aim was to 
constrain the parton densities in photon at high $x_{\gamma}$. Events were 
separated in to the  direct sample ($x_{\gamma}>$ 0.75) and the resolved photon sample ($x_{\gamma}<$ 0.75). 
The asymmetric cut is applied on the two highest transverse energy jets. 
The cross section is symmetrized in the pseudorapidities of these jets.
Events correspond to the highest $E_T$ jets with the threshold
 $E_{T leading}=$ 14  \gev and for the  second jet
the threshold is given by $E_{T second}=$ 11 \gev.  
The rapidity range $-1<\eta<2$ is considered.
 Two Monte Carlo generators  were used: HERWIG 5.9 and PYTHIA 5.7
with  the $k_T$ clustering method for the jet identification.

\begin{figure}[hb]
\vskip 0.in\relax\noindent\hskip 3.6cm
       \relax{\includegraphics{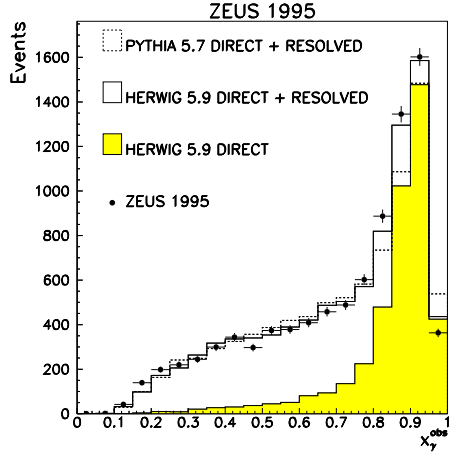}}
\vspace{0ex}
\caption{\small\sl The  uncorrected $x_{\gamma}$ distribution
for the direct (defined as 
$x_{\gamma}>$ 0.75) and the full (direct+resolved) contributions obtained 
in the PYTHIA 5.7 and HERWIG 5.9 generators for $E_T>$ 14 GeV
(from \cite{DESY-99-057}).}
\label{fig:jet49}
\end{figure}

The uncorrected  $x_{\gamma}$ distribution is shown in  
fig.~\ref{fig:jet49}, together with 
Monte Carlo predictions for a direct 
and the full (direct+resolved) contributions.
\newline
\vspace*{5.8cm}

\begin{figure}[hc]
\vskip 0.in\relax\noindent\hskip 2cm
       \relax{\includegraphics{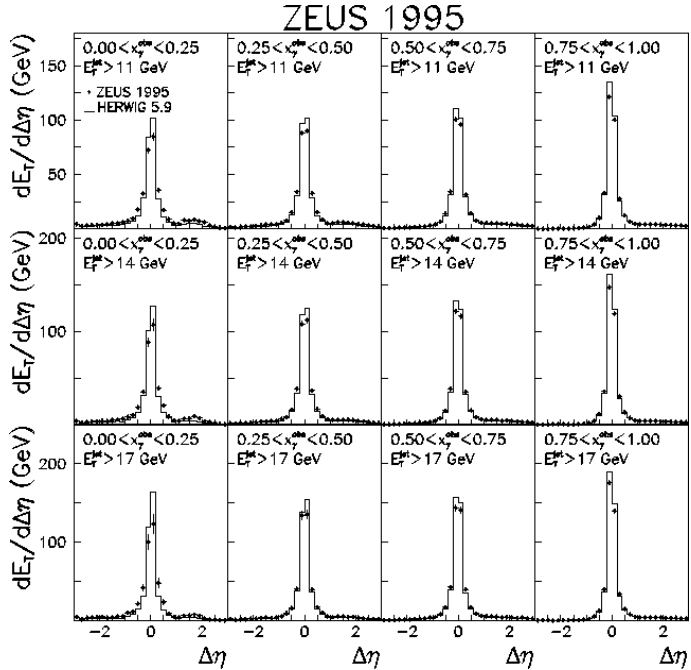}}
\vspace{0ex}
\caption{\small\sl The transverse energy flow around the jet axis (integrated
 over $|\delta \phi | <$ 1), in different ranges
 of transverse energy and $x_{\gamma}$
(from \cite{DESY-99-057}).}
\label{fig:jet50}
\end{figure}

The energy flow around the jet was studied for different regions of 
$x_{\gamma}$ and for different  $E_{T,leading}$ thresholds, see fig.~\ref{fig:jet50}.

\vspace*{9.3cm}

\begin{figure}[ht]
\vskip 0.cm\relax\noindent\hskip 1.5cm
       \relax{\includegraphics{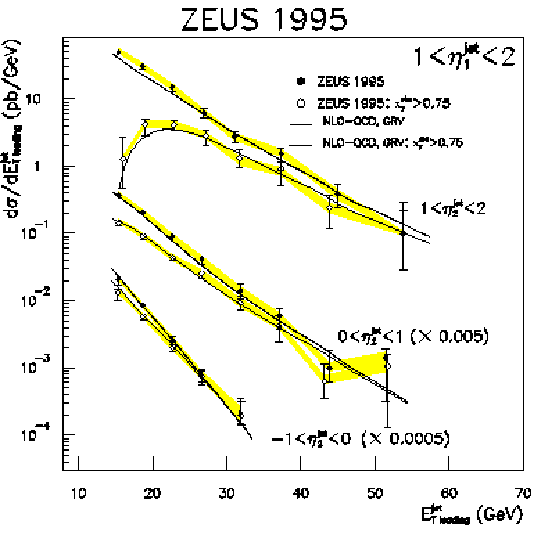}}
\vskip -0.48cm\relax\noindent\hskip 8.5cm
       \relax{\includegraphics{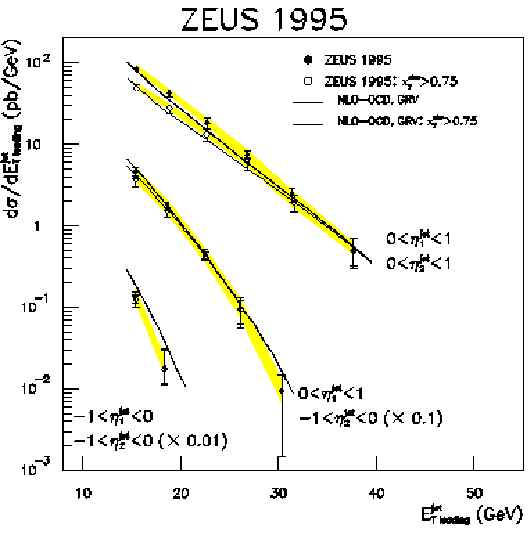}}
\vspace{-0.6cm}
\caption{\small\sl Left: Dijet cross section 
$d\sigma$/$dE_{Tleading}$ with one jet
within the rapidity range 1 $< \eta_1 <$ 2 and the other jet in 
three rapidity ranges, described  in the figure.
Right: Dijet cross section 
$d\sigma$/$dE_T$ with one jet
within rapidity range 0 $< \eta_1 <$ 1 and the other jet in 
three rapidity ranges, described  in the figure.
Comparison with the NLO calculation (with GRV HO) \cite{l10}
- thick line (thin line shows the direct contribution)
(from \cite{DESY-99-057}).}
\label{fig:6541}
\end{figure}

Below the results for the $E_T$ distribution in different regions of 
$\eta_{1,2}$ are presented (fig. \ref{fig:6541}), with the direct 
 contributions displayed separately. Comparison with the NLO calculation
(with CTEQ4M and the GRV HO, GS96 HO, AGF HO parton parametrizations for the proton and photon, respectively) is shown. 
\vspace*{5.7cm}

\begin{figure}[ht]
\vskip 0.in\relax\noindent\hskip 4.2cm
       \relax{\includegraphics{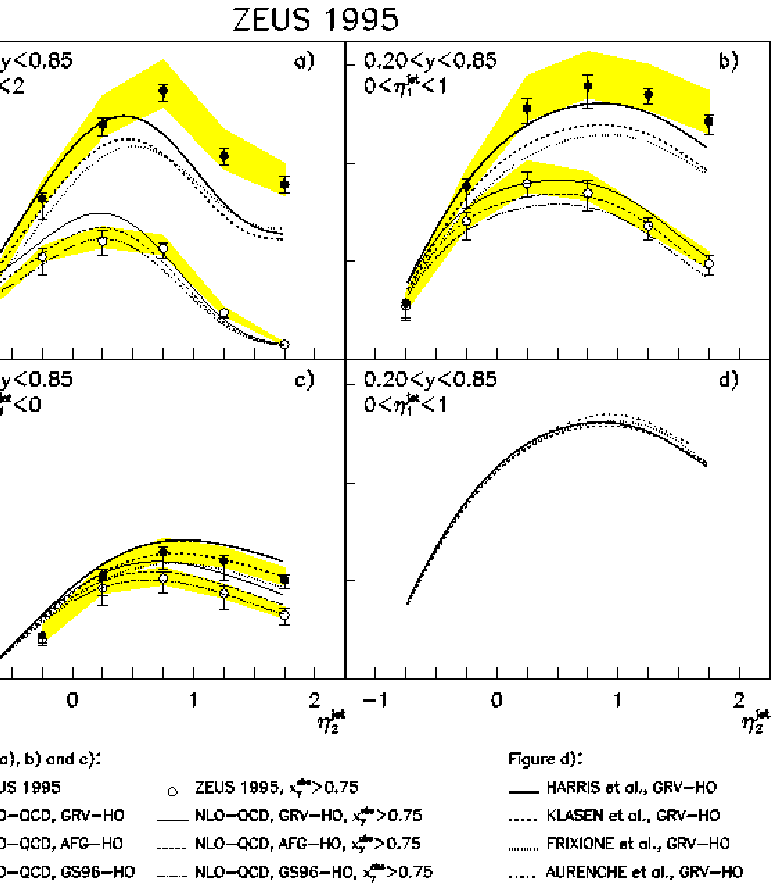}}
\vskip 8.8cm\relax\noindent\hskip 2.1cm
       \relax{\includegraphics{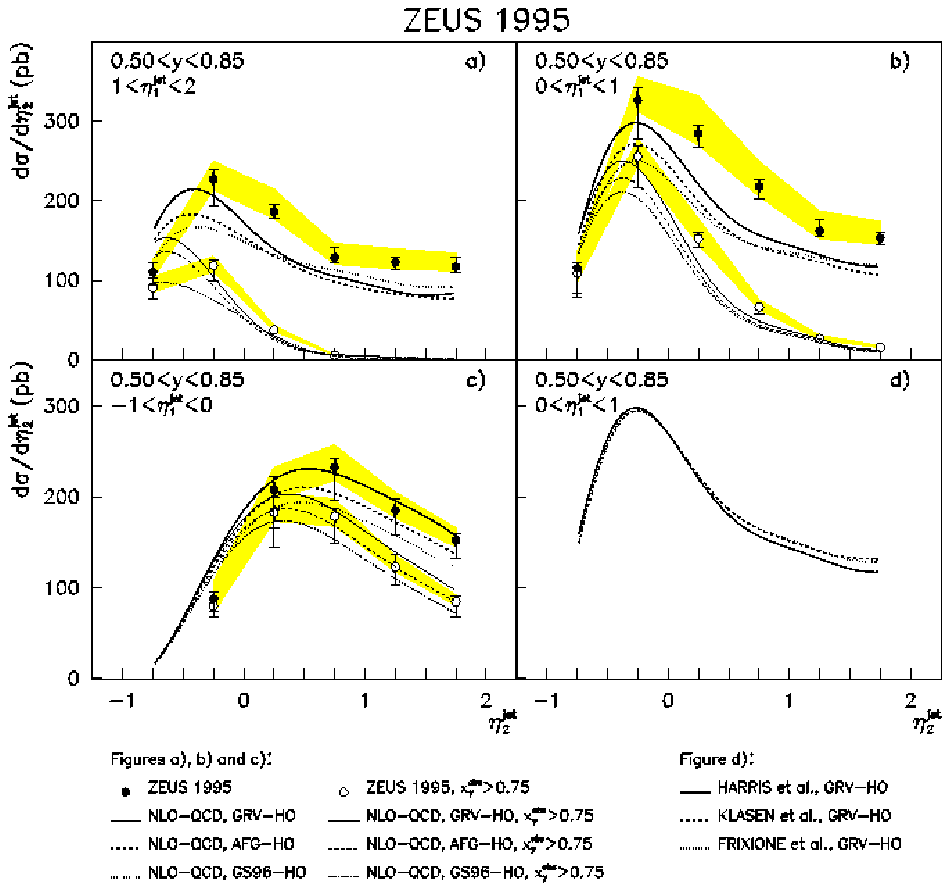}}
\vspace{-0.6cm}
\caption{\small\sl The rapidity distribution $d\sigma$/$d\eta_2$
for  0.2 $<y<$ 0.85 (upper set of figures) and  
0.5 $<y<$ 0.85 (lower set of  figures).
The extra cut for the leading jet, $E_{T leading}\ >$ 14 GeV, was applied.
Three $\eta_1$ jet rapidity ranges were considered. 
The comparison with the NLO calculation 
with different parton parametrization for the photon is made in a), b) 
and c). In d) comparison of different NLO calculations with the
 particular parton parametrization for the photon (GRV-HO) is shown, see text
(from \cite{DESY-99-057}).}
\label{fig:6543}
\end{figure}

For the leading jet,  with  $E_{T leading}\ >$ 14 GeV, 
the rapidity distribution $d\sigma$/$d\eta_2$ was measured as well.
It was done  for a full  $y$ region measured in this experiment and, 
to enhance the sensitivity to the choice of parton density in the photon, 
for a large $y$ region (\ie 0.5 $<y<$ 0.85 corresponding to 
212 $<W<$ 277 GeV$^2$).
Three $\eta_1$ jet rapidity ranges were studied, see 
fig.~\ref{fig:6543} for results. 
 Here the  comparison
was made with the NLO calculation \cite{l10}
 with the GRV HO, AGF HO  and GS96 HO parton parametrization for the photon
(the CTEQ4M parametrization for the proton). Also the comparison between the  
different NLO calculations (\cite{l4,npb507,l11,l13,kkk})
 is made using the  particular (GRV-HO) parton parametrization.\\
\newpage
\vspace*{8.9cm}

~\newline
Comment: {\it The underlying events ``play no role in the present 
kinematic regime.''\\
For the  full and for the  higher energy range 
NLO QCD calculations using the GRV HO, AGF HO and GS96 HO parametrizations
for the parton densities in the photon  can describe properly the distribution
in $E_{Tlaeding}$. However 
``at central and forward pseudorapidities, both
for the full and for the high $x_{\gamma}$ range, the data  lie above the
NLO calculations''.
It ``suggests that in this kinematic region the parton
 densities in the photon are too small in the available parametrizations''. 
``This region has not been studied in $F_2^{\gamma}$ measurements.''}

~\newline
$\bullet${\bf {ZEUS 99b,conf  \cite{ICHEP99-540} (HERA) }}\\
This is the  continuation of the previous analysis of the dijet
events for $E_T$ up to 75 GeV. The data were collected in 1996-1997
with the kinematical cuts   as in {\bf ZEUS  99a}.
The $k_T$ clustering algorithm was used for  jet identification.
The thresholds on $E_{T leading}$ from 14 to 55 GeV was introduced.

\begin{figure}[ht]
\vskip 0.in\relax\noindent\hskip -2.5cm
       \relax{\includegraphics{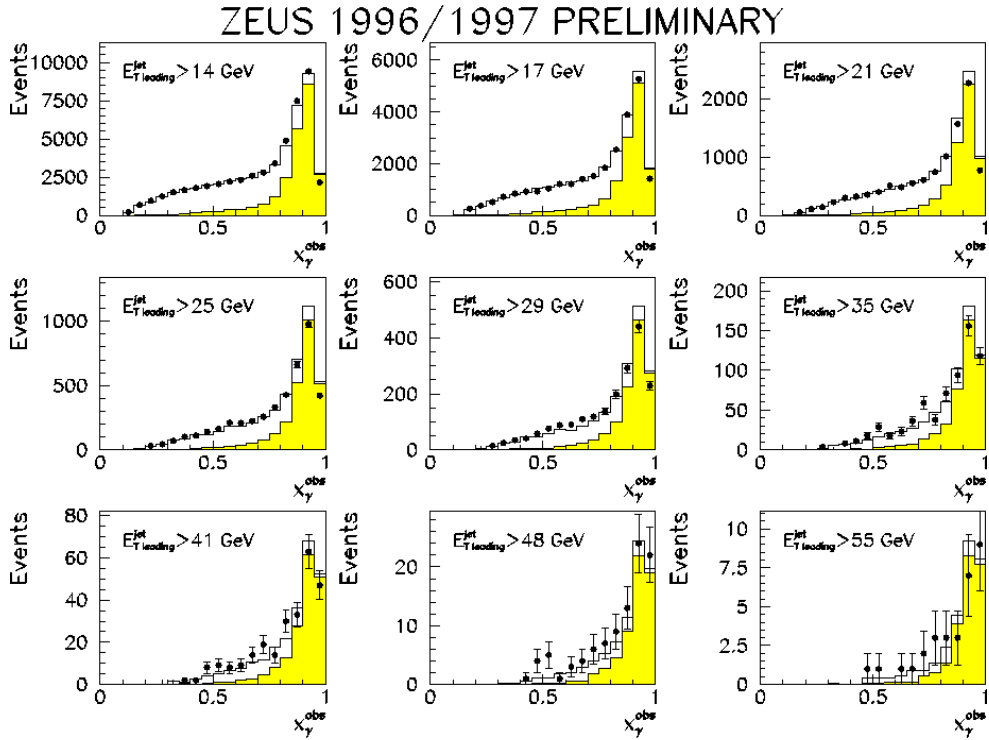}}
\vspace{0ex}
\caption{\small\sl The  $x_{\gamma}$ distribution 
(with the statistical errors only) for different thresholds 
on the highest transverse energy of the leading jet. 
The second one has $E_{Tsecond}>11$ GeV. The solid 
histograms represent  predictions of HERWIG 5.9, the shaded ones represent
 the direct contributions
(from \cite{ICHEP99-540}).}
\label{fig:540_1}
\end{figure}

The $x_{\gamma}$ distributions for different transverse energy thresholds
are presented in fig.~\ref{fig:540_1}; they are well described by 
Monte Carlo generators (also the transverse energy flows, not shown). 

\vspace*{9.5cm}

\begin{figure}[ht]
\vskip 0.cm\relax\noindent\hskip -1.9cm
       \relax{\includegraphics{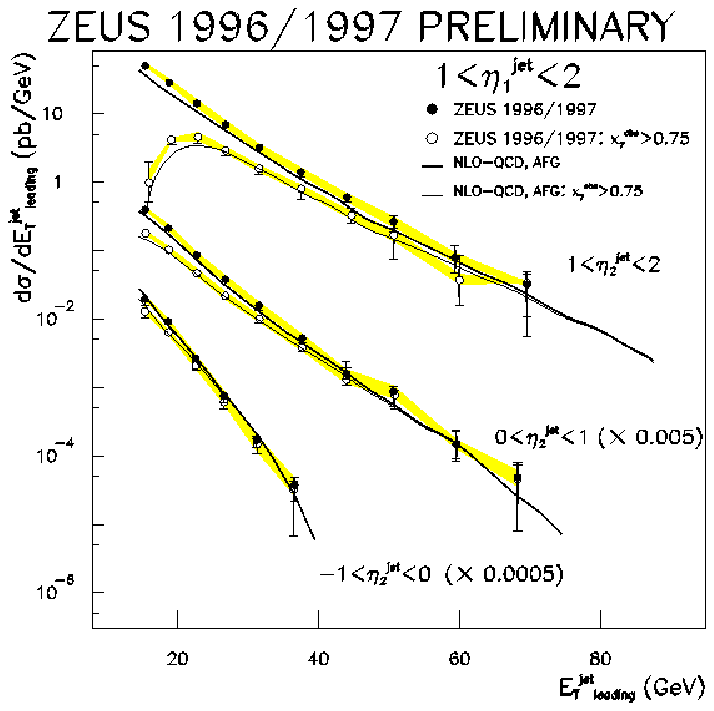}}
\vskip -0.48cm\relax\noindent\hskip 5.9cm
       \relax{\includegraphics{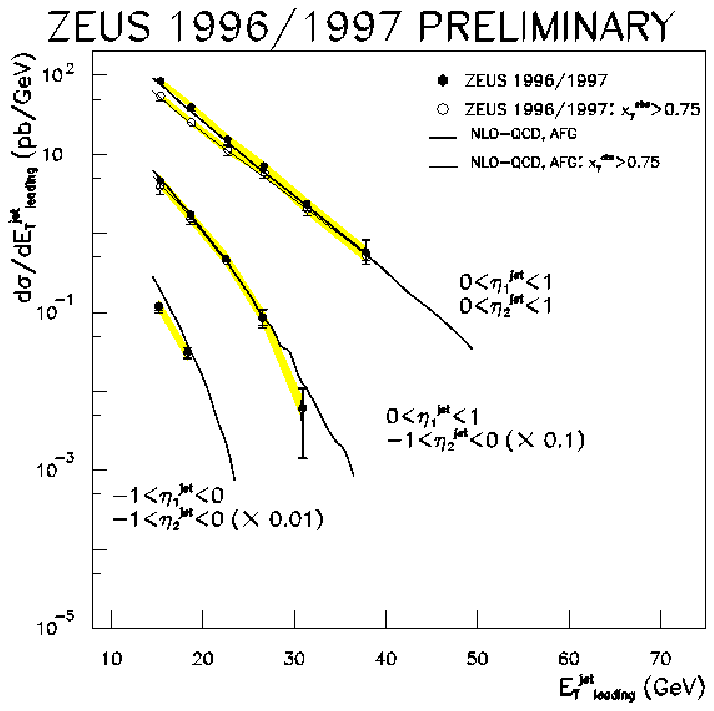}}
\vspace{-0.6cm}
\caption{\small\sl Left: Dijet cross section 
$d\sigma$/$dE_{Tleading}$ with one jet
within the rapidity range 1~$<\ \eta_1\ <$~2 and the other jet in 
three rapidity ranges, described  in the figure.
Right: Dijet cross section 
$d\sigma$/$dE_{Tleading}$ with one jet
within rapidity range 0~$<\ \eta_1\ <$~1 and the other jet in 
three rapidity ranges, described  in the figure.
Comparison with the NLO calculation \cite{l10} with the AGF HO parametrization
for the photon and the CTEQ4M parametrization for the proton
- solid line (dashed line shows the direct contribution)
(from \cite{ICHEP99-540}).}
\label{fig:540_2}
\end{figure}

\begin{figure}[ht]
\vskip 0.in\relax\noindent\hskip 0cm
       \relax{\includegraphics{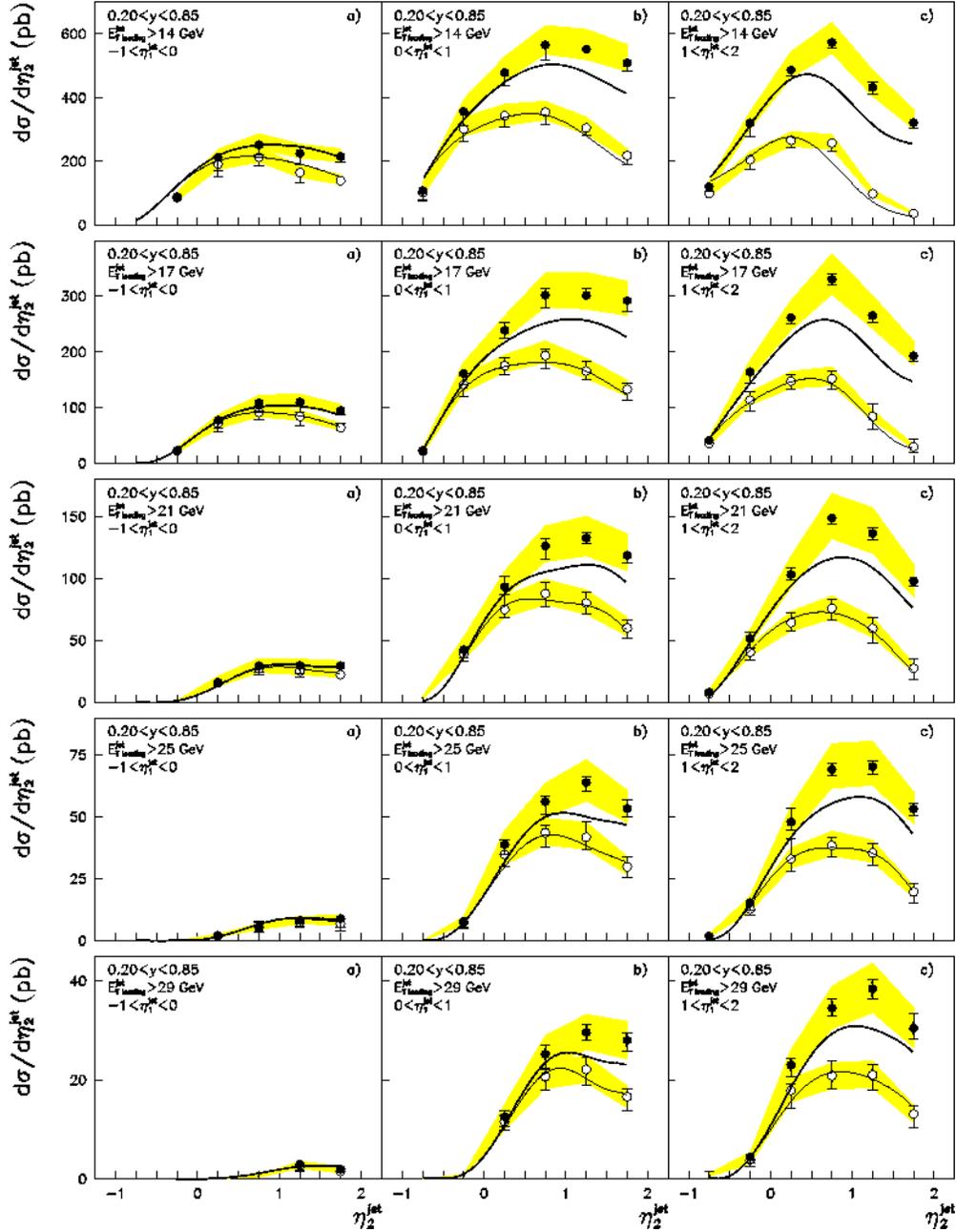}}
\vspace{0.1cm}
\caption{\small\sl The rapidity distribution $d\sigma$/$d\eta_2$
for  0.2 $<y<$ 0.85.
Five  different  cuts  for the leading jet $E_{T leading}\ $ were applied
from 14 to 29 GeV.
Three $\eta_1$ jet rapidity ranges were considered in a), b) and c),
 with the AGF HO parametrization
for the photon and the CTEQ4M parametrization for the proton
- solid line (dashed line shows the direct contribution). 
The  particular parton parametrization for the photon 
(AGF HO) was used (from \cite{ICHEP99-540}).}
\label{fig:540_4}
\end{figure}

Both the transverse energy distribution for $E_{Tleading}$ 
and the rapidity distribution were studied.
Below, the results for the $E_{Tleading}$ distribution in different regions of 
$\eta_{1,2}$ are presented (fig. \ref{fig:540_2}). The direct (defined as 
$x_{\gamma}>$ 0.75) contributions are displayed separately.
The results for the rapidity distributions for various
thresholds on $E_{Tleading}$  from 14 to 29 GeV are shown in
fig. \ref{fig:540_2}, and in fig. \ref{fig:540_4} 
for a narrower high $y$ region. The data are compared with the
NLO calculation using CTEQ4M (proton) and AGF HO (photon)
parton parametrizations, and some discrepancies with data are 
found. 
\newpage
\vspace*{9cm}

~\newline
Comment: {\it These results confirm those  in {\bf ZEUS 99a}.\\
``...the measured cross sections remain higher than the NLO QCD predictions
when the transverse energy threshold on the leading jet is raised up 
to 29 GeV'' suggesting ``that, in the kinematic region of the measurement
presented here [high $x_{\gamma}$ and high $E_T$], 
the available parametrizations of the parton densities 
in the photon are too small.''}

\newpage
\vspace*{15.2cm}

\vspace*{15.2cm}
\begin{figure}[ht]
\vskip 0.in\relax\noindent\hskip 0cm
        \relax{\includegraphics{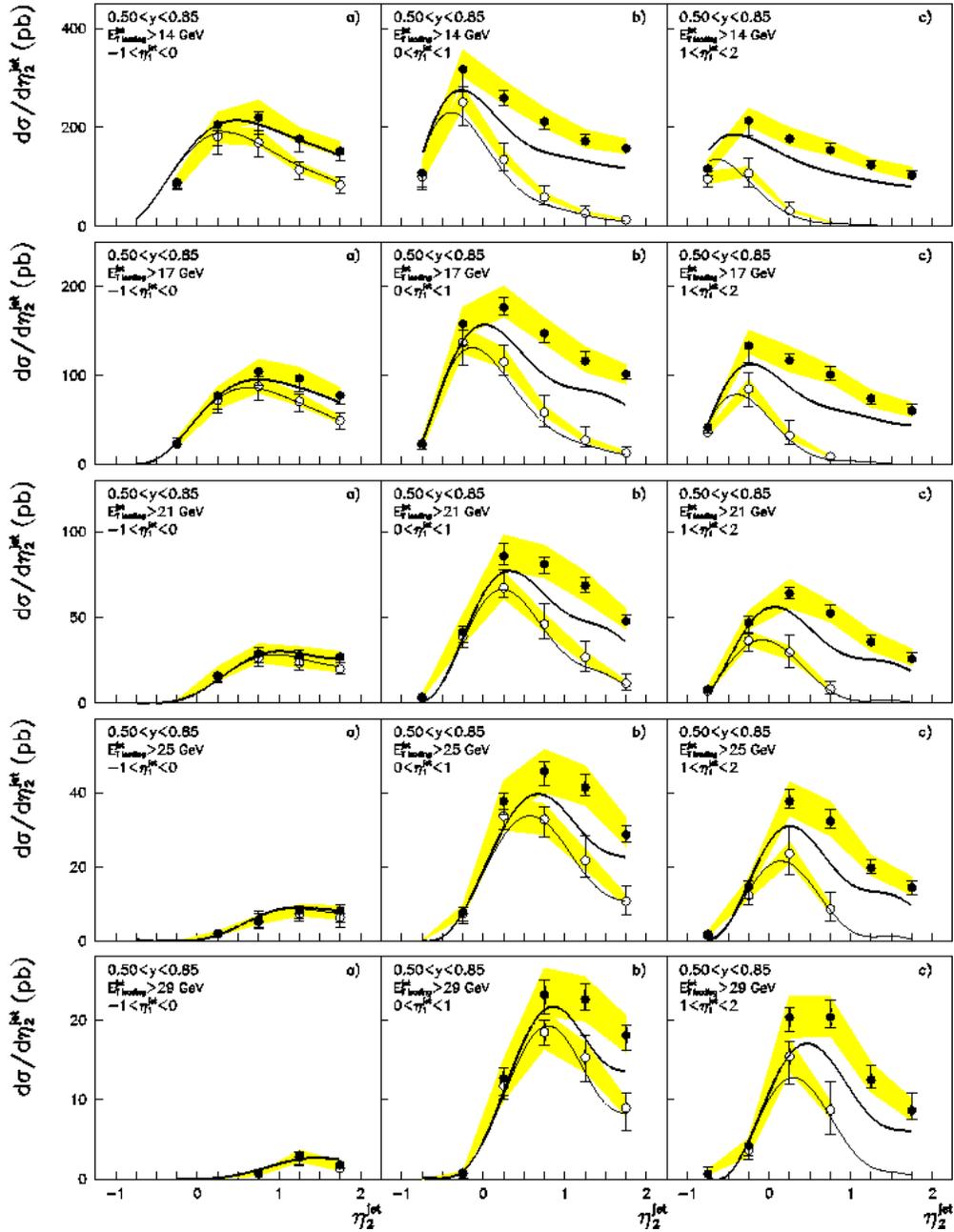}}
\vspace{-0.6cm}
\vspace{0.7cm}
\caption{\small\sl The rapidity distribution $d\sigma$/$d\eta_2$
for 0.5 $<y<$ 0.85, see fig.~\ref{fig:540_4}
 (from \cite{ICHEP99-540}).}
\label{fig:540_5}
\end{figure}

~\newline
$\bullet${\bf {ZEUS 98d,conf  \cite{ICHEP98-805} (HERA) }}\\
The high mass ($M_{jj}>$ 47 GeV, up to 140 GeV) dijet cross section was 
 measured for 134 $<W<277$ GeV (the  1995-1997 sample). 
Events with  $P^2$ smaller than 4 GeV$^2$ were collected 
and compared with the same NLO  calculation \cite{l10} as in
{\bf ZEUS 99b,conf}. In the present analysis 
the cone algorithm with $R$=1 was used  as well as $k_T$ algorithm. 

\begin{figure}[ht]
\vskip 0.in\relax\noindent\hskip 0.cm
       \relax{\includegraphics{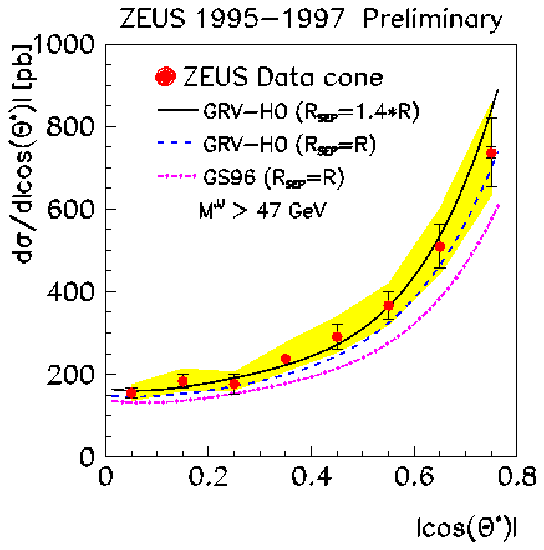}}
\vskip -0.48cm\relax\noindent\hskip 8.5cm
       \relax{\includegraphics{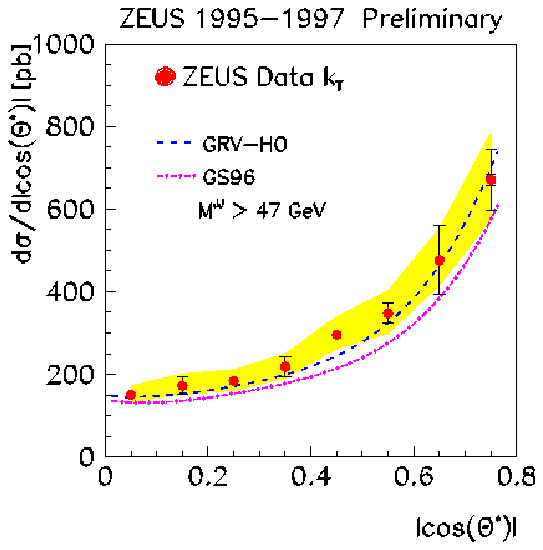}}
\vspace{-0.6cm}
\vspace{0ex}
\caption{\small\sl The $\cos\theta^*$ distribution 
for events with $M_{jj}$ above 47 GeV obtained using 
two-jet algorithms: the cone algorithm (left), the $k_T$ clustering (right). 
The comparison with the NLO calculation using the GRV HO parton densities
(with $R_{sep}=R, 1.4 R$ for the cone algorithm) and GS96 ($R_{sep}=R$
for the cone algorithm) is made
(from \cite{ICHEP98-805}).}
\label{fig:fig805_1}
\end{figure}

\begin{figure}[ht]
\vskip 0.in\relax\noindent\hskip 0.cm
       \relax{\includegraphics{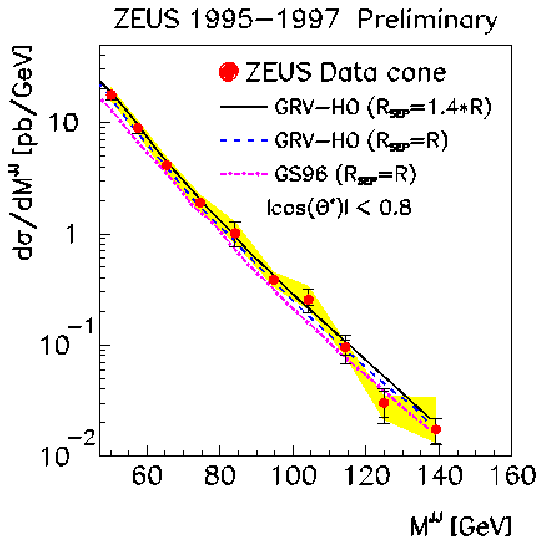}}
\vskip -0.48cm\relax\noindent\hskip 8.5cm
       \relax{\includegraphics{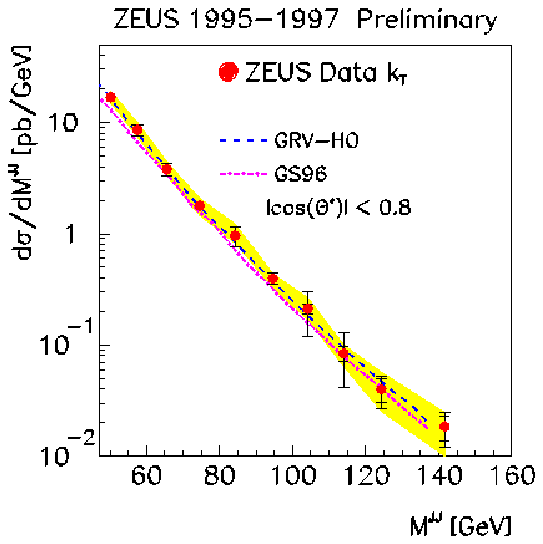}}
\vspace{-0.6cm}
\vspace{0ex}
\caption{\small\sl The $M_{jj}$  distribution 
for events with $|\cos\theta^*|<0.8$  obtained using 
two-jet algorithms: the cone algorithm (left), the $k_T$ clustering (right). 
The comparison with the NLO calculation as in fig. \ref{fig:fig805_1}
(from \cite{ICHEP98-805}).}
\label{fig:fig805_3}
\end{figure}

Results for  the $\cos\theta^*$ distribution obtained within these 
two-jet algorithms are presented in fig. \ref{fig:fig805_1}. In the 
fig. \ref{fig:fig805_3} the invariant mass distribution (integrated 
over the $|\cos\theta^*|<$ 0.8) is shown.\\
\vspace*{5.2cm}
\vspace*{5.7cm}

~\newline
Comment: {\it ``NLO QCD calculations account reasonably well for the shape 
and magnitude of the measured $d\sigma/d\cos\theta^*$ and 
$d\sigma/dM_{JJ}$''}.

~\newline
\newpage
\subsubsection{Prompt photon production in the resolved $\gamma p$ collision 
\label{sec245}}

The {prompt photons}. \ie large-$p_T$ photons (predominantly 
direct), produced in the $\gamma p$ collision \footnote{called also the
deep inelastic Compton (DIC) process, as it corresponds to the $\gamma q\ra \gamma q$ subprocess in the Born approximation} 
give an additional information as compared to jet(s)
production discussed in the previous sections. 
Although the corresponding rate is much smaller ($\sim\alpha^2$),
 the  advantage is the possible {\sl direct} access to the hard subprocess. 
However, there are also
{\sl indirect} 
contributions due to the quark and gluon {\sl fragmentation into photon}
\cite{DICold}. 
In practice the {\underline{isolated photon}} production is considered. 
It corresponds to the restriction of the hadronic production 
around the photon direction. Usually the cone algorithm, with R defined as for the jets, is used with a limit on the allowed hadronic 
energy (\eg $\sum E^h~<~0.1 E^\gamma$) within the cone  \cite{izolacja}.

The photoproduction of a hard direct
photon may occur due to the direct interaction of 
the initial $\gamma$, or due to the resolved initial $\gamma$. The sensitivity
to the particular subprocesses and to 
the form of parton density in the photon differs in such 
processes from that in the jet production, discussed in 
secs. \ref{sec243} and \ref{sec244}.
The domination of one contribution due to the gluonic content of the initial 
photon leading to the production of the forward, very energetic photons was
pointed out in \cite{DICold}g. 
This can be particularly useful in
extraction of the gluonic content of the real (and also slightly 
virtual photon) \cite{bawamkaz}. Also there is a chance to test a special 
 photon-gluon fusion process, namely the  
$\gamma g\rightarrow\gamma g$ box diagram contribution.

All these make  prompt photon processes a source of valuable independent
information on the photon-hadron interaction at high energies.
New data which appeared recently ({\bf ZEUS 2000a}) can be compared
to the two versions of the  NLO calculations \cite{NLO-DIC}.
The agreement of the MC models and the NLO calculation with data is  
reasonably good for rapidities of the final photon in the range
$0.1 < \eta < 0.9$, while for $-0.7 < \eta < 0.1$
the excess of the data over the predictions is seen.

~\newline
\centerline{\bf \huge DATA}

\begin{figure}[hb]
\vskip 0.in\relax\noindent\hskip 0.cm
       \relax{\includegraphics{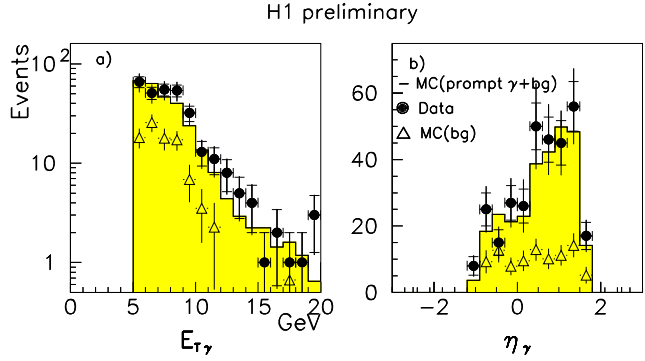}}
\vspace{0ex}
\caption{\small\sl Transverse energy $E_T^{\gamma}$ (a) and the pseudorapidity
$\eta^{\gamma}$ (b) distribution of the produced photon
(from \cite{prompth1}).}
\label{fig:prompth1_2}
\end{figure}

~\newline
$\bullet${\bf {H1 97a,conf \cite{prompth1} (HERA) }}\\
The measurement of the high $E_T$ photons ($E_T^{\gamma} > 5$ GeV
and $-1.2 < \eta_{\gamma} < 1.6$) in quasi-real $\gamma p$ collisions
was performed, based on 1996 data. The isolation of the photon within
a cone with R=0.8 around the photon  direction was
imposed. The total hadronic energy in this cone was restricted to be
not higher than 10\% of the photon energy. The balancing hadronic jet 
with transverse energy $E_T$ was also measured. The  PYTHIA generator
with the GRV LO parton densities in the proton and in the photon was used.
The results for the transverse energy and for the 
pseudorapidity of the final photon distributions are presented 
in fig. \ref{fig:prompth1_2}.
\newpage
\vspace*{3.5cm}

\begin{figure}[ht]
\vskip 0.in\relax\noindent\hskip 0.cm
       \relax{\includegraphics{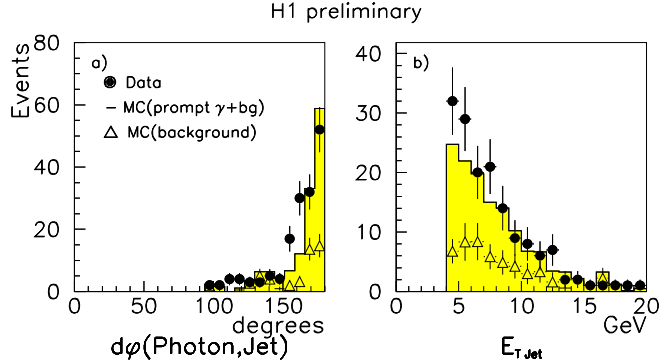}}
\vspace{0ex}
\caption{\small\sl a) The angle between the final photon and the jet,
$\Delta\phi_{(\gamma ,jet)}$, distribution, 
b) the transverse energy of the accompanying jet distribution 
(from \cite{prompth1}).}
\label{fig:prompth1_3}
\end{figure}

The number of events as a function of the azimuthal angle 
between the final photon and the jet ($\Delta\phi_{(\gamma ,jet)}$) 
and the transverse energy of the accompanying jet are presented
in fig.~\ref{fig:prompth1_3}.\\
\vspace*{4cm}

\begin{figure}[hc]
\vskip 0.in\relax\noindent\hskip 0.cm
       \relax{\includegraphics{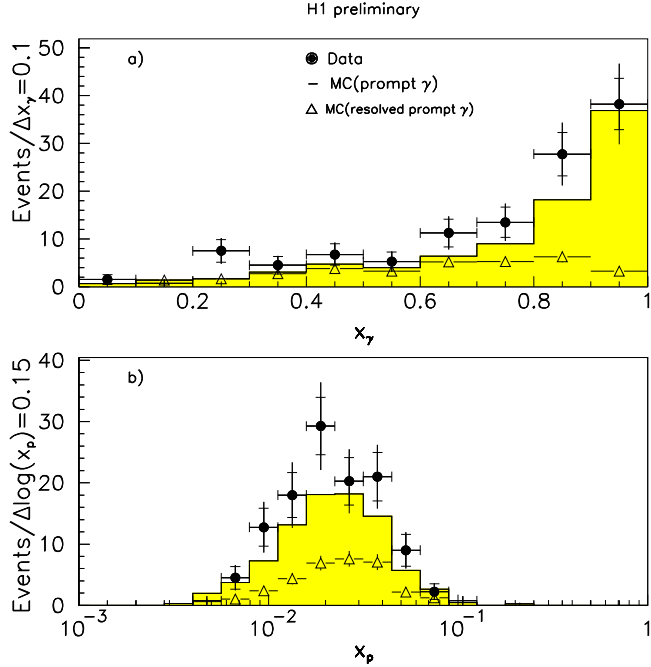}}
\vspace{0ex}
\caption{\small\sl The $x_{\gamma}$ (a) and $x_p$ (b) distributions 
compared with the MC predictions 
(from \cite{prompth1}).}
\label{fig:prompth1_4}
\end{figure}

The measurement of the photon and the jet allows to 
reconstruct the $x_{\gamma}$ and $x_p$ variables. 
Their distributions are shown in fig.~\ref{fig:prompth1_4}.
\newpage
\vspace*{8.cm}
~\newline
Comment: {\it ``It is shown that the prompt photon process at HERA
is dominated by the direct $\gamma p$ interaction as expected from QCD.''}

~\newline\newline
$\bullet${\bf {ZEUS 97 \cite{promptzeus} (HERA) }}\\
The photoproduction of the isolated prompt photon with high 
$E_T^{\gamma}$, together with a balancing jet (described by $E_T$ and $\eta$)
 has been observed for the first time at HERA. The data (collected in 1995)
correspond to $P^2\lsim 1$ GeV$^2$, $5 \leq E_T^{\gamma} < 10$ GeV
and $E_T\geq 5$ GeV, and in addition: $-0.7 < \eta^{\gamma} < 0.8$
and $-1.5 < \eta < 1.8$.

\begin{figure}[ht]
\vskip 0.in\relax\noindent\hskip -0.2cm
       \relax{\includegraphics{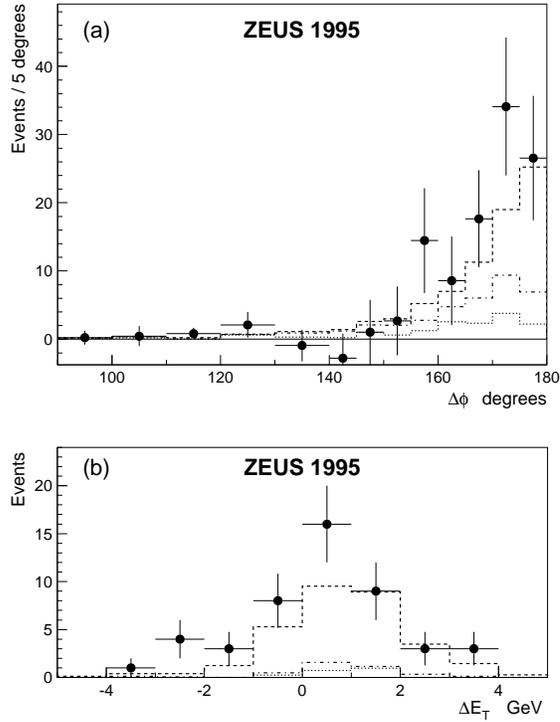}}
\vspace{0ex}
\caption{\small\sl The  $\Delta\phi (\gamma ,jet)$ (a)
and $\Delta E_T (\gamma ,jet)$ (b) distributions
(from \cite{promptzeus}).}
\label{fig:promptzeus3}
\end{figure}

\begin{figure}[ht]
\vskip 0.in\relax\noindent\hskip 0.cm
       \relax{\includegraphics{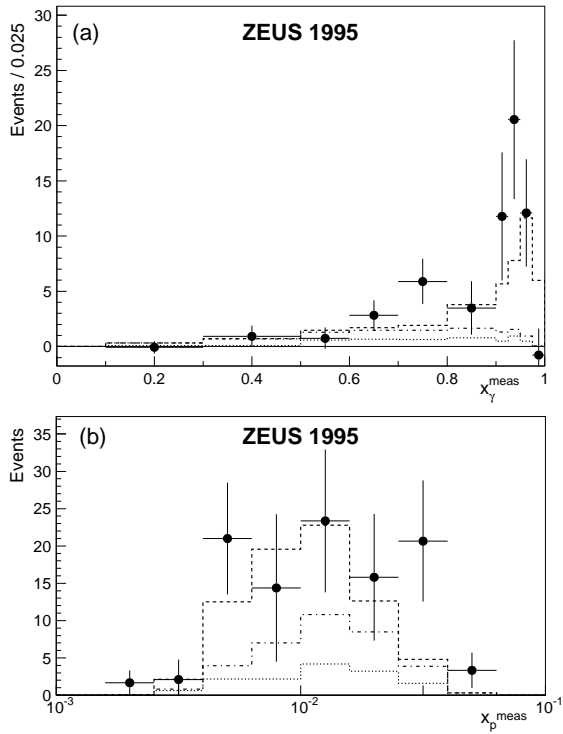}}
\vspace{0ex}
\caption{\small\sl The $x_{\gamma}$ and $x_p$ distributions
are shown for the isolated $\gamma$ plus jet production
(from \cite{promptzeus}).}
\label{fig:promptzeus4}
\end{figure}

The isolation cone was imposed around the photon candidate; within the cone
with R=1 the hadronic energy was assumed to be smaller than 10\% of the photon
 energy. The Monte Carlo PYTHIA program with the MRSA (proton) and GRV LO 
(photon) parton densities was applied. Results are presented in
fig. \ref{fig:promptzeus3} for $\Delta\phi (\gamma ,jet)$,
and $\Delta E_T (\gamma ,jet)$ distributions. In
fig. \ref{fig:promptzeus4} the $x_{\gamma}$ and $x_p$ distributions
are shown.\\
\vspace*{9cm}
\vspace*{9.cm}

~\newline
Comment: {\it For the first time the isolated high - $E_T$ photons
accompanied by balancing jets were observed at HERA. The $x_{\gamma}$
distribution is in good agreement with the LO QCD prediction of the PYTHIA 
generator.}

~\newline
$\bullet${\bf {ZEUS 2000a \cite{DESY-99-161} (HERA) }}\\
The first inclusive measurement of photoproduction of isolated prompt photon 
obtained from run 1996-1997 $(P^2<1 $GeV$^2$ and
$0.2<y<0.9)$ is reported. 
Events correspond to the following ranges of kinematical variables:
$E_{T}^{\gamma}>5~GeV$ and $-0.7<\eta^{\gamma}<0.9$ for the photon.
The PYTHIA 5.7 and HERWIG 5.9 generators, 
 with $p^{min}_T=2.5$ GeV and no  implementation of multi-parton interactions,
were used.  
The proton and photon parton parametrizations,
MRSA and GRV LO, respectively, were used in the analysis.
 
\begin{figure}[hc]
\vskip 0.in\relax\noindent\hskip 3.5cm
       \relax{\includegraphics{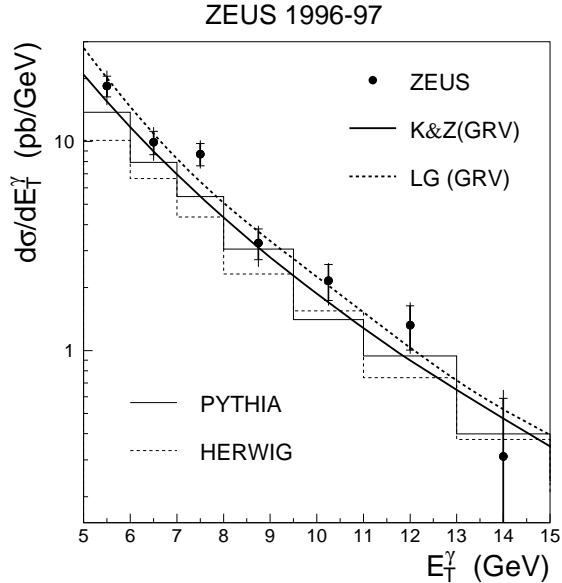}}
\vspace{0ex}
\caption{\small\sl The $E_T^{\gamma}$ distributions
for -0.7$<\eta^{\gamma}<$0.9 and 0.2$<y<$0.9,  is presented
together with the  MC predictions  and
the NLO calculations by Gordon and by Krawczyk\&Zembrzuski \cite{NLO-DIC}
with the GRV HO parton parametrizations
(from \cite{DESY-99-161}).}
\label{fig:DESY-99-161_2}
\end{figure}

\begin{figure}[ht]
\vskip 0.in\relax\noindent\hskip 1.7cm
       \relax{\includegraphics{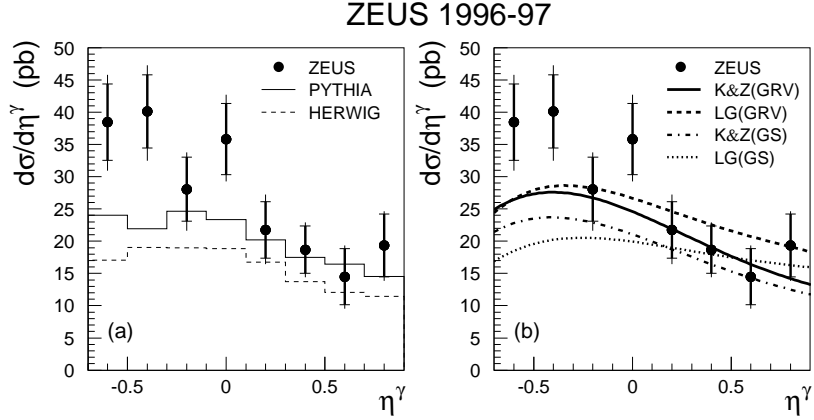}}
\vspace{-1.8cm}
\caption{\small\sl The $\eta^{\gamma}$ distribution
for 5$<E_T^{\gamma}<$10 GeV and 0.2$<y<$0.9, 
together a) with the predictions of the PYTHIA and HERWIG models, b) with
the NLO calculations by Gordon (LG) and by Krawczyk\&Zembrzuski (K\&Z)
 \cite{NLO-DIC} with the GRV HO and GS parton parametrizations for the photon.
(from \cite{DESY-99-161}).}
\label{fig:DESY-99-161_3}
\end{figure}

The isolation cut for the photon was imposed: in the cone  R=1
 the total hadronic energy was restricted to be below $0.1~E_{T}^{\gamma}$. 
The $E_T^{\gamma}$ distribution is presented in Fig. \ref{fig:DESY-99-161_2} 
together with the  MC predictions  and
the NLO calculations by Gordon and by Krawczyk \& Zembrzuski \cite{NLO-DIC}.
In Fig. \ref{fig:DESY-99-161_3} the data for the $\eta^{\gamma}$ distribution
together with the predictions of the 
MC models and of the NLO calculations \cite{NLO-DIC} are shown. 
In order to find an origin of the discrepancy between data and 
both MC and NLO calculations the rapidity distributions were 
studied in three $y$ bins (not shown). The disagreement is strongest for 
the smallest $y $ bin ($y$ between 0.2 and 0.32) 
\vspace*{8.cm}
\vspace*{7.cm}

~\newline
Comment: {\it 
``The [MC] models  are able to describe the data well for the forward (proton direction) photon pseudorapidities, but are low in the rear direction. The disagreement is  strongest in the W interval 134-170 GeV.
This result, together with the disagreements with the NLO predictions seen
 also in  recent dijet results at HERA [{\bf ZEUS 98a}],
would appear to indicate a need to review the present theoretical modelling of the parton structure of the photon.''}

\newpage
\section{Partonic content of the virtual  photon \label{sec3}}
The notion of partonic content of the virtual  photon
has appeared in  high energy interactions soon after
the related concept for the real one (\cite{rev}a, \cite{uw}). 
From the very beginning it has a status of a unique test of QCD,
as without additional experimental or model assumption 
the definite, singularity free  
predictions can be derived for both the $x$ and $Q^2$ dependence
with the LO and NLO accuracy
\footnote{However, the importance of nonperturbative components
of the virtual   photon was pointed out, see below (sec. \ref{sec321}).}. 
By virtue of the analytical continuation (dispersion relation)
one expects that a limit of zero virtuality
will lead to  the partonic content of the real photon.

Not only the DIS$_{e\gamma^*}$ can be performed
for the virtual photon, 
where the corresponding \underbar{structure functions of the $\gamma^*$}
are measured,
but also  the \underbar{resolved virtual photon} processes
sensitive to the individual parton densities in $\gamma^*$
can be studied.  

The deep inelastic scattering on the virtual photon with hadronic final state 
was studied experimentally already in 1984 ({\bf PLUTO 84}). 
Recently     new data on DIS$_{e\gamma^*}$ 
have appeared from the LEP accelerator ({\bf L3 2000}). Also,
the measurement of the leptonic final state in the deep inelastic
scattering on the virtual $\gamma^*$ has been performed 
at LEP ({\bf OPAL 99b}). 
It shows the importance of the
interference terms
which makes the extraction of the corresponding structure functions 
$F_i^{\gamma^*(QED)}$ unfeasible in 
some kinematical regions (see sec. \ref{sec4}).

The bulk of the resolved virtual photon data, with $P^2$ up to 80 GeV$^2$
is just arriving from the  HERA collider (single and double jet production),
see sec. \ref{sec33}. 
The early analyses treated these events only as a special class of DIS$_{ep}$
events - therefore the standard (for such analyses) types of 
Monte Carlo generators were used. In the recent analyses the 
generators incorporating resolved virtual photon interaction are used.
The recent measurement of jet production in the virtual photon-proton
scattering allowed to extract
{\underbar {the effective parton density in $\gamma^*$}} ({\bf H1 2000b}).
Still some discrepancies are observed, \eg in {\bf ZEUS 2000b}, where
 the dijet production in 
the $\gamma^* p$ collision  for $P^2$ below 4.5 GeV$^2$ 
cannot be properly described by Monte Carlo models nor by the 
NLO QCD.

A dedicated study of the \underbar{forward production} of particles and jets,
performed at HERA with the aim to establish a dynamics
of the parton radiation at small $x_{Bj}$ uses a large sample of
the Monte Carlo generators - those typical for DIS$_{ep}$ events
and those for processes with a resolved virtual photon.
Here the possible signal from the BFKL evolution for the  partons
from a proton is confronted with the importance of the
 contribution due to the resolved $\gamma^*$ interaction.
We discuss this topic in the separate
section (sec. \ref{sec34}). 

\vspace*{-0.1cm}
\subsection{The virtual photon-induced processes}
The description of hadronic high energy processes with the 
initial virtual photon depends on its  
characteristic  virtuality: on  how much it differs from zero
and how big it is as  compared to other relevant  scales,
\eg $\Lambda_{QCD}$ or $p_T$.
From these relations one can derive  the role of the virtual photon
in the considered process. 
However, the interpretation is highly nontrivial, as it may depend on the 
reference frame, and even on the chosen gauge.
We will not discuss these theoretical aspects any further. Below we try
to underline only the roles in which  
the photon appears in the experimental analyses. 

We start with the short discussion of the flux of virtual
photons.
\subsubsection{The flux of virtual photons}
As for today, we always deal with  virtual photons
arising from the {\sl leptonic} beam(s)\footnote{Protons, due to their larger mass,
are much weaker source of the virtual photons.}.
Assuming the factorization 
holds between the leptonic source and the hadronic hard subprocess, 
one can introduce the flux of the {\underbar{ 
transversely polarized virtual photons}}
(\eg \cite{rev}a, \cite{www})
\begin{eqnarray}
f_{\gamma^*/e}^T(y,P^2)={{\alpha}\over {2 \pi P^2}} [ {{1+(1-y)^2}\over {y}} 
- {{2 (1-y) P^2_{min}}\over {y P^2}} ].
\label{26}
\end{eqnarray}
The flux of {\underbar {longitudinal virtual photons}},
\begin{eqnarray}
f_{\gamma^*/e}^L(y,P^2)={{\alpha}\over {2 \pi P^2}} [ {{2(1-y)}\over {y}} ],
\label{266}
\end{eqnarray}
is expected to be small for the present experimental setup
and it is usually neglected  in the analyses \cite{Freiburg, Chyla}.

Both the flux of virtual photons emitted by the electron
(eqs. \ref{26}, \ref{266}), and the cross sections involving initial 
virtual photons (\eg $\sigma^{\gamma^*e}$ and $\sigma^{\gamma^*p}$) depend 
on convention used in the definitions\footnote{as they contain the 
convention - dependent flux of the initial virtual photons.}.
This ambiguity is however absent in the $ee$ or $ep$ cross sections,
where a convolution of the virtual photon flux and the virtual photon 
cross section appears (\eg $\sigma^{ep} \sim f_{\gamma^*/e} \otimes
\sigma^{\gamma^*p}$). 
Note that the concept of the \underbar{resolved electron} 
may happen to be very useful here, see sec. \ref{sec6}.

The virtual photons described by eqs. (\ref{26}, \ref{266})
undergo further interaction. They may  interact directly, 
or via their partonic agents, provided that in a process
that follows a  probe exists, hard enough to resolve the 
virtual photon.
 
\subsubsection{The virtual photon as a (non-elementary) probe}
The description of the inclusive processes with $\gamma^*$
as a probe of another  object, \eg of a real photon in  DIS$_{e\gamma}$,
is relatively simple if  global quantities, like total 
$\gamma^*\gamma$ cross sections or 
$F_2^{\gamma}(x_{Bj}, Q^2)$ for a real photon, are considered.
Here  the virtuality of the 
photon probe (we denote it  consequently through the survey  by $Q^2$) 
provides  the hardest scale in the process,  and  a  direct 
interaction of $\gamma^*$ with partonic content of the target dominates.
However, when one tries to reconstruct the
different classes of  hadronic final states
in the DIS-type of measurements,  among them  the rare events 
with $p_T^2\gg Q^2$, one should include the interaction due to 
the partonic content  of the probe - $\gamma^*$ (see also discussion in sec. 
\ref{sec23}).

Similarly, in the case of $ep$ collision the highly
virtual photon, being  a probe for the proton target 
(DIS$_{ep}$ events with standard variables $Q^2$ and  $x_{Bj}$), 
can  be resolved itself by the large $p_T$ jet (particle) 
production\footnote{This problem (``Who is probing who?'')
was underlined \eg in \cite{alevy}.}
(see below). 

\subsubsection{The virtual photon as a target}
Sec. \ref{sec3} covers the dedicated measurements
of the ``structure'' of the virtual photon.
Such a photon, with virtuality $P^2$, plays here a role of  a {\sl target} -
with the structure to be resolved.
The structure functions for such object
are measured in the DIS$_{e\gamma^*}$ experiments at $e^+e^-$ colliders
in the double-tag events,  with $Q^2\gg P^2$. 
Jet production resolving the initial virtual photon, 
\ie with $p_T^2 \gg P^2$, corresponds in $e^+e^-$ colliders to
 the single ($\gamma^*\gamma$) and double-tag  ($\gamma^*\gamma^*$) events,
and in the $ep$ case - to the single tagged events $\gamma^* p$.
In order to pin down the partonic content of the virtual photon
also large $p_T$ particles (among them a real photon \cite{bawamkaz})
can be used. 

In practice an admixture of the resolved virtual photon 
contribution is always present in measurements on the ``real'' 
photon discussed in the previous section.
Data are usually corrected for this effect (see sec. \ref{sec2}).

As it was mentioned above, there is a 
possibility to introduce here the structure function of 
the {\sl {electron}} (see sec. \ref{sec6}), 
given as a convolution of the 
flux for virtual photons (eqs. \ref{26}, \ref{266}) and the structure 
function of ${\gamma}^*$.

\subsection{DIS$_{e\gamma^*}$ experiments \label{sec32}}
Measurements of the virtual photon structure functions in the 
deep inelastic scattering at $e^+e^-$ colliders 
are performed using double-tag events.
These  events are selected in the kinematic 
region where one of the virtual photons (the probe) 
has, on the average, a large  virtuality $Q^2$ and the other, 
the target, a small one, $P^2\ll Q^2$.

For general double-tag events the hadronic state $X$ is produced
in the lepton beams collision, $e(p_1)e(p_2)\ra e(p_1')e(p_2')X$,
via the $\gamma(q) \gamma(p)$ collision with  definite helicity 
states of the photons. The corresponding cross section for the
unpolarized lepton beams is given by (see \cite{rev}a,b,c)
\bea
\label{ttll}
E_1'E_2'{{d\sigma(ee\ra ee X)}\over{d^3p_1'd^3p_2'}} =
~~~~~~~~~~~~~~~~~~~~~~~~~~~~~~~~~~~~~~~~~~~~~~~~~~~~~~~~~~~~~~~ \\  \nonumber 
={{\alpha^2}\over{16\pi^4q^2p^2}}
\left[  {{(q\cdot p)^2-q^2p^2}\over{(p_1\cdot p_2)^2-m_e^4}}\right] ^{1/2}
(4\rho_1^{++}\rho_2^{++}\sigma_{TT}+2|\rho_1^{+-}\rho_2^{+-}|\tau_{TT}
\cos2\bar \phi+ \\  \nonumber
+2\rho_1^{++}\rho_2^{00}\sigma_{TL}
+2\rho_1^{00}
\rho_2^{++}\sigma_{LT}
+\rho_1^{00}\rho_2^{00}\sigma_{LL}-8| \rho_1^{+0}\rho_2^{+0} |
\tau_{TL}\cos\bar \phi),
\eea
where T denotes transverse (+ or -) and L - longitudinal helicity states (0).
The $q$ and $p$ stand for the photons four-momenta: $q\equiv p_1-p_1'$,  
$p\equiv p_2-p_2'$.
The $\rho_{1}$ and $\rho_{2}$
are the photon (with $q$ and $p$ four-momentum, respectively)
 density matrices.
The $\sigma_{TT,TL,LT,LL}$ and $\tau_{TT,TL}$ denote 
the corresponding cross sections and interference terms
(the first subscript corresponds to the photon with four-momentum $q$). 
The $\bar \phi$ is the angle between two scattering planes 
of the scattered electrons in the $\gamma \gamma$ CM system.

For large virtualities of both photons,
$Q^2=|q^2|$, $P^2=|p^2|\gg 4 m_e^2$,
eq. \ref{ttll} can be written in the following 
form: 
\bea
\label{ttll2} \nonumber
E_1'E_2'{{d\sigma(ee\ra ee X)}\over{d^3p_1'd^3p_2'}}
=L_{TT}(\sigma_{TT}+\epsilon_1\sigma_{LT}+\epsilon_2\sigma_{TL}+
\epsilon_1\epsilon_2\sigma_{LL}+\\ \nonumber
+{1\over 2}\epsilon_1\epsilon_2\tau_{LL}\cos 2\bar{\phi}-
2\sqrt{\epsilon_1(1+\epsilon_1)}\sqrt{\epsilon_2(1+\epsilon_2)}
 \tau_{TL}\cos \bar{\phi}),
\eea
where 
\bea
\nonumber
L_{TT}={{\alpha^2}\over{16\pi^4q^2p^2}}
\left[  {{(q\cdot p)^2-q^2p^2}\over{(p_1\cdot p_2)^2-m_e^4}}\right] ^{1/2}
4\rho_1^{++}\rho_2^{++}
\eea
and 
\bea
\nonumber
\epsilon_i={1\over 2} \rho_i^{00}/\rho_i^{++}.
\eea
For $\epsilon_i$ close to 1, which is a typical value in present experiments,
one can furthermore simplify the above formulae, and obtain:
\bea
\nonumber
E_1'E_2'{{d\sigma(ee\ra ee X)}\over{d^3p_1'd^3p_2'}} =
L_{TT}(\sigma_{eff}
+{1\over 2}\tau_{LL}\cos 2\bar{\phi}-
4 \tau_{TL}\cos \bar{\phi}),
\eea
where  an effective
cross section $\sigma_{eff}=\sigma_{TT}+\sigma_{LT}+\sigma_{TL}+\sigma_{LT}$
was introduced.

For the deep inelastic scattering, where
$Q^2>P^2$, the corresponding structure functions 
for a polarization-averaged photon target 
with virtuality $P^2$ can be introduced
\bea
2xF_1^{\gamma^*}={-q^2\over 4\pi^2\alpha}
{\sqrt{(q\cdot p)^2-q^2p^2}\over q\cdot p}
\left( \sigma_{TT}(x,q^2,p^2)-{{1}\over{2}}\sigma_{TL}(x,q^2,p^2) \right),
\eea
\bea
F_2^{\gamma^*}={{-q^2}\over{4\pi^2\alpha}}{{q\cdot p}\over{\sqrt{(q\cdot p)^2
-q^2p^2}}}(\sigma_{TT}(x,q^2,p^2)+~~~~~~~~~~~~~~~~~\\ \nonumber
+\sigma_{LT}(x,q^2,p^2)
-{{1}\over{2}}\sigma_{LL}(x,q^2,p^2)-{{1}\over{2}}\sigma_{TL}(x,q^2,p^2)),
\eea
\be
\nonumber
{\rm with} ~~~~~~~~~~~~~~~~~ F_L^{\gamma^*} = F_2^{\gamma^*} - 2xF_1^{\gamma^*},
~~~~~~~~~~~~~~~~~~~
\ee
and $x = Q^2/2pq$.

\subsubsection{Theoretical description \label{sec321}}
The structure function of the virtual photon  can 
be obtained in the Parton Model assuming   
the production of the $q {\bar q}$ pairs in the 
$\gamma^*(Q^2) \gamma^*(P^2)$
collision. In the mass range
\begin{eqnarray}
 Q^2\gg P^2\gg m_q^2
\label{27}
\end{eqnarray}
it has the form \cite{uw}:
\begin{eqnarray}
\nonumber{F_2^{\gamma^*}(x,Q^2,P^2)=N_c N_f<Q^4>
{{{\alpha}}\over{\pi}}}
x\{ [x^2+(1-x)^2]\ln{{Q^2}\over{P^2x^2}}+6x(1-x)-2\},
\end{eqnarray}
where
\be
<Q^4>={1\over{N_f}}\sum^{N_f}_{i=1}Q_{i}^4,
\ee
to be compared with the eq.~(\ref{f2}).\newline 
One can clearly see  that the scale of the probe, $Q^2$,
has to differ from the $P^2$ in order to test the  structure of 
the virtual photon in the leading logarithmic accuracy (LL).
 The corresponding quark density
in the virtual photon defined in the LL approximation has the form:
\begin{eqnarray}
\nonumber
q_i^{\gamma}(x,Q^2,P^2)\mid_{PM}^{LL} = {\alpha\over 2\pi} N_c Q_i^2
[x^2+(1-x)^2] \ln {Q^2\over P^2}.
\end{eqnarray}
The QCD  
evolution equations for the virtual photon are analogous to those
for the real photon with the inhomogeneous term given by the 
corresponding splitting function $P_{q\gamma^*}$.
In the case of the virtual photon there is a hope  
that to solve evolution equations the initial conditions are not needed,
since for $Q^2\gg P^2\gg \Lambda^2_{QCD}$ the nonperturbative effects
should be absent (see ref.\cite{uw}).
Indeed, the virtual photon 
may play  a unique   role  in testing the QCD.

However, recently there appeared papers \cite{sas,grs,grs99}
showing that the parton content of the virtual photon is not solely
described by purely perturbative contributions in this region, in
contrast to the expectation from \cite{uw}.

Recently the LO \cite{grs,lovir,Freiburg}  
and the NLO QCD \cite{l14,l15,krapot,potter99}
calculations have appeared for the DIS$_{e\gamma^*}$ and the
jet production in the considered processes involving
virtual photon(s)
$\gamma^*\gamma$, $\gamma^* \gamma^*$ and $\gamma^* p$.

The existing parton parametrizations for the virtual photon
are described in the Appendix. 

\subsubsection{Measurements of $F_2^{\gamma^*}$\label{sec322}}
For the DIS$_{\gamma^*}$ measurements, the quantity 
$$x_{vis}=Q^2/(Q^2+P^2+W^2_{vis})$$  
needs to be converted to the true $x_{Bj}=Q^{2}/2p q$ variable
\footnote {At finite $P^2$ a modified variable $x_{Bj}$,
which extends over the whole range between 0 and 1 may be introduced.
For small ratio $P^2/Q^2 \ll1$ they coincide.}.

We start with the old data from PLUTO, the first
``DIS''-type  measurements for  the virtual photon.

\newpage
\centerline{\bf \huge DATA}

\begin{figure}[ht]
\vskip 0.in\relax\noindent\hskip 2.3cm
       \relax{\includegraphics{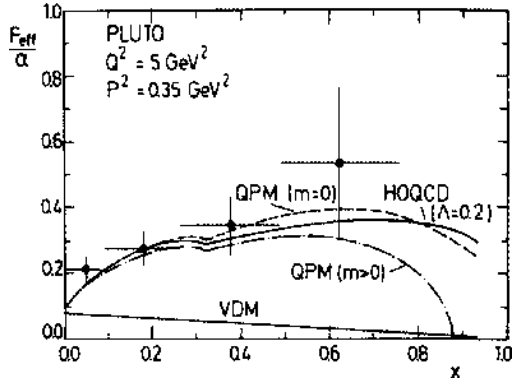}}
\vspace{-2cm}
\caption{\small\sl The  data for the effective 
structure function $F_{eff}/\alpha$ for the virtual photon.
For the  averaged $Q^2=5$ GeV$^2$, and $P^2=0.35$ GeV$^2$,
the dependence on $x_{Bj}$ is shown in comparison with the QPM 
(with massless and massive quarks) and QCD calculations,
both with the VMD for the target contributions added  (from \cite{pluto}).}
\label{fig:marysia2}
\end{figure}

\begin{figure}[hc]
\vskip 0.in\relax\noindent\hskip 1.5cm
       \relax{\includegraphics{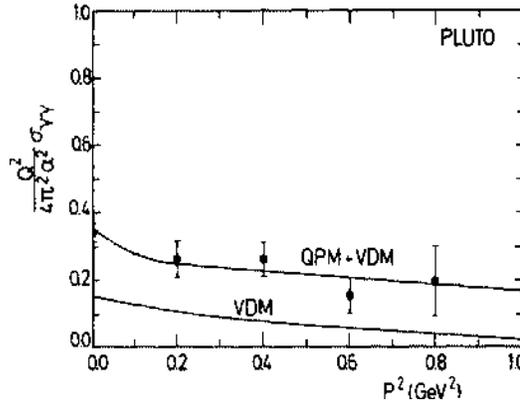}}
\vspace{-2.3cm}
\caption{ {\small\sl The  results for the 
${{Q^2}\over{4{\pi}^2{\alpha}^2}}{\sigma}_{{\gamma}{\gamma}}$
for virtual photon as a function of  $P^2$
(averaged over $x_{Bj}$ and $Q^2$), compared with QPM and VMD
contributions (from \cite{pluto}). }}
\label{fig:marysia1}
\end{figure}

~\newline
$\bullet${\bf {PLUTO 84 \cite{pluto} (PETRA) }}\\
The double-tag events   were measured
in $e^+e^-$ collision, with  the energy of the beam equal to  17.3 GeV.
One of the virtual photons (the probe) 
had, on the  average, virtuality  $<Q^2>=$5 GeV$^2$ and the other 
(the target)  $<P^2>=$0.35 GeV$^2$. 
For the cross sections, the assumptions $\sigma_{TL}\approx\sigma_{LT}$
and $\sigma_{LL}\approx 0$ were made.
The experiment was sensitive to the following combination of  
the virtual photon structure functions (the {\sl effective} structure 
function):\\
\centerline{$F_{eff}\equiv F_2+(3/2)F_L$.}
Results for the extracted $F_{eff}$ as a function of $x_{Bj}$ 
are presented in fig.~\ref{fig:marysia2} together with the theoretical 
predictions. In fig.~\ref{fig:marysia1}  the  quantity
$Q^2{\sigma}_{{\gamma}{\gamma}}/4{\pi}^2{\alpha}^2
\approx F_{eff}/\alpha$, averaged over both $x_{Bj}$ and $Q^2$,
is shown as a function of the measured $P^2$ (0.2 - 0.8 GeV$^2$),
in comparison with QPM and VMD (for the target) contributions.\\
\vspace*{6.5cm}
\vspace*{6.7cm}

~\newline
$\bullet${\bf {L3 2000 \cite{l3_highq2} (LEP 1) }}\\
The measurement of the structure function for the virtual
photon ($<P^2>=3.7$ GeV$^2$)
at $\sqrt s$= 89-92 GeV was performed based on the  1991-1995 data.
The large $Q^2$ region, between 40 and 500 GeV$^2$ ($<Q^2>$=120 GeV$^2$),
was analysed.  
The real photon structure function was studied as well, see sec. \ref{sec222}.

In the analysis the effective structure function:\\
\centerline{$F_{eff}\equiv {{Q^2}\over{4\pi^2\alpha}}(
\sigma_{TT}+\sigma_{LT}+\sigma_{TL}+\sigma_{LL})$} was measured.

The data were analysed using the JAMVG
generator modelling the QPM with $N_f$=4, the PHOJET 1.05c with a cutoff 
$p_T^{min}$=2.5 GeV and TWOGAM generating the three processes: QPM, 
VMD and QCD resolved photon contributions.
The hadronic final state was investigated  (see  section \ref{sec23} for details).

\begin{figure}[hc]
\vskip 0in\relax\noindent\hskip 0.cm
       \relax{\includegraphics{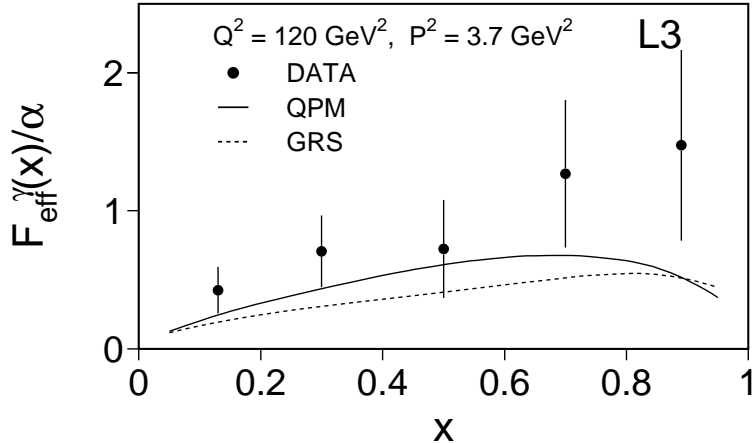}}
\vspace{0cm}
\caption{\small\sl   $F_{eff}^{\gamma}/\alpha$
 as a function of $x_{Bj}$ for $<Q^2>$=120 GeV$^2$ and $<P^2>$=3.7 GeV$^2$.
The solid line corresponds to the QPM prediction.
The QCD prediction (only for transverse target
photon states) using GRS LO parton parametrization 
is given by the  dashed line.
(from \cite{l3_highq2}).}
\label{fig:l3_highq2_6b}
\end{figure}

The $F_{eff}^{\gamma}$ data are presented in table \ref{table_l3high_v} 
and fig.~\ref{fig:l3_highq2_6b},
where the results are compared with the predictions of the QPM and 
QCD calculations performed for the structure function of 
transverse photon only. The QCD contribution is needed 
to describe the $x_{Bj}$ region  below 0.5.  

\begin{table}[ht]
\caption{}
\label{table_l3high_v}
$$
\begin{array}{|r|r|c|}
\hline
\,\,<Q^2>~~\!\!&x_{Bj}~~~~~~&F_{eff}^{\gamma}/{\alpha}\\  
~[GeV^2]~~~&&(stat. + syst.)\\
\hline
~~~120~~~~~&0.05-0.20&0.42\pm0.16\pm 0.05\\
            &0.20-0.40&0.71\pm0.24\pm 0.09\\
            &0.40-0.60&0.72\pm0.34\pm 0.09\\
            &0.60-0.80&1.27\pm0.51\pm 0.16\\
            &0.80-0.98&1.48\pm0.66\pm 0.19\\
\hline
\end{array} 
$$
\end{table}
\vspace*{6cm}

\begin{figure}[hc]
\vskip 0in\relax\noindent\hskip 0.cm
       \relax{\includegraphics{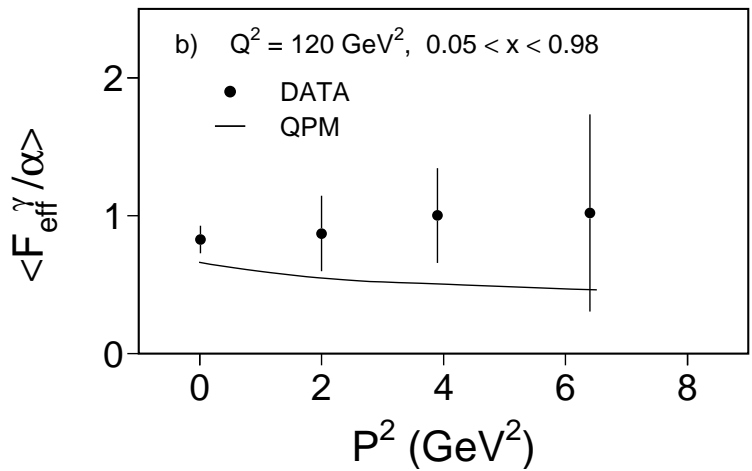}}
\vspace{0cm}
\caption{\small\sl The $P^2$ dependence of   $F_{eff}^{\gamma}/\alpha$
 averaged over $x_{Bj}=0.05-0.98$
for the double-tag data, compared with the QPM prediction (solid line) 
(from \cite{l3_highq2}).}
\label{fig:l3_highq2_8b}
\end{figure}

The averaged values of $F_{eff}^{\gamma}/\alpha$ as a function 
of $<P^2>$ are presented in table \ref{table_l3_highq2_av_v} 
and in fig.~\ref{fig:l3_highq2_8b}.

\begin{table}[ht]
\caption{}
\label{table_l3_highq2_av_v}
$$
\begin{array}{|c|c|c|c|}
\hline
<P^2>&<F_{eff}^{\gamma}/\alpha>\\
~[GeV^2]~& (stat.+syst.)\\
&x_{Bj}=0.05-0.98\\
\hline
0&0.83\pm0.06\pm0.08\\
2.0&0.87\pm0.25\pm0.11\\
3.9&1.00\pm0.32\pm0.13\\
6.4&1.02\pm0.70\pm0.13\\
\hline
\end{array}
$$
\end{table}
\vspace*{5.6cm}

\subsection{Measurements of resolved virtual photon(s) processes\label{sec33}}
As in the case of the real photon, the large $p_T$ jets or particles may resolve 
the virtual photon(s).
Provided the corresponding mass relation
$\tilde Q^2\sim p_T^2 \gg P^2_1 ( P^2_2 ) \gg \Lambda^2_{QCD}$ holds,
one  introduces the contribution due to a partonic content of $\gamma^*$ 
and  a direct $\gamma^*$ subprocesses, as for the real photon
(eqs.~(\ref{twojets}, \ref{xpm})). For the same reason as in sec. \ref{sec2}
we will limit ourselves to the jet(s) production. 
The exception is the hard forward production, where  a single particle
production is also considered, see sec. \ref{sec34}.

As we already discussed in sec. \ref{sec23}, in analyses of hadronic final
state, accompanying the $F_2^{\gamma}$ measurements, jet production
in $\gamma^*\gamma$ collision was studied, \eg {\bf TOPAZ 94},
{\bf DELPHI 96b,conf}, {\bf OPAL 97d,conf}.
In some of these analyses, \eg {\bf TOPAZ 94}, there were
events with the hard scale $\tilde{Q}^2\sim p_T^2$ much larger than
the virtuality of $\gamma^*$, so $\gamma^*$ can be considered as 
being resolved. 

However, so far there are {\sl no} data from {\sl dedicated}
experiments on the resolved virtual photon processes 
in $e^+e^-$ collisions at LEP
and TRISTAN, whereas
a large amount of data from the $ep$ collider HERA have
appeared recently. The single and double jets produced in
the resolved virtual photon processes are being studied in detail
at HERA. Three and four jets including remnant jets, and jet shapes
were also studied, however we will omit these results in our presentation.
For the first time the {\sl effective parton density}, based
on polarization-averaged parton densities in the virtual photon,
was extracted from the data ({\bf H1 2000b}). 

The {\underbar {transition region}} 
between the interaction of an almost real  photon and of a virtual photon
with the proton 
is studied by  the H1 and ZEUS collaborations in the $ep$ collisions at HERA.
In such analysis various Monte Carlo generators (LEPTO, ARIADNE, MEPS, DISJET,
PROJET, RAPGAP
and PHOJET, HERWIG), used to describe the photoproduction
and the DIS$_{ep}$ events as well as the rapidity gap events at HERA, are used.

There are some discrepancies observed between the 
virtual photon cross sections for the dijets production
at HERA and the NLO QCD \cite{l14,l15,krapot,potter99}
calculations, see {\bf ZEUS 2000b}.

\subsubsection{Jet production in $\gamma^*\gamma$ and 
$\gamma^* \gamma^*$ collisions}
The jet production in $\gamma^*\gamma$ collision
were studied in DIS$_{e\gamma^*}$ experiments, \eg
{\bf ALEPH 97a,conf}, {\bf TOPAZ 94}, see sec. \ref{sec23}. 
There are no data so far for the jet production in the double-tag events.\\

\subsubsection{Jet production in $\gamma^*p$ collisions}
~\newline
In the experimental analyses at HERA the flux  of virtual photons is 
introduced ({\bf H1 97b})
in the  form  of the integral  
of the transversely polarized virtual photons over the 
relevant range of $y$, 
and over its  positive squared mass (virtuality), $P^2$:
\begin{equation}
F_{\gamma^*/e}^T=\int^{y_{max}}_{y_{min}}dy \int^{P^2_{max}}_{P^2_{min}}d P^2
f_{\gamma^*/e}^T(y,P^2).
\label{34} 
\end{equation}

The kinematical variables used in the analysis of the single or
double jet production are defined as for the real photon case
(secs. \ref{sec242} and \ref{sec244}).
In the resolved {virtual photon} processes 
some kinematical variables are defined also in the $\gamma^*p$
CM system, and are denoted below by  a star, \eg $E_T^*$. 
Note that here we use the notation $P^2$ for the  virtuality
of the photon although in the context of the DIS$_{ep}$ events (on the 
{\sl {proton}}) at HERA it plays the role of the $Q^2$.
The hard scale $\tilde{Q}^2$ is usually provided by the transverse
energy or transverse momentum of jets.

\newpage
\centerline{\bf \huge DATA}
~\newline\newline
$\bullet${\bf {H1 94 \cite{h194} (HERA)}}\\
The jet rate measurement in DIS$_{ep}$ events with $Q^2 \gg p_T^2$ was 
reported. The JADE jet clustering algorithm was used.

~\newline
$\bullet${\bf {H1 97b \cite{rick} (HERA)}}\\
The single jet cross section for the events with 
0.3 $<y<$ 0.6 was studied in  the transition between
the photoproduction and the standard DIS$_{ep}$ regime at 
HERA. 
The three ranges of the squared mass of the virtual photon 
were studied: $P^2<$ 10$^{-2}$ GeV$^2$ (1994 data),
0.65 $<P^2<$ 20 GeV$^2$ (the 1995 shifted vertex data) 
and 9 $<P^2<$ 49 GeV$^2$ (1994 data).
The jets with $E_T^*>$ 4 GeV (for the photoproduction  
$E_T^*>$ 5 GeV only) for $-2.5<\eta^*<-0.5$ were measured.
The $k_T$ - clustering algorithm was used. For the photoproduction
PHOJET 1.03, for $DIS_{ep}$ events - LEPTO 6.5 and ARIADNE 4.08
were used in simulation.
RAPGAP and HERWIG 5.9 (with $p_T^{min}$ = 1.5 GeV) were applied
to model direct and resolved real and virtual photon processes. 
The GRV94 HO parton parametrization was taken for the proton.
GRV HO with the Drees - Godbole parametrization 
($\omega^2$ = 1 GeV$^2$), denoted GRV HO/Drees-Godbole, and SaS2D 
parametrizations for the virtual photon were used.

\begin{figure}[ht]
\vskip 0.in\relax\noindent\hskip 0.cm
       \relax{\includegraphics{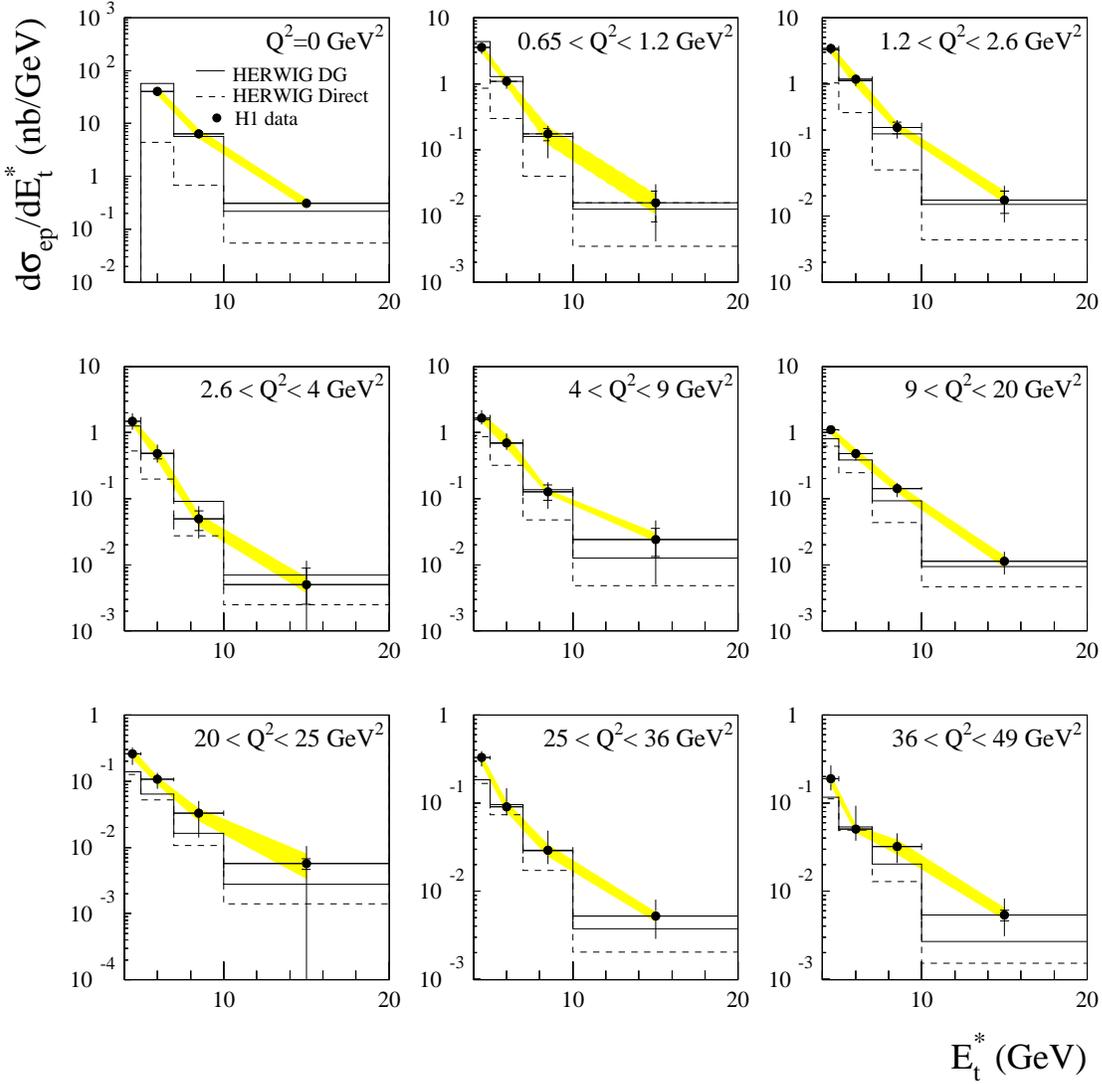}}
\vspace{0ex}
\caption{\small\sl The inclusive $d\sigma_{ep}/dE_T^*$ jet cross section
as a function of the transverse energy $E_T^*$ for
various initial photon virtuality $P^2$ (denoted as $Q^2$) ranges and for 
$-2.5<\eta^*<-0.5$. The  HERWIG GRV HO/Drees-Godbole
prediction is denoted by 
the solid line, the dashed line corresponds to the direct contribution to 
this prediction (from \cite{rick}).}
\label{fig:97091} 
\end{figure}

\begin{figure}[ht]
\vskip 0.in\relax\noindent\hskip 0.cm
       \relax{\includegraphics{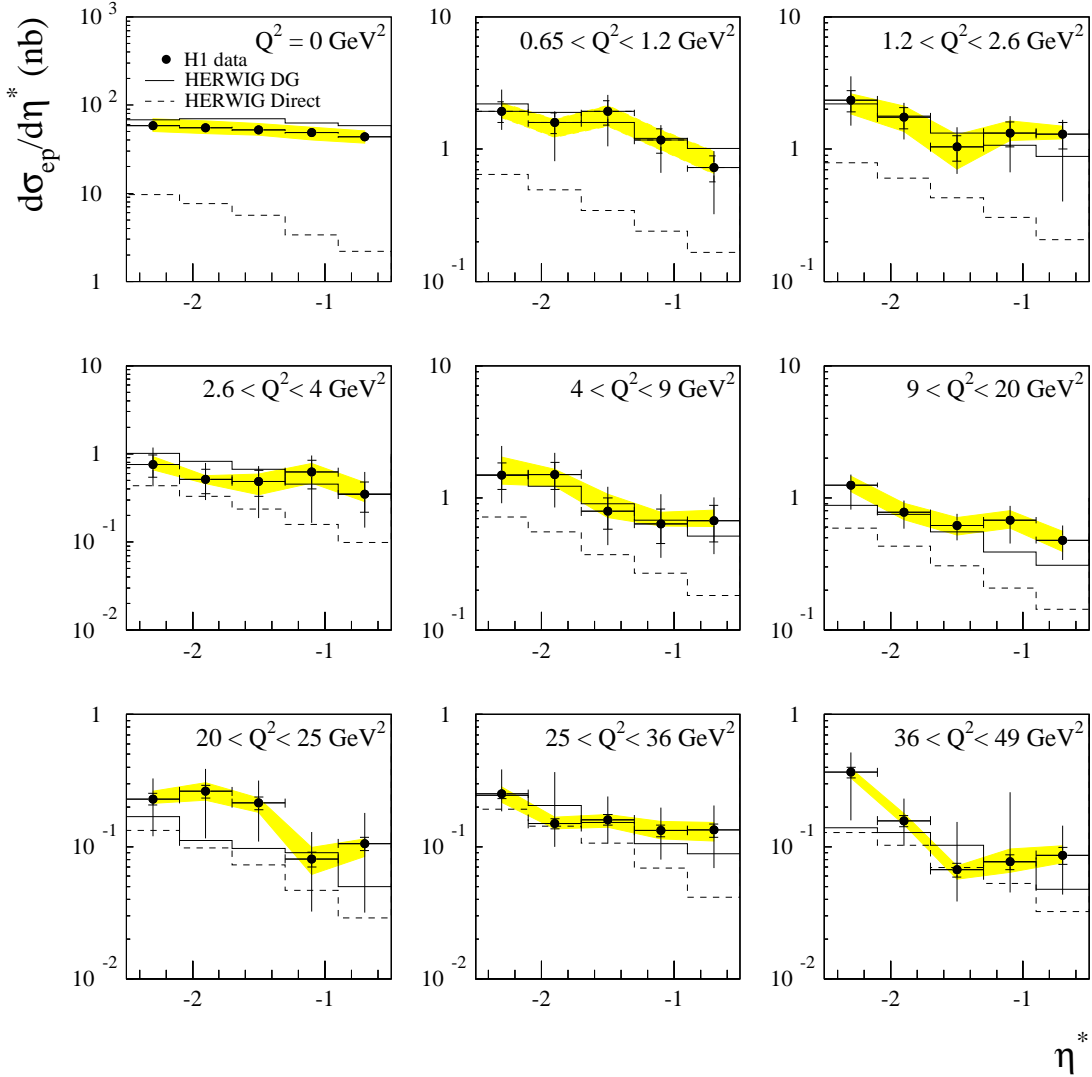}}
\vspace{0ex}
\caption{\small\sl The inclusive $d\sigma_{ep}/d\eta^*$ jet cross section
as a function of the rapidity $\eta^*$ for various initial photon 
virtuality $P^2$ (denoted as $Q^2$) ranges for the $E_T^*>$ 5 GeV. 
The  HERWIG (with the GRV HO/Drees-Godbole parton parametrization) 
prediction is denoted by 
the solid line, the dashed line denotes the direct contribution 
(from \cite{rick}).}
\label{fig:97092} 
\end{figure}

The measured $d\sigma_{ep}/dE_T^*$ as a function of the transverse 
energy of the jet for various $P^2$ ranges integrated over $y$
and $\eta^*$ ranges is presented in fig.~\ref{fig:97091}. 
It was found to be in agreement with the HERWIG model (using 
the GRV HO/Drees-Godbole parton parametrization).
In fig.~\ref{fig:97092}   the  corresponding 
data for the rapidity distribution
for jets with $E_T^*>$ 5 GeV are shown in the various virtuality 
ranges.\\
\newpage
\vspace*{14cm}

\newpage
\vspace*{13.1cm}

To study the dependence of the virtuality of the photon,
the flux (\ref{34}) is introduced and the
cross section $\sigma_{\gamma^* p}$ is calculated,
\be
\sigma_{\gamma^*p\rightarrow jet+X}=  {{\sigma_{ep\rightarrow jet+X}
\over{F^T_{\gamma^*/e}}}}.
\ee
Note that it is not certain that the implied factorization really holds
for the whole range of kinematical variables.

\begin{figure}[ht]
\vskip 0.in\relax\noindent\hskip 0.cm
       \relax{\includegraphics{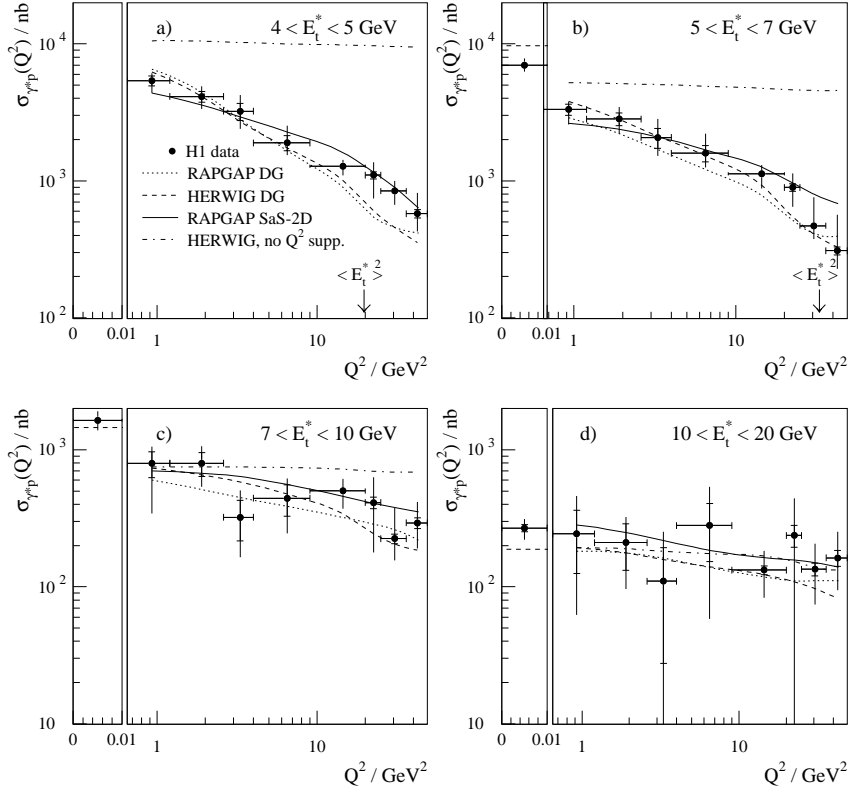}}
\vspace{-0.1cm}
\caption{\small\sl The inclusive $\gamma^* p$ jet cross section
as a function of $P^2$ (denoted as $Q^2$) for
various ranges of the transverse jet energy $E_T^{*2}$, for  
$-2.5<\eta^*<-0.5$. The HERWIG (with the GRV HO/Drees-Godbole 
distribution) prediction is denoted by the dashed line, 
RAPGAP (with the GRV HO/Drees-Godbole distribution) 
- the dotted line, RAPGAP (with the SaS2D distribution) - 
the solid line, and the dot-dashed line corresponds to  HERWIG 
with the GRV HO parametrization as for the real photon 
("no $P^2$ suppression") (from \cite{rick}).}
\label{fig:97093} 
\end{figure}

The results for $\sigma_{\gamma^* p}$ agree with the HERWIG and 
RAPGAP predictions based on the
GRV HO/Drees-Godbole and SaS2D parton parametrizations in the virtual photon,
as shown in fig. \ref{fig:97093}. 
Similar comparison with LEPTO and ARIADNE was made (not shown) and
``neither model can describe the data when $Q^2<E_T^{*2}$ and the virtual
photon can be resolved''.\\

\vspace*{9.4cm}
~\newline
Comment: {\it ``The inclusive jet cross-section can therefore be 
understood if a partonic structure is ascribed to the virtual photon.\\
The data are best described by the RAPGAP model using the SaS2D 
parametrization of the virtual photon.''}

~\newline
$\bullet${\bf {H1 2000b} \cite{h19806029} (HERA) }\\
The dijet event rates have been measured for $5 \lsim P^2 \lsim 100$
GeV$^2$ and $10^{-4} \lsim x_{Bj} $$\lsim$ $10^{-2}$ (\ie for the DIS$_{ep}$
events at small $x_{Bj}$), and for jet transverse momenta squared
$p_T^2 \gsim P^2$ (see also {\bf H1 94}). 
The data collected in 1994 correspond to $y > 0.05$.
The condition $\mid\Delta\eta^*\mid < 2$ was imposed. 
The jet $p_T$ was assumed to be at least 5 GeV, with the following
requirements: for the symmetric case $p_{T 1,2}^* \geq 5$ GeV,
for the asymmetric case $p_{T 1}^* \geq 5$ GeV and $p_{T 2}^* \geq 7$ GeV,
and for the sum scenario $p_{T 1}^* + p_{T 2}^* \geq 13$ GeV.
The cone algorithm with $R$=1 was applied in the $\eta^*$ - $\phi^*$ plane.

The standard DIS$_{ep}$ Monte Carlo generators:
LEPTO and ARIADNE were used to describe in LO the direct
$\gamma^*q$ contribution with two sets of 
parton parametrization for the proton: MRS-H and GRV94 HO.
The additional 
mechanism based on the resolved virtual photon interaction 
neglecting the longitudinal $\gamma^*$ was
introduced in the analysis. Its contribution 
was obtained using the  RAPGAP Monte Carlo model
with the factorization and renormalization scale 
$\tilde Q^2=Q^2+p_T^2$. To describe this contribution also 
the NLO calculations (for partons) in  JeTViP program with
SaS1D parton parametrization for $\gamma^*$ were studied.
For comparison
the DISENT program was used also to calculate the NLO direct virtual 
photon  contribution. The factorization and renormalization scale
$\tilde Q^2=Q^2+50$ GeV$^2$ was used in both NLO calculations.

The transverse energy flow with respect to the jet axis was studied. 
There is a good agreement between the data and the events
simulated with LEPTO and ARIADNE, except for the $\eta^*$ distribution
of the jets (not shown).

\begin{figure}[ht]
\vskip 0.in\relax\noindent\hskip 0.cm
       \relax{\includegraphics{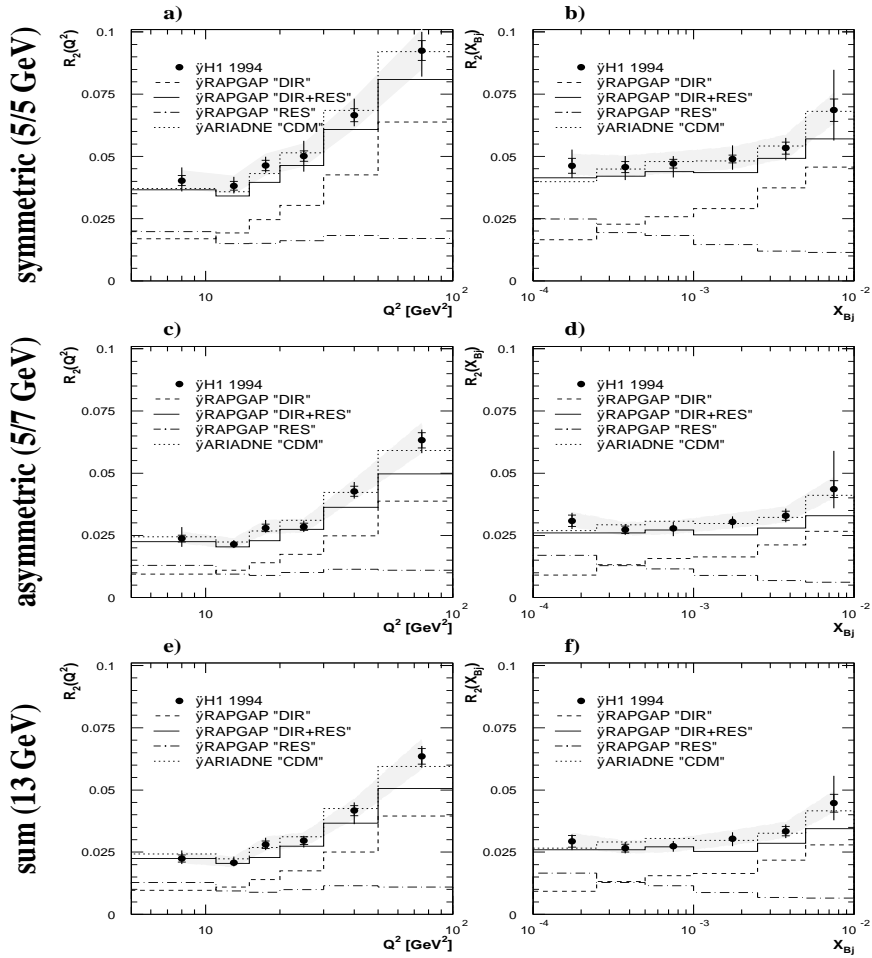}}
\vspace{0.cm}
\caption{\small\sl The dijet rate as a function of $P^2$ 
(denoted as $Q^2$) (a,c,e), integrated
over $x_{Bj}$, and as a function of $x_{Bj}$ (b,d,f), integrated over $P^2$.
The symmetric (a,b), asymmetric (c,d) and sum (e,f) cut scenarios on
the $p_T^*$ of the two jets are presented. The comparison
is made with RAPGAP and ARIADNE predictions for the direct and 
resolved virtual photon contributions (from \cite{h19806029}).}
\label{fig:9806029_3} 
\end{figure}

\begin{figure}[ht]
\vskip 0.in\relax\noindent\hskip 0.cm
       \relax{\includegraphics{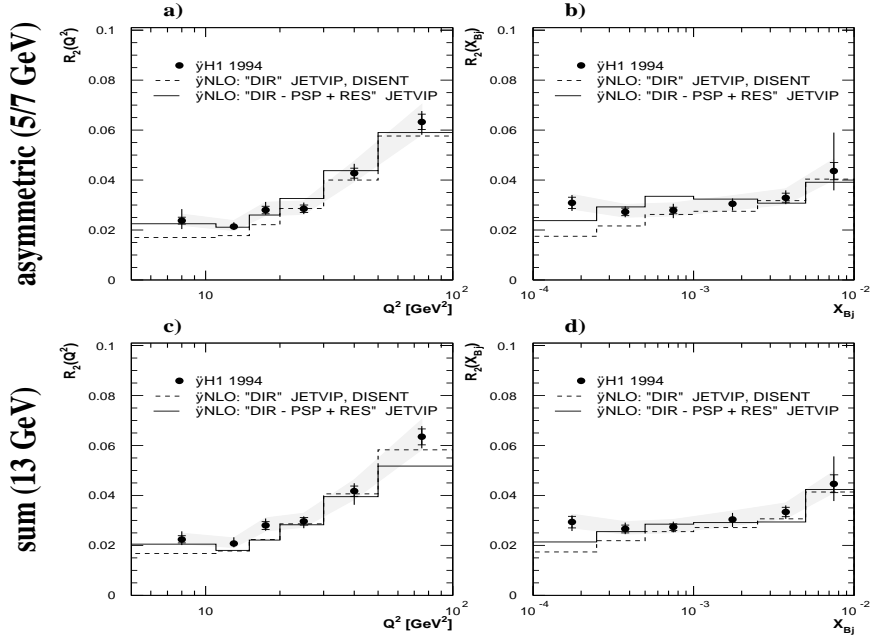}}
\vspace{0.cm}
\caption{\small\sl The dijet rate as a function of $P^2$
(denoted as $Q^2$) (a,c), integrated
over $x_{Bj}$, and as a function of $x_{Bj}$ (b,d), integrated over $P^2$.
The data (the same as in fig. \ref{fig:9806029_3}c-f) for the asymmetric
(a,b) and the sum (c,d) cut scenario are compared to different
NLO calculations for the direct and resolved virtual photon
contributions (from \cite{h19806029}).}
\label{fig:9806029_4} 
\end{figure}

\begin{figure}[ht]
\vskip 0.in\relax\noindent\hskip 0.cm
       \relax{\includegraphics{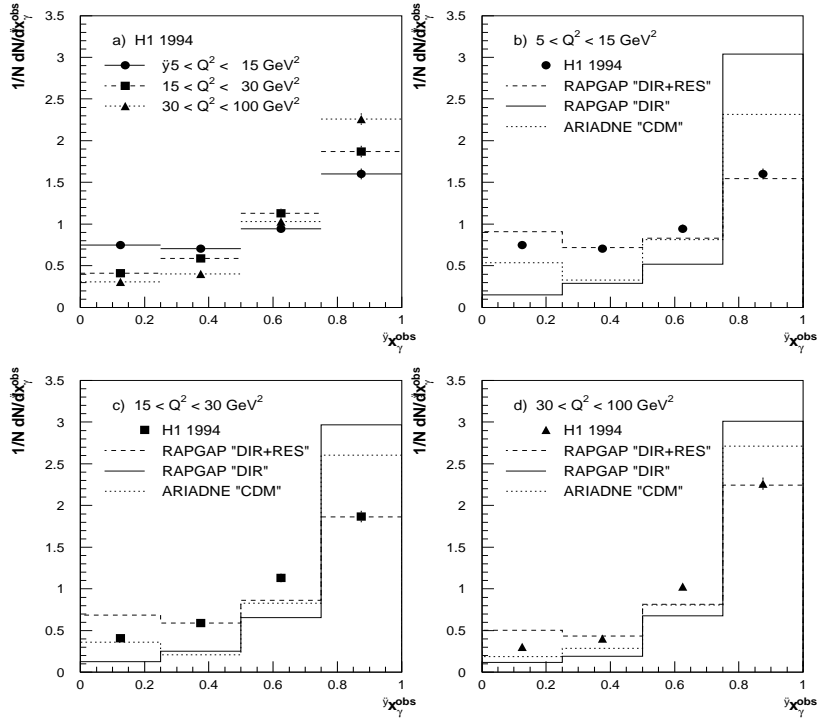}}
\vspace{0.cm}
\caption{\small\sl The uncorrected distribution of $x_{\gamma}^{obs}$
in three different $P^2$ (denoted as $Q^2$) bins (a). In b), c) and d) the data 
in different $P^2$ bins are compared to the RAPGAP and ARIADNE models
for the direct and resolved virtual photon
contributions (from \cite{h19806029}.}
\label{fig:9806029_5} 
\end{figure}

The comparison of the fraction  of dijet events in all DIS$_{ep}$
events with the RAPGAP and ARIADNE models prediction can be found
in fig. \ref{fig:9806029_3},
where the rates as a function of $P^2$ (integrated over $x_{Bj}$)
and $x_{Bj}$ (integrated over $P^2$) are shown.
The same distributions are shown in 
fig. \ref{fig:9806029_4} for the NLO calculations
obtained using DISENT (the direct photon contribution)
and JeTViP, where also the resolved photon component is included. 
The $x_{\gamma}^{obs}$ distribution is shown in fig. \ref{fig:9806029_5}.

~\newline
Comment: {\it The agreement with the data at low $x_{Bj}$ and moderate 
$P^2$ is improved when beside the direct also contributions 
from the resolved virtual photon  are 
included in the Monte Carlo model (RAPGAP) and in the NLO calculations.
``The CDM model, as implemented in ARIADNE, is also able
to describe the dijet rate well.''}\\
\newpage
\vspace*{11.5cm}
\newpage
\vspace*{7.5cm}
\vspace*{0.3cm}
\vspace*{8.5cm}

\newpage
~\newline
$\bullet${\bf {H1 2000c} \cite{h1eff98} (HERA) }\\
The dijet cross sections were measured in 1996 for 
$1.6<P^2<80$ GeV$^2$, $0.1<y<0.7$, and 30 $< \bar E_T^2 <300$ GeV$^2$.
For jets the ranges $|\Delta\eta |<$ 1, -3.0 $<\bar{\eta}<$ -0.5,
and 2~$|\Delta E|/\bar{E} <$ 0.25 were assumed. 
Two Monte Carlo generators were used, HERWIG and RAPGAP,
with different (LO and NLO) virtual photon parton densities:
GRV HO/Drees-Godbole ($p_T^{min}$ = 3 GeV),
GRV LO/Drees-Godbole ($p_T^{min}$ = 2 GeV),
and SaS1,2D ($p_T^{min}$ = 2 GeV).
For proton, the GRV HO and GRV LO parton parametrizations
were used.
The $k_T$ - clustering algorithm for jets was applied.
Only the transversely polarized photon flux was assumed in the analysis.

The data were used to extract (for the first time) an effective LO parton
density for the virtual photon.

\begin{figure}[ht]
\vskip 0.in\relax\noindent\hskip 0.cm
       \relax{\includegraphics{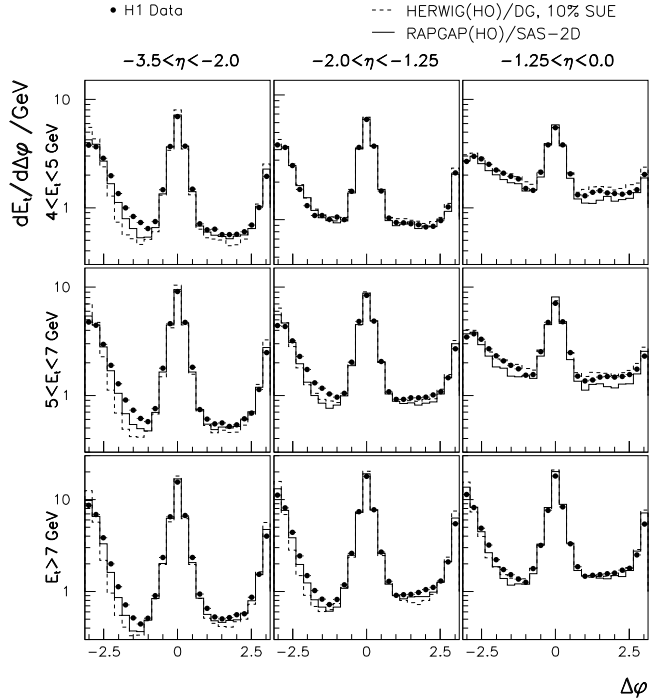}}
\vspace{-0.1cm}
\caption{\small\sl The transverse energy flow as a function of
$\delta\phi$ for different $\eta$ bins. Comparison with predictions of
HERWIG (HO) using the GRV HO/Drees-Godbole 
parton distribution (dashed line) and RAPGAP (HO) using
the SaS2D parton distribution (solid line) is shown (from \cite{h1eff98}).}
\label{fig:9812024_2} 
\end{figure}

The transverse energy flow was studied as a function of
$\delta\phi$ and compared with the Monte Carlo predictions
with and without soft underlying event (HERWIG and RAPGAP,
respectively), see fig.~\ref{fig:9812024_2}.
\newline

\vspace*{8.5cm}

\begin{figure}[ht]
\vskip 0.in\relax\noindent\hskip 0.cm
       \relax{\includegraphics{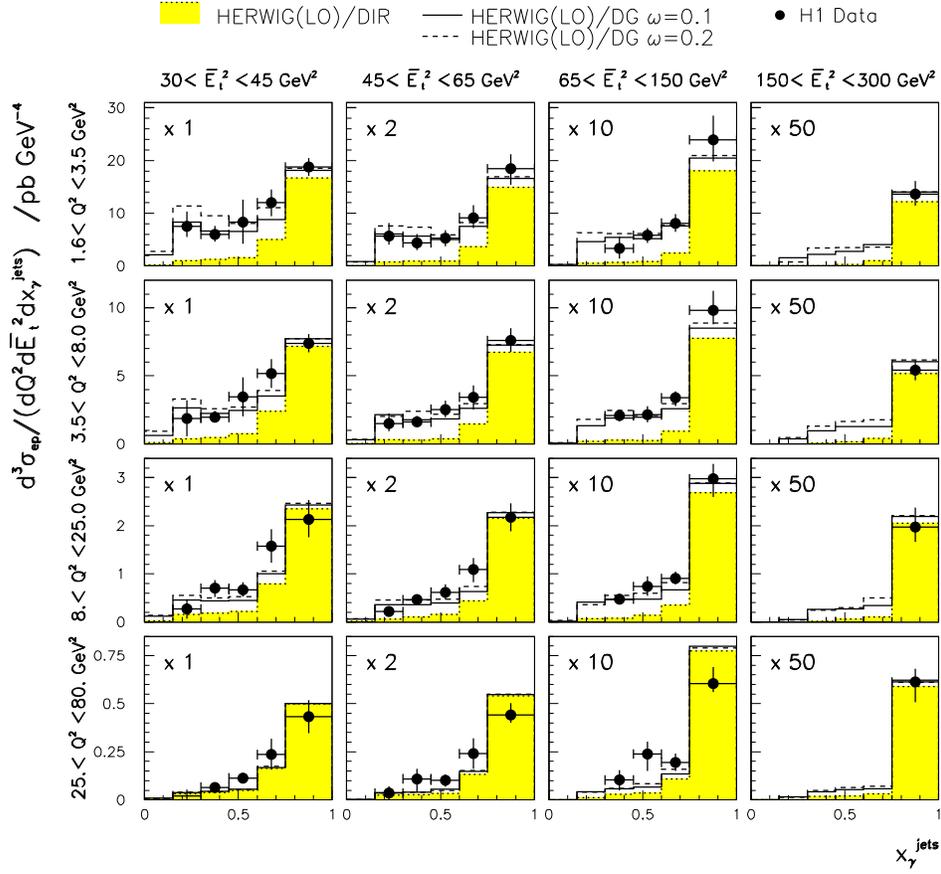}}
\vspace{-0.1cm}
\caption{\small\sl The differential cross section as a function
of $x_{\gamma}$ in four bins of the virtuality $P^2$
(denoted as $Q^2$) and four bins of
$\bar E_T^2$. The comparison with the LO prediction
of the HERWIG Monte Carlo based on the GRV LO/Drees-Godbole parametrization
with $\omega$=0.1 (solid line) and $\omega$=0.2 (dashed line) is shown.
Also the direct contribution is displayed  (dotted line) (from \cite{h1eff98}).
}
\label{fig:9812024_3} 
\end{figure}

\begin{figure}[ht]
\vskip 0.in\relax\noindent\hskip 0.cm
       \relax{\includegraphics{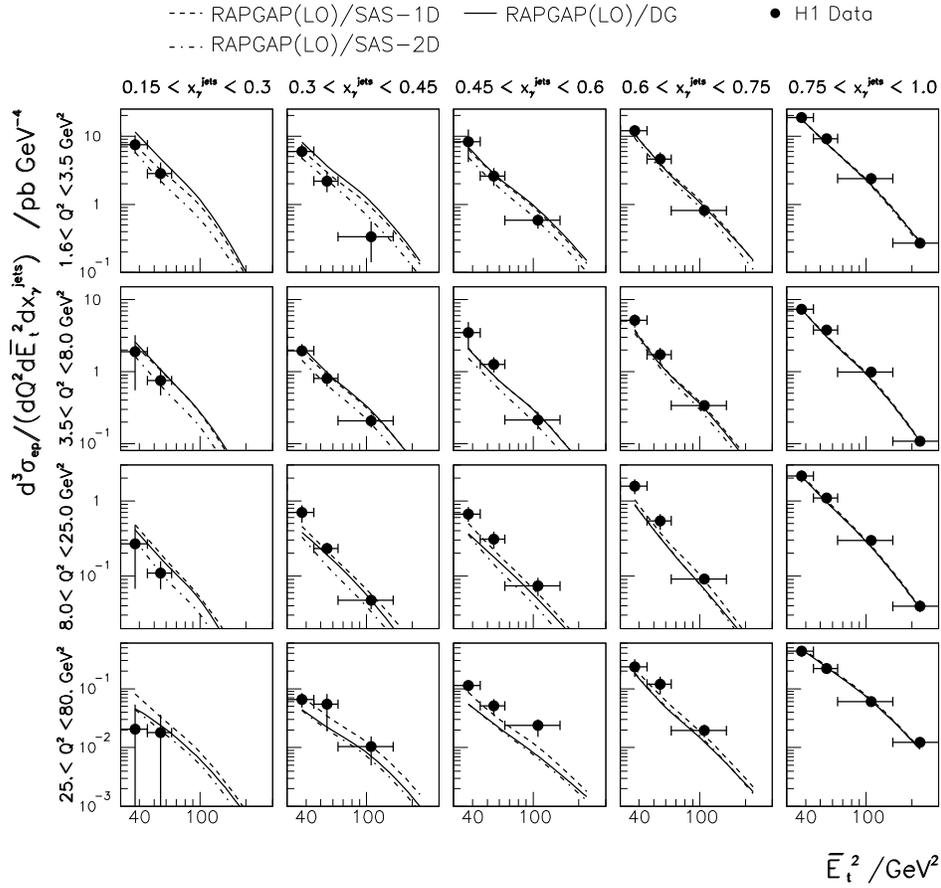}}
\vspace{-0.1cm}
\caption{\small\sl The differential cross section as a function
of $\bar E_T^2$ in four bins of the virtuality $P^2$
(denoted as $Q^2$) and five bins of 
$x_{\gamma}$. The comparison with the LO prediction
of the RAPGAP Monte Carlo using the SaS1D (dashed line), SaS2D 
(dash-dotted line) and GRV LO/Drees-Godbole (solid line) parton distributions
in the photon is shown (from \cite{h1eff98}).}
\label{fig:9812024_4} 
\end{figure}

\begin{figure}[ht]
\vskip 0.in\relax\noindent\hskip 0.cm
       \relax{\includegraphics{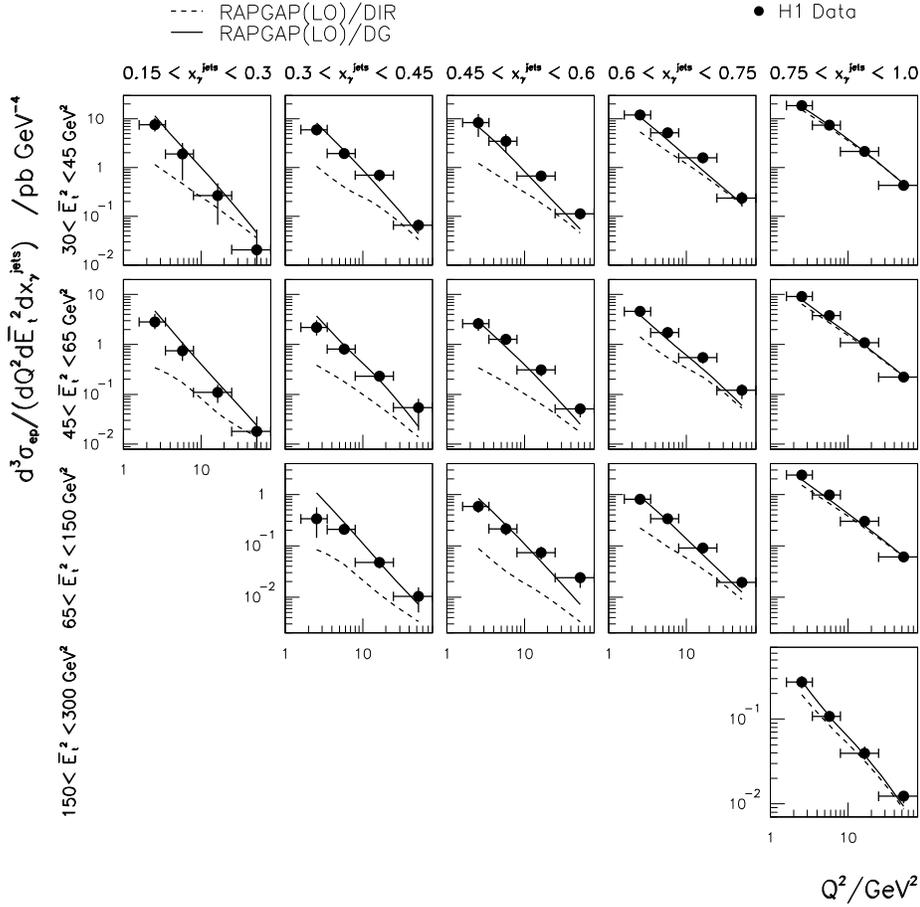}}
\vspace{-0.1cm}
\caption{\small\sl The differential cross section as a function
of virtuality $P^2$ (denoted as $Q^2$) in four bins of
$\bar E_T^2$ and five bins of the 
$x_{\gamma}$. The comparison with the LO prediction
of the RAPGAP Monte Carlo: using the GRV LO/Drees-Godbole parton distributions
in the photon (solid line) and the the direct contribution only
(dashed line) is shown (from \cite{h1eff98}).}
\label{fig:9812024_5} 
\end{figure}

The triple differential cross sections as a function of 
$x_{\gamma}$, $\bar E_T^2$ and $P^2 (=Q^2)$ are presented in 
figs.~\ref{fig:9812024_3}, \ref{fig:9812024_4} and 
\ref{fig:9812024_5}, respectively, together with predictions of the HERWIG and
RAPGAP Monte Carlo programs.
\newpage
\vspace*{11cm}

\newpage
\begin{table}[ht]
\caption{}
\label{tablenew} 
$$
\begin{array}{|c|c|c|c|}
\hline
P^2&\tilde{Q}^2 (=p_t^2)&x_{\gamma}&f(x_{\gamma})/{\alpha}\\
~[GeV^2]~&~[GeV^2]~&&(stat. + syst.)\\
\hline
2.4&40.0&0.275&0.55\pm 0.02^{+0.23}_{-0.19}\\
   &    &0.425&0.60\pm 0.02^{+0.15}_{-0.12}\\
   &    & 0.6 &0.95\pm 0.03^{+0.17}_{-0.29}\\
\cline{2-4}
   &52.0&0.275&0.59\pm 0.02^{+0.32}_{-0.19}\\
   &    &0.425&0.57\pm 0.02^{+0.20}_{-0.15}\\
   &    &0.6  &0.93\pm 0.03^{+0.19}_{-0.18}\\
\cline{2-4}
   &85.0&0.425&0.53\pm 0.02^{+0.29}_{-0.18}\\
   &    &0.6  &0.98\pm 0.03^{+0.21}_{-0.26}\\
\hline
5.3&40.0&0.275&0.33\pm 0.01^{+0.15}_{-0.15}\\
   &    &0.425&0.49\pm 0.02^{+0.14}_{-0.13}\\
   &    &0.6  &0.82\pm 0.03^{+0.13}_{-0.30}\\
\cline{2-4}
   &52.0&0.275&0.36\pm 0.01^{+0.20}_{-0.16}\\
   &    &0.425&0.50\pm 0.02^{+0.14}_{-0.15}\\
   &    &0.6  &0.85\pm 0.03^{+0.14}_{-0.25}\\
\cline{2-4}
   &85.0&0.425&0.64\pm 0.02^{+0.19}_{-0.21}\\
   &    &0.6  &0.87\pm 0.03^{+0.15}_{-0.23}\\
\hline
12.7&40.0&0.275&0.16\pm 0.01^{+0.07}_{-0.08}\\
    &    &0.425&0.42\pm 0.02^{+0.09}_{-0.19}\\
    &    &0.6  &0.54\pm 0.02^{+0.10}_{-0.25}\\
\cline{2-4}
   &52.0&0.275&0.18\pm 0.01^{+0.09}_{-0.09}\\
   &    &0.425&0.45\pm 0.02^{+0.09}_{-0.20}\\
   &    &0.6  &0.62\pm 0.03^{+0.10}_{-0.23}\\
\cline{2-4}
   &85.0&0.425&0.46\pm 0.02^{+0.12}_{-0.21}\\
   &    &0.6  &0.74\pm 0.03^{+0.14}_{-0.27}\\
\hline
40.0&85.0&0.425&0.43\pm 0.04^{+0.09}_{-0.27}\\
    &    & 0.6 &0.65\pm 0.04^{+0.19}_{-0.41}\\
\hline
\end{array}
$$
\end{table}

\newpage
\vspace*{11.5cm}

\newpage
\vspace*{11.7cm}

\begin{figure}[ht]
\vskip 0.in\relax\noindent\hskip 0.cm
       \relax{\includegraphics{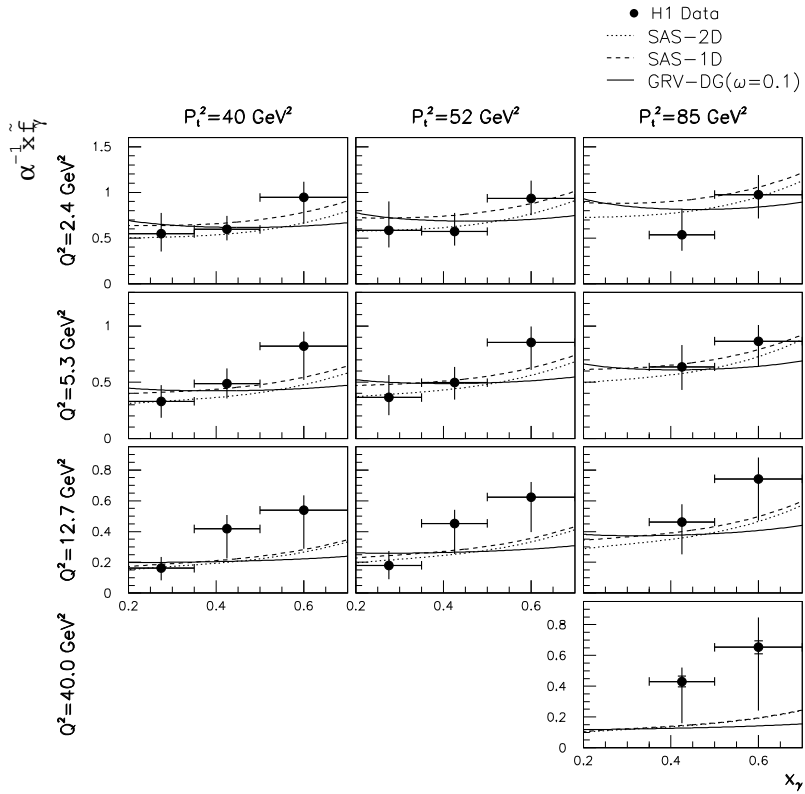}}
\vspace{-0.1cm}
\caption{\small\sl The effective parton density as a function of
$x_{\gamma}$ for various values of the parton transverse momentum 
squared $\tilde{Q}^2$ (= $p_t^2$)  and
virtuality $P^2$ (denoted as $Q^2$). The comparison with predictions
of the SaS2D (dotted line), SaS1D (dashed line) and 
Drees-Godbole (with GRV LO parametrization for the
real photon and using $\omega$ = 0.1, solid line)
parton distributions in the photon is shown (from \cite{h1eff98}).}
\label{fig:9812024_7} 
\end{figure}

\begin{figure}[ht]
\vskip 0.in\relax\noindent\hskip 0.cm
       \relax{\includegraphics{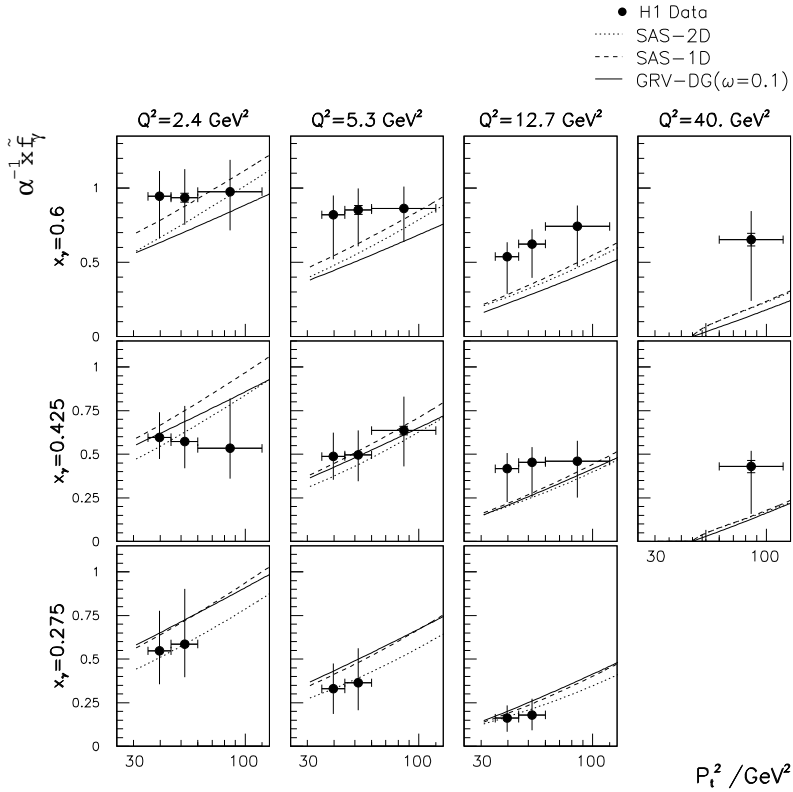}}
\vspace{0.cm}
\caption{\small\sl The effective parton density as a function of
$\tilde{Q}^2$ (= $p_T^2$) for various values of $x_{\gamma}$ and
virtuality $P^2$ (denoted as $Q^2$). Comparison with predictions as in
fig. \ref{fig:9812024_7} (from \cite{h1eff98}).}
\label{fig:9812024_8} 
\end{figure}

\begin{figure}[ht]
\vskip 0.in\relax\noindent\hskip 0.cm
       \relax{\includegraphics{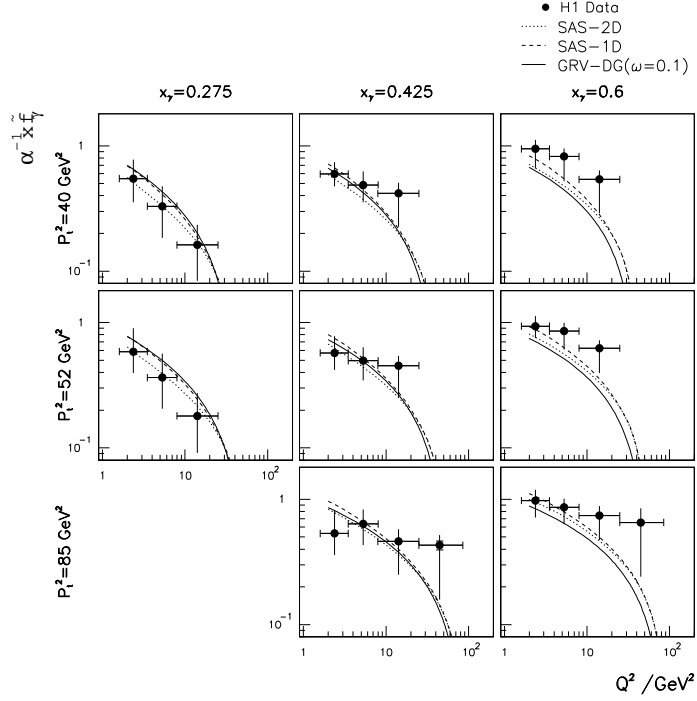}}
\vspace{0.cm}
\caption{\small\sl The effective parton density as a function of
$P^2$ (denoted as $Q^2$) for various 
$\tilde{Q}^2$ (= $p_t^2$) and
$x_{\gamma}$. Comparison with predictions as in
fig. \ref{fig:9812024_7} (from \cite{h1eff98}).}
\label{fig:9812024_9} 
\end{figure}

The Single Effective Subprocess Approximation \cite{com-max}
was assumed in the analysis
and  the effective (LO) parton density of the virtual photon
\be
{\alpha}^{-1}x_{\gamma}(\tilde q^{\gamma}+{9\over 4}G^{\gamma}),
\ee 
with $\tilde q^{\gamma} = \sum (q^{\gamma} + {\bar q^{\gamma}})$,
was extracted for the first time. 
Here  $q^{\gamma} (\bar{q}^{\gamma})$ and $G^{\gamma}$ are photon
polarization-averaged parton densities.
This effective density is presented in table \ref{tablenew}. In 
figs.~\ref{fig:9812024_7}, \ref{fig:9812024_8} and \ref{fig:9812024_9}
it is displayed as a function of $x_{\gamma}$, $p_t^2$ (playing the role of the hard
scale $\tilde{Q}^2$) and $P^2$, respectively. Only points where
averaged $p_t^2$ is larger than $<P^2>$ are shown.\\

\begin{figure}[ht]
\vskip 0.in\relax\noindent\hskip 0.5cm
       \relax{\includegraphics{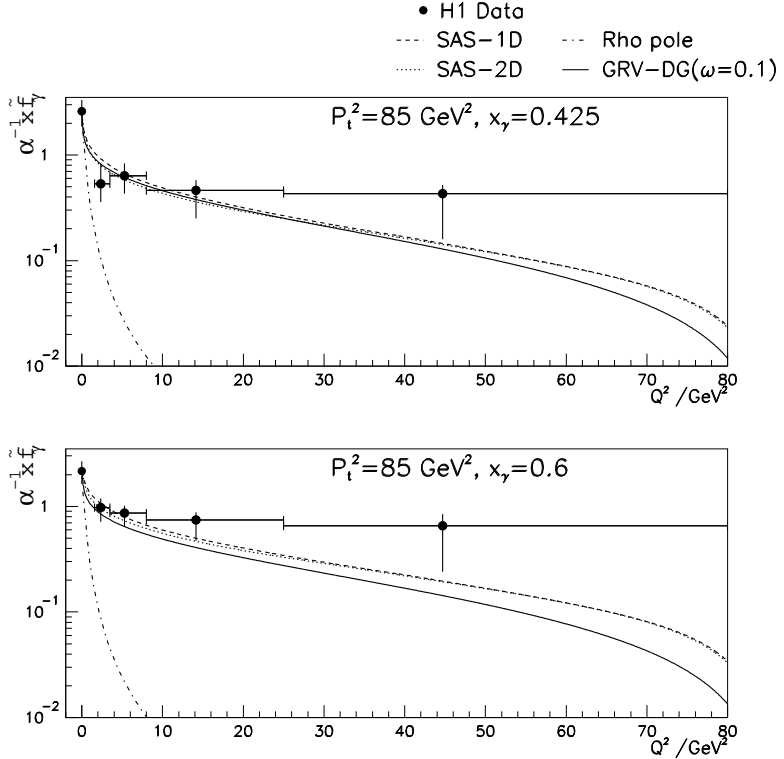}}
\vspace{0.cm}
\caption{\small\sl The LO effective parton density as a function of
$P^2$ (denoted as $Q^2$) for $\tilde{Q}^2$ (= $p_t^2$)
= 85 GeV$^2$,
$x_{\gamma}$=0.425 and 0.60. Comparison with predictions as in
fig. \ref{fig:9812024_7} (from \cite{h1eff98}).}
\label{fig:9812024_10} 
\end{figure}

In fig.~\ref{fig:9812024_10} the obtained LO effective parton density 
as a function of the virtuality $P^2$ is shown for $x_{\gamma}$=0.425 and 0.6, 
at $\tilde{Q}^2$=$p_t^2$. 
\newpage
\vspace*{8cm}
\vspace*{9.8cm}
\vspace*{8.2cm}
\vspace*{10cm}

\noindent
Comment: {\it ``Both simulations give a  
good description of the energy flow in the core of the jets'', neither Monte Carlo model
is able to describe the pedestal for all ranges of $E_T$ and $\eta$ 
(fig.~\ref{fig:9812024_2}).}

~\newline
$\bullet${\bf {ZEUS 93 \cite{plb306} (HERA) }}\\
The evidence for events with two or more jets in
DIS$_{ep}$ (neutral current interaction) is presented
for $Q^2>$ 4 GeV$^2$ (1992 data).

~\newline
$\bullet${\bf {ZEUS 95d \cite{zpc67} (HERA) }}\\
The two-jet production in DIS$_{ep}$ for $160 < Q^2 < 1280$ GeV$^2$,
$0.01 < x_{Bj} < 0.1$ and $0.04 < y < 0.95$ was studied.
In reconstructing jets the JADE algorithm was used (in the
HERA laboratory frame). 
The LEPTO 6.1 generator was used in the analysis.
The comparison was made with the NLO calculations.

~\newline
${\bullet}$ {\bf ZEUS 2000b \cite{STEFAN_NEW} (HERA)}\\
Dijet production for  photon virtualities $P^2$ from 0 up to $4.5$
GeV$^2$ was measured, based on the data  collected in 1995.
Events corresponding  to 0.2 $<y<$ 0.55, $E_{T1,2}>5.5$ GeV and 
$-1.125<\eta<2.2$ were studied in three $P^2$ bins: $P^2\approx 0$
(a quasi-real photon sample obtained by antitagging condition,
which corresponds to $P^2<1.0$ GeV$^2$ with median 10$^{-3}$ GeV$^2$),
0.1 - 0.55 GeV$^2$ and 1.5 - 4.5 GeV$^2$. 
The $k_T$ algorithm  and  the HERWIG 5.9  program (with $p_T^{min}$=2.5 GeV),
 with the GRV LO (photon) and MRSA (proton) parton parametrizations 
were used in the analysis of data. The possible multiple 
parton interaction ('underlying event') was simulated for the quasi-real photon
sample (denoted MI). To estimate uncertainties due to modelling of 
the jet fragmentation the PYTHIA  model was used in addition. The LEPTO model 
for the direct $\gamma^*q$ events
and the NLO prediction for the resolved $\gamma^*$ processes 
(JeTViP) are compared with data.

\begin{figure}[ht]
\vskip 0.cm\relax\noindent\hskip 0.cm
       \relax{\includegraphics{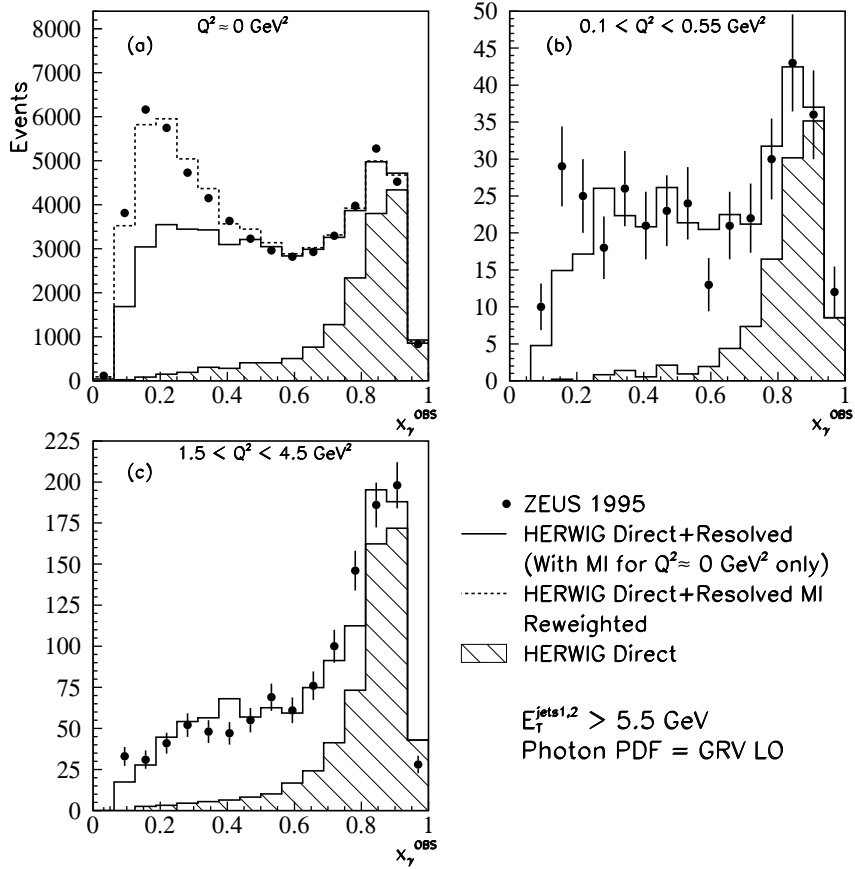}} 
\vspace{0.cm}
\caption{\small\sl Uncorrected  $x_{\gamma}^{obs}$ distributions for 
various  photon virtualities: a) $P^2$ (denoted here as $Q^2$)=0, 
b) 0.1 $\leq P^2 \leq$ 0.55 GeV$^2$, and
c) 1.5 $\leq P^2 \leq$ 4.5 GeV$^2$. Events with $E_{T1,2}>$ 5.5 GeV, 
$-1.125<\eta<2.2$ were included. The predictions of the HERWIG program
(GRV LO) are shown for: a  direct contribution (shaded histogram), 
a direct + resolved + MI (only for  $P^2=0$ sample) contribution 
(solid histogram), 
in a) a reweighted prediction for a direct + resolved + MI
contribution (dashed histogram) (from \cite{STEFAN_NEW}).}
\label{fig:DESY-00-017_1}
\end{figure}

Uncorrected $x_{\gamma}$ distributions are presented in 
fig.~\ref{fig:DESY-00-017_1} for real and for virtual photons (in two bins),
and compared to the HERWIG predictions for 
the direct and resolved (LO) photon contributions.

In the low $x_{\gamma}$ region the data for quasi-real photons disagree with 
MC prediction. ``Disagreement is also observed in the $\eta$, $y_{Bj}$, 
and $E_T$ 
distributions (not shown) and can be attributed to the presence of 
underlying event effects or uncertainty in the PDFs of the photon.'' 
The inclusion of MI 
is not enough to reproduce the shape of the low $x_{\gamma}$
 data (fig.~\ref{fig:DESY-00-017_1}a), so a reweighting has to be performed 
to include this fact in the correction of the data 
for migrations and acceptance.  The obtained results
 (showing an agreement with data)
are presented in fig.~\ref{fig:DESY-00-017_1}a (dashed histogram), 
also the $\eta$, $y_{BJ}$, 
and $E_T$ distributions  (not shown) ``agree well with data''.
\newpage
\vspace*{10.6cm}

\begin{figure}[ht]
\vskip 0.cm\relax\noindent\hskip 0.cm
       \relax{\includegraphics{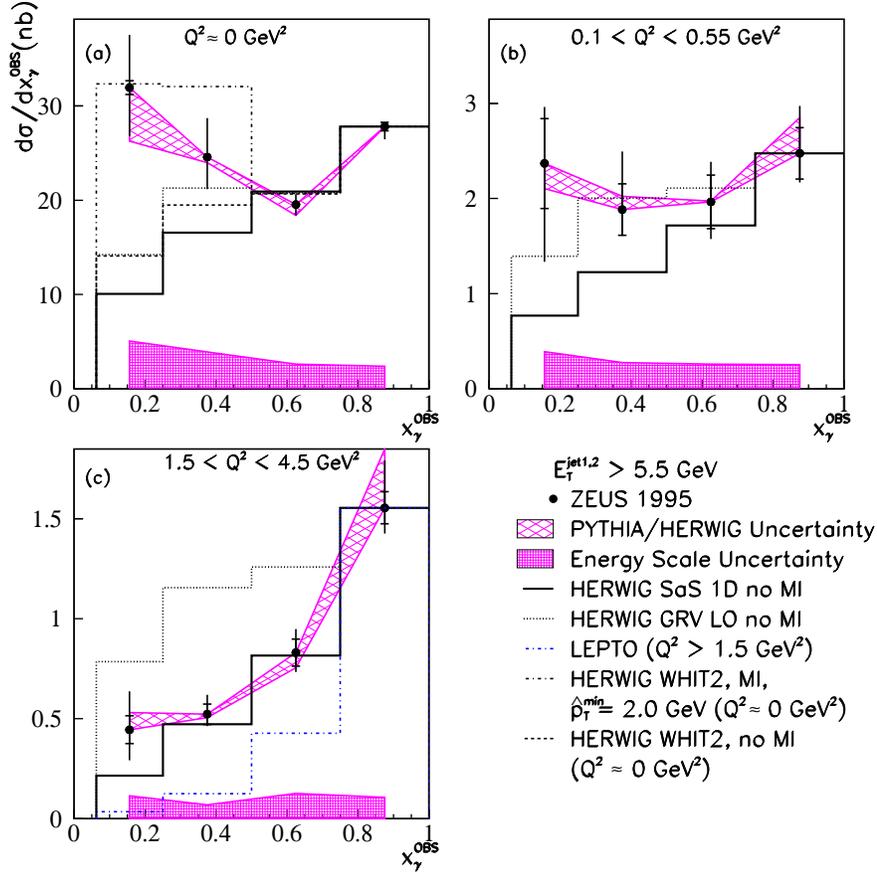}} 
\vspace{0.cm}
\caption{\small\sl The $d\sigma/dx_{\gamma}^{obs}$ cross section 
for various photon virtualities: a) $P^2$ (denoted here as $Q^2$)=0, b) 
0.1 $\leq P^2 \leq$ 0.55 GeV$^2$, and c)
1.5 $\leq P^2 \leq$ 4.5 GeV$^2$, for low $E_T$ sample:
$E_{T1,2}>$ 5.5 GeV, and $-1.125<\eta<2.2$. The predictions of the HERWIG 
program with GRV LO (SaS1D) are shown by dotted (solid) histograms.
In a) the HERWIG predictions obtained using the WHIT2 parton paramerization
with  and without MI  (dot-dashed and dashed histograms) are shown,
in c) the LEPTO prediction (dot-dashed) is given.
The shaded bound represents uncertainties in the simulation
(from \cite{STEFAN_NEW}).}
\label{fig:DESY-00-017_2}
\end{figure}

The dijet cross section $d\sigma/dx_{\gamma}$, corrected to the hadron level,
is presented  in fig.~\ref{fig:DESY-00-017_2} for a low $E_T$ sample, 
where, as above, $E_{T1,2}>$ 5.5 GeV and $-1.125<\eta<2.2$. 
The data are compared with prediction of the HERWIG  program,
 with different parton parametrizations for the photon: 
GRV LO and  WHIT2 (without MI and with MI, then assuming 
$p_T^{mi}$ = 2 GeV), which are  valid for a real photon, 
and SaS1D. In addition a
comparison with prediction of the LEPTO  program is shown for events with 
$P^2>1.5$ GeV$^2$.
\newline
\vspace*{10.6cm}

\begin{figure}[ht]
\vskip 0.cm\relax\noindent\hskip 0.cm
       \relax{\includegraphics{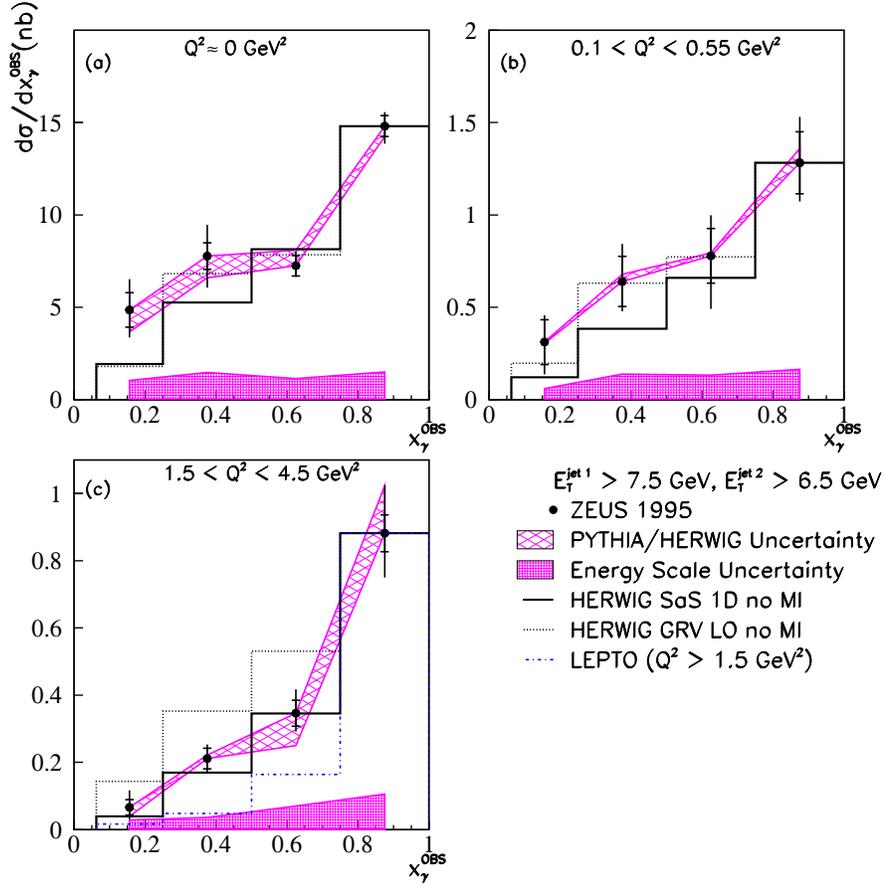}} 
\vspace{0.cm}
\caption{\small\sl
The $d\sigma/dx_{\gamma}^{obs}$ cross section 
for various photon virtualities: a) $P^2$ (denoted here as  $Q^2$)=0, b) 
0.1 $\leq P^2 \leq$ 0.55 GeV$^2$, and c)
1.5 $\leq P^2 \leq$ 4.5 GeV$^2$, for high $E_T$ sample:
$E_{T1}>$ 7.5 GeV, $E_{T2}>$ 6.5 GeV, and $-1.125<\eta<1.875$. 
The predictions of the HERWIG 
program with GRV LO (SaS1D) are shown by dotted (solid) histograms.
In c) the LEPTO prediction (dot-dashed histogram)
is given. The shaded bound represents uncertainties of the simulation
(from \cite{STEFAN_NEW}).} 
\label{fig:DESY-00-017_3}
\end{figure}

Similar analysis has been performed for the high  $E_T$ sample, 
where $E_{T1}>$ 7.5 GeV, $E_{T2}>$ 6.5 GeV and $-1.125<\eta<1.875$,
 see fig.~\ref{fig:DESY-00-017_3} for results.
\newpage
\vspace*{10.6cm}

\begin{figure}[ht]
\vskip 0.cm\relax\noindent\hskip 0.cm
       \relax{\includegraphics{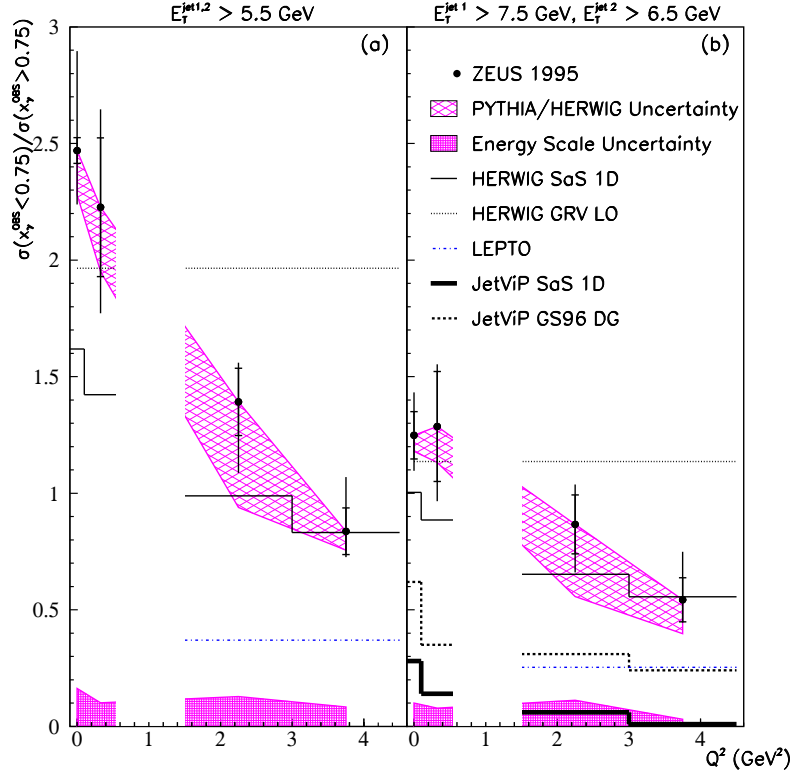}} 
\vspace{0.cm}
\caption{\small\sl The ratio 
$\sigma(x_{\gamma}^{obs}<0.75)/\sigma(x_{\gamma}^{obs}>0.75)$
as a function of the photon virtuality $P^2$ (denoted as $Q^2$) for both $E_T$ cuts.
The predictions of the HERWIG program with the GRV LO(SaS1D) parton 
parametrization are presented as  dotted (solid) histograms.
The LEPTO program predictions are shown for $P^2>1.5$ GeV 
(dot-dashed histograms). In b) the NLO QCD prediction of the JeTViP program  
with the SaS1D  (GS96/Drees-Godbole) parton parametrization with  the squared 
hard scale $P^2+p_T^2$ is shown as thick solid (dashed) histogram
(from \cite{STEFAN_NEW}).}
\label{fig:DESY-00-017_4}
\end{figure}

The ratio  $\sigma(x_{\gamma}<0.75$)/  $\sigma(x_{\gamma}>0.75)$
as a function of the photon virtuality $P^2$ is shown in
fig.~\ref{fig:DESY-00-017_4} for both sets of  $E_T$ cuts.
Here a comparison with the prediction of the HERWIG program with the 
 GRV LO and SaS1D parton parametrizations is shown.
The small $P^2$ region is not properly modelled.
Also the LEPTO program predictions are shown for $P^2>1.5$ GeV.
The NLO QCD predictions obtained by using the JeTViP program  
with the SaS1D and GS96/Drees-Godbole parton parametrizations 
and the squared hard
scale equal to $P^2+p_T^2$, are compared to high $E_T$ data.
 They ``lie well below the data.'' 
\newpage
\vspace*{9.5cm}

~\newline
Comment: {\it ``The shape of the dijet cross sections, 
$d\sigma/dx_{\gamma}$, is compared to the prediction of HERWIG MC for a variety of photon PDFs. None of these models is able to explain the data for both high- and low-$E_T$ cuts in all $P^2$.''\\
For the dependence of the ratio $\sigma(x_{\gamma}<0.75$)/
$\sigma(x_{\gamma}>0.75)$ on the photon virtuality $P^2$...'' none of 
the LO model, or the NLO calculation examined here, gives a good 
description of the data across the full kinematic region.''
\newline
Note, that data were corrected for acceptance, smearing and kinematical
cut using HERWIG 5.9 with the GRV LO parton parametrization, valid
for a real $\gamma$.} 

\newpage
\subsection{Forward jet (particle) production in $\gamma^*p$ collisions and the
resolved virtual photon contribution\label{sec34}}
The production of high $E_T$ jets (particles) in the direction of the proton 
in events with large virtualities, \ie {\sl DIS$_{ep}$} events, small $x_{Bj}$ 
and $E_T^2 \sim Q^2$ was studied by both the H1 and ZEUS groups at the HERA 
collider.

The contribution due the standard DGLAP  
evolution based on the {\sl  strong} $p_t$ ordering 
of parton emission along
the whole chain between the proton and the photon is expected
to be strongly suppressed for such events. In contrast, 
 signals from the {\sl non-strongly ordered} in transverse
momenta partons from the {\sl {proton}} (the
BFKL \cite{bfkl} or CCFM \cite{ccfm} evolution) are
expected to show up \cite{F1}. At first the BFKL calculation for the 
forward jet (particle) production was based on the LO approach, recently  
the modified LO BFKL approach has appeared including the kinematical 
constraint, called the {\underbar{consistency constraint}}, on the gluon emission.
This constraint  
presumably embodies a major part of the NLO 1/$x$ corrections \cite{kmo},
see also discussion on this point in \cite{note}.
 Besides the BFKL, the  mechanism with
the two DGLAP evolution chains corresponding to the resolved virtual 
photon-proton interaction may also lead to forward jets (or particles),
 as was pointed out in \cite{F2}.
The NLO prediction of this contribution, based on 
the parton level generator JeTViP \cite{krapot}, was performed in
\cite{9901314}. Several Monte Carlo programs were used to 
model the radiation characteristics for a particular evolution. 
The HERWIG and LEPTO models are based on DGLAP, while LDC 
(called also LDCMC) - on the 
CCFM equation. In the ARIADNE program, which  implements the Color Dipole 
Model, there is {\sl no $p_t$ ordering}, so it has the properties of the BFKL evolution. 
The RAPGAP generator (also JeTViP) contains the resolved virtual photon 
contribution with the DGLAP cascades.

Both forward jet and forward  single particle production are measured.
The advantage of studying single particles is the potential to reach a smaller 
angle than it is  possible with jets.

The scale for the deep inelastic scattering on the proton
(DIS$_{ep}$), $Q^2$,
plays the role of $P^2$, if the contribution due to the resolved
virtual photon is considered. For the sake of transparency,
in this section we
 keep $Q^2$ notation for the virtuality of the photon.

\newpage
\centerline{\bf \huge DATA}
~\newline
$\bullet${\bf {H1 95b \cite{forh195} (HERA) }}\\ 
The first analysis of the forward jets in DIS$_{ep}$  was done.

~\newline
$\bullet${\bf {H1 97c \cite{forh197} (HERA) }}\\
The $p_T$ distribution of charged particles in DIS$_{ep}$
events was studied. 
The hadronic final state was analysed
in order to test the different schemes for parton evolution in the
proton (DGLAP and BFKL).

~\newline 
Comment: {\it The large parton radiation between the current and the proton 
remnant system was found.}

~\newline
$\bullet${\bf {H1 99b \cite{forh198} (HERA) }}\\ 
The production of forward jets and forward particles 
(the charged particles and the neutral 
pions) with large $p_T$ is analysed in the DIS$_{ep}$ events at small $x_{Bj}$
(1994 data).
The forward particles were measured for $5^0<\theta<25^0$ (LAB),
and for the energy ratio $x_{cp(\pi)} = E_{cp(\pi)}/E_p$ between
0.01 and 0.15 (for charged particles below 0.015 only). 
Two $p_T$ thresholds were used, 1 and 2 GeV (only for $\pi^0$), 
with additional requirement 0.0002 $< x_{Bj} < $ 0.00235.
Forward jets with the energy ratio
$x_{jet} = E_{jet}/E_p > 0.035$, and 
$0.5 < p_T^2/Q^2 < 2$, $7^{\circ} < \theta_{jet} < 20^{\circ}$ and
$p_T > 3.5$ and 5 GeV were studied
$(10^{-4}<x_{Bj}< 4 \cdot 10^{-3})$. 
A simple cone algorithm with R=1 was used for jets. 

\begin{figure}[hb]
\vskip 0.cm\relax\noindent\hskip 0.cm
       \relax{\includegraphics{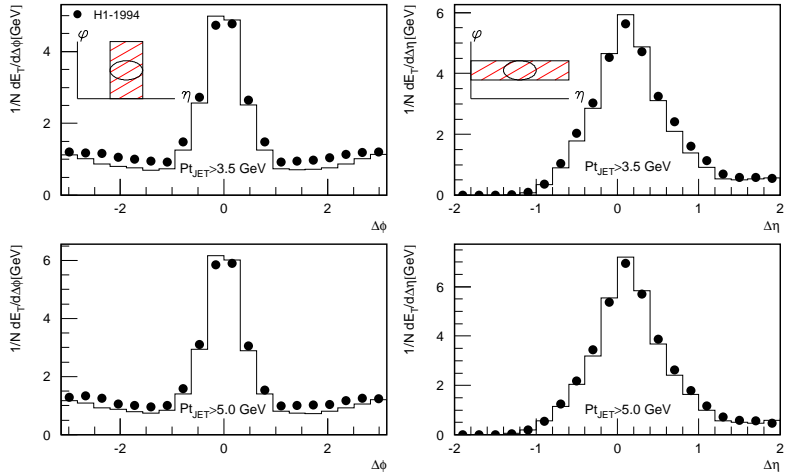}} 
\vspace{0.cm}
\caption{\small\sl The transverse energy flow as a function of $\delta\phi$
and $\delta\eta$ for two $p_T$ cuts for jets (from \cite{forh198}).}
\label{fig:9809028_2}
\end{figure}

The predictions of the LO BFKL calculation \cite{klm} and of the
Monte Carlo generators: LEPTO 6.5 (with renormalization scale $Q^2$),
LDC 1.0 and {ARIADNE} 4.08 (scale $p_T^2$) were compared with
data. The RAPGAP 2.06 generator with resolved photon processes
(with SaS2D parton parametrization for the photon, assuming
scale $Q^2 + p_T^2$) was also used in comparison. 
To describe a proton the GRV HO parton densities were used
in all the programs. The transverse energy flow as a function of 
$\delta\eta$ was measured for jets and results are presented in 
fig.~\ref{fig:9809028_2}.

\vspace*{6.cm}

\begin{figure}[ht]
\vskip 0.cm\relax\noindent\hskip 0.cm
       \relax{\includegraphics{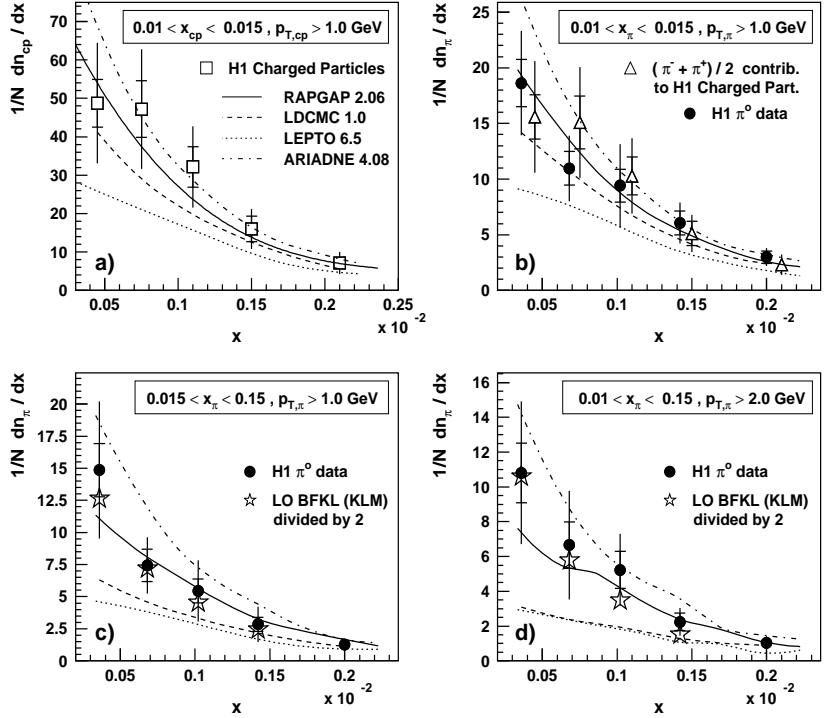}} 
\vspace{0.cm}
\caption{\small\sl The single particle spectra as a function of
$x_{Bj}$: a) for charged particles, b) for charged pions and $\pi^0$'s;
c), d) for $\pi^0$'s with different cuts on $x_{\pi}$ and $p_T^{\pi}$. 
Comparison with the Monte Carlo predictions, and in panels (c, d)
in addition with the LO BFKL
calculation \cite{klm}, is presented (from \cite{forh198}).}
\label{fig:9809028_3}
\end{figure}

The comparison of the measured $x_{Bj}$ distribution  with 
prediction of different types of Monte Carlo 
models for single particle spectra
is  shown in fig.~\ref{fig:9809028_3}. 
Here also  the 
comparison with the 
results of the analytical calculation based on the BFKL approach
\cite{klm}, with the factor
${1\over 2}$ to describe the data, is made.
\newline
\vspace*{8.7cm}

\begin{figure}[ht]
\vskip 0.cm\relax\noindent\hskip 0.cm
       \relax{\includegraphics{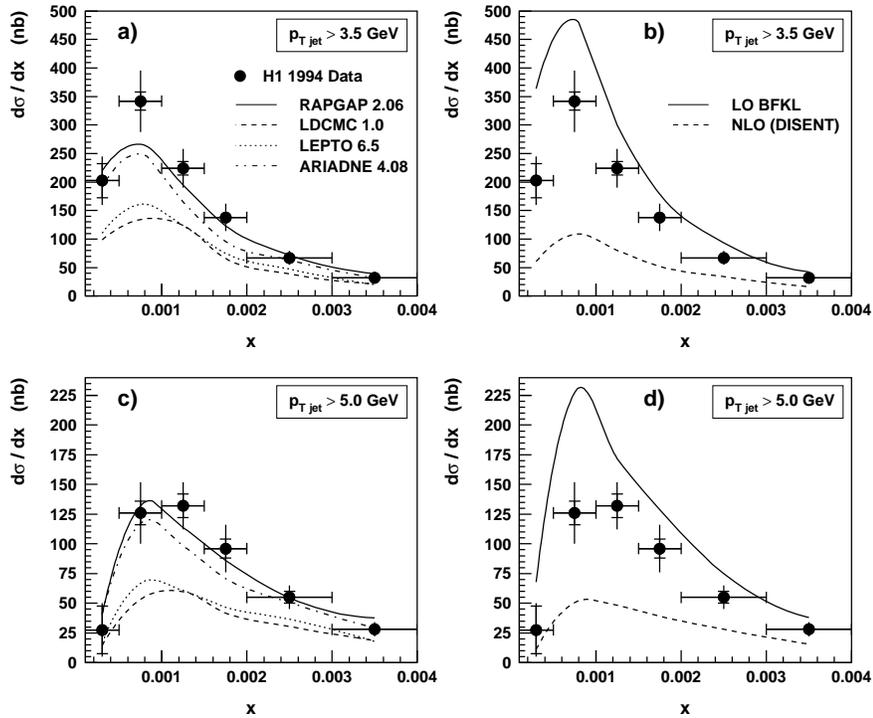}} 
\vspace{0.cm}
\caption{\small\sl The forward jet cross section as a function
of $x_{Bj}$ for $p_T$ cuts: 3.5 GeV (a, b) and 5 GeV (c, d). 
a) and c) Comparison with the Monte Carlo predictions: RAPGAP, LDCMC,
LEPTO and ARIADNE. b) and d) Comparison with  the LO BFKL calculation 
\cite{bart96} and NLO direct prediction (DISENT) (from \cite{forh198}).}
\label{fig:9809028_4}
\end{figure}

In fig.~\ref{fig:9809028_4}
the forward jet data for $p_T > 3.5$ GeV and 
$p_T > 5$ GeV are shown 
and compared to the Monte Carlo predictions 
and analytical calculations, LO BFKL and the NLO direct
contribution (implemented in the DISENT program).
\newpage
\vspace*{8.2cm}

\begin{figure}[ht]
\vskip 0.cm\relax\noindent\hskip 0.cm
       \relax{\includegraphics{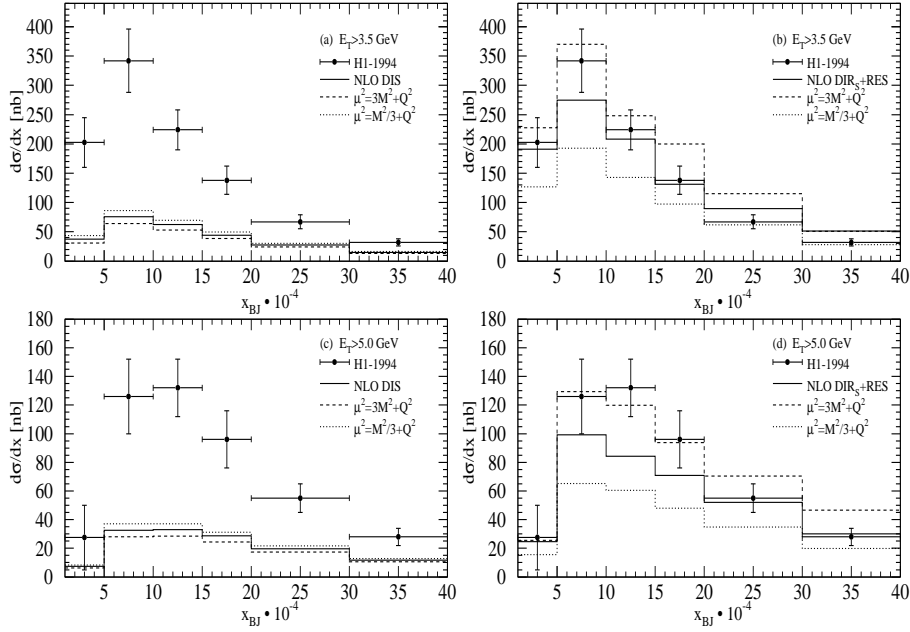}} 
\vspace{0.cm}
\caption{\small\sl Results of the NLO calculations of the forward 
jet cross section as a function of $x_{Bj}$,
compared to the H1 data for $E_T >$ 3.5 GeV (a, b)
and $E_T >$ 5 GeV (c, d). The NLO description within the DIS$_{ep}$ approach
is presented in a) and c), while b) and d) show the NLO description
based on the direct (DIR) and the resolved (RES)
virtual photon contributions. Three different
renormalization scales are used: 
$\tilde{Q}^2 (= \mu^2$) = $ M^2$ + $Q^2$ (solid line),
$\tilde{Q}^2 (= \mu^2$) = $3 M^2$ + $Q^2$ (dashed line) and
$\tilde{Q}^2 (= \mu^2$) = $M^2/3$ + $Q^2$ (dotted line), 
with $M^2=<E_T>^2$=50 GeV$^2$ (from \cite{9901314}).}
\label{fig:9901314_3}
\end{figure}

Fig.~\ref{fig:9901314_3} (taken from \cite{9901314})
shows the data together with the standard DIS$_{ep}$ NLO 
predictions, and the NLO calculation by Kramer and P\"otter 
\cite{9901314}, where the interaction due to partonic content of 
the virtual photon is included. Note that in this calculation the 
hard scale $\tilde{Q}^2$ is assumed in the form 
$\tilde{Q}^2 = k M^2 + P^2$, with $M^2 = <E_T>^2$ fixed to be 50 GeV$^2$
and parameter $k$ between $1\over 3$ and 3. 
\newpage
\vspace*{7.2cm}

The azimuthal correlation $\Delta\phi$ between the jet and
the scattered electron was also studied (not shown).

~\newline
Comment: {\it "The conclusions obtained using the  forward jet and
 the single particle measurements are in agreement''. 
\newline Predictions
based on the DGLAP approach fail to describe the data, except
for those which allow for resolved virtual photon contributions.
The RAPGAP model gives a good description of data. The BFKL calculation
is above the measurements for the data. ARIADNE model reproduces
the rise towards low $x$.
\newline ``Predictions of a CCFM evolution,
which should smoothly interpolate between the DGLAP and BFKL regimes,
give a poor description of all measured distributions.''}

~\newline
$\bullet${\bf H1 99c \cite{DESY99-094} (HERA)} \\
The forward production of large $p_T$ $\pi^0$-mesons was measured
in 1996 in DIS$_{ep}$ events at small $x_{Bj}$.
The events correspond to 2.0 $<Q^2<$ 70 GeV$^2$, 0.1 $<y<0.6$ 
and small polar angle between 5$^{\circ}$ and 25$^{\circ}$ (LAB).
Events with  forward $\pi^{0}$ production were collected
for $p_T^{* \pi}>$ 2.5 GeV (in the hadron CMS) and 
$x_{\pi}=E_{{\pi}}/E_p>$  0.01.

The comparison of the data is made with 
the prediction of the modified LO BFKL calculation \cite{kmo} and with
the Monte Carlo models: LEPTO 6.5 based on the DGLAP description of 
the direct $\gamma^* q$ scattering, and RAPGAP 2.06, which includes 
the resolved photon contributions and uses the DGLAP evolution for 
$\gamma^*$ (SaS1D with $\tilde{Q}^2=Q^2+p_T^2$). For the proton all 
models use the CTEQ4M parton parametrization.

\begin{figure}[ht]
\vskip 0.cm\relax\noindent\hskip 0.2cm
       \relax{\includegraphics{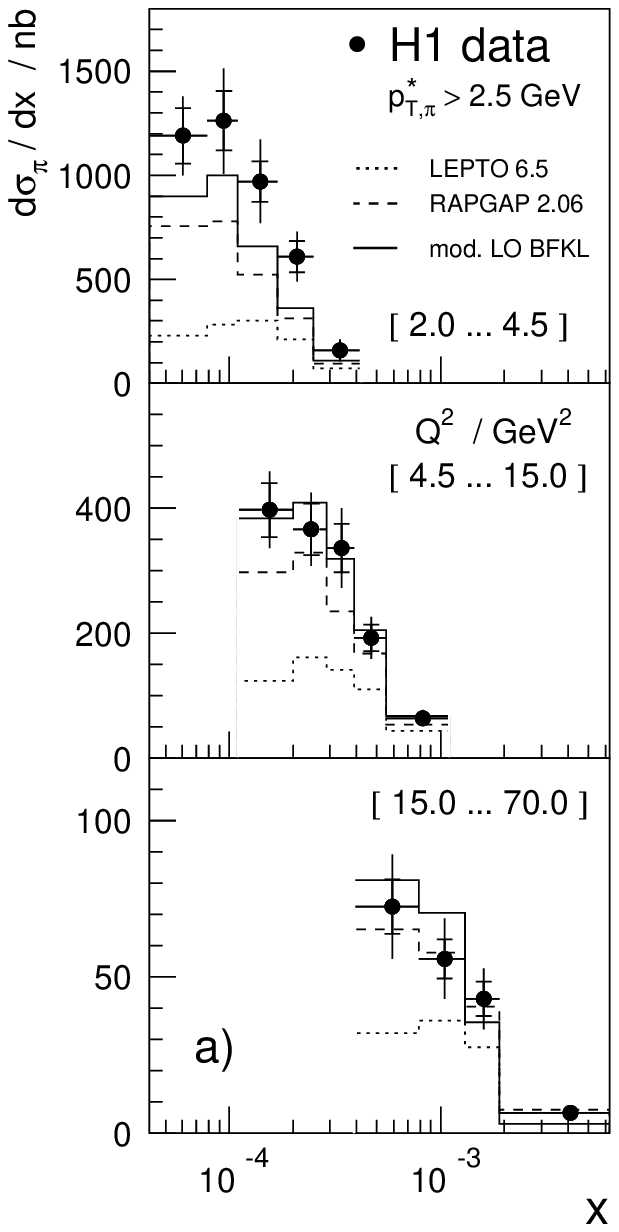}}
\vskip -0.5cm\relax\noindent\hskip 7.6cm
       \relax{\includegraphics{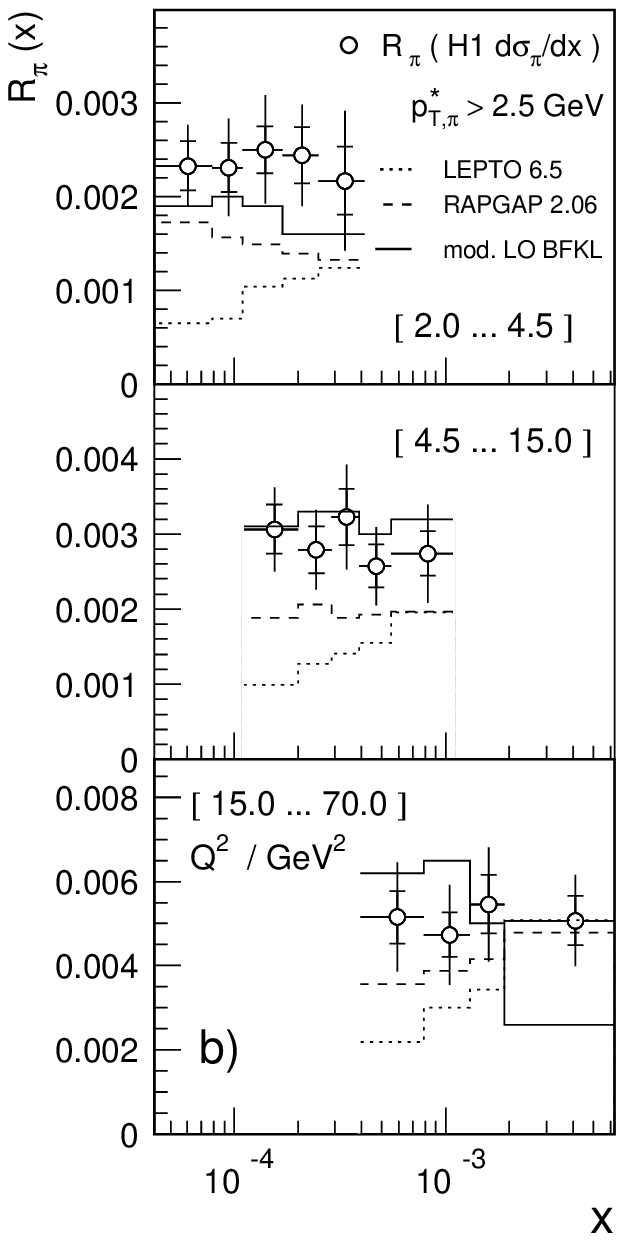}} 
\vspace{-0.5cm}
\caption{\small\sl The inclusive forward $\pi^0$ cross section
for $p_T^{*\pi} >$ 2.5 GeV
as a function of $x_{Bj}$ in bins of $Q^2$. Comparison with the modified 
LO BFKL calculation (solid line) \cite{kmo}, 
with the LEPTO (dotted line) and RAPGAP (dashed line)
Monte Carlo models. a) The $x_{Bj}$
distributions for $\pi^0$. b) The ratio of the $x_{Bj}$
distributions presented in a) and the corresponding ones for
the full sample of $ep$ events (from \cite{DESY99-094}).}
\label{fig:d99-0941}
\end{figure}

In fig.~\ref{fig:d99-0941}a the $x_{Bj}$ dependence is presented for 
 $p_T^{*\pi} >$ 2.5 GeV as a function of $x_{Bj}$ in bins of $Q^2$. 
The observed  strong rise of the cross section towards small $x_{Bj}$ 
in the forward $\pi^0$ production is identical to the rise of the 
inclusive $ep$ cross section for all events, see fig.~\ref{fig:d99-0941}b.
Comparison with the modified LO BFKL calculation \cite{kmo}, 
and with LEPTO and RAPGAP Monte Carlo models are also present in the figure.
\vspace*{13cm}

\begin{figure}[ht]
\vskip 0.cm\relax\noindent\hskip 0.2cm
       \relax{\includegraphics{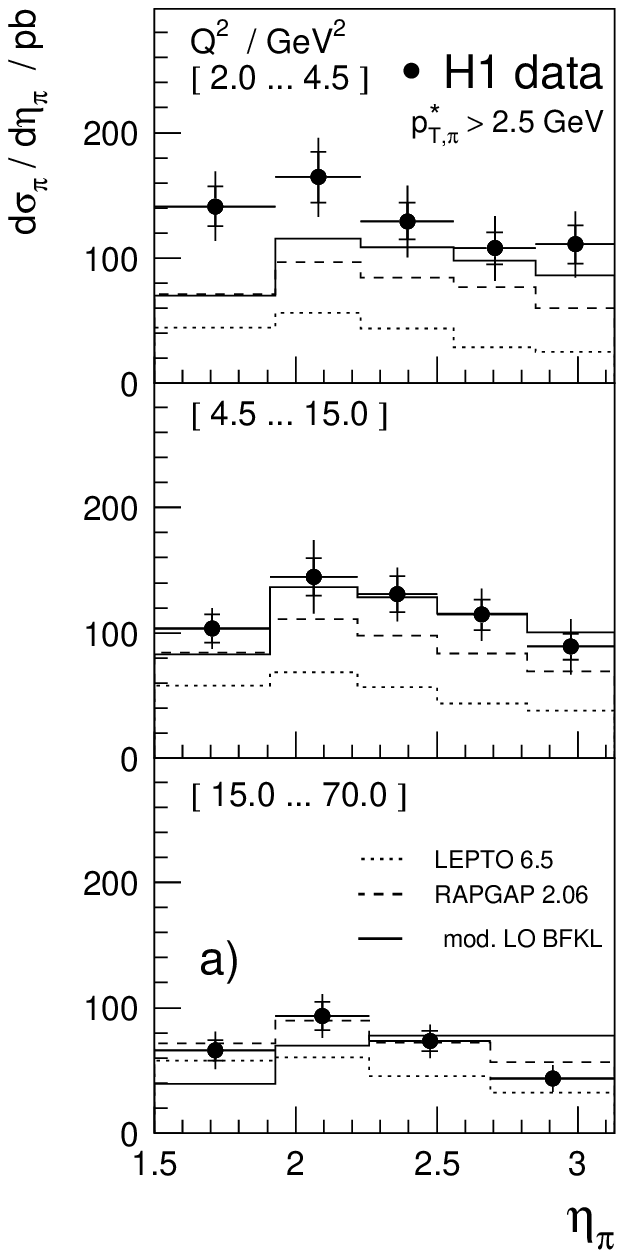}}
\vskip -0.5cm\relax\noindent\hskip 7.6cm
       \relax{\includegraphics{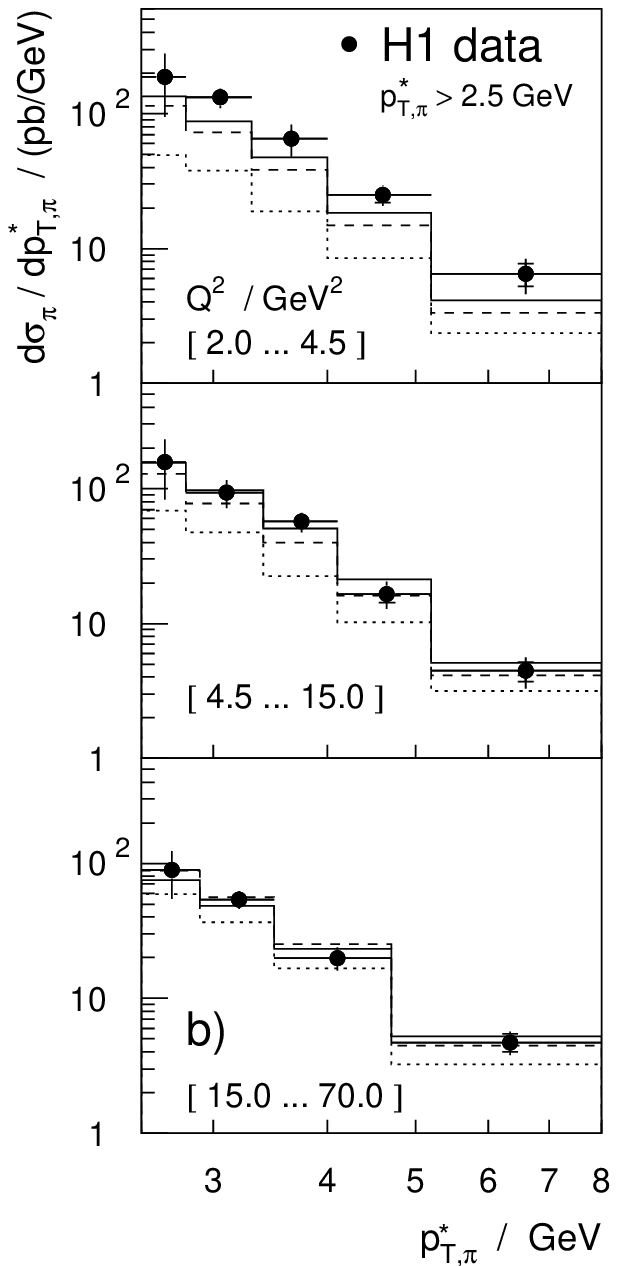}} 
\vspace{-0.3cm}
\caption{\small\sl The inclusive forward $\pi^0$ cross section
for $p_T^{*\pi} >$ 2.5 GeV
as a function
of $\eta_{\pi}$ (a) and of $p_T^{*\pi}$ (b), in bins of $Q^2$. 
Comparison with model predictions as in fig.~\ref{fig:d99-0941}
(from \cite{DESY99-094}).}
\label{fig:d99-0942}
\end{figure}

The analysis of the  $p_T^{*\pi}$ and $\eta$ distributions in the  $Q^2$ bins
was performed and the results are presented in fig.~\ref{fig:d99-0942}.
\newpage
\vspace*{12cm}

\begin{figure}[ht]
\vskip 0.cm\relax\noindent\hskip 1.5cm
       \relax{\includegraphics{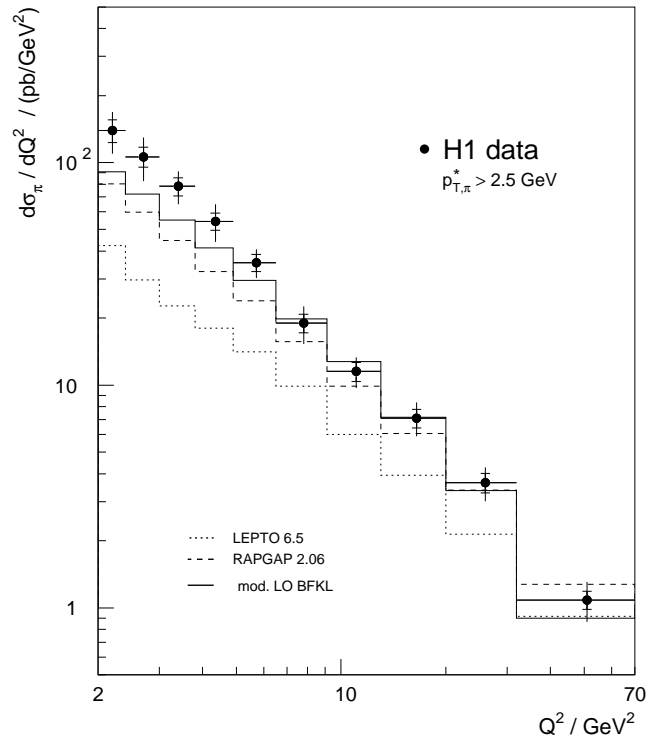}}
\vspace{0.cm}
\caption{\small\sl The inclusive forward $\pi^0$ cross section
for $p_T^{*\pi} >$ 2.5 GeV
as a function of $Q^2$.
Comparison with model predictions as in fig.~\ref{fig:d99-0941}
(from \cite{DESY99-094}).}
\label{fig:d99-0943}
\end{figure}

\newpage
In fig.~\ref{fig:d99-0943} the $Q^2$ dependence is presented 
for events with $p_T^{*\pi} >$ 2.5 GeV, in comparison with the
modified LO BFKL and Monte Carlo predictions. 
\vspace*{9.7cm}

\begin{figure}[ht]
\vskip 0.cm\relax\noindent\hskip 0.cm
       \relax{\includegraphics{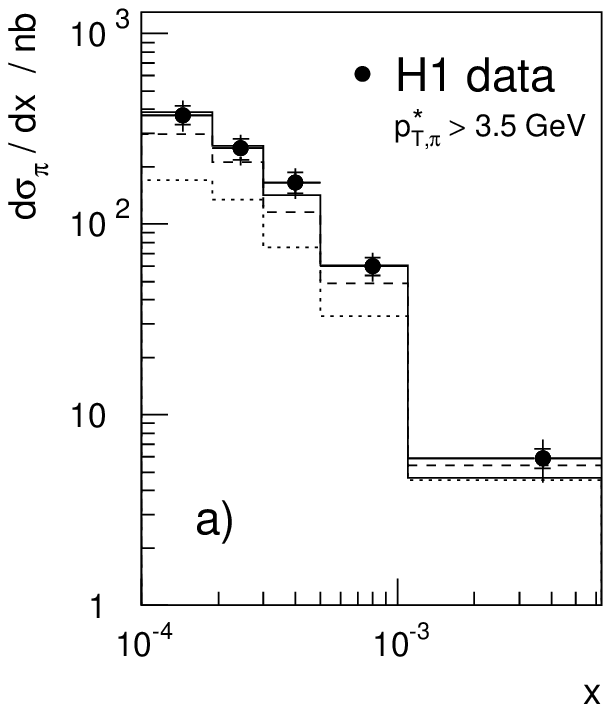}}
\vskip -0.5cm\relax\noindent\hskip 8.cm
       \relax{\includegraphics{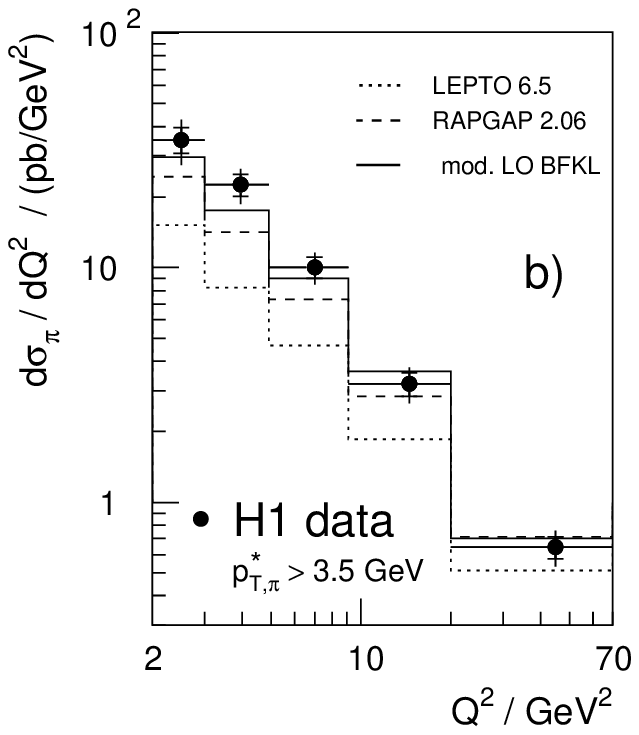}} 
\vspace{-0.6cm}
\caption{\small\sl The inclusive forward $\pi^0$ cross section
for $p_T^{*\pi} >$ 3.5 GeV as a function
of (a) $x_{Bj}$ and (b) $Q^2$. 
Comparison with model predictions as in fig.~\ref{fig:d99-0941}
(from \cite{DESY99-094}).}
\label{fig:d99-0944}
\end{figure}

\newpage
For comparison also the $x_{Bj}$ and $Q^2$ dependences are presented for 
 the $p_T^{*\pi} $ threshold equal to  3.5 GeV, see fig.~\ref{fig:d99-0944}. 
\vspace*{7.5cm}

~\newline
Comment: {\it The measurement extends down to $x_{Bj}=5\cdot 10^{-5}$,
lower than the previous values. It covers lower polar angles and
larger $p_T^{*\pi}$ range, as compared with {\bf H1 99b}.

``Inclusion of processes in which the virtual photon is resolved 
improves the agreement with the data,
but does not provide a satisfactory description in the full $x_{Bj}$ and $Q^2$
range. A calculation based on the BFKL formalism is in good agreement
with  the data, particularly for the shape description, but the absolute 
normalization remains strongly affected by the scale uncertainty.''}

~\newline
$\bullet${\bf {ZEUS 99c \cite{forzeus99} (HERA) }}\\ 
In the $e^+p$ scattering the forward jets produced
 with $y>0.1$, $E_T>$ 5 GeV and $x_{Bj}$ between $4.5 \cdot 10^{-4}$ 
and $4.5 \cdot 10^{-2}$ are studied  for the 0.5 $<E_T^2/Q^2<2$. 
Data were taken in 1995.
The motivation was the search for effects of the BFKL type parton 
shower evolution. 
The jet was defined according to the cone algorithm (R=1).
The LO BFKL and
few Monte Carlo programs (ARIADNE 4.08, LEPTO 6.5, HERWIG 5.9, 
LDC 1.0 and MEPJET for the NLO direct contribution) 
predictions were compared with data.

\begin{figure}[ht]
\vskip 0.cm\relax\noindent\hskip -3cm
       \relax{\includegraphics{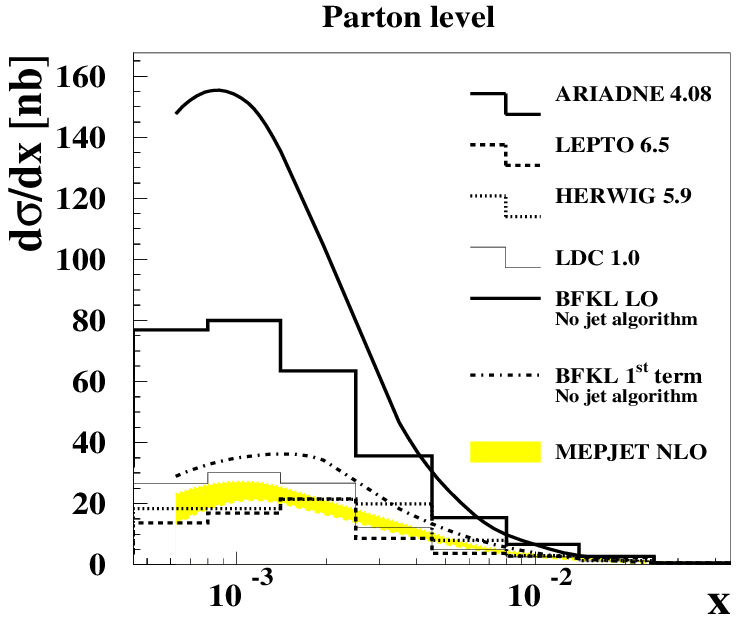}} 
\vskip -0.48cm\relax\noindent\hskip 4.7cm
       \relax{\includegraphics{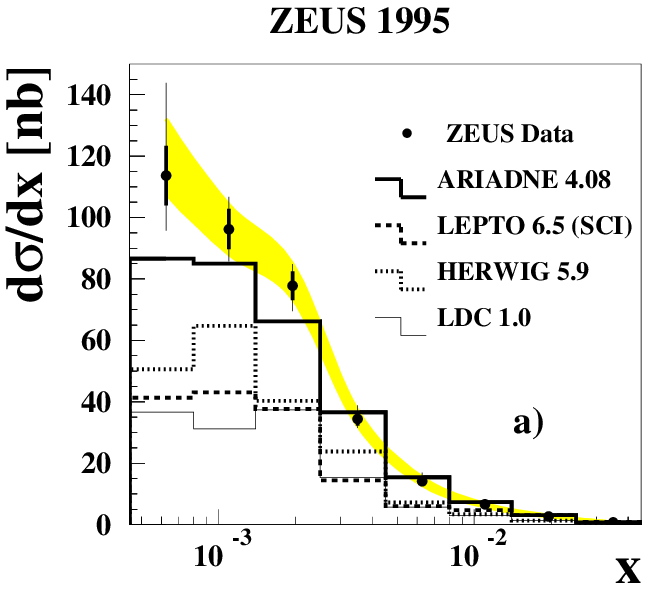}} 
\vspace{0.cm}
\caption{\small\sl Left: The $x_{Bj}$ distribution at the parton level
obtained in various approaches (Monte Carlo and QCD analytical
calculations). The shaded band corresponds to the range of results obtained with MEPJET using a renormalisation scale between $0.25 k_T^2$ and $2.0 k_T^2$.
Right: The $x_{Bj}$ distribution at the hadron level:
data points and various Monte Carlo predictions (from \cite{forzeus99}).}
\label{fig:9805016_7_10a}
\end{figure}

The parton-level predictions for forward jet
cross sections as a function of $x_{Bj}$ are shown in 
fig.~\ref{fig:9805016_7_10a} (left) and for the hadron level in 
fig.~\ref{fig:9805016_7_10a} (right). The comparison with the LO BFKL and 
LO BFKL $1^{st}$ term (\ie with one parton rung in the
gluon ladder) and the Monte Carlo predictions is also given. 

The $E_{T}^2/Q^2$ distribution was studied as well (see
{\bf ZEUS 2000c} for the new results).
\newline
\vspace*{5.3cm}
~\newline
Comment: {\it ``Three regions are identified in the $E_{T}^2/Q^2$
distribution:
i) the standard DGLAP region with $E_{T}^2\ll Q^2$, where all Monte Carlo
models are in agreement with the data;
ii) the region of phase space where BFKL dynamics is expected to
contribute significantly with $E_{T}^2\approx Q^2$, where only
the Colour Dipole model describes the data well;
iii) the region with $E_{T}^2\gg Q^2$, where none of the models
reproduces the data.''\newline
No contribution due to partonic content of the virtual photon
was considered.}

~\newline
$\bullet${\bf {ZEUS 2000c \cite{DESY-99-162} (HERA) }}\\ 
 The forward jet production (1995 data) was studied based on 
in DIS$_{ep}$ events with $Q^2>10$ GeV$^2$, $y>$0.1.
Events with $E_T>$5 GeV, 0.00025 $< x_{Bj} <$ 8.0 $\cdot$ 10$^{-2}$
and $E^2/Q^2$ between 10$^{-2}$ and 10$^2$ were considered,
in addition the cut $x_{jet}$=$p_{z,jet}/p_{proton}>$
0.036 was used to select forward jets.
The cone algorithm was applied to jets in LAB frame with R=1.

\begin{figure}[ht]
\vskip 0.cm\relax\noindent\hskip 0.7cm
       \relax{\includegraphics{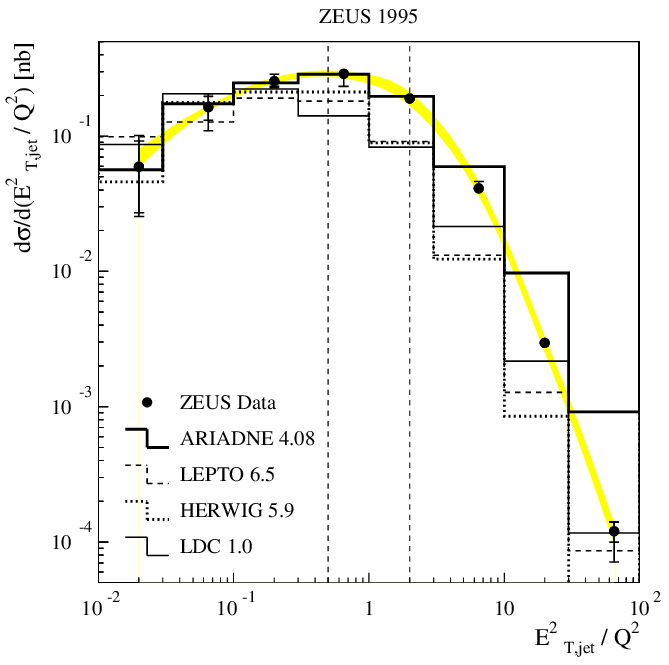}} 
\vskip -0.51cm\relax\noindent\hskip 7.5cm
       \relax{\includegraphics{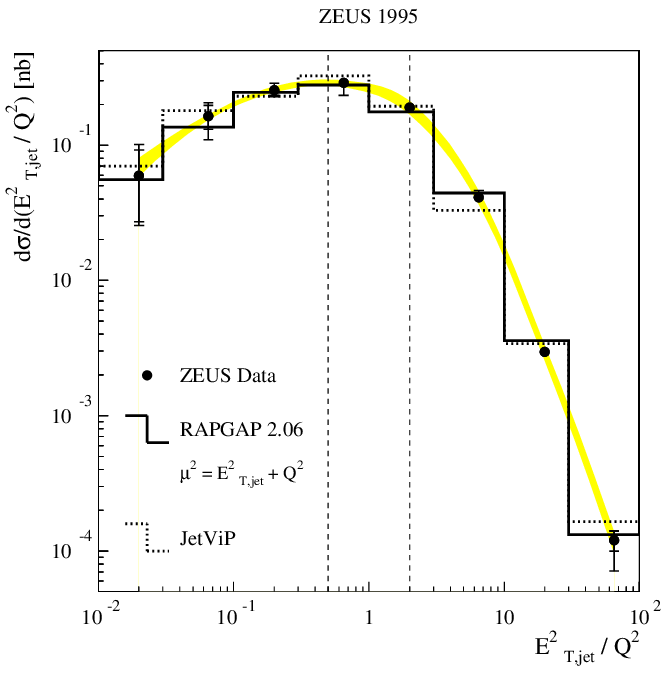}} 
\vspace{-0.6cm}
\caption{\small\sl  Forward jet production cross section as a function
of $E_T^2/Q^2$. 
Comparison with predictions of the various MC models, left: ARIADNE 4.08, LEPTO 6.5, HERWIG 5.9, LDC 1.0, and  right: RAPGAP 2.06, JeTViP
(from \cite{DESY-99-162}).}
\label{fig:DESY-99-162_1}
\end{figure}

\begin{figure}[ht]
\vskip 0.cm\relax\noindent\hskip 3.7cm
       \relax{\includegraphics{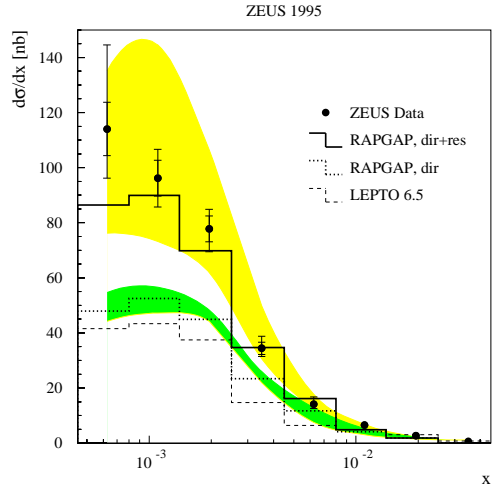}} 
\vspace{-0.7cm}
\caption{\small\sl Forward jet production cross section as a function
of $x_{Bj}$: data compared to the various Monte Carlo predictions:
LEPTO based on the standard DGLAP evolution (dashed histogram),
RAPGAP with the  direct virtual photon contribution (dotted histogram)
and RAPGAP with the  direct and the  resolved photon contribution (solid histogram)
(from \cite{DESY-99-162}).}
\label{fig:DESY-99-162_3}
\end{figure}

The LO MC models, standard for the description of the DIS$_{ep}$ events: 
LEPTO 6.5, HERWIG 5.9, LDC 1.0 (the linked-dipole chain as in CCFM),
and ARIADNE 4.08 were used. In addition  
the RAPGAP 2.06 Monte Carlo program  and the JeTViP (NLO calculation) program
were included in the analysis 
to describe the  resolved virtual photon interaction, with
SaS2D and SaS1D parton parametrization
for the photon, respectively.
Both introduce the renormalization/factorization scale $\tilde Q^2=E_T^2+Q^2$. In   
fig.~\ref{fig:DESY-99-162_1} the $E_{T}^2/Q^2$ distribution, 
and in fig.~\ref{fig:DESY-99-162_3} the $x_{\gamma}$ distribution
is shown
and compared with the Monte Carlo models, with and without the resolved
photon contribution.\\

\vspace*{6.1cm}
\vspace*{6.4cm}

\begin{figure}[ht]
\vskip 0.cm\relax\noindent\hskip 0.cm
       \relax{\includegraphics{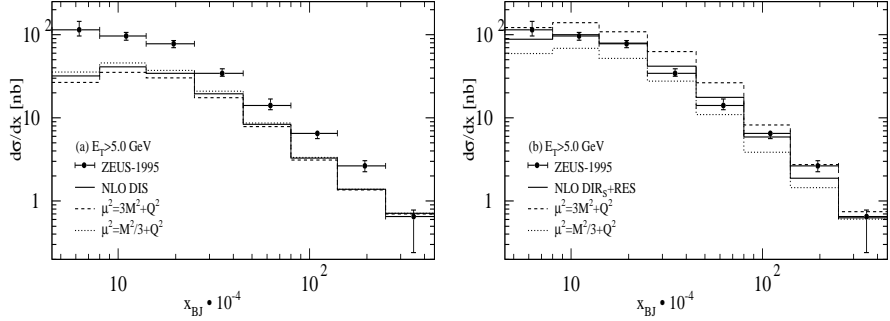}} 
\vspace{0.cm}
\caption{\small\sl Results of the NLO calculations of the forward 
jet cross section
for $E_T > 5$ GeV,
without (left) and with (right) resolved virtual photon interaction,
compared to the ZEUS data. 
Three different
renormalization scales are used: 
$\mu^2$ = $ M^2$ + $Q^2$ (solid line),
$\mu^2$ = $3 M^2$ + $Q^2$ (dashed line) and
$\mu^2$ = $M^2/3$ + $Q^2$ (dotted line), with $M^2=<E_T>^2$=50 GeV$^2$
(from \cite{9901314}).}
\label{fig:9901314_2}
\end{figure}

\newpage
In fig. \ref{fig:9901314_2} (taken from \cite{9901314}) the data 
are compared to the  NLO calculation \cite{9901314}, with
and without resolved virtual photon interaction (and with 
different renormalization scales).\\

\vspace*{3.5cm}
~\newline
Comment: {\it ``All models describe DIS regime.
In the BFKL regime ($Q^2\approx E_T^2$) ARIADNE reproduces the 
measured distributions well. Only model with resolved photon component, 
RAPGAP, describes data everywhere.''}

\newpage
\section{ Leptonic structure functions of the photon \label{sec4}}
The lepton pair production by two photons in $e^+$$e^-$
scattering, $e^+e^-\ra e^+e^- l^+l^-$, is  one of the basic test of 
QED\footnote{See ref. \cite{courau} for the 
review of the early experiments.}. 
It serves  also as  a test of experimental 
methods applied to more complicated two-photon
reactions involving hadrons, especially to the process 
$e^+$$e^-\rightarrow hadrons$, where the hadronic 
structure functions of the photon are measured.
To study the reliability of unfolding and tagging applied in these 
measurements\footnote{See discussion in secs. \ref{sec22} and \ref{sec23}}, 
including the effect of the (small) virtuality $P^2$
of the quasi-real target photon,
the lepton pair productions can be used as an ideal tool.

Single-tag events with lepton pair in the final state can
be described in the unpolarized $e^+e^-$ collision in terms of 
{\underbar{leptonic structure functions}}, $F_1^{\gamma (QED)}$ (=$F_T^{\gamma (QED)}$), 
$F_2^{\gamma (QED)}$ or $F_L^{\gamma (QED)}$,
in analogy to the hadronic case.
The leptonic structure functions were measured already in early 
eighties\footnote{The earliest measurements are summed up in 
\cite{fraz} (see also \cite{rev}j).}.
Although the production of final states with electrons, muons and
even taons have been measured, only the muonic case allows for the
extraction of the structure function from the data. 
Also nowadays experiments at LEP measure only muonic structure
functions.
The invariant mass of the $\mu^+ \mu^-$ pair, and hence 
$x_{Bj}$, can be determined very accurately and the measurements
of the muonic structure function $F_2^{\gamma}$, denoted below
as $F_2^{\gamma (QED)}$, are only statistically limited 
in contrast to the hadronic $F_2^{\gamma}$.
Recently also two other (muonic) structure functions, 
$F_A^{\gamma (QED)}$ and $F_B^{\gamma (QED)}$,
have been measured at LEP.

Recently, for the first time, a measurement of the leptonic final state
 mediated 
by two highly virtual photons (double-tag events) was performed at LEP.
For both photons with high virtualities the presence of {\sl large 
interference} 
terms was established ({\bf OPAL 99b}) even for events with very different 
virtualities $Q^2\gg P^2$. Thus in this case the (effective) leptonic 
structure function for a virtual photon 
cannot be introduced, 
nevertheless we report these results together with leptonic structure
functions of the real photon.

\subsection{Theoretical description \label{sec41}}

The QED structure functions for photon:
$F_2^{\gamma (QED)}$, $F_1^{\gamma (QED)}$, $F_L^{\gamma (QED)}$,
etc., unlike in the hadronic case, 
can be reliably calculated in QED.
Moreover all the particles in the 
final state can, in principle,  be directly observed.

The $F_L^{\gamma (QED)}$ structure function is much harder to
measure than $F_2^{\gamma (QED)}$, because its contribution to the 
relevant cross section is  
weighted by the small factor $y^2$ (as for hadronic
$F_L^{\gamma}$, see eq.~(\ref{si2})).
However, this longitudinal structure function
is not the only structure function that
contains additional information about the structure of the photon. 
It has been shown that
there are {\underbar{azimuthal correlations}} in the final state particles
from two-photon collisions which are sensitive to
additional structure functions \cite{aret,aure}. 
Below we discuss some of them for the real photon.

Beside  the cross section
${d\sigma^{e\gamma\ra eX}\over dx_{Bj}dy}$
(eqs.~(\ref{si}), (\ref{si2})), one can also measure in the two-body 
$\gamma^*\gamma$ process  one 
final state particle $a$. Then additional structure functions 
$\tilde {F}_A^{\gamma}$ (resulting from interference between the
 amplitudes involving both 
 longitudinal and transverse photons)   and $\tilde{F}_B^{\gamma}$ 
(describing interference between amplitudes for transverse photons only)
 appear \cite{aure}:
\begin{eqnarray}
\nonumber {d\sigma (e\gamma\ra eaX)\over dx_{Bj}dyd\Omega_a/4\pi}=
{2\pi\alpha^2\over Q^2}{1+(1-y)^2\over x_{Bj}y}
[(2x_{Bj}\tilde{F}_T^{\gamma}+\epsilon (y)\tilde{F}_L^{\gamma})\\
-\rho (y)\tilde{F}_A^{\gamma}\cos 
\phi_a+{1\over 2}\epsilon (y)\tilde{F}_B^{\gamma}\cos 2\phi_a]. 
\label{ea}
\end{eqnarray}
~\newline
Here $\Omega_a$ describes the solid angle for  the  particle $a$ in the
$\gamma^*\gamma$ centre of mass frame, and $\phi_a$ is its 
azimuthal angle around the $\gamma^*\gamma$ axis, relative to
the electron (tag) plane. The corresponding polar angle 
relative to the $\gamma^*\gamma$ collision axis is denoted by $\theta$. 
The functions $\epsilon (y)=2(1-y)/(1+(1-y)^2)$ and
$\rho (y)=(2-y)\sqrt{1-y}/(1+(1-y)^2)$ are  close to 1, 
as they are $\sim (1-{\cal {O}}(y^2))$. The standard functions
$F_T^{\gamma}$ and $F_L^{\gamma}$ (and also $F_{A,B}^{\gamma}$) are 
obtained from the corresponding 
$\tilde{F}_{T,L}^{\gamma}$  ($\tilde{F}_{A,B}^{\gamma}$)
by integration over the solid angle $\Omega_a$. Note that the
formula (\ref{ea}) holds for two leptons (or two partons)
produced in the final state.

The expression for $F_T^{\gamma (QED)}$ and
$F_L^{\gamma (QED)}$ (or $F_2^{\gamma (QED)}$ =
$F_L^{\gamma (QED)}$ + $2 x_{Bj} F_T^{\gamma (QED)}$)
in the lowest order QED is given by the Bethe-Heitler type formula
(see eq.~\ref{mh}).
The functions $F_A^{ \gamma (QED)}$ and $F_B^{ \gamma (QED)}$ 
calculated in the same approximation and with the full dependence on the muon mass
up to terms of order $m_{\mu}^2/W^2$ \cite{ns}, have the
following form:
\bea
F_A^{ \gamma (QED)}(x_{Bj},\beta)=
{{4 \alpha}\over{\pi}} \xb \sqrt{\xb (1-\xb)}
(1-2\xb)\{ \beta[1+(1-\beta)^2{{1-\xb}\over{1-2\xb}}]\\ \nonumber
+{{3\xb-2}\over{1-2\xb}}\sqrt{1-\beta^2}\arccos(\sqrt{1-\beta^2})\},\\
F_B^{ \gamma (QED)}(\xb,\beta)={{4 \alpha}\over{ \pi}}\xb^2(1-\xb)
\{ \beta[1-(1-\beta^2){{1-\xb }\over{2\xb}}]\\  \nonumber
+{{1\over2}}(1-\beta^2)[{{1-2\xb}\over{ \xb}}
-{{1-\xb}\over{2\xb}}(1-\beta^2)]\log({{1+\beta}\over{1-\beta}})\},
\eea
where $$\beta=\sqrt{1-{{4m_{\mu}^2}\over{W^2}}}.$$

The collection of other useful formulae for the above structure 
functions in the lowest order QED approximation with the full 
dependence on the fermion mass can be  found in
\cite{ns}, see also \cite{pez} for the first calculation
in the lowest order QED.

The function $F_B^{\gamma (QED)}$ is, 
in a specific logarithmic approximation \cite{brew} and zero muon
mass limit, {\sl accidentally} equal  to $F_L^{\gamma (QED)}$, 
although it involves quite 
different photon helicity structures. Thus extracting 
$F_B^{\gamma (QED)}$ can give us indirectly
information on $F_L^{\gamma (QED)}$ (in this approximation).

The above description concerns only the leptonic production
in $\gamma^*\gamma$ collisions, so with one (almost) real photon.
Similar consideration can be performed for $\gamma^*\gamma^*$ collisions,
when the notion of leptonic structure functions of the virtual photon  
can in principle be applied if $Q^2\gg P^2$, in full analogy to
hadronic structure functions for the virtual photon (sec. \ref{sec3}).
The extraction of such functions from the corresponding cross section,
\eg $F_2^{\gamma^* (QED)}$, is feasible if the
interference terms can be neglected.

\subsection{Measurements of single and double-tag events \label{sec42}}

A measurement of single-tag events allows to extract
$F_2^{ \gamma (QED)}$, as well as $F_A^{ \gamma (QED)}$ and
$F_B^{ \gamma (QED)}$.
The function $\tilde{F}^{\gamma (QED)}_A$ is antisymmetric in $\cos \theta^*$
\footnote{Often variables referring to the $\gamma^*\gamma$
centre of mass frame are denoted by adding a star}.
To extract it from the data, the $x_{Bj}$-dependence of the term 
in the differential cross 
section proportional to $F^{\gamma (QED)}_A$ has to be studied separately 
for the ranges $\cos \theta^* <0$ and $\cos \theta^* >0$
in order to avoid cancellation.

In the last few years measurements of the leptonic structure function 
$F_2^{\gamma (QED)}$ have been performed at LEP 1 by the 
four collaborations; 
the ratios of  structure functions $F_{A,B}^{\gamma (QED)}$ to $F_2^{\gamma (QED)}$ 
have also been measured by all the  groups. Only the L3 and OPAL groups
provided also the results for individual structure functions 
$F_{A}^{\gamma (QED)}$ and $F_{B}^{\gamma (QED)}$.

The Vermaseren, Galuga and BDK Monte Carlo generators, which have
implemented full dependence on both photons' virtualities, 
were used in analyses (for background simulation also
other generators were used).

Results of  all $F_2^{\gamma (QED)}$ measurements 
discussed here will be shown together in figs.~\ref{fig:chap6_02}
and \ref{fig:chap6_03}. A similar compilation is also presented 
for the ratios $F_{A, B}^{\gamma (QED)}/F_2^{\gamma (QED)}$
(fig. \ref{fig:chap6_05}) 
and for individual functions $F_A^{\gamma (QED)}$, $F_B^{\gamma (QED)}$
(fig. \ref{fig:chap6_06}).

\begin{figure}[hb]
\vskip 0.in\relax\noindent\hskip 0.cm
       \relax{\includegraphics{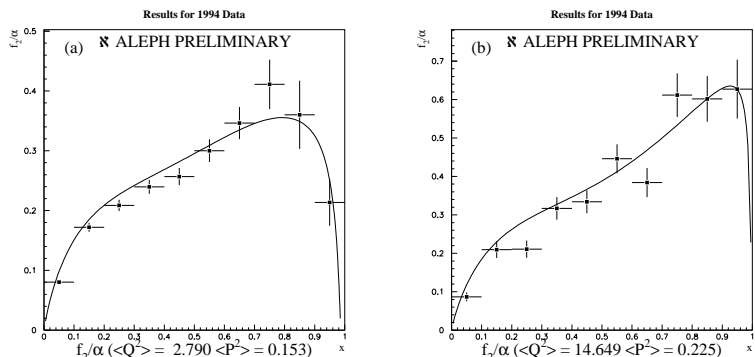}}
\vspace{0cm}
\caption{\small\sl The structure functions $F_2^{\gamma (QED)}$/$\alpha$ 
as a function $x_{Bj}$ for
a) $<Q^2>$=2.79 GeV$^2$, $<P^2>$=0.153 GeV$^2$ and  
b) $<Q^2>$=14.65 GeV$^2$, $<P^2>$=0.225 GeV$^2$.
The solid lines are the QED expectations (from \cite{brew}).}
\label{fig:brew1}
\end{figure}

The relevant cross section for the double-tag events,
\ie for the production of the leptonic ($\mu^+\mu^-$) final state
in $\gamma^*\gamma^*$ collision,
is as for the hadronic final state (eq. \ref{ttll}),
with the same definition of kinematical variables.
Note that for these events
 the additional angle  $\bar \phi$ 
between two scattering planes for the 
two scattered electrons appears.
~\newline\newline
\centerline{\bf \huge DATA}
\newline\newline
$\bullet${\bf ALEPH 97b,conf \cite{brew} (LEP 1)}
\newline
$F_2^{\gamma (QED)}$ has been measured (data from 1994) for
two samples: 0.6$<Q^2<$6.3 GeV$^2$ ($<Q^2>$=2.8 GeV$^2$) with $<P^2>$
equal to 0.15 and 3.0$<Q^2<$60.0 GeV$^2$ ($<Q^2>$= 14.6 GeV$^2$) with
$<P^2>$ = 0.22 GeV$^2$. Results for these two samples are shown 
in fig.~\ref{fig:brew1}.\\
\vspace*{4.2cm}

\begin{figure}[ht]
\vskip 0.in\relax\noindent\hskip 0.cm
       \relax{\includegraphics{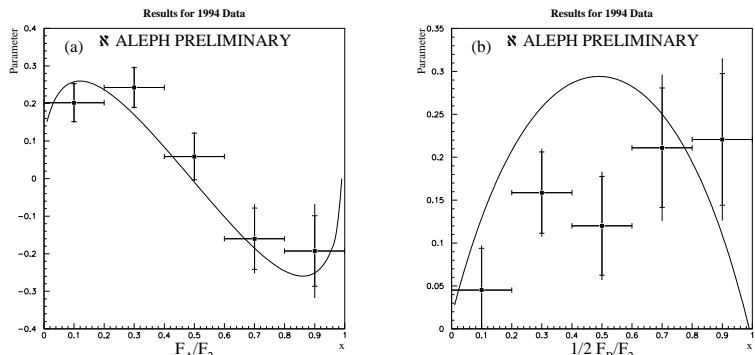}}
\vspace{0.ex}
\caption{\small\sl Results for a) $F_A^{\gamma (QED)}/F_2^{\gamma (QED)}$
and b) ${1\over 2}F_B^{\gamma (QED)}/F_2^{\gamma (QED)}$
as a function $x_{Bj}$ at $<Q^2>$=8.8 GeV$^2$.
The solid lines are the QED expectations (from \cite{brew}).}
\label{fig:brew2}
\end{figure}

In this experiment also the azimuthal angle distributions have been 
measured and ratios $F_A^{\gamma (QED)}/F_2^{\gamma (QED)}$ and 
$F_B^{\gamma (QED)}/F_2^{\gamma (QED)}$ extracted 
(fig.~\ref{fig:brew2}).

\vspace*{4.5cm}

~\newline
$\bullet${\bf DELPHI 96a \cite{delphi2} (LEP 1)}
\newline
$F_2^{\gamma (QED)}$ has been measured at $<Q^2>$=12 GeV$^2$,
as a test for the unfolding and tagging methods in extraction 
of the hadronic $F_2^{\gamma}$ (see sec. \ref{sec222}). 
The effect of nonzero
target virtuality has been studied. 
 A satisfactory fit to the
measured $F_2^{\gamma (QED)}$ is obtained for the fixed value
of $P^2$=0.04 GeV$^2$.

~\newline\newline
$\bullet${\bf DELPHI 99,conf \cite{da-silva} (LEP 1)}
\newline
The single-tag events with the muon pairs (data from 1992-1995) 
for $Q^2$ between 2.5 and 750 GeV$^2$
 are used to extract the leptonic photon structure functions. 
The SAT (1991-1993) and the STIC (after 1994) samples were 
studied corresponding to $<Q^2>$=12.5 GeV$^2$, and the FEMC sample
for $<Q^2>$=120 GeV$^2$.

\begin{figure}[ht]
\vskip 0.in\relax\noindent\hskip 0.cm
       \relax{\includegraphics{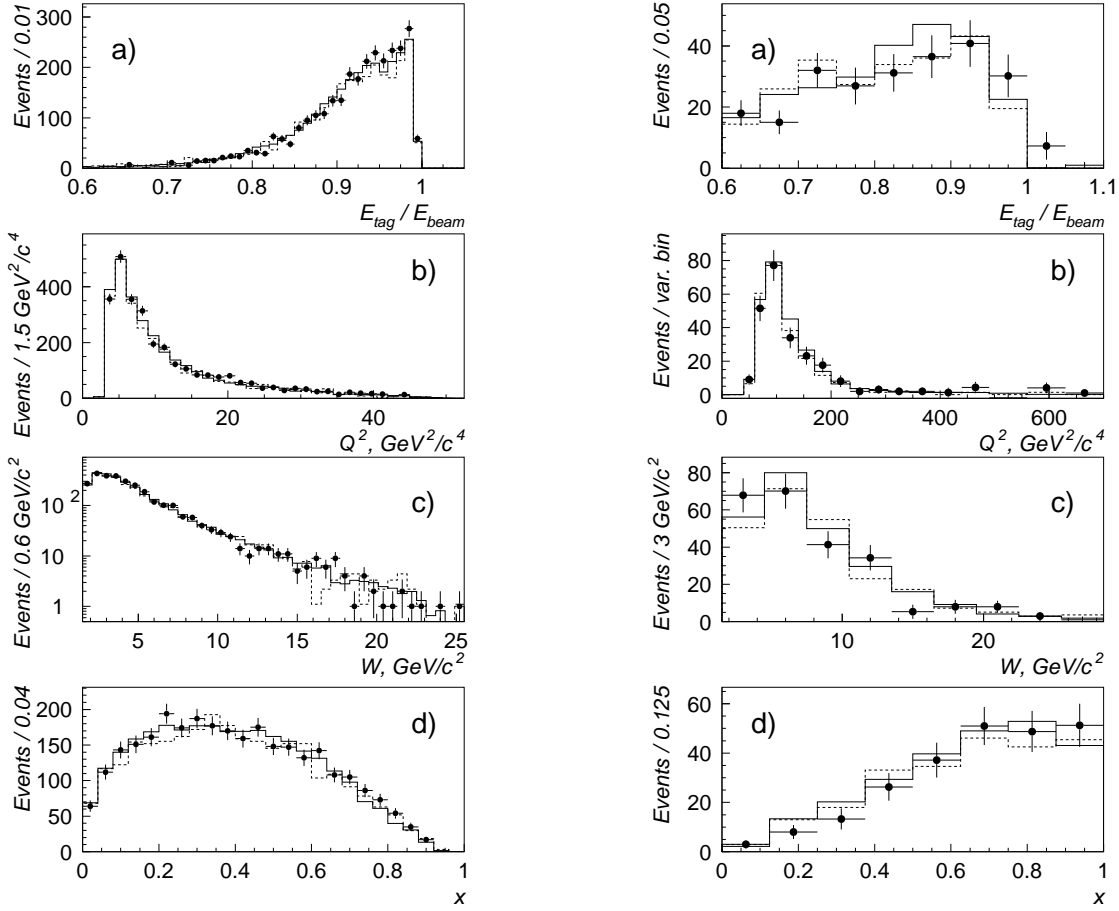}}
\vspace{0cm}
\caption{\small\sl
Left: Distributions for STIC sample: a) $E_{tag}/E_{beam}$, b) $Q^2$, c) $W$,
d) $x_{Bj}$, and a comparison with BDKRC and DIAG36 generators.
Right: The same for the FEMC sample
(from \cite{da-silva}).}
\label{fig:da-silva_23}
\end{figure}

 The effect of nonzero
target virtuality has been studied, a fit to $F_2^{\gamma}$ data
gives for SAT (STIC) events $P^2=0.032\pm0.007$ ($0.022\pm0.007$) GeV$^2$. 
The results for various distributions for SAT and FEMC samples
are presented in fig.~\ref{fig:da-silva_23}.
\newpage
\vspace*{11cm}

\begin{figure}[ht]
\vskip 0.in\relax\noindent\hskip 0.cm
       \relax{\includegraphics{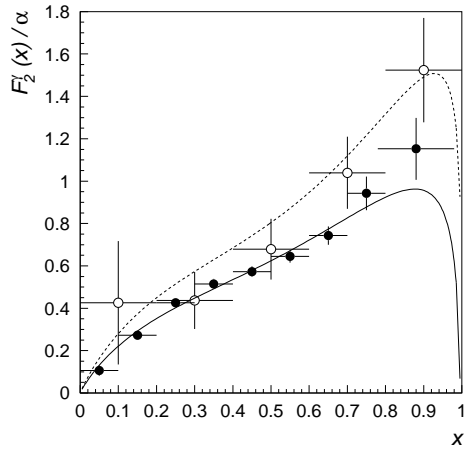}}
\vspace{0cm}
\caption{\small\sl
The results for $F_2^{\gamma (QED)}/\alpha$ as a function of $x_{Bj}$
for the SAT and STIC combined data ($<Q^2>$=12.5 GeV$^2$)
and for FEMC sample
($<Q^2>$=120 GeV$^2$), together with QED predictions
(from \cite{da-silva}).}
\label{fig:da-silva_4}
\end{figure}

The results for $F_2^{\gamma(QED)}/\alpha$ as a function of $x_{Bj}$
for the  combined SAT and STIC data ($<Q^2>$=12.5 GeV$^2$)
and for FEMC sample
($<Q^2>$=120 GeV$^2$), together with QED predictions are shown in fig.
\ref{fig:da-silva_4}.

\vspace*{5.7cm}
\begin{figure}[hb]
\vskip 0.in\relax\noindent\hskip 0.cm
       \relax{\includegraphics{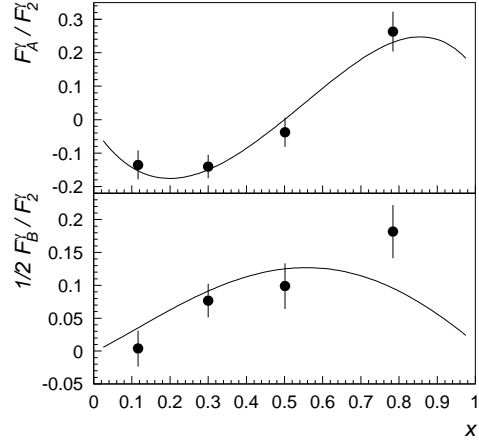}}
\vspace{0cm}
\caption{\small\sl
The results for 
$F_A^{\gamma(QED)}/F_2^{\gamma(QED)}$ (top) and 
$F_B^{\gamma(QED)}/2 ~F_2^{\gamma(QED)}$ (bottom)
as a function of $x_{Bj}$. The QED predictions are also shown
(from \cite{da-silva}).}
\label{fig:da-silva_6}
\end{figure}

\newpage
Azimuthal correlations have been  studied to derive 
$F^{\gamma (QED)}_A$ and $F^{\gamma(QED)}_B$. 
In order to increase the observed correlation only events with 
20$^{\circ}<\theta^*<160^{\circ}$ were taken into account. 
The azimuthal angle distributions (not shown) were fitted to obtain the factors multiplying
the $\cos\phi_a$ and  $\cos2\phi_a$ contributions. Next the extrapolation in these factors was made and the final results for ratios  
$F_A^{\gamma(QED)}/F_2^{\gamma(QED)}$ and 
$F_B^{\gamma(QED)}/F_2^{\gamma(QED)}$ obtained,
see fig.~\ref{fig:da-silva_6}.

\vspace*{6cm}
 
~\newline
$\bullet${\bf L3 98b \cite{l3qed_98} (LEP 1)}\\
The production of $e^+e^-$ and  $\mu^+\mu^-$
pairs for 1.4 GeV$^2 \leq Q^2 \leq 7.6$ GeV$^2$ was studied,
based on the 1991-1994 run. For a muonic final state, data on  
$F_2^{\gamma (QED)}$,
and on the ratios $F_A^{\gamma (QED)}/F_2^{\gamma (QED)}$  and
${1\over 2}F_B^{\gamma (QED)}/F_2^{\gamma (QED)}$, as well as   
on $F_A^{\gamma (QED)}$ and $F_B^{\gamma (QED)}$
were obtained.

\begin{figure}[ht]
\vskip -0.3cm\relax\noindent\hskip 4.2cm
       \relax{\includegraphics{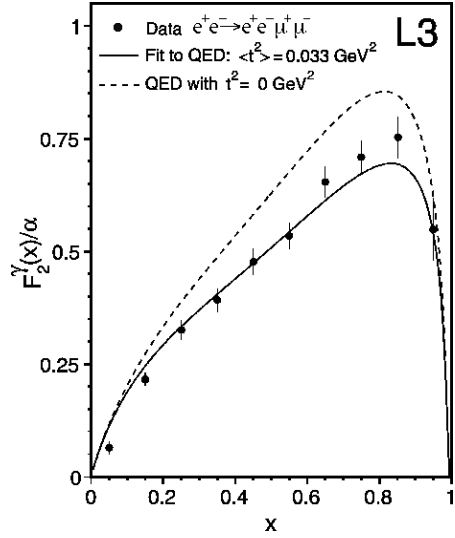}}
\vspace{0.3cm}
\caption{\small\sl $F_2^{\gamma (QED)}(x_{Bj},Q^2,P^2)/\alpha $
for $1.4 < Q^2 < 7.6 $ GeV$^2$ as a function of $x_{Bj}$
compared to QED predictions at $P^2$=0 (dashed line)
and at $<P^2>_{fit}=0.033$ GeV$^2$ (solid line)
(from \cite{l3qed_98}).}
\label{fig:tnis_6c}
\end{figure}

The effect of the photon target virtuality has been studied 
and is clearly seen in all the three structure functions (see
fig.~\ref{fig:tnis_6c} for $F_2^{\gamma (QED)}$ only).\\
\newpage
\vspace*{6.5cm}
\vspace*{6.2cm}
\begin{figure}[ht]
\vskip 0.in\relax\noindent\hskip 4.cm
       \relax{\includegraphics{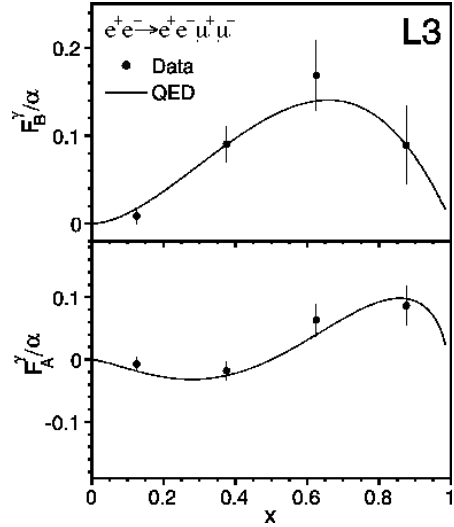}}
\vspace{0.cm}
\caption{\small\sl Results for 
$F_B^{\gamma (QED)}/\alpha$ and $F_A^{\gamma (QED)}/\alpha$ 
measured at 1.4 $<Q^2<$ 7.6 GeV $^2$. The curves
are the QED predictions for $P^2$=0 (from \cite{l3qed_98}).}
\label{fig:tnis_8a}
\end{figure}
~\newline
$\bullet${\bf OPAL 93 \cite{opal} (LEP 1)}
~\newline 
The QED structure function $F_2^{\gamma (QED)}$ for $<Q^2>$=8.0 GeV$^2$
was extracted from single-tag events at CM energy $\sim M_Z$, for the 
first time at LEP.

~\newline 
$\bullet${\bf OPAL 97f \cite{opalqed} (LEP 1)}
\newline
Here the extraction of $F_B^{\gamma (QED)}$ was performed for
0.85$<Q^2<$31 GeV$^2$ ($<Q^2>$ =5.2 GeV$^2$).

~\newline
$\bullet${\bf OPAL 99b \cite{opal99qed} {\bf (LEP 1)}}
\newline
The complete OPAL dataset taken in years 1990-95 at $e^+e^-$ CM
energies close to the mass of the $Z$ boson has been used to
determine the QED structure functions $F_2^{\gamma (QED)}$, 
$F_A^{\gamma (QED)}$ and $F_B^{\gamma (QED)}$ of quasi-real photon
(\ie single tagged events, with $y <$ 0.5 and $P^2 <$ 1.4 GeV$^2$, with
$<P^2>$=0.05 GeV$^2$) and, for the first time, the differential
cross section ${{d\sigma}\over{d\xb}}$ for highly virtual photons with 
$y <$ 0.5 and 1.5 $<P^2<$ 20 GeV$^2$.
For single-tag events the different  samples 
(SW, FD, EE) were collected, depending on the $Q^2$ range (from 1.5
till 30 GeV$^2$) and the calorimeter
used to detect the scattered electron. The double-tag sample is
denoted by DB and corresponds to $Q^2$ between 1.5 and 300 GeV$^2$.

\begin{figure}[hb]
\vskip 0.in\relax\noindent\hskip 0.cm
       \relax{\includegraphics{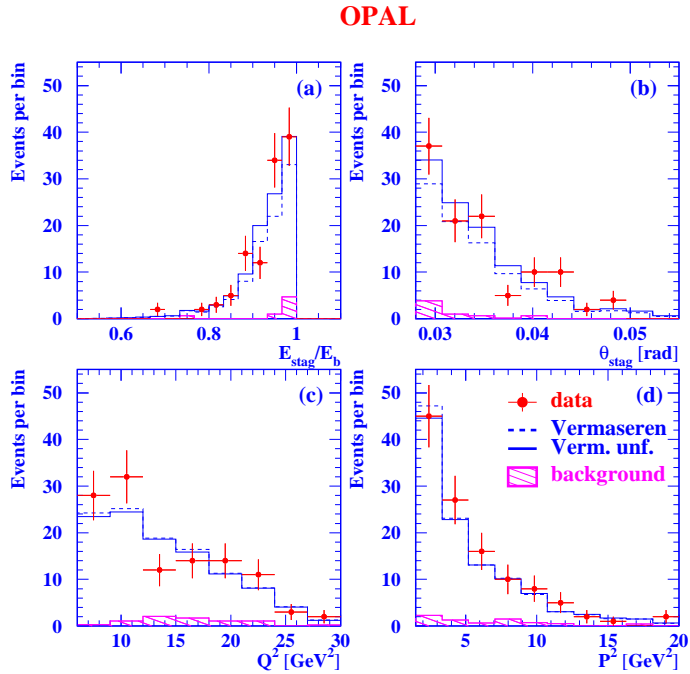}}
\vspace{0.cm}
\caption{\small\sl
Various distributions for  double-tag events
(from \cite{opal99qed}).}
\label{fig:pr271_4}
\end{figure}

The various distributions for the single and double-tag events were measured
and compared with MC (Vermaseren generator) predictions. In 
fig.~\ref{fig:pr271_4}
the results for DB sample are presented. Here the energy and the polar angle 
of the second electron, as well as the $Q^2$ and $P^2$
distributions, are shown.
\vspace*{8.5cm}

\begin{figure}[hb]
\vskip 0.in\relax\noindent\hskip 0.cm
       \relax{\includegraphics{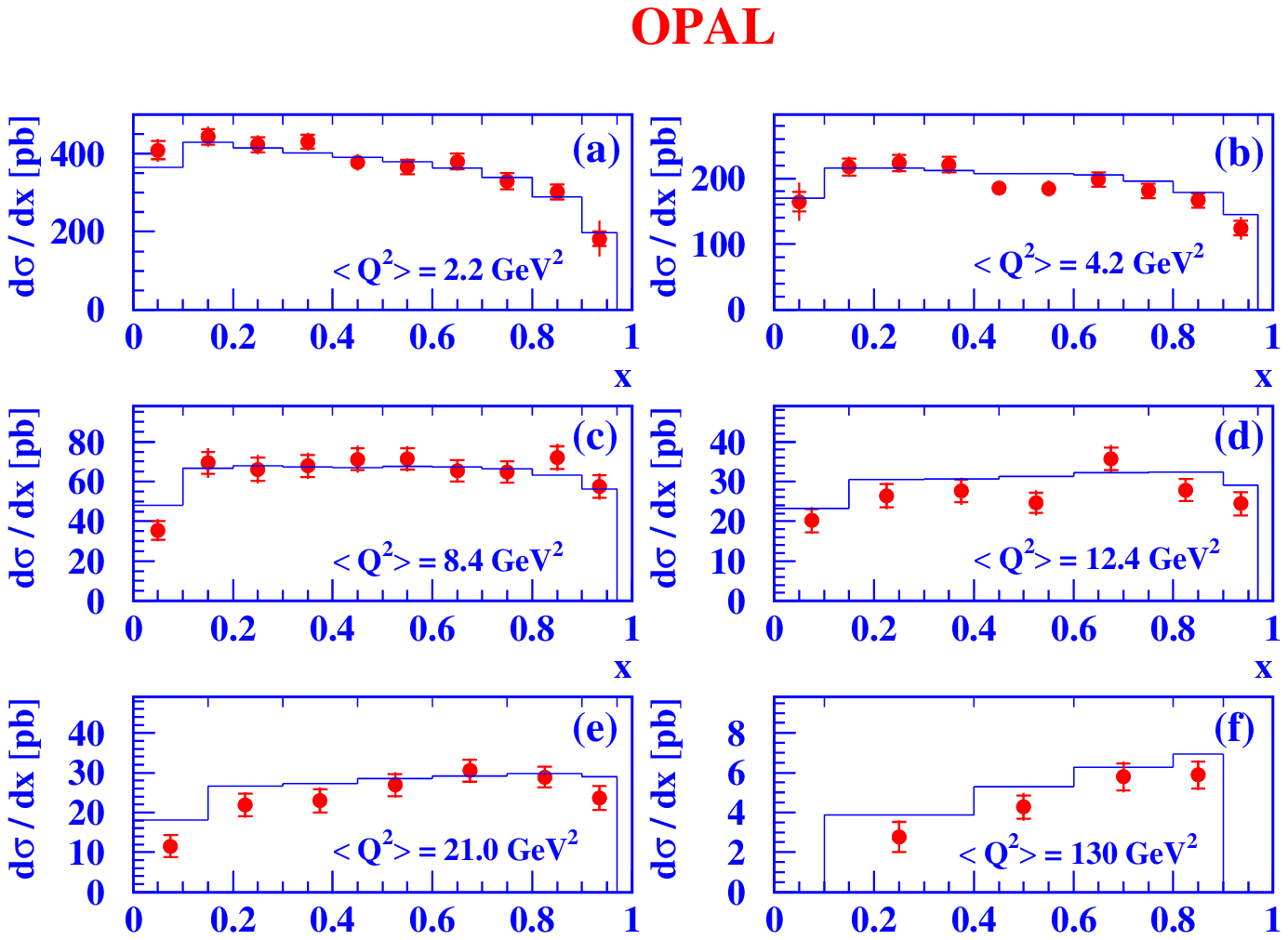}}
\vspace{0.cm}
\caption{\small\sl Differential cross section $d\sigma /dx_{Bj}$ for
various $<Q^2>$ samples of singly tagged events. The full line is 
the differential cross section as predicted by QED within Vermaseren 
Monte Carlo (from \cite{opal99qed}).}
\label{fig:opal99_7}
\end{figure}

\begin{figure}[hb]
\vskip 0.in\relax\noindent\hskip 0.cm
       \relax{\includegraphics{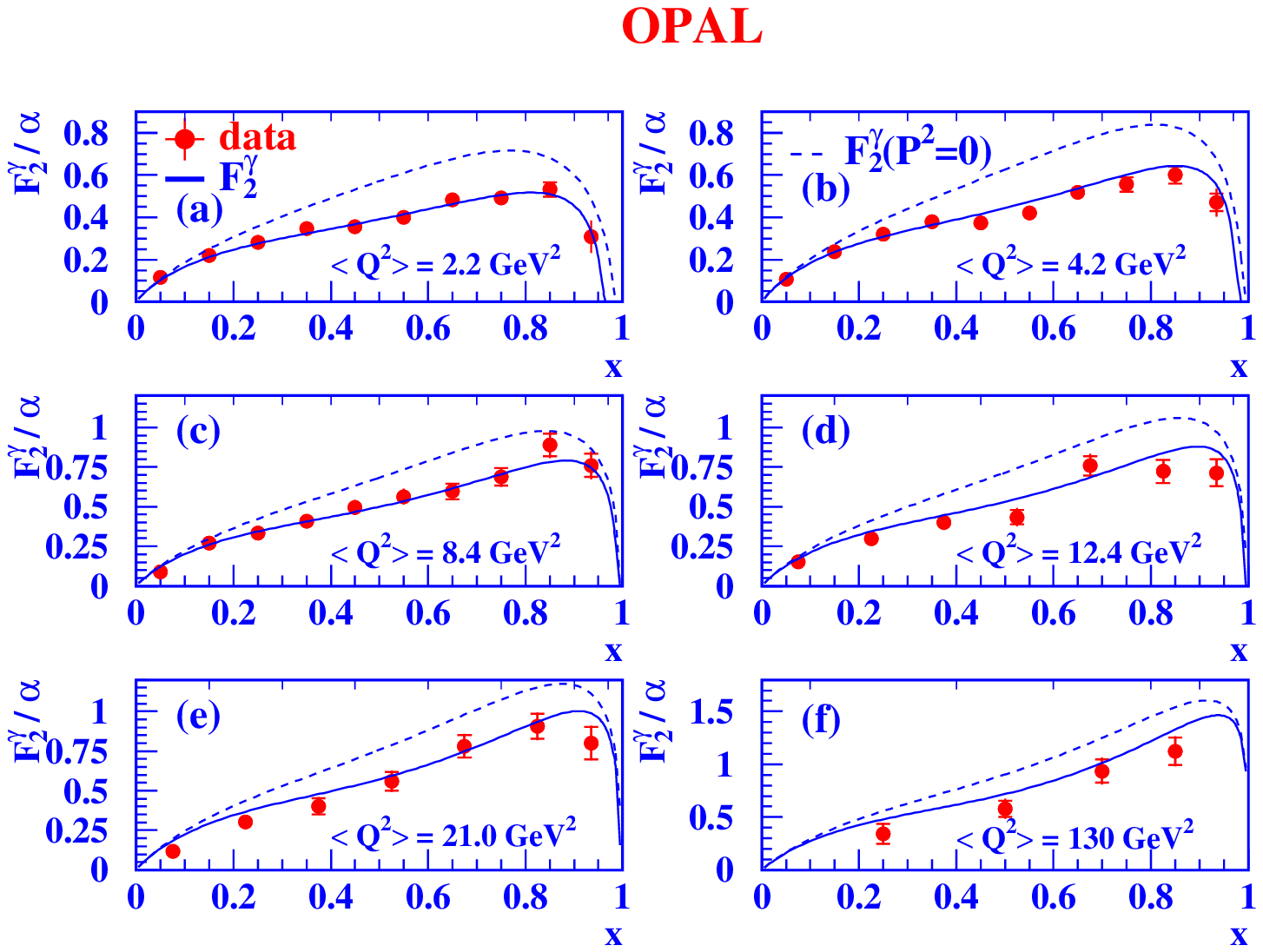}}
\vspace{0.cm}
\caption{\small\sl Structure functions $F_2^{\gamma (QED)}/\alpha$
for various $<Q^2>$ samples of singly tagged events. The full line is
the QED prediction for $<P^2> = 0.05$ GeV$^2$, the dashed line - 
for $<P^2> = 0$ (from \cite{opal99qed}).}
\label{fig:opal99_9}
\end{figure}

For single tagged events (with $<P^2> = 0.05 $ GeV$^2$) 
the differential cross section 
${{d\sigma}\over{d\xb}}$ have 
been measured and the $F_2^{\gamma (QED)}$ was extracted.
Results for various samples are shown in fig.~\ref{fig:opal99_7}
(${{d\sigma}\over{d\xb}}$) and \ref{fig:opal99_9} ($F_2^{\gamma (QED)}$),
together with the corresponding QED predictions.
\newpage
\vspace*{9.5cm}

\newpage
\vspace*{10.5cm}

For the first time the differential cross section 
has been measured for two highly virtual photons
and the importance of the interference terms
$\tau_{TT}$ and $\tau_{TL}$ (see eq. \ref{ttll}) has been shown.

\begin{figure}[ht]
\vskip 0.in\relax\noindent\hskip 0.cm
       \relax{\includegraphics{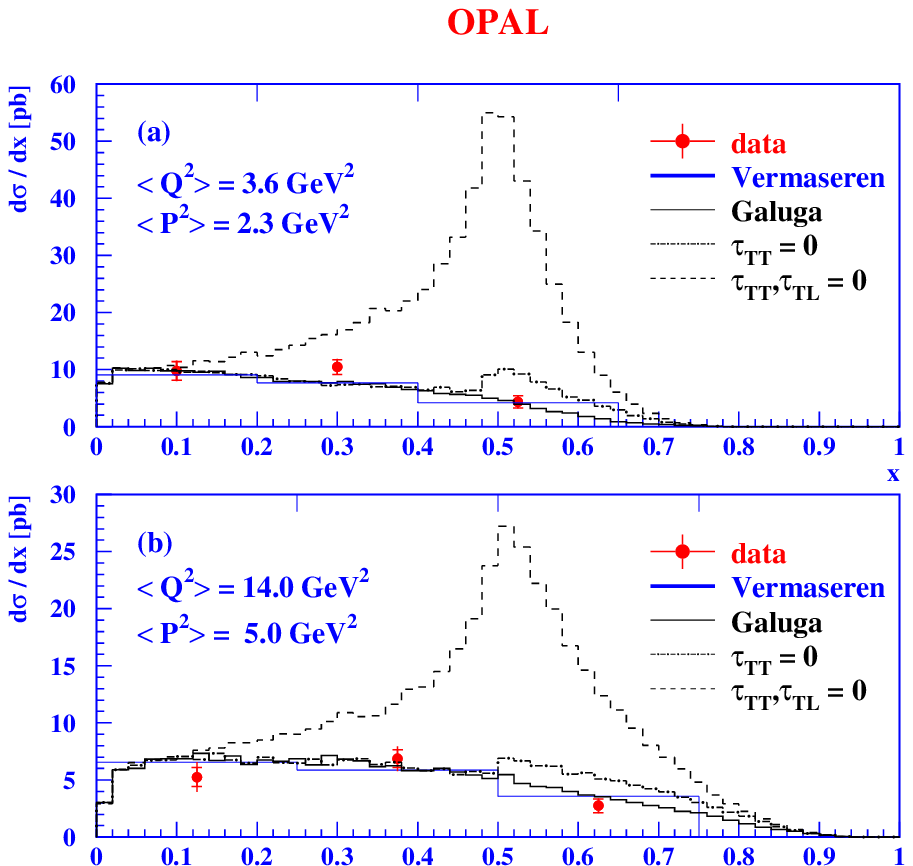}}
\vspace{0.cm}
\caption{\small\sl Differential cross section for $\mu^+\mu^-$
production with two virtual photons, for two $<Q^2>$, $<P^2>$ samples,
as a function of $x_{Bj}$.
The full line is the differential cross section predicted by the
Vermaseren Monte Carlo. The additional  three histograms 
represent the cross section calculations with the GALUGA Monte
Carlo for the full cross section (solid line), cross section obtained
for $\tau_{TT}=0$ (dot-dashed line) and that for $\tau_{TT}$ =
$\tau_{TL}=0$ (dashed line) (from \cite{opal99qed}).}
\label{fig:opal99_11}
\end{figure}

If, after integration of the differential cross section
(eq. \ref{ttll}) over $\bar {\phi}$, the contributions of
the terms  $\tau_{TT}$ and $\tau_{TL}$ vanish, 
one can introduce an effective structure function of virtual
photon (see sec. \ref{sec322}, {\bf PLUTO 84} and {\bf L3 2000}). 
This is {\sl not} the case
in the present measurement,
as can be seen in fig.~\ref{fig:opal99_11}
where Monte Carlo predictions 
also for the options with  $\tau_{TT}$=0 and $\tau_{TT}$ =
$\tau_{TL}$=0 are displayed and compared with the data.

\begin{figure}[ht]
\vskip 0.in\relax\noindent\hskip 0.cm
       \relax{\includegraphics{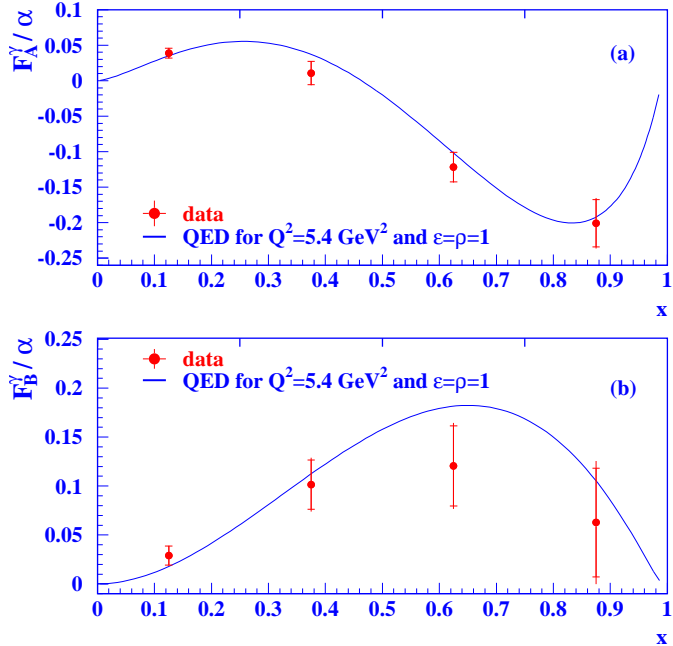}}
\vspace{0cm}
\caption{\small\sl The measured structure functions 
$F_A^{\gamma (QED)}/\alpha$ and $F_B^{\gamma (QED)}/\alpha$
as a function of $x_{Bj}$ for $<Q^2>$ = 5.4 GeV$^2$. The solid lines are 
the QED predictions for $Q^2=5.4$ GeV$^2$ and $\epsilon = \rho =1$
(from \cite{opal99qed}).}
\label{fig:opal99_16}
\end{figure}

\begin{figure}[ht]
\vskip 0.in\relax\noindent\hskip 1.3cm
       \relax{\includegraphics{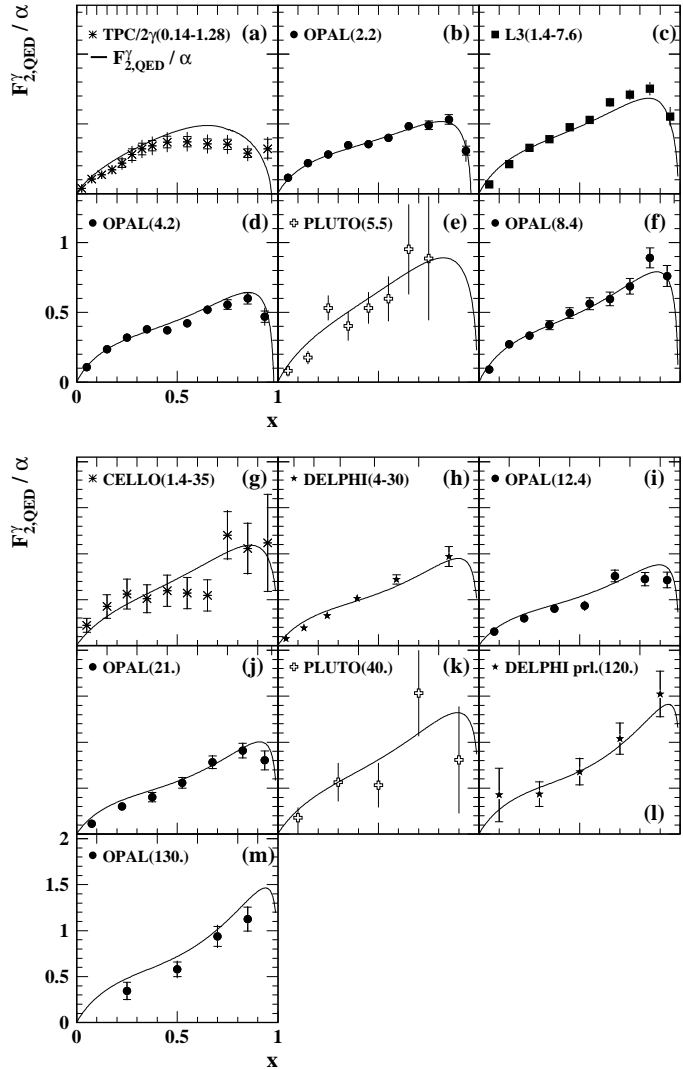}}
\vspace{0.3cm}
\caption{\small\sl Summary of existing $F_2^{\gamma (QED)}/\alpha$
data as a function of $x_{Bj}$
for broad $<Q^2>$ range shown with QED predictions for $P^2$=0
(from \cite{rev}j).}
\label{fig:chap6_02}
\end{figure}

\begin{figure}[ht]
\vskip 0.in\relax\noindent\hskip 1.8cm
       \relax{\includegraphics{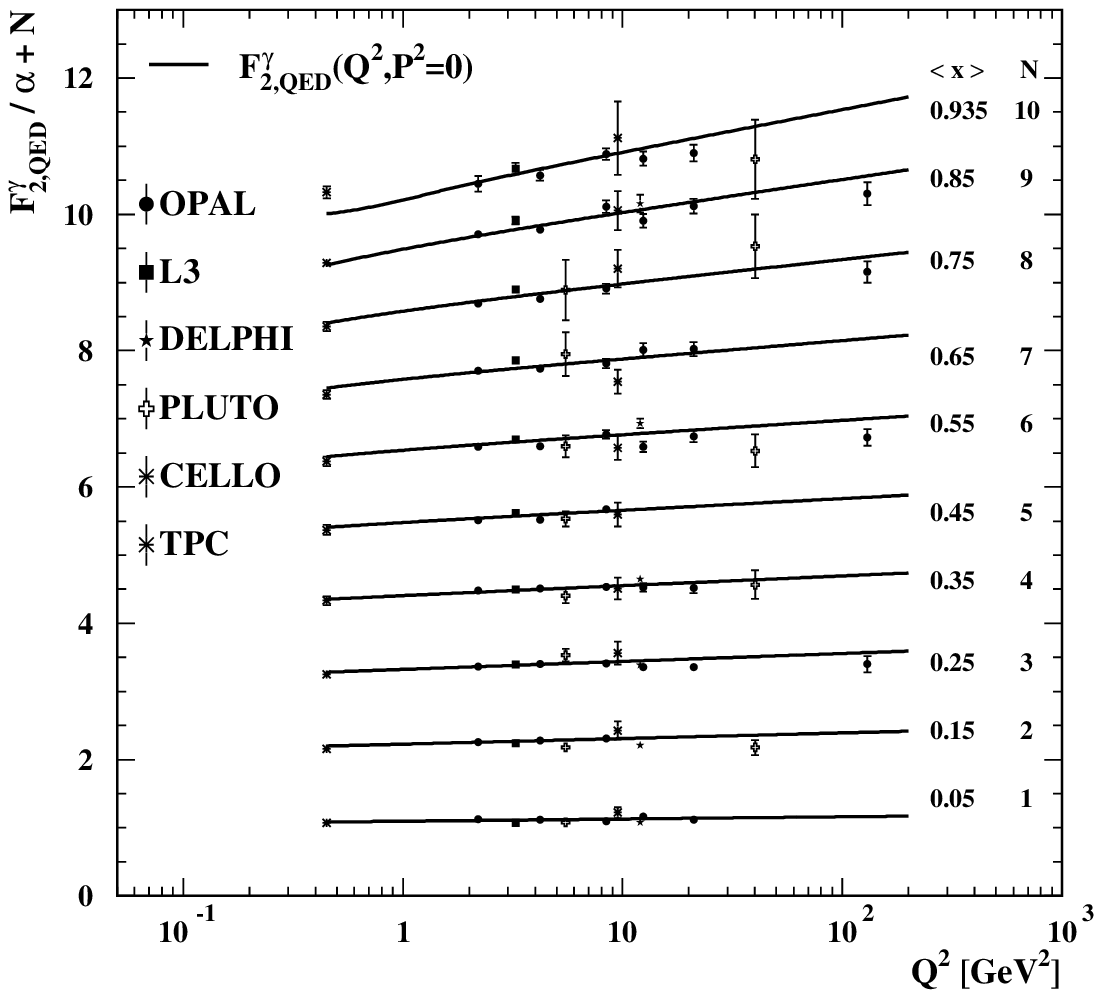}}
\vspace{-0.4cm}
\caption{\small\sl Summary of existing $F_2^{\gamma (QED)}/\alpha$
data as a function of $Q^2$
for fixed $<x_{Bj}>$ bins shown with QED predictions for $P^2$=0
(from \cite{rev}j).}
\label{fig:chap6_03}
\end{figure}

\begin{figure}[ht]
\vskip 0.in\relax\noindent\hskip 3cm
       \relax{\includegraphics{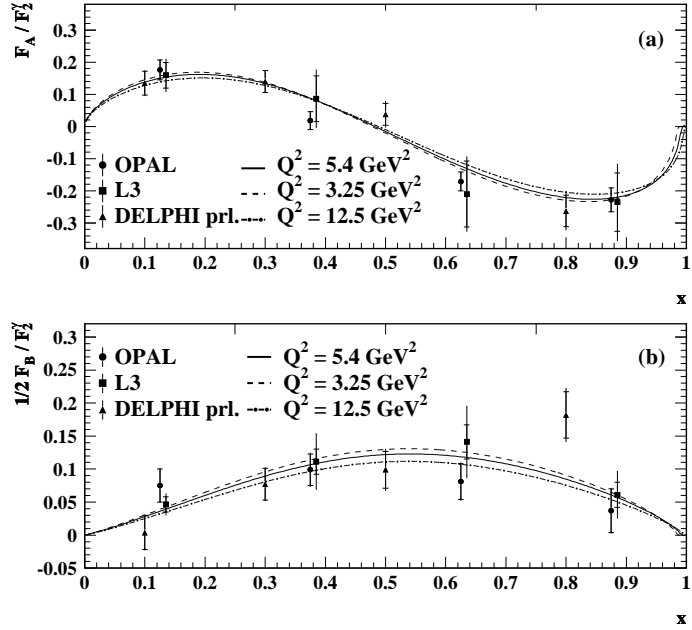}}
\vspace{-0.ex}
\caption{\small\sl Summary of existing data: a)
$F_A^{ (QED)}/F_2^{\gamma (QED)}$ and b)
$F_B^{ (QED)}/F_2^{\gamma (QED)}$ as a function of $x_{Bj}$
for $<Q^2>$ = 3.25, 5.4, 12.5 GeV$^2$, shown with QED predictions for $P^2$=0
(from \cite{rev}j).}
\label{fig:chap6_05}
\end{figure}

\begin{figure}[ht]
\vskip 0.in\relax\noindent\hskip 3cm
       \relax{\includegraphics{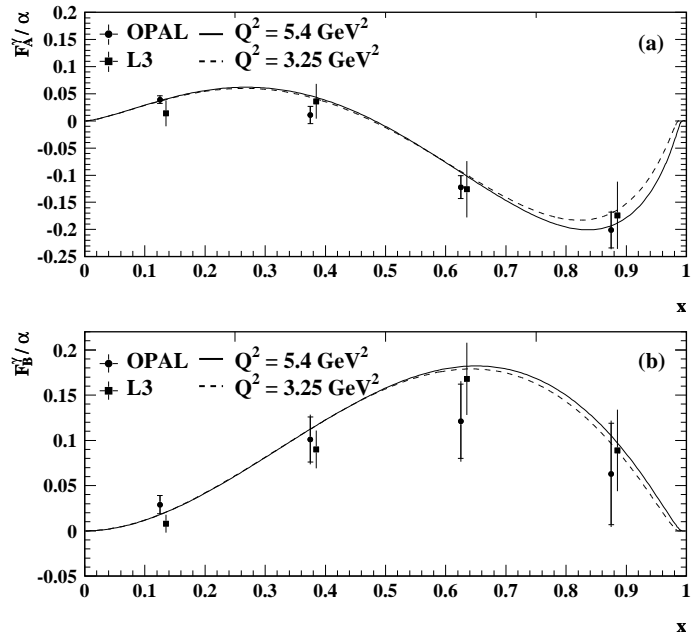}}
\vspace{-0.ex}
\caption{\small\sl Summary of existing data: a) $F_A^{\gamma (QED)}/\alpha$
and b) $F_B^{\gamma (QED)}/\alpha$ as a function of $x_{Bj}$.
Data for $<Q^2>$=3.25 and 5.4 GeV$^2$ 
are shown with corresponding QED predictions for $P^2$=0
(from \cite{rev}j).}
\label{fig:chap6_06}
\end{figure}

Based on azimuthal correlations, the QED structure functions 
$F_A^{\gamma (QED)}$ and $F_B^{\gamma (QED)}$ for quasi-real photons have
been determined for $<Q^2>$=5.4 GeV$^2$.
First, ratios $F_A^{ (QED)}/F_2^{\gamma (QED)}$ 
and $F_B^{ (QED)}/F_2^{\gamma (QED)}$
were obtained from a fit to the $\phi_a$ distribution.
Then from the
measurements of $F_2^{\gamma (QED)}$ also the 
individual structure functions
$F_A^{ \gamma (QED)}$ and
 $F_B^{\gamma (QED)}$
were calculated. 
The obtained results are shown in fig.~\ref{fig:opal99_16}.\\

\newpage
\vspace*{7.5cm}
\vspace*{8cm}
~\newline
Comment: {\it For the single-tag events leptonic structure
functions were extracted in the wide range
of $Q^2$. The effect of $P^2$ dependence was studied.

For the first time the leptonic final state was studied
for the deep inelastic scattering on the highly virtual photon.
Due to large interference terms there is no clear relation between 
structure functions and cross sections.

 The definition of $F_A^{\gamma (QED)}$ adopted here
differs by sign and a factor 2 from that used by {\bf L3 98b}.
It is worth noting that the analytical formulae used here with
the muon mass dependence, give noticeably different values from those with
zero mass.

``The measurements presented here supersede the earlier structure function
results'' ({\bf OPAL 93, 97f}).}

\vspace*{0.1cm}

\centerline{*****}
\vspace*{0.4cm}

For an overall comparison  we present below  collective figures 
(from ref. \cite{rev}j) of the
 QED structure functions for the real photon
together with corresponding QED predictions for
$P^2$ = 0. Data on $F_2^{\gamma (QED)}$ from old and new experiments
are presented in  fig.~\ref{fig:chap6_02} 
as a function of $x_{Bj}$, for $<Q^2>$ between 0.14 and 130 GeV$^2$,
and in fig.~\ref{fig:chap6_03} as a function of $Q^2$ 
for various $<x_{Bj}>$ bins. Also the summary of structure functions ratios 
$F_A^{ (QED)}/F_2^{\gamma (QED)}$  and $F_B^{ (QED)}/F_2^{\gamma (QED)}$ 
 and of the individual structure functions
$F_A^{ (QED)}$  and $F_B^{ (QED)}$ is shown
in fig.~\ref{fig:chap6_05} and ~\ref{fig:chap6_06}, respectively.  
\newpage 
\vspace*{13.5cm}
\newpage
\vspace*{12.3cm}
\newpage
\vspace*{8cm}
\vspace*{8.5cm}

\newpage
\section{Selected results on the heavy quark content of the photon\label{sec5}}
An additional  information on the ``structure'' of the photon
is coming from the production of  heavy quarks
 in a photon-induced processes ( 
{\sl open heavy-flavor} or   {\sl heavy-flavor  hadron} production).
In such processes heavy quark mass provides in a natural way 
the hard scale  allowing
 to apply the perturbative QCD.
However, the question how to describe within the perturbative QCD the 
heavy quark (hadron) production in {\sl hadronic} 
processes  is still open. Both the
so called ``massive'' ({\sl no} heavy parton) and
``massless''(with heavy quark treated as a {\sl massless parton}) 
schemes are being applied \cite{OlnRiem}-\cite{BuzMatSmNe}.
 For hard processes initiated by photons,
 where direct and resolved photon contributions are possible,
this problem is particularly  complex  (see ref. 
\cite{DrKrZuZer}- \cite{cacciari-HERA-massless}). 
In both approaches  calculations for the  heavy quark production
in  $\gamma \gamma$ 
and $\gamma p$ collisions
were performed up to the NLO accuracy. 
The  fragmentation of  heavy quarks into heavy-flavor hadrons is
 usually described by a simple Peterson function \cite{peter} 
(sometimes with  QCD evolution (DGLAP), see \eg \cite{kniehl-HERA-massless}).
The NLO calculation of the charm quark production in the DIS$_{e\gamma}$
 (contributions to $F_2^{\gamma}$ and $F_L^{\gamma}$)  was done
 in \cite{LRSN}. 
The progress in the theoretical description of heavy quark processes 
is necessary as  large discrepancies in the description of the data
involving heavy quarks are observed.
This progress  is needed  not only for understanding of hadronic interaction. 
It is also crucial in  searches of ``new physics'', \eg for 
reliable estimation of the background.

On the experimental side 
- new results   from LEP and HERA colliders  have appeared recently
on the production of $c$-quark (and $D^{*\pm}$ and $D^{\pm}_s$ mesons) as 
well as $b$ -quark,  in the DIS$_{e\gamma}$  and in the resolved photon 
processes in $\gamma \gamma$ and $\gamma p$ collisions.
 
\begin{figure}[hc]
\vskip 0.in\relax\noindent\hskip 2cm
       \relax{\includegraphics{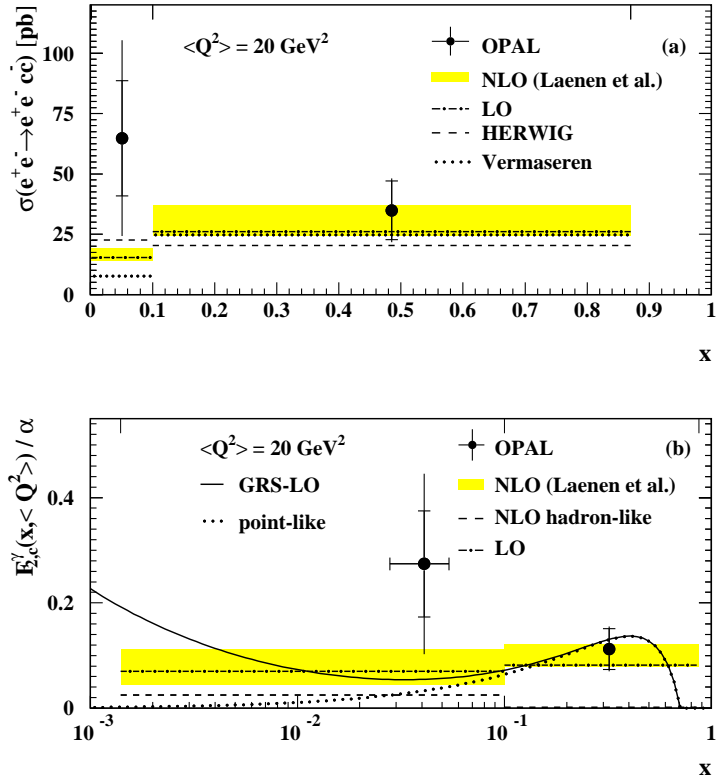}}
\vspace{-0.3cm}
\caption{\small\sl OPAL data at $<Q^2>$=20 GeV$^2$
 in comparison  with predictions
of LO and NLO calculations  and Monte Carlo models. 
a) The cross section for $c \bar c$ production
as a function of $x_{Bj}$.
b) The extracted  
structure function $F_{2,c}^{\gamma}(x_{Bj},<Q^2>)/\alpha$ 
(from \cite{OPALheavy}).}
\label{fig:pr294_11}
\end{figure}

As it was already mentioned in sec. \ref{sec2},
in early $F_2^{\gamma}$ experiments the heavy 
quark distributions, mainly $c$-quark, were subtracted from
$F_2^{\gamma}$ data, see \eg \cite{amy2}, {\bf AMY 95} and {\bf TOPAZ 94}
and references in \cite{whalley}.
	The recent DIS-type measurement at LEP \cite{OPALheavy}
led to the extraction, for the first time,
of the  charm contribution to $F_2^{\gamma}$, denoted  $F^{\gamma}_{2,c}$.
In  fig.~\ref{fig:pr294_11} the 
 OPAL data in comparison with the NLO calculation \cite{LRSN} are shown.\\
\vspace*{10.2cm}

\begin{figure}[ht]
\vskip 0.in\relax\noindent\hskip 1.6cm
       \relax{\includegraphics{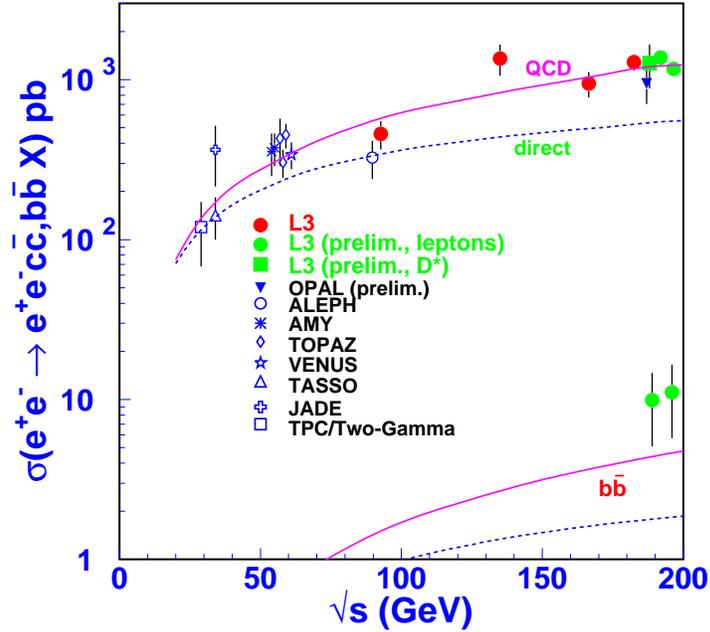}}
\vspace{-0.ex}
\caption{\small\sl Summary of charm and beauty production in 
$\gamma \gamma$ collision at $e^+e^-$ collisions (from \cite{LEPC}).}
\label{fig:bbb}
\end{figure}

\begin{figure}[ht]
\vskip 0.in\relax\noindent\hskip 0cm
       \relax{\includegraphics{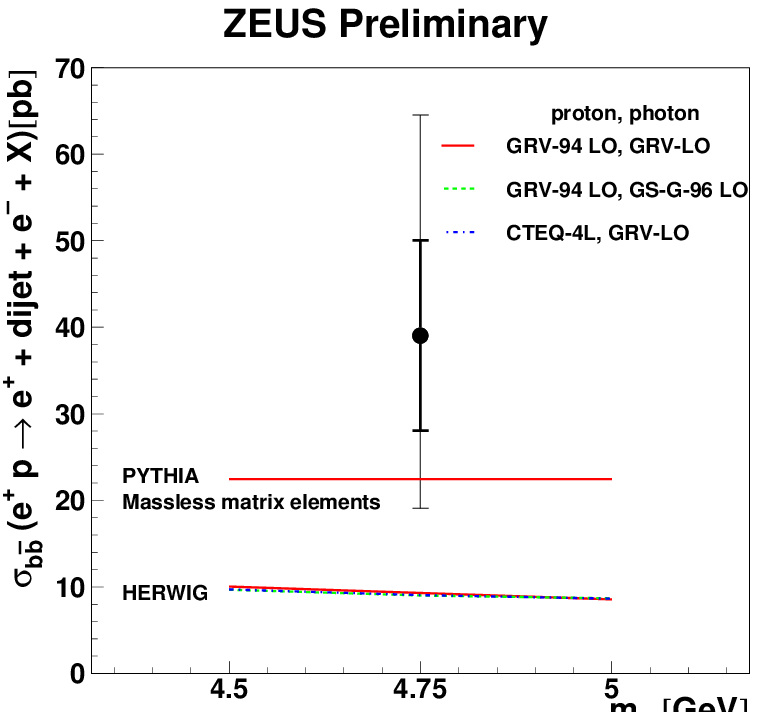}}
\vskip -0.20in\relax\noindent\hskip 0cm
       \relax{\includegraphics{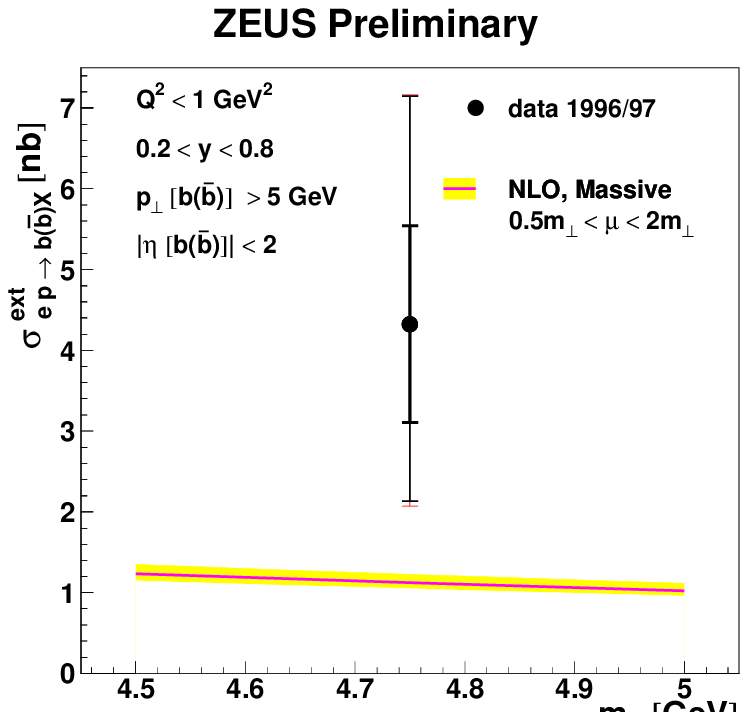}}
\vspace{-0.ex}
\caption{\small\sl ZEUS preliminary data. 
Left: Measured $b\bar b$ cross section for $ep$ collision
with $P^2 (=Q^2) < 1$ GeV$^2$,
0.2$<y<$0.8 and for $E_{T 1,2}>$ 7,6 GeV, $|\eta|<$2.4, $p_T^{e^-}>$ 1.6 GeV,
$|\eta^{e^-}|<$ 1.1, compared 
to HERWIG and PYTHIA MC predictions.
Right: Extrapolated $b(\bar b)$ cross section in $ep$ collisions
for events with $P^2< 1$ GeV$^2$,
0.2$<y<$0.8, $p_{T b(\bar b)}>$ 5 GeV, $|\eta_{b(\bar b)}|<$2,
compared to the ``massive'' NLO  predictions
(from \cite{zeushq}).}
\label{fig:herabb}
\end{figure}

The first hint that the charm-quark
production in $\gamma\gamma$ collisions requires the resolved photon
processes can be found in {\bf AMY 1.5 94}, \cite{plb341}, \cite{prd50};
for other early results see ref. \cite{jadeh} - \cite{alephh}.
The new measurements of this type were performed recently
at LEP by L3 group \cite{LEPC}, for the first time with $b$-quarks production,
see the collective figure, fig. \ref{fig:bbb}. 
 The data arrive also from the $c$-quark (also $D^{*\pm}$ 
and $D_s^{\pm}$ mesons) 
and $b$-quark photoproduction 
measurements at HERA \cite{h1hq,zeushq}. 
Some disagreement with the Monte Carlo predictions and the
NLO QCD calculations, especially for the
$b$-quark production, is observed, see fig.~\ref{fig:herabb}
for ZEUS preliminary results.\\
\vspace*{7.5cm}
\vspace*{7.cm}
\newpage
\section{Related topics\label{sec6}}
\subsection{On the polarized photon structure functions \label{sec61}}
The polarized photon structure function $g_1^{\gamma}$ (another name is the 
{\sl spin dependent} structure function, \ie {\sl not} spin-averaged
 like $F_1^{\gamma}$ and $F_2^{\gamma}$)  both for a real 
and virtual photon can be studied in the DIS-type  experiments 
 with polarized electron beams and polarized 
 photon target. The other spin dependent 
function, $F_3^{\gamma}$, can be studied in the conventional experimental 
setup with unpolarized electron beams, see sec. 2.2.
 The  structure function $g_1^{\gamma}$ is of great importance, since its  
first moments ({\sl a sum rule for the 
``spin `` of the photon}) involve  strong and electromagnetic anomalies,
and it is deeply connected with the chiral 
properties of QCD \cite{anomaly}.

The calculation of the structure function  $g_1^{\gamma}$
for  polarized real and virtual 
photon 
has been done in NLO in \cite{g1-sv} and \cite{mr,pol3}, respectively. 
The NLO calculations of the jet production in the
 polarized real $\gamma$-hadron collision can be found in \cite{sv}-
\cite{pol2}.

There is no data on the polarized photon 
structure functions $F_3^{\gamma}$, $g_1^{\gamma}$
 nor on the {\sl polarized partonic content} of the real and virtual 
photons from resolved photon processes.
The future measurements in linear colliders,
both in $e^+e^-$ and $\gamma e$ or $\gamma\gamma$ options , may be crucial 
for testing the structure of polarized photons, see \eg \cite{kziaja,stratman}.

\subsection{On the structure function of the electron}
In the DIS experiments in $e^+e^-$ collisions 
the inclusive hadron production can be ascribed  not to the 
photon target, but rather to the parent
electron or positron target. This in some
cases may be more straightforward 
than pinning down the structure of the virtual photon,
as was mentioned before.

This topic is discussed in \eg Ref.~\cite{rev}f,~\cite{mr}. 
See also the results in \cite{ws}, where the structure function of the electron
(in general - lepton) is related not only to the structure function of photon 
but also  to the electroweak gauge bosons $W$ and $Z$.
The {\em structure of weak bosons} which appears in this approach 
was introduced and discussed in Ref.~\cite{ws1}.

\subsection{On the photon content of the proton}

The {\sl photonic content} of the proton
relevant for the elastic and inelastic processes with an initial { proton}
has been studied in \cite{kn-pcp} and \cite{gsv-pcp,rv-pcp},
respectively.

%
%
\newpage
\section{Summary and outlook}
The status of recent measurements ($\sim$ 1990 and later) of
the ``structure'' of unpolarized   photon in 
``DIS$_{e\gamma}$'' experiments as well as in large $p_T$ jet
production processes involving resolved photon(s) at $e^+e^-$ and 
$e^{\pm}p$ colliders is presented.
In updating our previous work \cite{rev}i we have kept its form of the collection of selected data, presenting the main results of individual measurements.
The final content of the survey has been dictated by the large amount of
 pioneering data that have appeared lately.

For the real photon the data from LEP 1
and  higher energy data from LEP 1.5 and 2,
with the higher  statistics and improved unfolding methods
are now published in final or preliminary form.
Final results from TRISTAN appeared.
The basic features predicted by the QCD are observed
with higher than previous accuracy, as  qualitative change has appeared.
The hadronic structure functions for real photon have been measured 
in the $\gamma^* \gamma$ processes in a wide range of $Q^2$ and down to very 
small $x_{Bj}$. 
On the other hand impressive progress has been obtained in 
pinning down the parton densities, in particular the gluonic 
content, in large $p_{T}$
resolved photon processes in $ep$ collisions. The data are not only
 from the larger
than before kinematical range but are also more precise, which allowed to pose 
more quantitative questions, \eg on the rise of the $F_2^{\gamma}$ 
or gluon density at the small $x_{Bj}$.
The prompt photon production has been measured in the resolved $\gamma p$
collisions at HERA (and also at TRISTAN in $\gamma \gamma$
collisions). This process can give an additional information on the
partonic content of the photon as compared 
to the jet(s) production.
The heavy ($c$ and $b$) quark contribution to the photon structure functions has been measured in both $e^+e^-$ and $e^{\pm}p$ collisions.

The survey contains a large amount of data related to 
the "structure" of the virtual photon 
from the hard photon-proton 
collisions at HERA, where 
the effective parton distribution has been extracted from the dijet production.
More than 15 years after the first measurement at PLUTO
the corresponding new precise DIS$_{e\gamma^*}$ measurement of the structure 
functions of virtual photon  have been performed at LEP.

The new measurements of the partonic content of  the  photon are 
accompanied by the impressive progress  made in the NLO QCD 
calculations for the resolved real and for virtual photon processes.
 
The  discrepancies in describing the  hadronic final states in the 
DIS$_{e\gamma}$ experiments as well as in the resolved photon processes
in $\gamma \gamma$ and $\gamma p$ collisions may have  
common origin. The need of additional $p_T$
in the standard QCD picture in the distribution of produced hadrons and jets
may suggest an extra interaction involving the constituents of 
photon(s) - a multiple interaction. The newest data seem to give us a 
message that the partonic content of the photon is not properly described
by existing parton parametrizations.

The future high-energy linear $e^+e^-$ colliders (LC) \cite{NLC} as well as
related $e{\gamma}$ and ${\gamma}{\gamma}$ colliders, based on 
the backward Compton scattering on the laser light, will offer 
a unique opportunity to measure the structure of photon in a new 
kinematical regime \cite{vogt}. Moreover, at these colliders measurements of  
structure functions for photon with a definite polarization should 
become feasible with a good accuracy.
To what extent the new generation of $ep$ or $\gamma p$ accelerators
under discussion (\eg THERA \cite{klein}) may extend our present knowledge, 
remains to be seen.

\vskip 2cm
\noindent{\large \bf {Acknowledgments}} 
\newline
We express our special thanks to Aharon Levy for his 
encouragement, critical reading of the early version of the 
manuscript and  important comments. We would like to thank R. Nisius
for reading the earlier version of the updated paper, and his 
very useful, critical and constructive comments.
We thank also I. Tyapkin, M. Kienzle, F. Kapusta, A. Finch and
S.S\"{o}ldner-Rembold for many important explanations and suggestions.
We are very grateful to Stefan S\"{o}ldner-Rembold for  
compilations of the newest results on $F_2^{\gamma}$.
We are grateful to our colleagues from all collaborations carrying the
$\gamma p$ and $\gamma\gamma$ analyses, especially to A. Finch, 
I. Tyapkin, F. Kapusta, M. Kienzle,
M. Chamizo, J. A. Lauber, B. Surrow, S. Schlenstedt, P. Bussey
and S.W. Lee for  discussion and 
 for providing us with their
new data. Thanks are due to 
H. Hayashii and T. Nozaki for
making available to us the postscript versions of older TRISTAN figures.

We thank T. Sj\"ostrand for illuminating discussion, and 
also D. Miller and L. Jonsson for important information.

We appreciate the
collaboration with Jerzy Rowicki, Jan \.Zochowski, Pawe\l ~Jan\-ko\-wski
and Urszula Jezuita-D\c{a}browska.
We would like to thank Mike Whalley for sending us
copies of their "Compilation..." \cite{whalley}.
\newline\newline
Supported in part by Polish Committee for Research, grant number
2P03B18209 (1996-1997), 2P03B18410 (1998-1999) and
2P03B11414 (2000), and by Interdisciplinary
Centre for Mathematical and Computational Modelling, Warsaw University,
Grant No G16-10 (1999).

%
%
\newpage
\section{Appendix}
\vspace*{0.5cm}
\subsection{Parton parametrizations for the  real photon \label{sec81}}
Parton densities are given for fixed number
of massless flavours (DO,LAC,GRSch,GIKO,IO) or for the number of flavours
dependent on the scale $Q^2$ (DG,GRV,GS,AGF,WHIT,SaS).
\newline
{\bf{Duke - Owens (DO)}} \cite{do} 
\newline
A leading logarithmic parametrization of the parton distributions 
in an asymptotic 
form. Quarks with equal charges have the same
distribution functions: $f_u = f_c$, $f_d = f_s$ ($N_f=4$).
\newline
{\bf{Drees - Grassie (DG)}} \cite{dg}
\newline
A parametrization of the  full solution of the leading order evolution
equations. The input parton distributions with free parameters are
assumed at $Q_0^2 = 1$ GeV$^2$ and fitted to the only data on
$F_2^{\gamma}$ existing at that
time, at $Q^2 = 5.9$ GeV$^2$, from PLUTO.
\newline
{\bf{Field - Kapusta - Poggioli (FKP)}} \cite{fkp} 
\newline
In this approach, there are two contributions to $F_2^{\gamma}$: 
the nonperturbative (``hadronic'') part,
$F_2^{\gamma HAD}$, and the perturbative (pointlike) one, $F_2^{\gamma PL}$.  
The $F_2^{\gamma PL}$ 
arises from the basic direct ${\gamma}^*\rightarrow q\bar{q}$ coupling 
(and higher order QCD corrections), and it is included if the
intrinsic quark $p_T$ is above a cutoff $p_T^0$. 
If $p_T$ is smaller than $p_T^0$ then the $q\bar{q}$ 
pair creates a bound state leading to $F_2^{\gamma HAD}$, usually
taken from the VMD model. The perturbative $F_2^{\gamma PL}$ is
calculated using the first order splitting functions and the 
one-loop $\alpha_S$. 
\newline
{\bf{Levy - Abramowicz - Charchu\l a (LAC)}} \cite{lac}
\newline
A parametrization of the   full solution of the leading order 
evolution equation fitted to all available in 1991 measurements of 
$F_2^{\gamma}$ for $Q^2\geq Q_0^2$. The number of flavours is fixed to $N_f = 4$.
Three sets are provided with the
different choices of an input scale $Q_0^2$, and the $x\ra 0$ 
behaviour of a gluon distribution $G(x)$, namely:\newline
$\bullet$ LAC1: $Q_0^2 = 4$ GeV$^2$\newline
$\bullet$ LAC2: $Q_0^2 = 4$ GeV$^2$, $xG(x)\ra const.$\newline
$\bullet$ LAC3: $Q_0^2 = 1$ GeV$^2$.
\newline
{\bf{Gordon - Storrow (GS)}} \cite{gs} 
\newline
The LO and NLO parametrizations. The input structure function at the 
scale $Q_0^2 = 5.3$ GeV$^2$ \cite{gs}a and $Q_0^2 = 3$ GeV$^2$ 
\cite{gs}b  is chosen in the LO analysis as a sum of a hadronic part 
from the VMD model and of a pointlike part based on the Parton Model. 
Free parameters (also light quarks masses) are fitted to the data with 
$Q^2\geq Q_0^2$. 
The input gluon distribution in LO \cite{gs}a is assumed in two
different forms (set GS1 and GS2).
The NLO distributions in the $\overline{\rm MS}$ scheme are 
obtained by matching of the $F_2^{\gamma}$ in the LO and the NLO approaches 
at the $Q_0^2$ scale.
$N_f =$ 3, 4, and 5 is used in \cite{gs}a. In \cite{gs}b the number
of flavours equal to $N_f =$ 3 for $Q_0^2\leq Q^2\leq$ 50 GeV$^2$ and
 $N_f =$ 4 for 50 GeV$^2 \leq Q^2$ was used.
\newline
{\bf{Gl\"{u}ck - Reya - Vogt (GRV)}} \cite{grv}
\newline
The LO and NLO parametrizations of the parton distributions generated 
dynamically 
with the boundary conditions at $Q_0^2$ given by a VMD input. 
The physical photon is then assumed to be a coherent superposition
of vector mesons, whose parton distributions are further assumed 
to be similar to those of a pion.
The low initial scale $Q_0^2$(LO) = 0.25, $Q_0^2$(NLO) = 0.3 GeV$^2$ 
is universal for structure functions of $p$, $\pi$, $\gamma$ etc. 
The DIS$_{\gamma}$ scheme is introduced 
to avoid the large-$x_{Bj}$ instability problems. The one free 
parameter, which is the VMD input normalization constant relative 
to $\pi$, is fixed by the data. 
\newline
{\bf{Aurenche - Chiapetta - Fontannaz - Guillet - Pilon (ACFGP)}} \cite{acfgp}
\newline
A solution of the NLO evolution equation with the boundary condition
taken at $Q_0^2$ = 0.25~GeV$^2$. The input parton distributions were
obtained from the VMD model at $Q^2$ = 2~GeV$^2$ and evaluated down 
to $Q_0^2$.
\newline
{\bf{Aurenche - Guillet - Fontannaz (AGF)}} \cite{agf}
\newline
The NLO parton distributions obtained with the input distributions 
(shown to be scheme-dependent) at $Q_0^2 = 0.5$ GeV$^2$. The input 
distributions are based on the VMD model modified to agree with the 
$\overline{\rm MS}$ scheme used in this analysis. 
\newline
{\bf{Hagiwara - Izubuchi - Tanaka - Watanabe (WHIT)}} 
\cite{whit}\newline
A set of six  LO parametrizations obtained by fitting the input 
distributions to all available data for $4$ GeV$^2\leq Q^2\leq 100$ 
GeV$^2$. The parametrizations WHIT1-WHIT6 are based on different 
input gluon distributions. A massive charm contribution is calculated 
from the Quark Parton Model below  $Q^2=100$ GeV$^2$, above this scale
 from the 
massive quark evolution equations. 
\newline
{\bf{Schuler - Sj\"{o}strand (SaS)}} 
\cite{sas}\newline
Four sets of the LO parametrizations. The nonperturbative input 
distributions at $Q_0 = 0.6$ GeV in the SaS1D, SaS1M and $Q_0 = 2$ 
GeV in the SaS2D, SaS2M sets are based on the VMD model (their 
normalization is fixed by the momentum sum rule,
and the $x$ - dependence is obtained from 
the fits to the data). The fully calculable pointlike contribution 
to  $F_2^{\gamma}$ is expressed as an integral over the 
$\gamma^*\rightarrow q\bar{q}$  virtuality $k^2$.\newline 
The nonleading term $C_{\gamma}$ is included into the $F_2^{\gamma}$, 
leading to the $\overline{\rm MS}$ distributions (SaS1M, SaS2M). 
$C_{\gamma} = 0$ gives distributions in the DIS scheme (SaS1D, SaS2D). 
\newline
{\bf{Gl\"{u}ck - Reya - Schienbein} (GRSch)} \cite{grs99}
\newline
The LO and NLO (DIS$_{\gamma}$ scheme) parameter-free 
parton distributions obtained 
from the VMD input based on a coherent superposition of
vector mesons applying the new Gl\"{u}ck - Reya - Schienbein pionic
distributions \cite{grspi}. The low initial scale (universal for the 
proton, pion and photon distributions) $Q_0^2$(LO) = 0.26,
$Q_0^2$(NLO) = 0.40 GeV$^2$ is assumed. The parton distributions are 
calculated for light quarks, $N_f$ = 3 (contribution of heavy quarks 
is taken into account in the structure function $F_2^{\gamma}$).
\newline
{\bf Laenen - Riemersma - Smith - Neerven (LRSN)} \cite{LRSN}
\newline
The NLO calculation of the structure function $F_2^{\gamma}$ and 
$F_L^{\gamma}$, including the heavy quark contribution.
\newline 
{\bf Gorski - Ioffe - Khodjamirian - Oganesian (GIKO)} \cite{ioffe}
\newline
The structure function $F_2^{\gamma}$ of the real  photon, obtained from 
the results for the virtual photon,
 where beside PM also a gluon condensate contribution is included (see below). 
Only the ligh quarks contributions are considered (\ie $N_f$=3).
This approach is valid for $Q^2$ below 10 GeV$^2$ due to the lack of $Q^2$
evolution, and for intermediate $x_{Bj}$ values: 0.2 - 0.7.
\newline
{\bf Ioffe - Oganesian (IO)} \cite{ioog}
\newline
Gluon distributions in the real photon calculated using 
the LO and NLO evolution equations, 
with the light quark densities input from GIKO. Two sets of gluon density
are studied: 1) without the intrinsic gluons at the input 
scale $Q^2_0$=1 GeV$^2$, 2) with the gluon density at the input scale
equal to gluonic content of vector mesons $\rho,\omega,\phi$.
The range of applicability:  $x_{Bj}$  below 0.6 down to 0.03-0.05, and 
1 $<Q^2<$ 100 GeV$^2$.\\
Here also  results for  $F_2^{\gamma}$ of the real  photon, 
 with the LO and NLO evolutions in $Q^2$  are given (with $N_f$=3), valid
for 0.2 $<x_{Bj}<$ 0.7 and 1 $<Q^2<$ 50 GeV$^2$.
 \newline
{\bf H. Abramowicz, E. Gurvich, A. Levy (GAL)} \cite{gal}
\newline
The LO \cite{gal}a and NLO \cite{gal}b parton densities in the photon 
are extracted from the existing $F_2^{\gamma}$ measurements and the low-$x$
proton structure function (related to the $F_2^{\gamma}$ by Gribov
factorization at low-$x$).

\subsection{Parton parametrizations for the virtual photon}
The notion of the partonic content of the virtual photon
($P^2\neq 0$) can be applied if $Q^2$ scale is larger than $P^2$
(typically $P^2 \lsim 0.2~Q^2$).
The parton distributions in the virtual 
photon can be calculated in the perturbative QCD
for $\Lambda^2\ll P^2\ll Q^2$ without any input, for lower $P^2$
extra assumptions are needed.
The parametrizations described below  
are valid for $0\leq P^2$ and become the corresponding
parametrizations for the real photon in the limit $P^2\rightarrow 0$. 
In the low-$P^2$ region some of them are based on interpolation
between parton distributions in the real photon ($P^2=0$)
and the QCD predictions in high-$P^2$ region ($\Lambda^2\ll P^2$).
All the  parametrizations presented below  deal with the 
{\underline {transverse}} polarized virtual photon, with one 
exception (Ch\'yla parametrization). 
~\newline
{\bf{Drees - Godbole}} \cite{rev}f 
\newline
The parametrization of the parton distributions in the electron
and in the virtual photon. The parton distributions in the virtual photon 
are equal to the corresponding distributions in the real photon
(\eg DG, GRV) multiplied by a factor dependent on virtuality
$P^2$. In the limit $P^2\rightarrow Q^2 \gg\Lambda^2$ the quark  
distributions in the virtual photon
approach the leading logarithmic (ln${Q^2\over P^2}$) parton model
predictions arising from 
$\gamma^*(Q^2)\gamma^*(P^2)\rightarrow q\bar{q}$ subprocess.
The gluon distributions in this limit approach the $Q^2$ - behaviour
($\sim \ln^2 {Q^2\over P^2}$) following from the 
$\gamma^*(Q^2)\gamma^*(P^2)\rightarrow g q\bar{q}$ cross section.
The parametrization contains one free parameter $P_c^2$,
a typical hadronic scale  
(in applications denoted also as $\omega^2$).
\newline
{\bf{Gl\"{u}ck - Reya - Stratmann} (GRS)} \cite{grs} 
\newline
The LO and NLO (DIS$_{\gamma}$ scheme)
distributions obtained by solving the $Q^2$-evolution 
equation with boundary conditions being a smooth interpolation
between boundary conditions valid at high and low $P^2$.
For high $P^2\rightarrow Q^2 \gg\Lambda^2$ the solutions correspond to
the cross section for
$\gamma^*(Q^2)\gamma^*(P^2)\rightarrow q\bar{q}$ subprocess.
For $P^2 = 0$ they are given by VMD input
parton distributions at the scale $Q^2_0$(LO) = 0.25 or 
$Q^2_0$(NLO) = 0.3 GeV$^2$, as for the real photon in \cite{grv}.
The applicability of the virtual photon parametrizations is assumed for 
$0 \leq P^2 \leq 10$ GeV$^2$ and $5 P^2\lsim Q^2$. The number of 
active flavours is fixed to $N_f$ = 3.
\newline
{\bf{Schuler - Sj\"{o}strand (SaS)}} \cite{sas} 
\newline
An extension of the SaS parton distributions in the real photon
to the virtual photon case, valid for $0\leq P^2\leq Q^2$. 
In the pointlike contribution, the
integral over the virtuality $k^2$ of the $\gamma^*\rightarrow q\bar{q}$
state, is modified by a factor $({1\over k^2})^2$
with the lower limit of the integration equal to
a cutoff scale $P^2_0$ (several choices for $P^2_0$ have been
considered, \eg $P^2_0 = max\{P^2, Q_0^2\}$).
In the hadronic contribution a factor $({m_V^2\over m_V^2+P^2})^2$
is introduced, where $m_V$ is a vector-meson mass. The 
notation SaS1D, SaS2D, SaS1M, and SaS2M
is the same as in the real photon case.
\newline
{\bf{Gl\"{u}ck - Reya - Schienbein} (GRSch)} \cite{grs99} 
\newline
The LO and NLO parton distributions valid for $0\leq P^2\ll Q^2$ and $N_f$ = 3. 
It is argued that whenever the virtual photon is probed at $Q^2\gg P^2$
it may be treated as the \underline{real} one: the longitudinal polarization
should be neglected and a partonic cross section involving
$\gamma (P^2)$ should be calculated as if $P^2 = 0$ (in this approach
the virtuality $P^2 \neq 0$ is present in the flux $e\rightarrow e\gamma^*$
and in the virtual photon parton distributions).
The boundary conditions in the limit $P^2\rightarrow Q^2\gg\Lambda^2$
should be calculated from 
$\gamma^*(Q^2)\gamma (P^2=0)\rightarrow q\bar{q}$ subprocess
instead of $\gamma^*(Q^2)\gamma^*(P^2)\rightarrow q\bar{q}$;
this implies changing the boundary conditions  compared to
the previous approach \cite{grs}. 
Now, at $P^2 = Q^2\gg\Lambda^2$ the parton distributions vanish in both LO and
NLO in DIS$_{\gamma}$ scheme.
The input distributions at the initial scale $Q^2 = 
max\{P^2, Q_0^2\}$ are assumed to be equal to the VMD parton
distributions in the real photon (see sec. \ref{sec81}) 
multiplied by the factor 
$({m_{\rho}^2\over m_{\rho}^2+P^2})^2$, where $m_{\rho}$ = 0.59 GeV$^2$,
$Q^2_0$(LO) = 0.26 GeV$^2$ and $Q^2_0$(NLO) = 0.4 GeV$^2$.
\newline 
{\bf Gorski - Ioffe - Khodjamirian - Oganesian (GIKO)} \cite{ioffe}
\newline
The structure function $F_2^{\gamma}$ of the virtual  photon, 
for $P^2>$ 0.5 GeV$^2$, calculated using the Operator Product Expansion 
as a series in $1/P^2$,
 where beside PM also a gluon condensate contribution is included. 
Using a  dispersion relation the structure function is then presented 
in terms of the  
contributions from  physical states ($\rho,\omega,\phi$) and continuum.
The proper continuation to $P^2$=0 can be performed, leading to
 prediction for the  real photon (see sec. \ref{sec81}). 
Only the light quark contributions are considered.
This approach is valid for intermediate $x_{Bj}$ values: 0.2 - 0.7,
and for $Q^2$ larger than $P^2$ but below 10 GeV$^2$,
due to the lack of $Q^2$ evolution.
\newline
{\bf Ioffe - Oganesian (IO)} \cite{ioog}
\newline
Gluon distributions in the (transverse) virtual photon calculated using 
the LO and NLO evolution equations, 
with the light quark densities input from GIKO. Two types of gluon density
are studied: 1) without the intrinsic gluons at the input 
scale, $Q^2_0$=1 GeV$^2$, 2) with the gluon density at the input scale,
$Q^2_0$=1 GeV$^2$,
equal to the gluonic content of the vector mesons $\rho,\omega,\phi$.
The range of applicability:  $x_{Bj}$  below 0.6 down to 0.03-0.05 and 
1 $<Q^2<$ 100 GeV$^2$.\\
Here also  results for  $F_2^{\gamma}$ of the (unpolarized and transverse)
virtual  photon, for 50 GeV$^2$ $>P^2>$ 0.5 GeV$^2$ with the LO and NLO 
evolutions in $Q^2$ are given (with $N_f$=3). Results for $F_2^{\gamma}$
for a real photon are also given (see above).
\newline
{\bf Ch\'yla} \cite{chylaL}
\newline
The LO parametrization of the {\sl pointlike} quark and gluon densities 
in the {\sl longitudinal} virtual photon, $1$ GeV$^2 < P^2 < 0.2~Q^2$, 
obtained from the homogeneous evolution equation, without extra input.
For $P^2\rightarrow Q^2$ the gluon distribution $G_L$ in the longitudinal 
virtual photon vanishes, while the quark densities
$q_L^S = q_L^{NS}$ approach the PM formula for 
$\gamma^*(Q^2)\gamma^*(P^2)\rightarrow q\bar{q}$.
Parton densities satisfy the momentum sum rule.

%
%

\end{document}